\documentclass[twocolumn,aps,epsfig,nofootinbib]{revtex4}
\usepackage{graphicx}
\usepackage{dcolumn}
\usepackage{amsmath}
\usepackage{amssymb}
\usepackage{epsfig}
\usepackage[colorlinks,linkcolor=blue,citecolor=blue,urlcolor=blue]{hyperref}

\begin{document}

\title{Preinflationary dynamics of $\alpha-$attractor in loop quantum cosmology}
\author{M. Shahalam$^1$ \footnote{E-mail address: shahalam@zjut.edu.cn}}
\author{M. Sami$^2$ \footnote{ E-mail address: samijamia@gmail.com}}
\author{Anzhong Wang$^{1,3}$ \footnote{E-mail address: Anzhong$\_$Wang@baylor.edu}}
\affiliation{$^{1}$Institute for Advanced Physics $\&$ Mathematics,
Zhejiang University of Technology, Hangzhou, 310032, China\\
$^2$Centre for Theoretical Physics, Jamia Millia Islamia, New Delhi, 110025, India\\
$^3$GCAP-CASPER, Department of Physics, Baylor University, Waco, TX, 76798-7316, USA }

\date{\today}

\begin{abstract}

We systematically study the preinflationary dynamics of the spatially flat  Friedmann-Lemaitre-Robertson-Walker universe filled with a single scalar field that has the generalized $\alpha-$attractor potentials, in the framework of loop quantum 
cosmology, in which the big bang singularity is replaced generically by a non-singular quantum bounce due to purely quantum geometric effects. The evolution can be divided into two different classes, one is dominated initially (at the quantum bounce) by 
the kinetic energy of the scalar field, and one is not. In both cases, we identify numerically the physically viable initial conditions that lead to not only a slow-roll inflationary phase, but also enough $e$-folds to be consistent with observations, and find that 
the output of such a viable slow-roll inflationary phase is generic. In addition, we also show that in the case when the evolution of the universe is dominated initially  by the kinetic energy of the scalar field (except for a very small set in the phase space), the evolution before reheating is aways divided into three different phases: {\em  bouncing, transition and  slow-roll inflation}. This universal feature does not depend on the initial conditions of the system nor on the specific potentials of the scalar field, as long as  it is dominated initially by the kinetic energy of the scalar field at the bounce. Moreover, we carry out  phase space analyses for the models under consideration and compare our results with the power-law and Starobinsky potentials. 

\end{abstract}
\maketitle

\section{Introduction}
\label{sec:intro}

In the early 1980's, the cosmic inflation stood out as a popular paradigm for resolving various problems in the standard model of cosmology, such as the horizon and flatness, etc. It also explains the origin of inhomogeneities that are observed in the cosmic microwave background and the formation of the large scale structure of the universe \cite{guth1981}. During the last three decades, a wide range of inflationary models have been proposed, including  conformal attractors \cite{conformal}, $\alpha-$attractors \cite{alpha,alpha1,alpha2,alpha3,alpha4}, Starobinsky and the chaotic inflation in supergravity,  which is known as Goncharov and Linde (GL) model \cite{staro1980,staro1,staro2,staro3,staro4,GL}. These models provide  very similar cosmological predictions with respect to the significant differences  in their potentials, and have an excellent fit with the current observations. According to Planck 2015 results \cite{Planck2015}, in the case of a single field inflation, the potentials of $\alpha-$attractors and Starobinsky are consistent with the observations, while the quadratic potential is not equally favorable. 

Despite the  triumph of the standard inflationary models, which are based on the classical theory of general relativity (GR), their past is inadequate due to the existence of a big bang singularity. All scalar field models of inflation suffer from this initial and inevitable singularity \cite{borde1994,borde2003}. Clearly,  with this it is  difficult to know when and how to set the initial conditions. Moreover, to be consistent with the current observations, the universe should have expanded at least 60 $e$-folds during the inflation. Meanwhile, in a large class of inflationary models, it is often more than 70 $e$-folds \cite{martin2014}. However, in these models the size of the current universe is smaller than the Planck at the onset of inflation. Consequently, the semi-classical treatments are questionable during inflation. This is the so-called  trans-Planckian problem \cite{martin2001,berger2013}.

To address the above issues, in this paper we shall study the preinflationary dynamics of the generalized $\alpha-$attractor model in the context of loop quantum cosmology (LQC),
 in which the big bang singularity is generically replaced by a quantum bounce  \cite{agullo2013a,agullo2013b,agullo2015,ashtekar2011,ashtekar2015,barrau2016}, 
and investigate whether following the quantum bounce a desired slow-roll inflation generically exists or not  \cite{ashtekar2010,psingh2006,zhang2007,chen2015,bolliet2015,schander2016,bolliet2016,Bonga2016,Mielczareka}.

In the literature, there are mainly two distinct approaches for   the preinflationary universe, the dressed metric \cite{agullo2013b,metrica,metricb,metricc} and the deformed algebra \cite{algebraa,algebrab,algebrac,algebrad,algebrae,algebraf}. Although both approaches give rise to the same set of dynamical equations in the case of the background evolution of the universe, their perturbations are different \cite{bolliet2016}. The corresponding non-Gaussianities were also studied both numerically \cite{agullo15,ABS17} and analytically \cite{ZWKCS18} recently, and found that it is consistent with current observations. 

However, in this paper since we are mainly concerned with the background evolution of the universe,  the results to be presented in this work will be valid to both approaches.  Keeping this in mind,  we shall compare our results with the power-law and Starobinsky potentials obtained in \cite{psingh2006,chen2015,Bonga2016,alam2017,Tao2017a,Tao2017b}. In particular, we shall show that, when the kinetic energy of the inflaton initially dominates at the bounce (except for a very small set in the phase space), the evolution of the universe before reheating can be divided universally into three different phases \cite{alam2017,Tao2017a,Tao2017b}: {\em bouncing, transition and slow-roll inflation}. During these phases, the evolutions of both background and linear perturbations of the universe are all known analytically \cite{Tao2017a,Tao2017b}. In the small exceptional region of the phase space, we find that the potential energy first evolutes almost  as a constant in the bouncing phase, but oscillating afterward, in contrast to the rest
of regions in which the kinetic energy of the inflaton dominates the evolution of the universe at the quantum bounce. As a result,  in this exceptional region,    a slow-roll inflation is not resulted. 

The rest of the paper is organized as follows. In Sec. \ref{sec:EOM}, we briefly discuss the basic equations of the background evolution of the universe in the framework of LQC. In Sec. \ref{sec:alpha}, we examine the generalized $\alpha-$attractor model, and shall divide it into three models, namely $T$, $E$ and $\alpha-$attractor with $n=2$ in the sub-sections \ref{sec:Tmodel}, \ref{sec:Emodel} and \ref{sec:n=2}, respectively. These sub-sections are devoted to the detailed analysis of the background evolution in the framework of the positive inflaton velocity (PIV, $\dot\phi > 0$) and negative inflaton velocity (NIV, $\dot\phi < 0$), and also in the form of kinetic energy dominated (KED) and potential energy dominated (PED) cases at the bounce. The phase portraits for the models under consideration are presented in Sec. \ref{sec:phase}. In Sec. \ref{sec:compare}, we compare our results with the ones obtained previously  for the power-law and Starobinsky potentials. Our main conclusions are summarized in Sec. \ref{sec:conc}.  

Before turning to the next section, it is interesting to note that pre-inflationary universe has been also studied recently in the framework of loop quantum gravity (LQG) \cite{yang2009,DL17,adlp,lsw2018a,lsw2018b,agullo18} by using Thiemann's quantization scheme for the Lorentz part of the Hamiltonian \cite{thiemann}, and among other things, it was shown that the resolution of the big bang singularity (replaced by a quantum bounce) is robust, although the details near the bounce depend on the ways to regularize the Hamiltonian  \cite{lsw2018a,lsw2018b}. When the kinetic energy of the inflaton dominates at the bounce, the evolution of the universe before reheating can be also divided universally into three different phases, {\em bouncing, transition and slow-roll inflation} \cite{lsw2018a,lsw2018b}. During these phases, the evolution of the background   of the universe is also known analytically \cite{Tao2017a,Tao2017b}. 

We would also like to note that  recently inflation  with different  potentials have been studied in Einstein's theory of gravity and string-inspired models \cite{HISY,BG15,sahni18,SW08,nozari}, and various interesting results were obtained.
In addition, in the framework of  quantum reduced loop gravity (QRLG), its Hamiltonian   ($H_{QRLG}$) is almost identical to the Hamiltonian of LQC ($H_{LQC}$), except for a leading term \cite{QRLG1,QRLG2}. If this term is zero, then $H_{QRLG}$ exactly coincides with $H_{LQC}$. Similar to LQC, bounce occurs in QRLG. If we use the modified Friedmann equation of QRLG, then we can also obtain three different phases, bouncing, transition and the slow-roll inflation. However, the background dynamics would not be exactly the same as in LQC due to the dependence on the parameters of QRLG. Yet,  the modified Friedmann equation of group field theory (GFT) is also almost the same as in LQC, except with the last term of energy $E_{j0}$ \cite{GFT1,GFT2}, The geometric interpretation of $E_{j0}$ is not transparent, but its effect on the dynamics is as following: (a) For $E_{j0}=0$, the effective dynamics is same as in LQC. (b) For $E_{j0}>0$, the bounce will take place at a higher space-time curvature.
(c) For $E_{j0}<0$, the bounce will occur at a lower space-time curvature.

 The issue of estimating the duration of the slow-roll inflation with effective isotropic, anisotropic and Bianch I Universe in LQC have been investigated in \cite{LQC1,LQC2,LQC3}, in which it was found  that the probability distribution function during the slow-roll inflation is peaked at the values of e-folds which are consistent with observations. Moreover, the duration of the slow-roll inflation does not depend crucially on the modified background evolution \cite{LQC4}. A new related study for the probability of inflation has been discussed in \cite{LQC5}, in which the existence of the quantum bounce affects the probability of inflation.

\begin{figure*}[tbp]
\begin{center}
\begin{tabular}{cc}
{\includegraphics[width=2.1in,height=1.65in,angle=0]{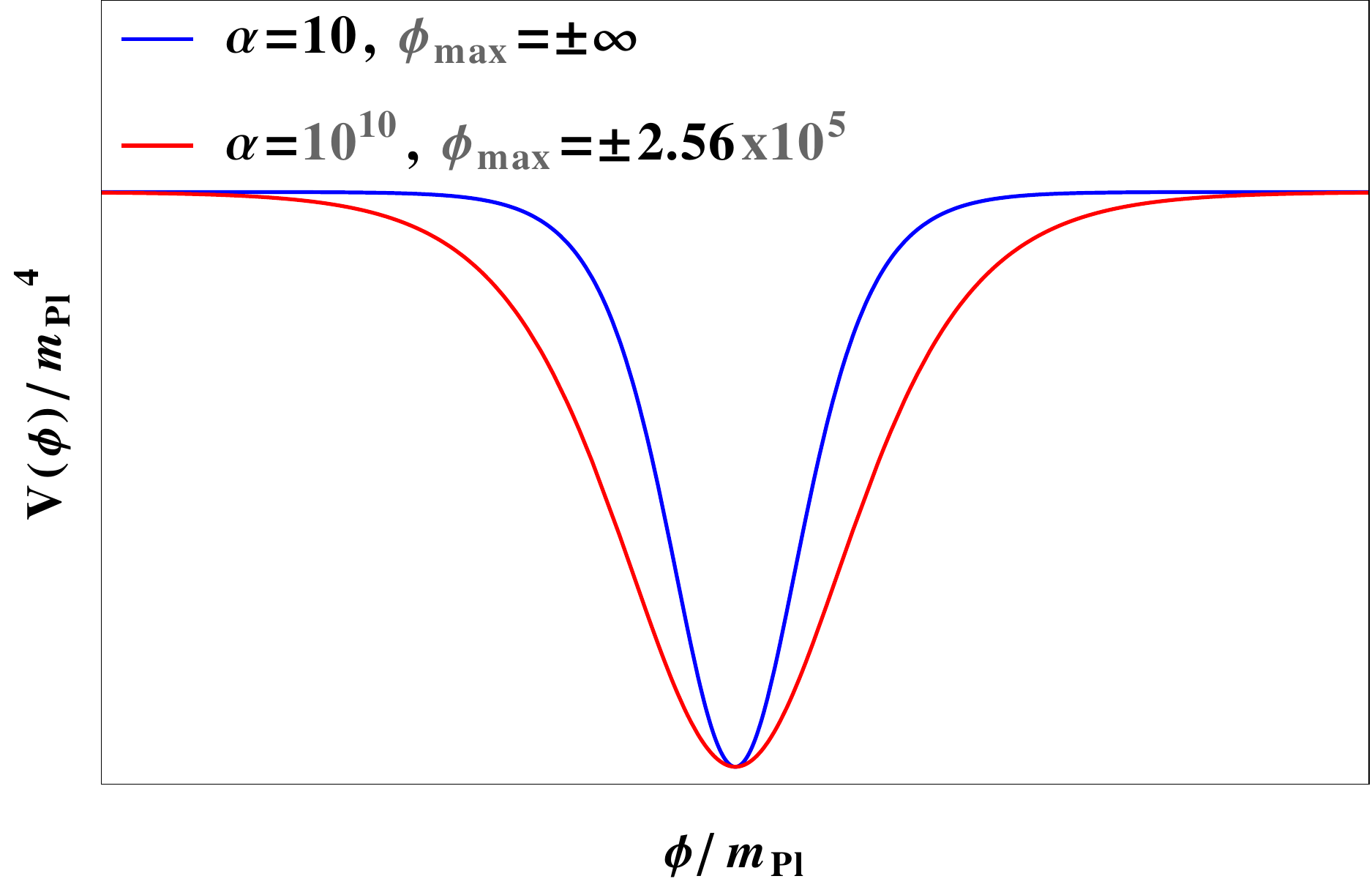}} &
{\includegraphics[width=2.1in,height=1.65in,angle=0]{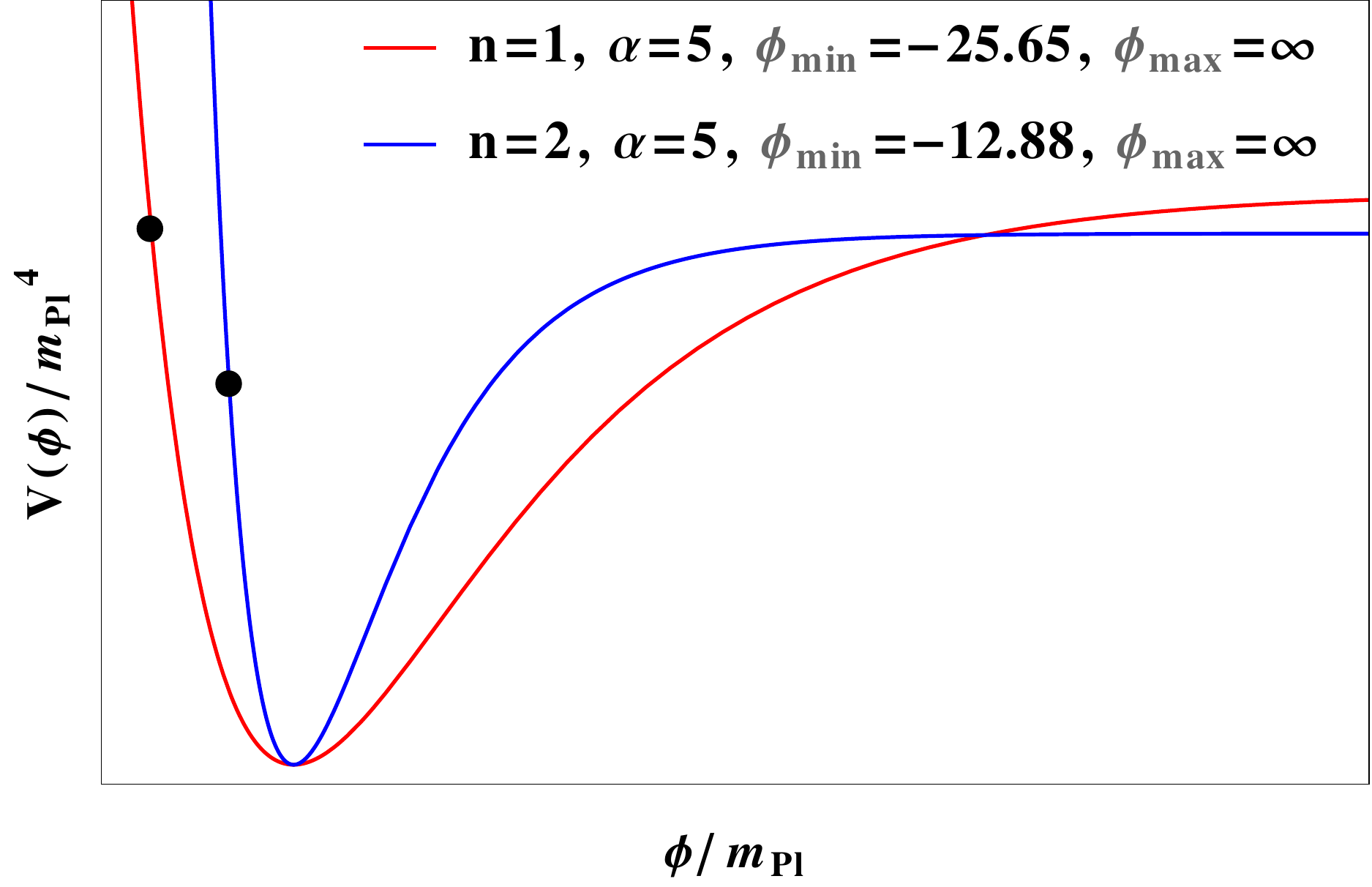}} 
\end{tabular}
\end{center}
\caption{ Left panel shows the evolution of the $T-model$ potential (\ref{eq:Tpot}) for $\alpha=10 m_{Pl}^2$ and $10^{10} m_{Pl}^2$. This potential is symmetric about $\phi=0$, and bounded by $V(\phi) \leq \alpha c^2$ (above, as $\phi \rightarrow \pm \infty$) and $V(\phi) \geq 0$ (below). In LQC, maximum energy density is $\rho_c$ that constrains $\phi_{max}$, see  Eq.(\ref{eq:TNphimax2}). Therefore, $\rho \leq \rho_c$ at the bounce. In the case of $\alpha=10 m_{Pl}^2$, $\rho \leq \rho_c$ as $\phi \rightarrow \pm \infty$. However, for $\alpha=10^{10} m_{Pl}^2$, $\rho \leq \rho_c$ as $\phi \rightarrow \pm 2.56 \times 10^5$. Right panel exhibits the evolution of the potentials (\ref{eq:Epot}) and (\ref{eq:n2pot}). Both potentials are bounded below by zero i.e. $V(\phi) \geq 0$. For  $\phi \rightarrow \infty$, the potentials are bounded above by $V(\phi) \simeq \alpha c^2/4 \simeq 3 \times 10^{-9} m_{Pl}^4 (n=1, \alpha=5 m_{Pl}^2)$ and $V(\phi) \simeq 2.8 \times 10^{-9} m_{Pl}^4 (n=2, \alpha=5 m_{Pl}^2)$, whereas for $\phi \rightarrow -\infty$, both potentials are unbounded from above. However, at the bounce, the critical energy density restricts $\phi_B$ to $(\phi_{min}, \infty)$, where $\phi_{min}$ are represented by the black dots, and given by Eqs.(\ref{eq:Ephimin}) ($n=1$) and (\ref{eq:n2phimin}) ($n=2$). }
\label{fig:pot}
\end{figure*}
\begin{figure*}[tbp]
\begin{center}
\begin{tabular}{ccc}
{\includegraphics[width=2.1in,height=1.65in,angle=0]{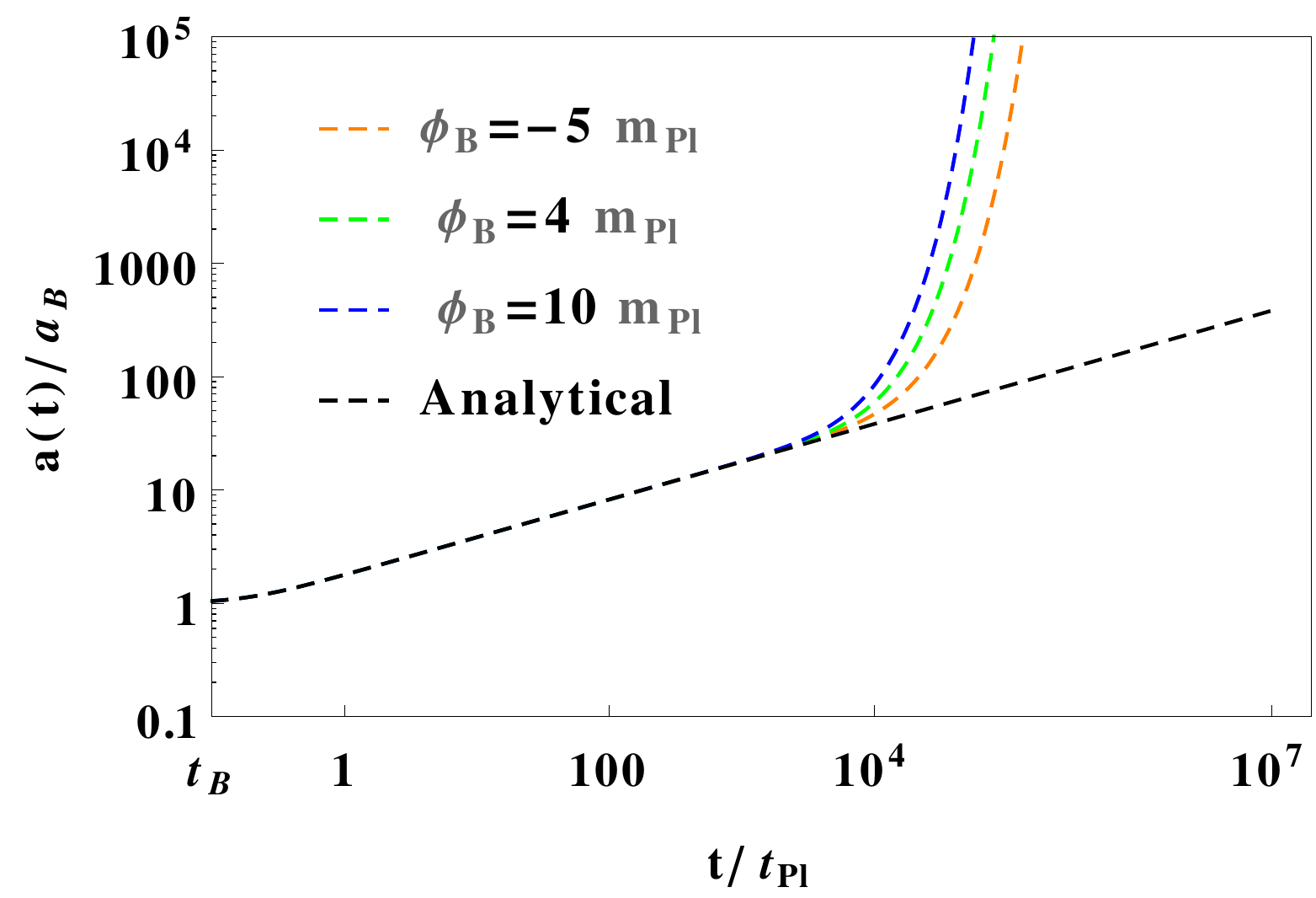}} &
{\includegraphics[width=2.1in,height=1.6in,angle=0]{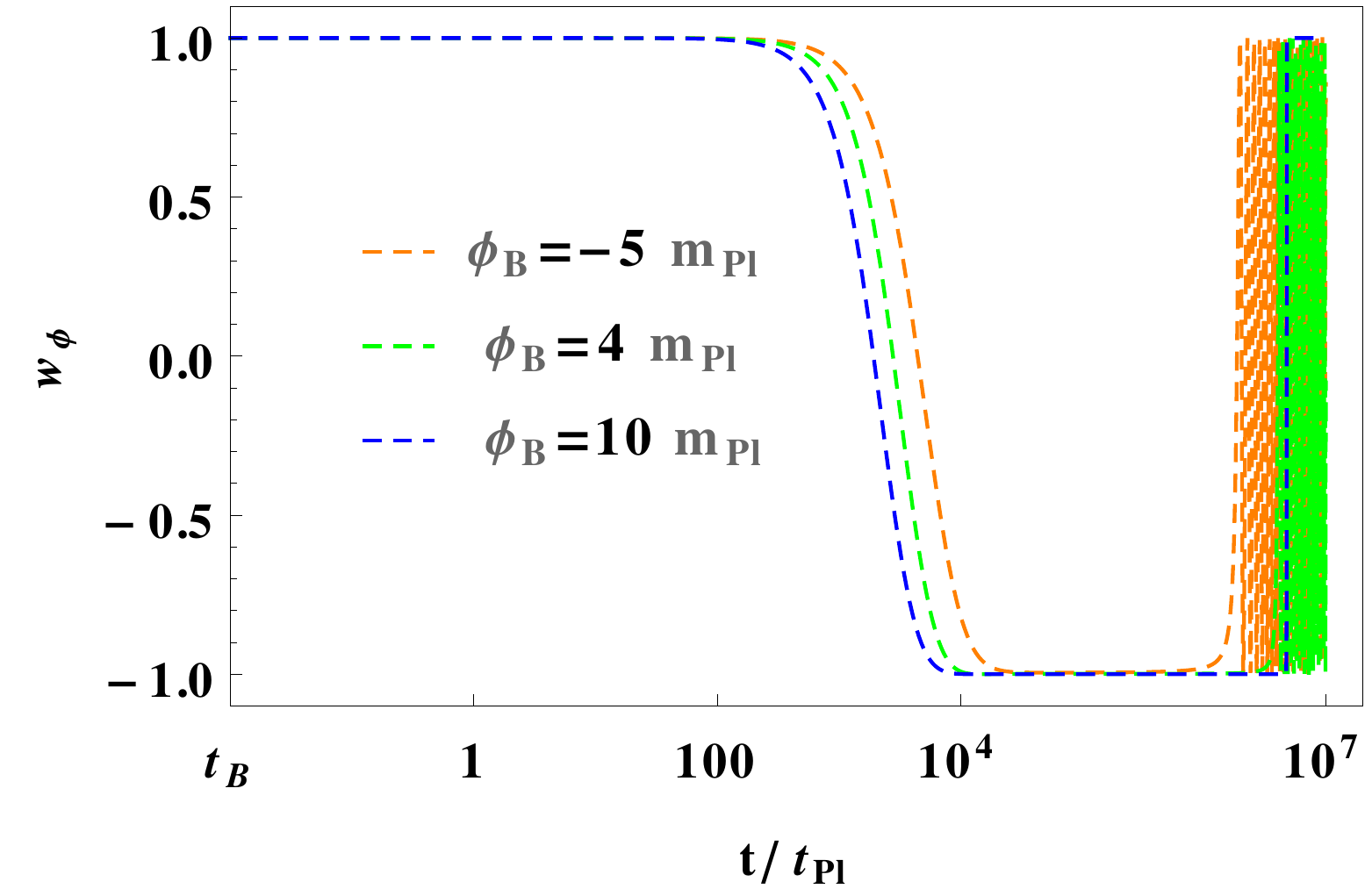}} &
{\includegraphics[width=2.0in,height=1.6in,angle=0]{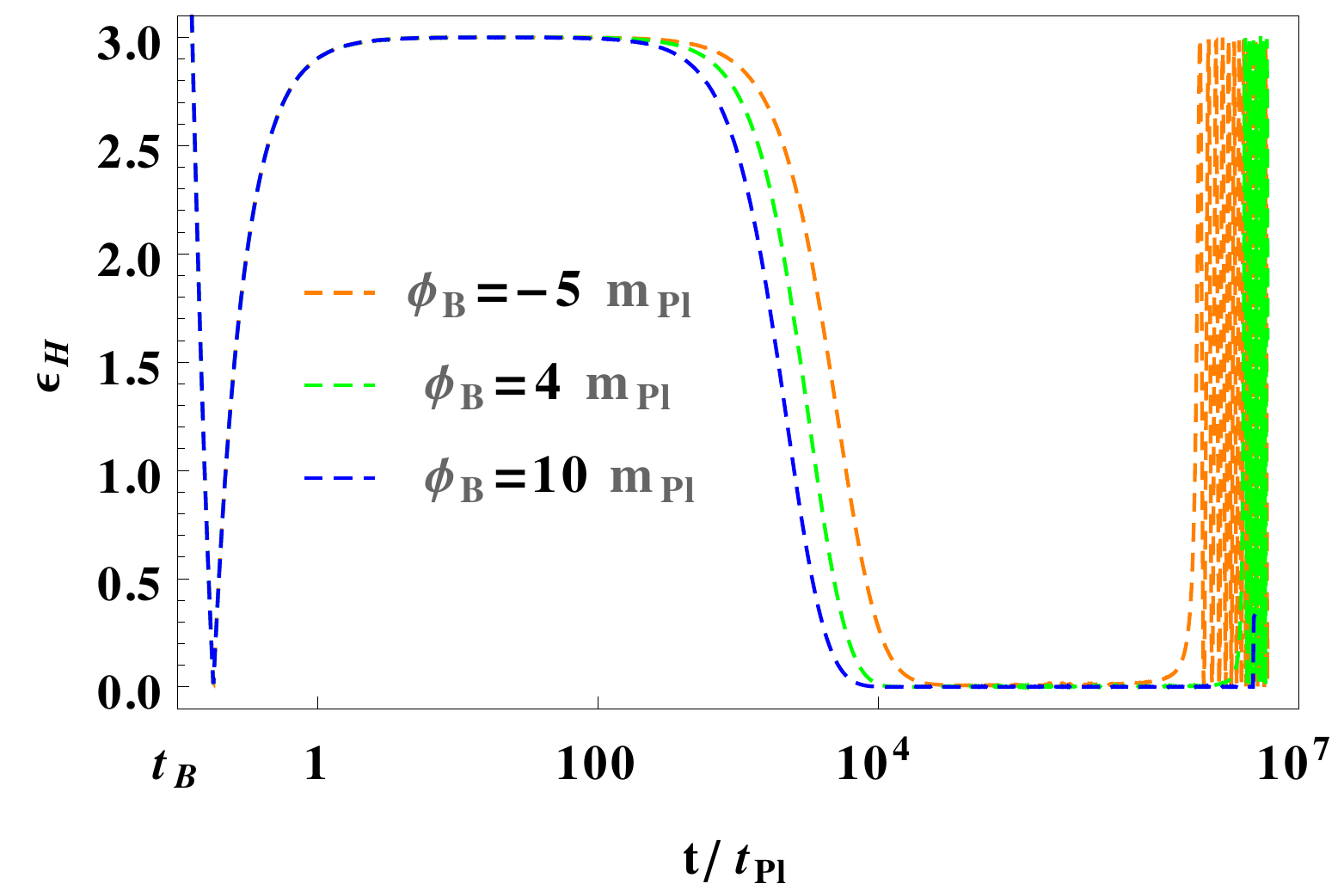}}
\\
{\includegraphics[width=2.1in,height=1.6in,angle=0]{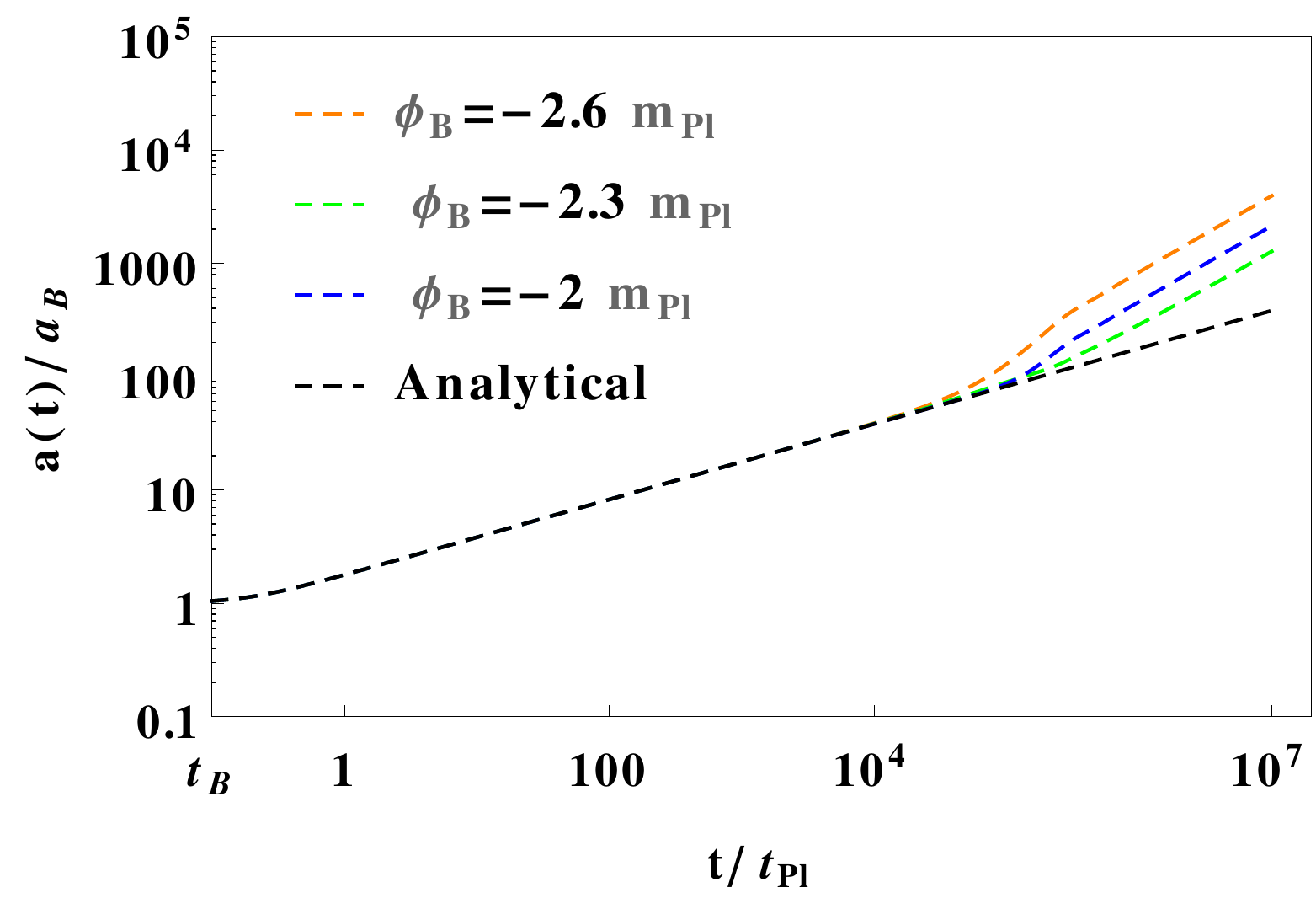}} & 
{\includegraphics[width=2.1in,height=1.6in,angle=0]{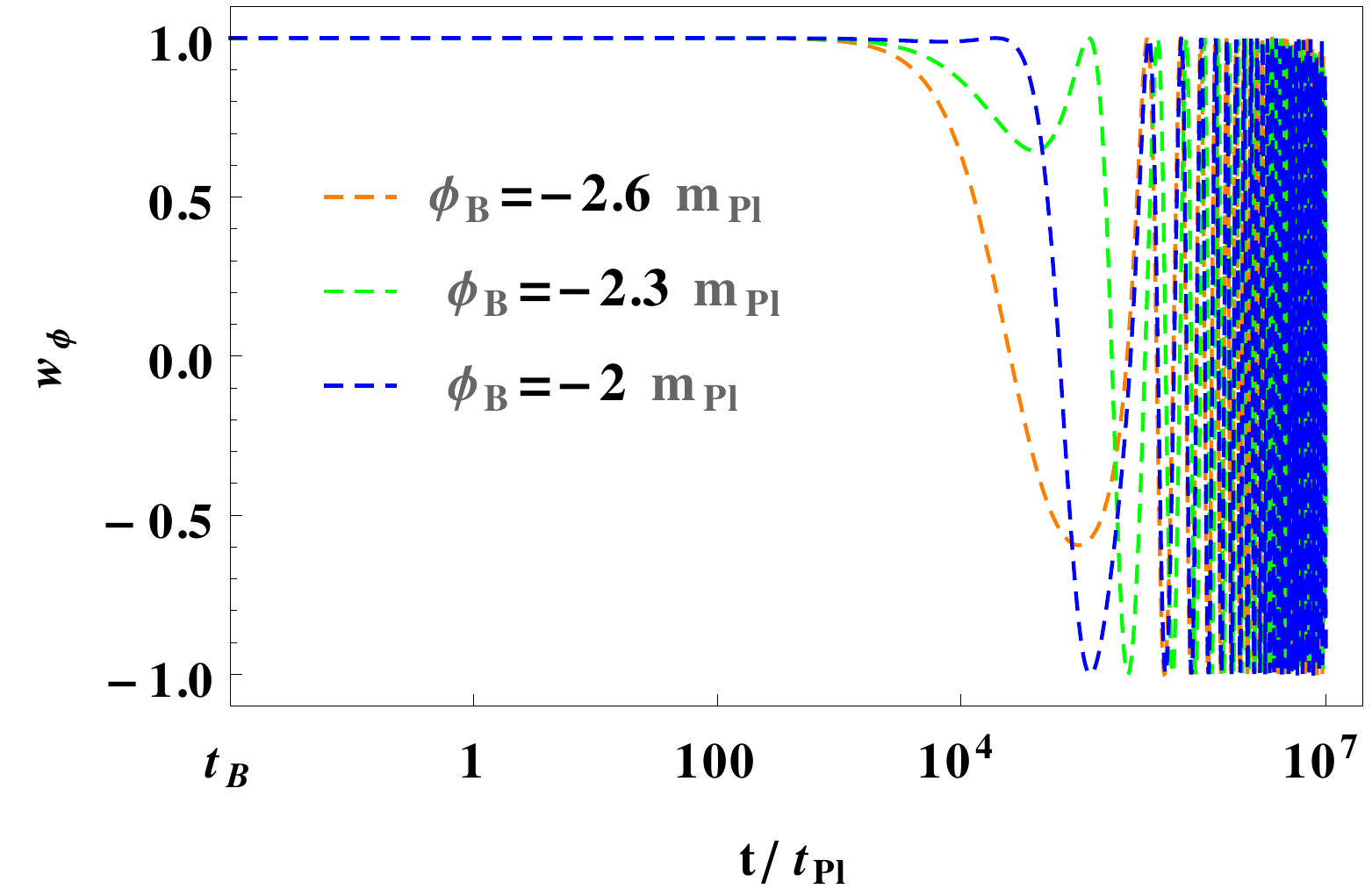}} & 
{\includegraphics[width=2.0in,height=1.6in,angle=0]{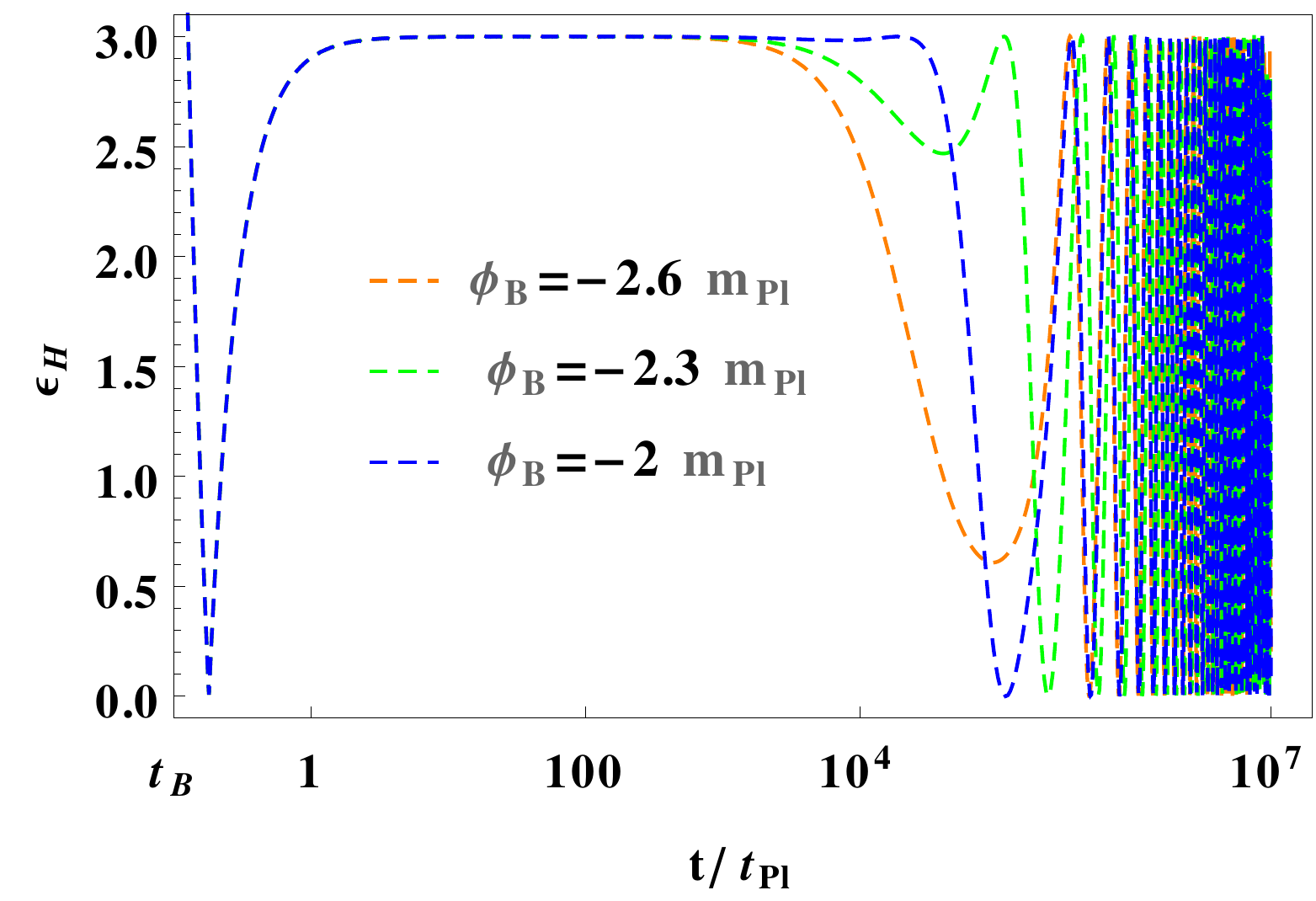}} 
\end{tabular}
\end{center}
\caption{ The numerical results for $T-model$ (\ref{eq:Tpot}) with $\dot{\phi}_B>0$. The figure shows the evolution of $a(t)$, $w(\phi)$ and $\epsilon_H$ for the KED with slow-roll inflation (top) and without slow-roll inflation (bottom) initial conditions at the quantum bounce. The analytical solution of $a(t)$ [Eq.(\ref{eq:a})] is also displayed in order to compare it with the numerical solutions. We use $\alpha=10 m_{Pl}^2$, $c=1.8 \times 10^{-5}m_{Pl}$, and $m_{Pl}=1$. Due to the symmetric nature of the $T-model$, similar results can be obtained for $\dot{\phi}_B<0$.}
\label{fig:n0alpha10_dphp}
\end{figure*}
\begin{figure*}[tbp]
\begin{center}
\begin{tabular}{ccc}
{\includegraphics[width=2.1in,height=1.65in,angle=0]{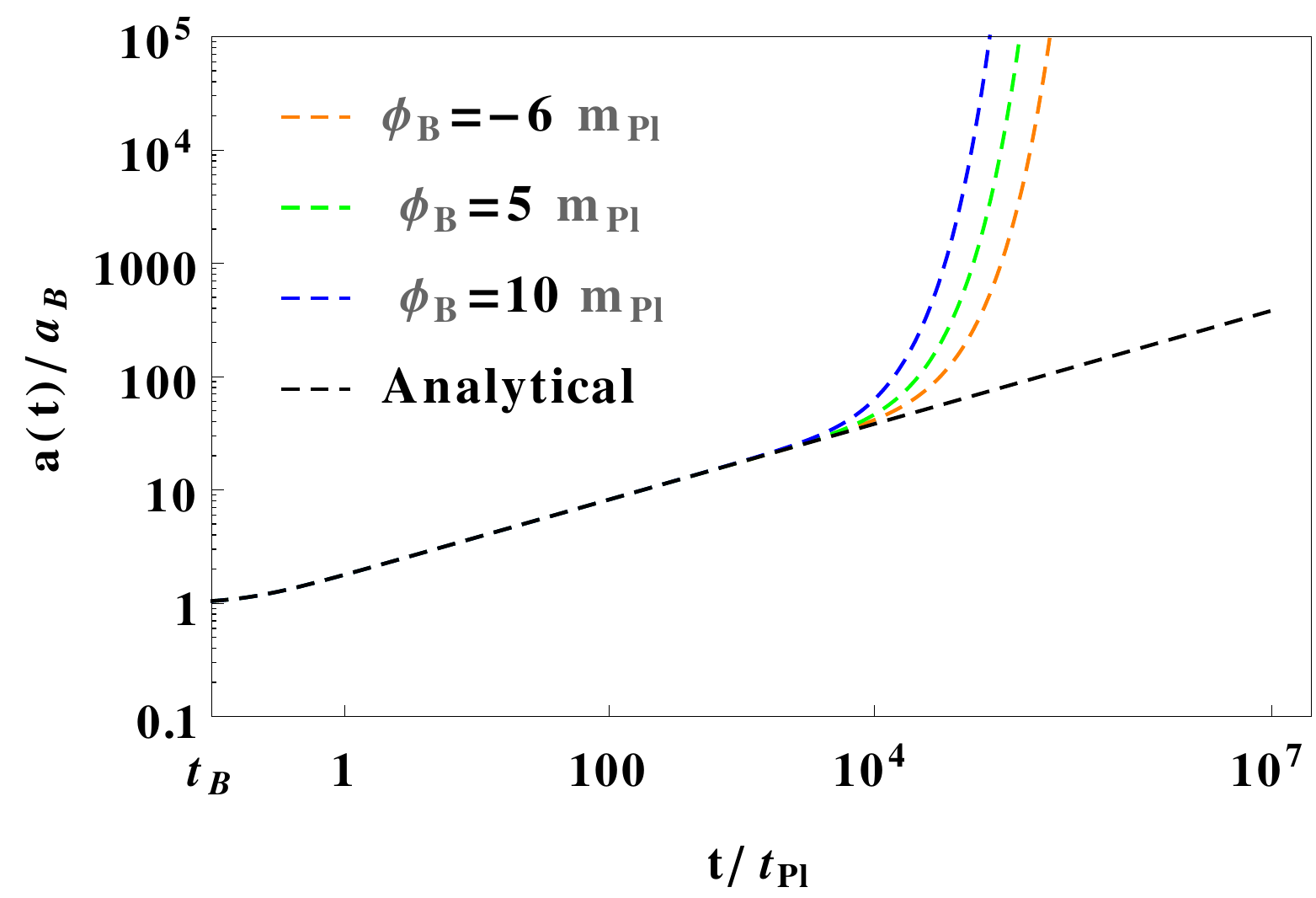}} &
{\includegraphics[width=2.1in,height=1.6in,angle=0]{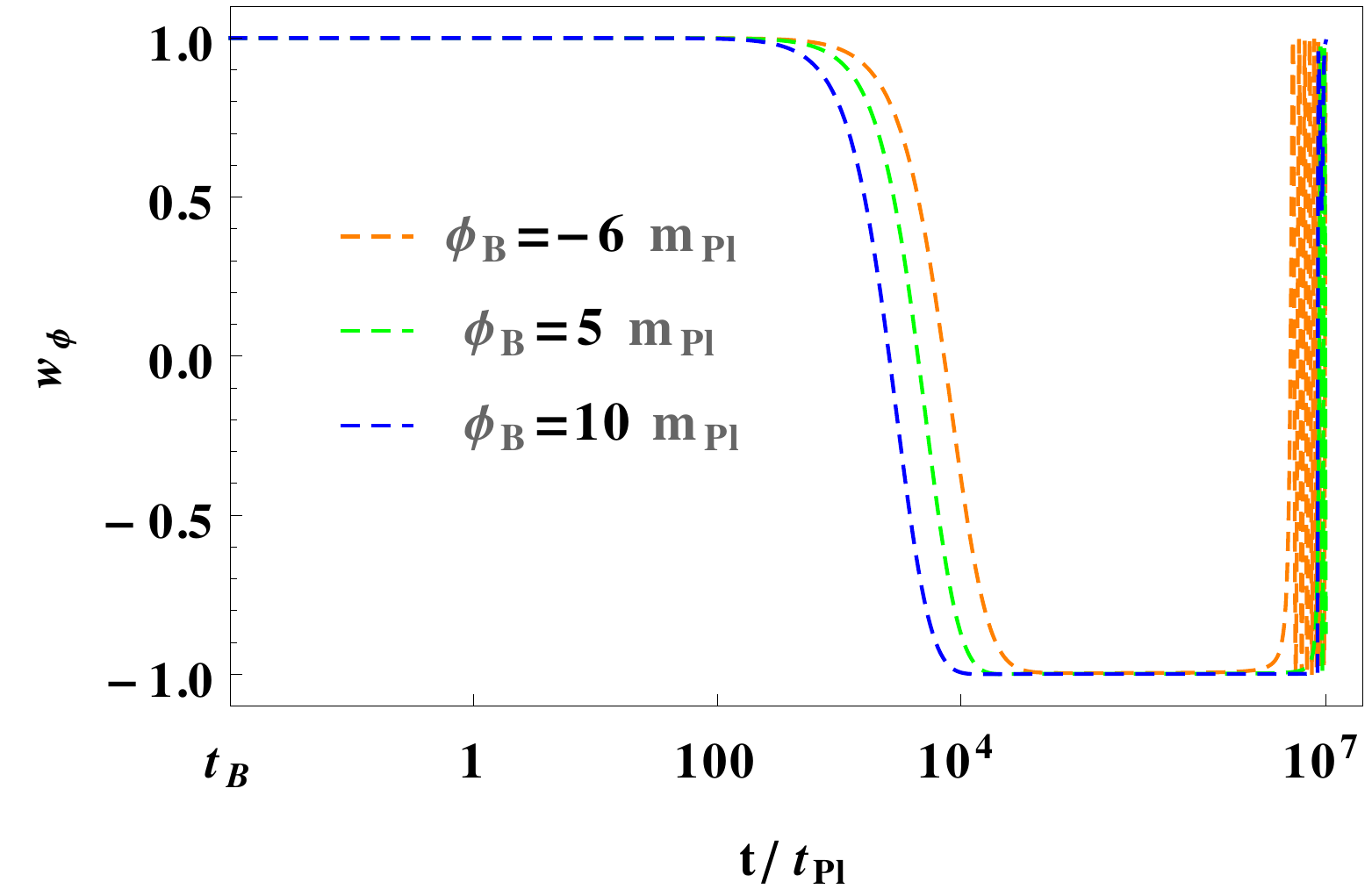}} &
{\includegraphics[width=2.0in,height=1.6in,angle=0]{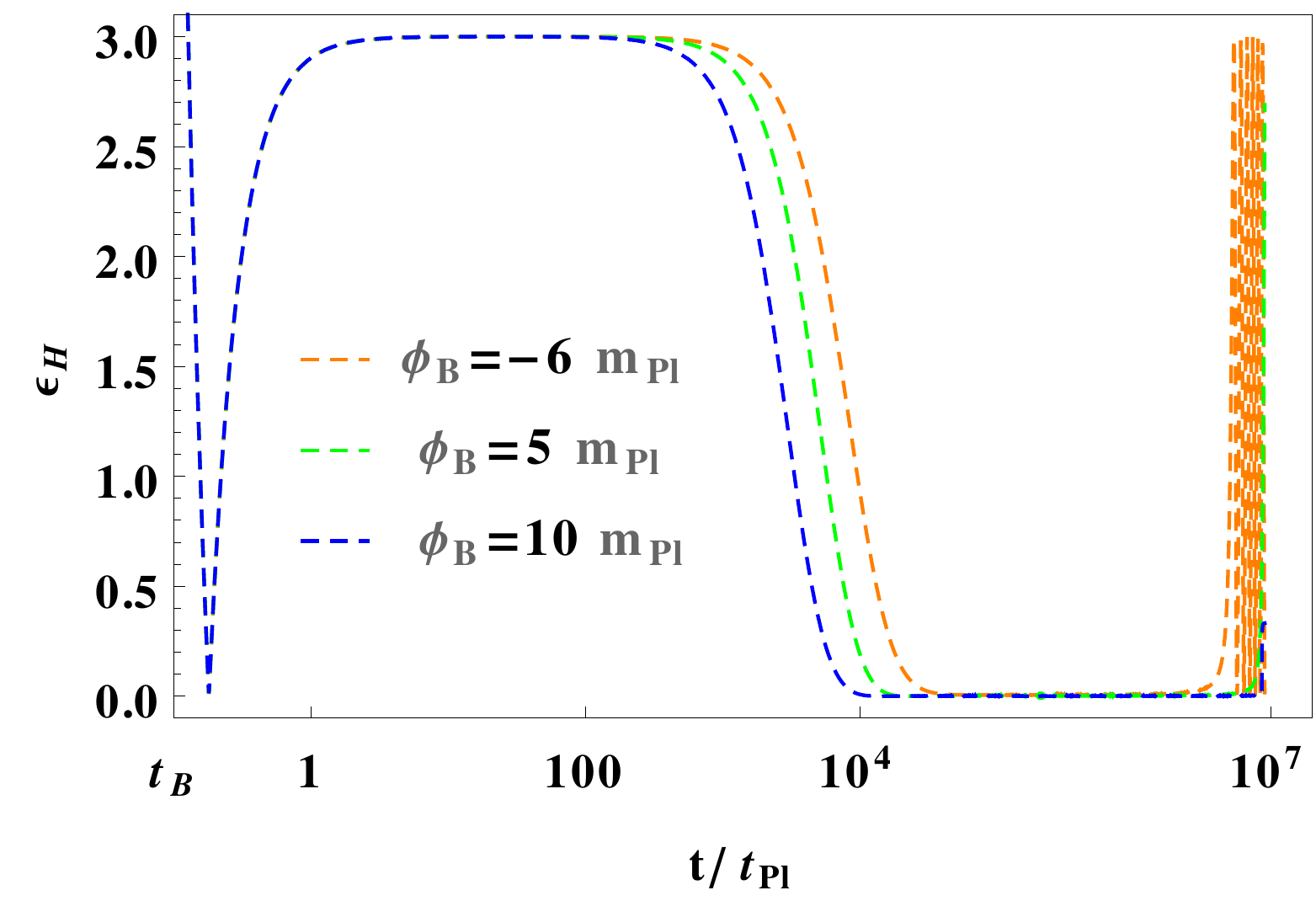}}
\\
{\includegraphics[width=2.1in,height=1.65in,angle=0]{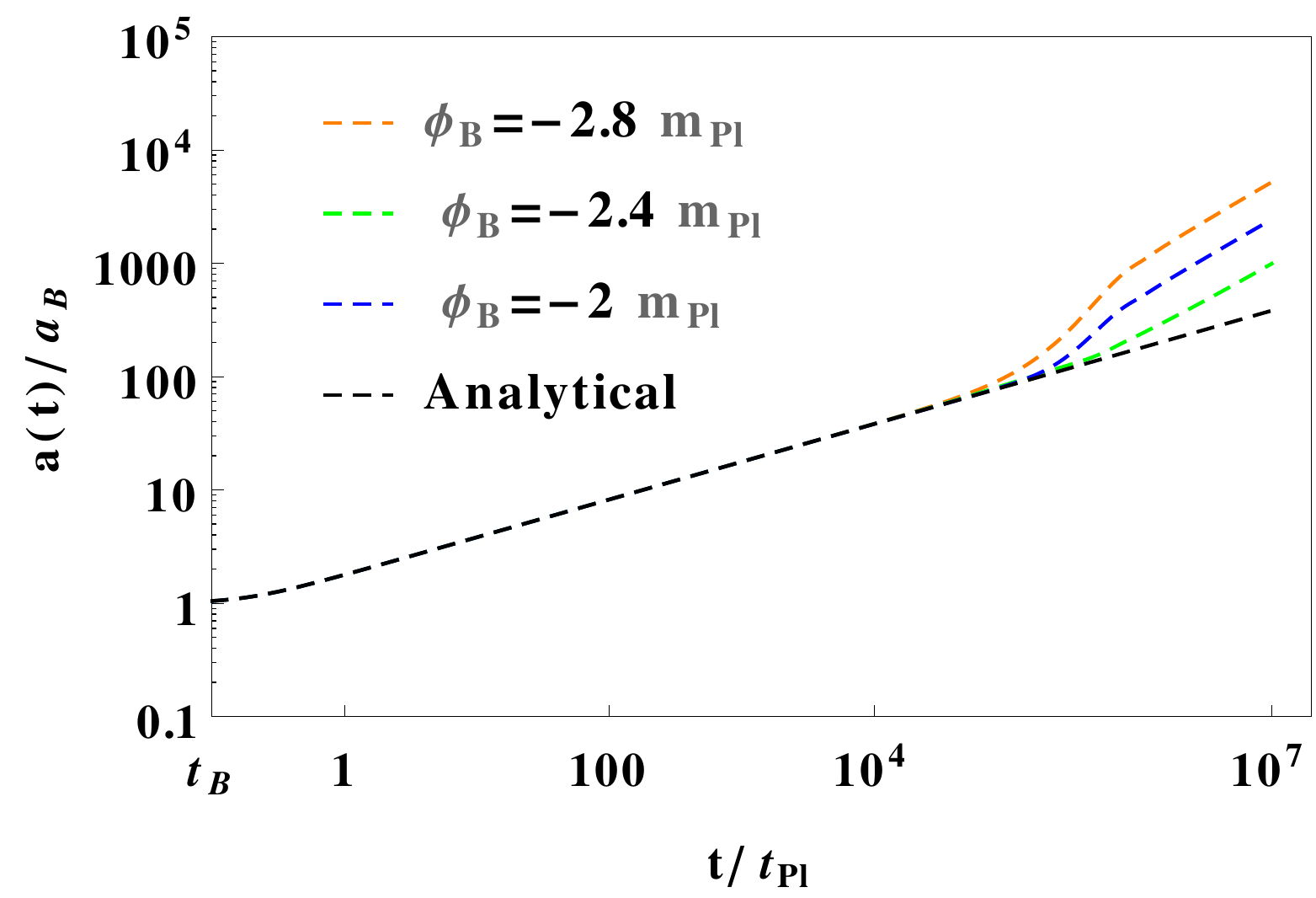}} &
{\includegraphics[width=2.1in,height=1.6in,angle=0]{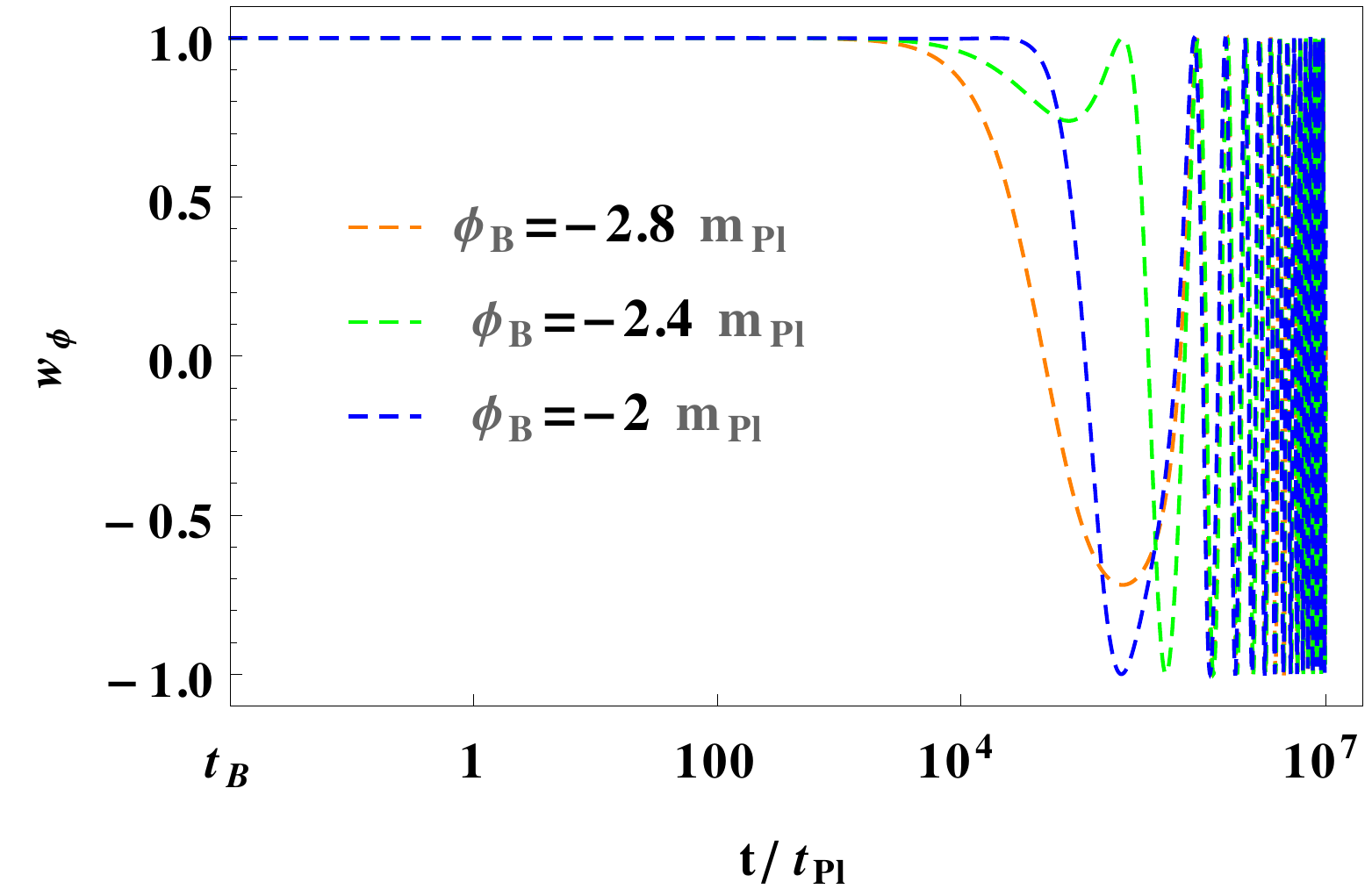}} &
{\includegraphics[width=2.0in,height=1.6in,angle=0]{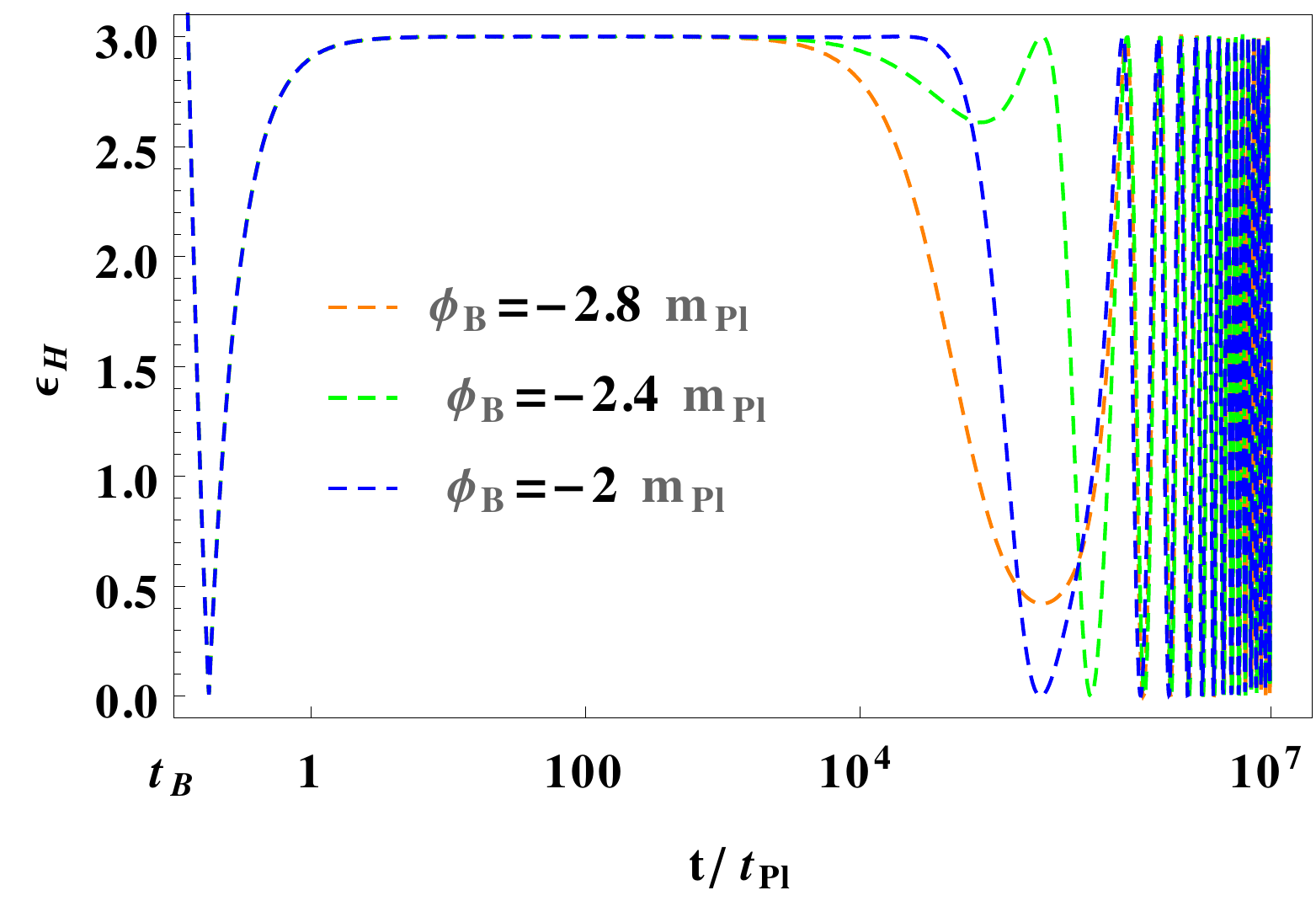}}
\\
{\includegraphics[width=2.1in,height=1.6in,angle=0]{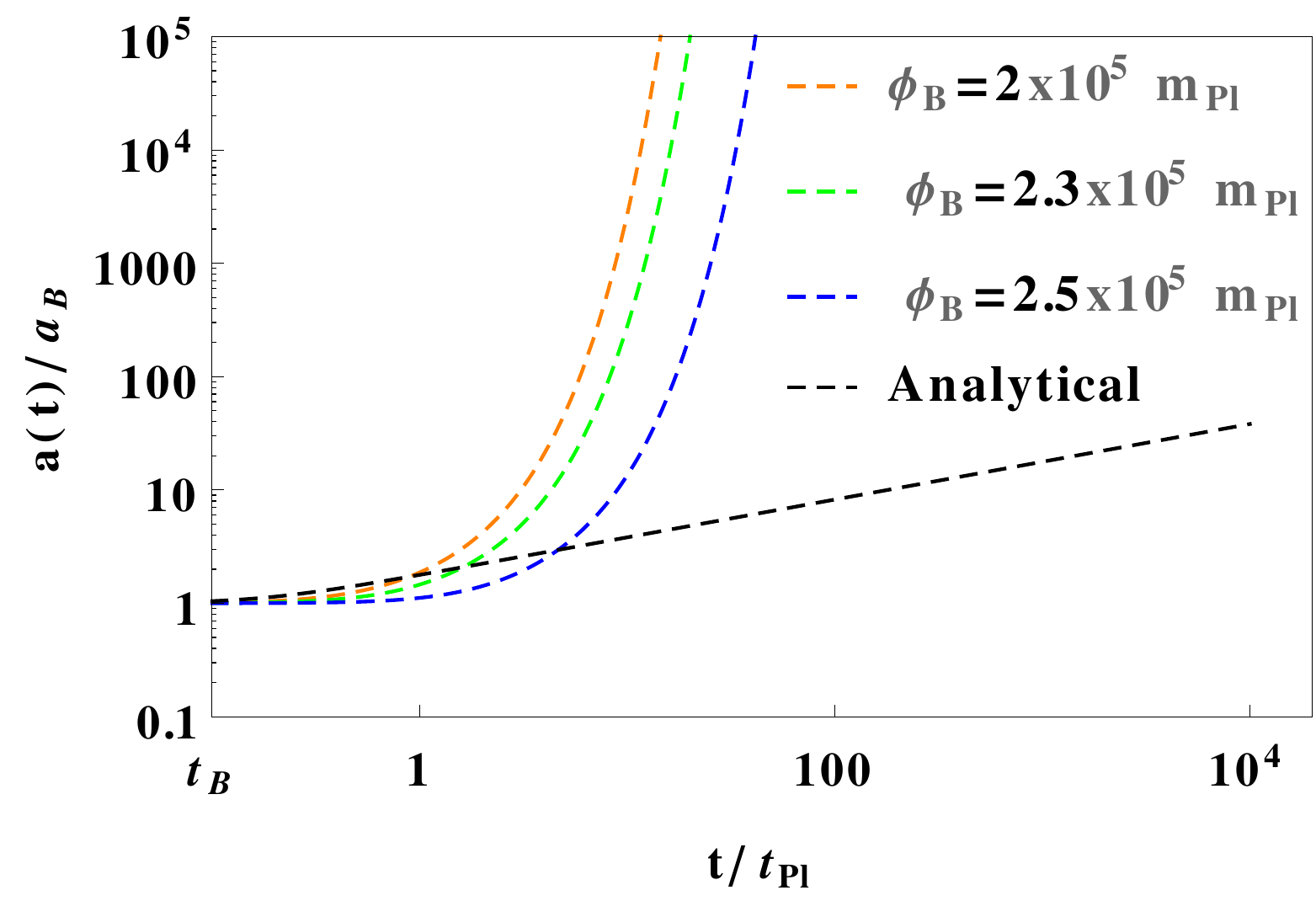}} & 
{\includegraphics[width=2.1in,height=1.6in,angle=0]{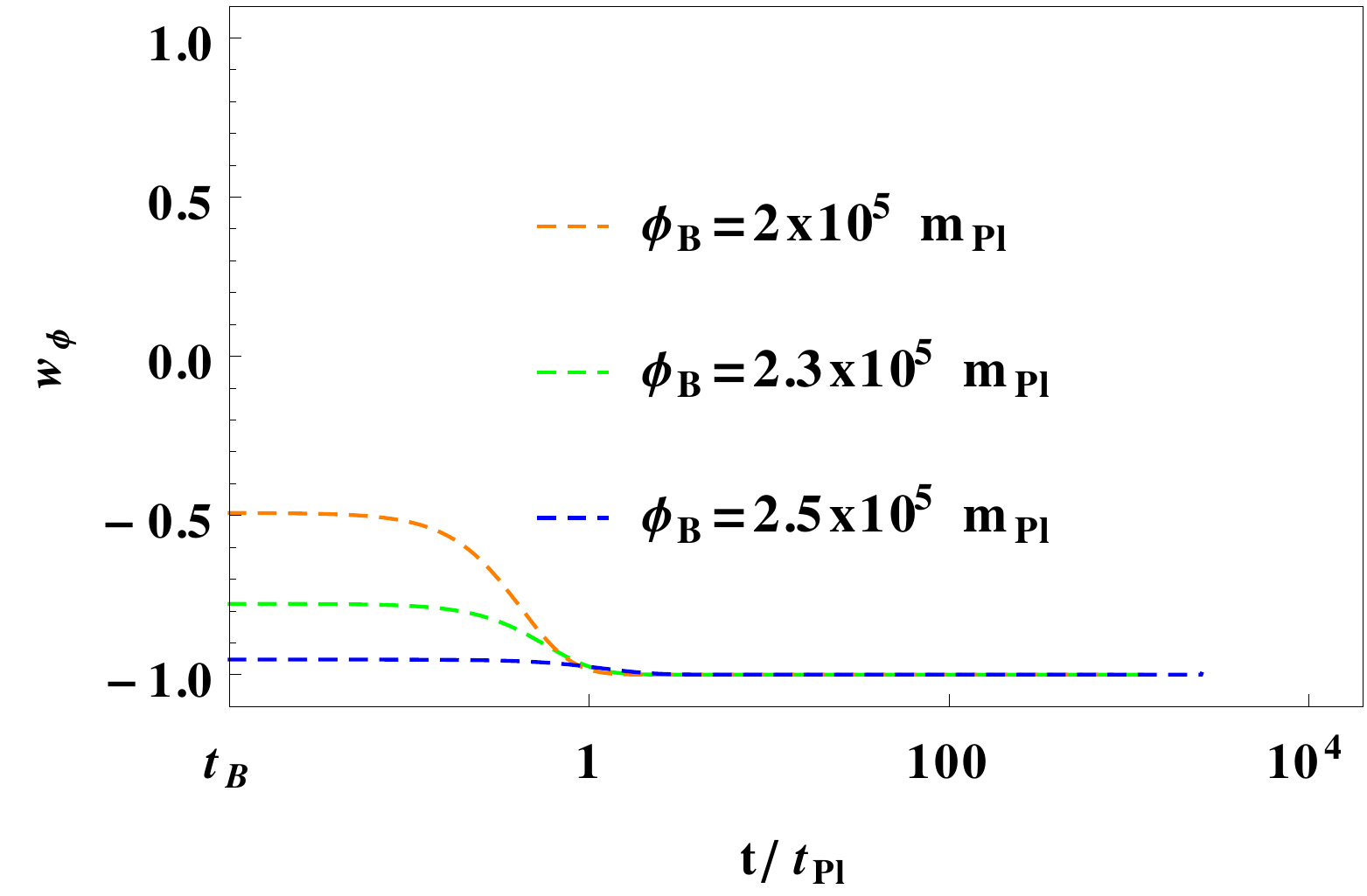}} & 
{\includegraphics[width=2.0in,height=1.6in,angle=0]{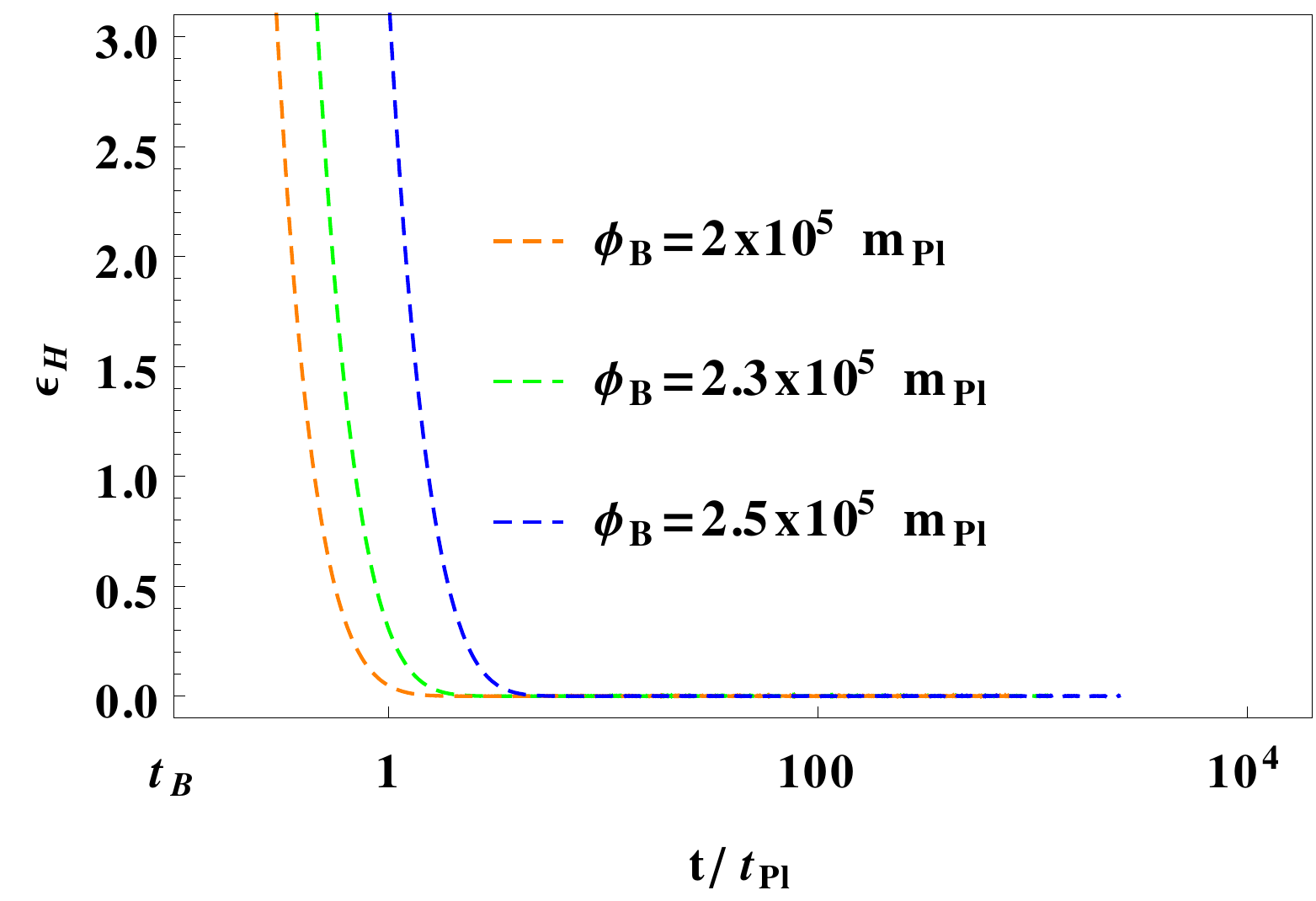}} 
\end{tabular}
\end{center}
\caption{ This figure represents the evolution of $a(t)$, $w(\phi)$ and $\epsilon_H$  for $T-model$ with $\dot{\phi}_B>0$. Top and middle panels correspond to the KED initial conditions with and without slow-roll inflation, whereas bottom ones are for PED initial conditions having slow-roll inflation. The numerical evolution of $a(t)$ with the analytical solution (\ref{eq:a}) is universal in the case of KED initial conditions (top and middle), while it is lost in the PED case (bottom).  We take $\alpha=10^{10} m_{Pl}^2$, $c=8.2 \times 10^{-6}m_{Pl}$, and $m_{Pl}=1$. In the case of $\dot{\phi}_B<0$, one can obtain similar results as the potential (\ref{eq:Tpot}) is symmetric.}
\label{fig:n0alpha10p10_dphp}
\end{figure*}
\begin{figure*}[tbp]
\begin{center}
\begin{tabular}{ccc}
{\includegraphics[width=2.1in,height=1.65in,angle=0]{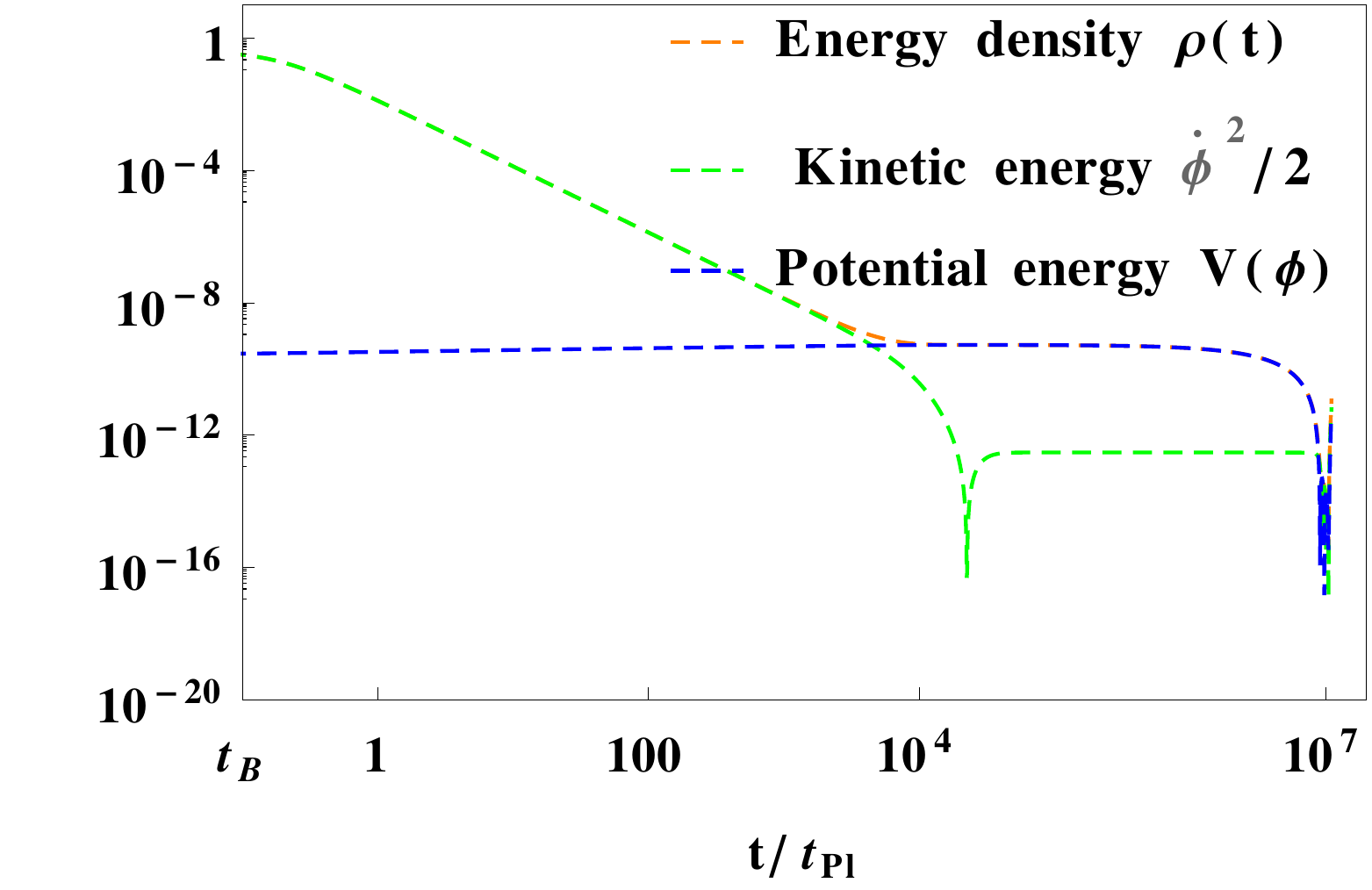}} &
{\includegraphics[width=2.1in,height=1.6in,angle=0]{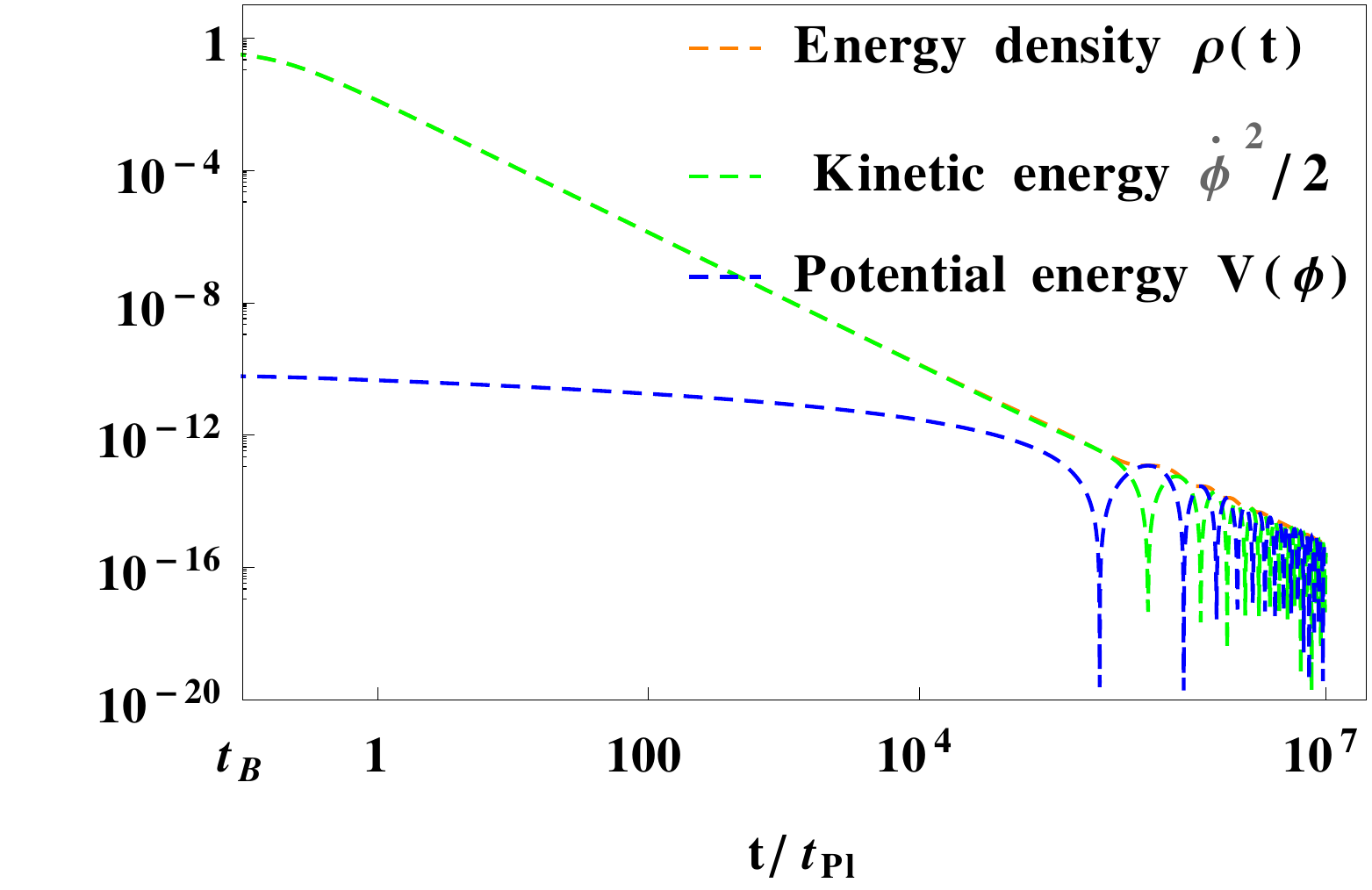}} &
{\includegraphics[width=2.0in,height=1.6in,angle=0]{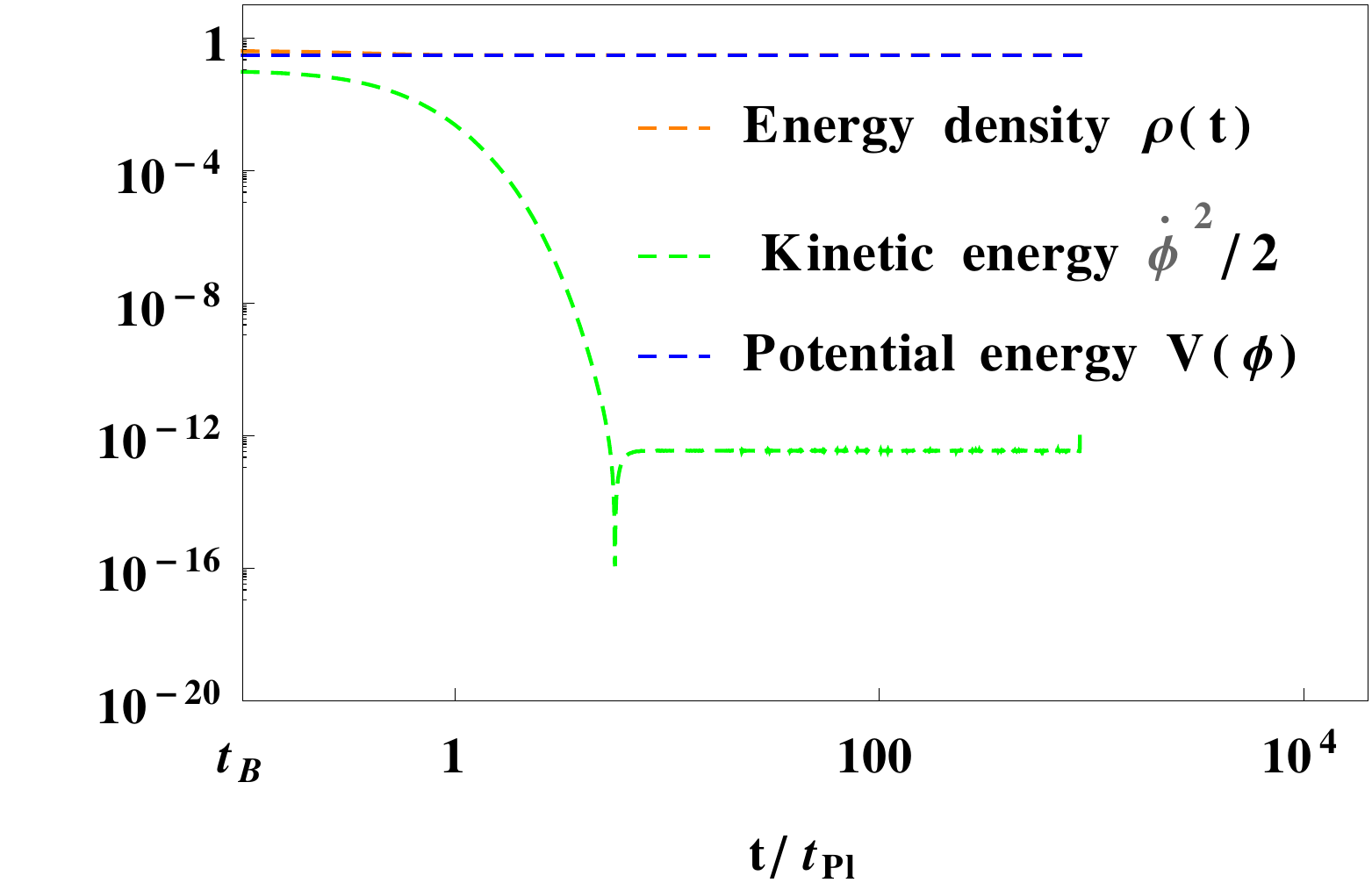}}
\end{tabular}
\end{center}
\caption{ This figure shows the evolutions of the kinetic $\dot{\phi}^2/2$ and potential  $V(\phi)$ energies of the inflaton for  $T-model$ with $\alpha=10^{10}m_{Pl}^2$ for $\dot{\phi}_B>0$. The energy density $\rho=\dot{\phi}^2/2+V(\phi)$ is also displayed. Left (KED case, $\phi_B=5 m_{Pl}$): during the bouncing phase, PE is sub-dominant and remains so until the transition phase. Thereafter, it overtakes KE and starts to dominate the evolution of the universe, whereby a slow-roll inflationary phase is resulted. Middle (KED case, $\phi_B=-2.4 m_{Pl}$): This panel shows that the PE remains sub-dominant and is never able to overtake the KE. Hence, it does not give rise to a slow-roll inflationary phase. Right (PED case, $\phi_B=2 \times 10^5 m_{Pl}$): in this case, PE dominates throughout the whole evolution, and provides a slow-roll inflationary phase for a long period.}
\label{fig:n0rho}
\end{figure*}
\begin{table*}
\caption{Table for the range of $\alpha$ having slow-roll (SR)$/$no slow-roll (NSR) inflation in the cases of both KED and PED initial conditions of the inflaton field at the bounce. For the sake of simplicity, we numerically choose $w_{\phi} \simeq -0.999$ and $\epsilon \simeq 10^{-2}-10^{-6}$ in the case of SR whereas $w_{\phi} \simeq -0.94$ and $\epsilon \simeq 0.10 $ for NSR. The solid line $(-)$ corresponds to the PED range that does not exist throughout the bouncing regime.  }
\begin{center}
\begin{tabular}{l  l cc l}
\hline\hline\\
Model & \qquad~~ $\alpha$ && Slow-roll inflation depending on the range of $\phi_B$  & 
\\  \cline{3-5} \\
& & KED (SR) & Existence of KED  & PED (SR)\\
& & (except subset) & subset (NSR) & \\
\hline\hline\\
$T$ & $0 < \alpha < 4.3 \times 10^9 $ & All & Yes & $-$ \\
    & $4.3 \times 10^9 \leq \alpha $  & All & Yes & All \\\\
$E$ & $0 < \alpha < 0.02$ & All & No (for $\dot{\phi}>0$) & All\\
 &  &  & Yes (for $\dot{\phi}<0$) & \\   
& $0.02 \leq \alpha < 0.6$  &  All & Yes & None\\
& $ 0.6 \leq \alpha $ & All & Yes & All\\\\
$n=2$ & $0 < \alpha < 0.1$  & All & No (for $\dot{\phi}>0$) & All\\ 
&  &  & Yes (for $\dot{\phi}<0$) & \\ 
& $0.1 \leq \alpha < 2.4$ & All & Yes & None\\
& $2.4 \leq \alpha $ & All & Yes & All\\ \\
\hline\hline
\end{tabular}
\label{tab:n012_alpha}
\end{center}
\end{table*}
\begin{table*}
\caption{Table for SR$/$NSR with the range of $\phi_B$ for $\dot{\phi}_B>0$ and $\dot{\phi}_B<0$ having different values of $\alpha$. For each value of $\alpha$, the corresponding values of $c$ are given by  Eqs.(\ref{eq:Talphac}), (\ref{eq:Ealphac}) and (\ref{eq:n2alphac}). The sign $(-)$ represents the case in which a single initial value of $\phi_B$ does not provide KED=PED $(w^B = 0)$ and PED $(w^B < 0)$ at the bounce. Here, $\phi_{max}$ corresponds to $T-model$ and given by  Eq.(\ref{eq:TNphimax2}). In the case of $E-model$ and $\alpha-$model with $n=2$, $\phi_{min}$ are given by Eqs.(\ref{eq:Ephimin}) and (\ref{eq:n2phimin}), respectively. }
\begin{center}
\resizebox{\textwidth}{!}{%
\begin{tabular}{lcccccc}
\hline\hline
\\
Model ~~& $\dot{\phi}_B$ ~~& $\alpha$~~ & KED (SR) ~~& Subset of KED (NSR) ~~& KED=PED (SR/NSR) ~~& PED (SR)\\
~~& ~~& ~~ & $(w^B > 0)$ ~~& $(w^B > 0)$ ~~& $(w^B = 0)$ ~~& $(w^B < 0)$\\\\
\hline
\\
$T$~~& $>0$ ~~& 10 ~~&  $ -\infty \leq \phi_B \leq -3.1 $ ~~& $-3.1 < \phi_B \leq -1.6$ ~~& $-$ ~~& $-$\\
~~& ~~& ~~&  $ -1.61 \leq \phi_B \leq +\infty $ ~~& ~~& ~~& \\\\
~~& ~~& $10^{10}$ ~~&  $ -1.52 \times 10^5 < \phi_B \leq -3.21 $ ~~& $-3.2 \leq \phi_B < -1.8$ ~~& $\phi_B = \pm 1.52 \times 10^5 $(SR) ~~& $-\phi_{max} \leq \phi_B < -1.52 \times 10^5 $\\
~~& ~~& ~~&  $ -1.8 \leq \phi_B < 1.52 \times 10^5 $ ~~& ~~& ~~& $1.52 \times 10^5 < \phi_B \leq +\phi_{max} $\\\\
$E$~~& $>0$ ~~& 0.1 ~~&  $-1.51 \leq \phi_B \leq \infty$ ~~& $-3.52 < \phi_B \leq -1.5$ ~~& $\phi_B = -3.52$ (NSR) ~~& None \\\\
~~& ~~& 5 ~~&  $-24.70 \leq \phi_B \leq -3.11$ ~~& $-3.1 \leq \phi_B < -1.4$ ~~& $\phi_B = -24.71$ (SR) ~~& $\phi_{min} \leq \phi_B \leq -24.72$ \\
~~&~~& ~~& \& $-1.4 \leq \phi_B \leq \infty$ ~~&  ~~&  ~~&  \\\\
~~& $<0$ ~~& 0.1 ~~&  $2.4 \leq \phi_B \leq \infty$ ~~& $-3.51 < \phi_B < 2.4$ ~~& $\phi_B = -3.51$ (NSR) ~~& None \\\\
~~& ~~& 5 ~~&  $-24.69 \leq \phi_B \leq 1.4$ ~~& $1.4 < \phi_B < 3$ ~~& $\phi_B = -24.7$ (SR) ~~& $\phi_{min} \leq \phi_B \leq -24.71$ \\
~~&~~& ~~& \& $3 \leq \phi_B \leq \infty$ ~~&  ~~&  ~~&  \\\\
$n=2$~~& $>0$ ~~& 0.5 ~~&  $-1.39 \leq \phi_B \leq \infty $ ~~& $-3.93 < \phi_B \leq -1.4$ ~~& $\phi_B = -3.93$ (NSR) ~~& None \\\\
~~&~~& 5 ~~&  $-12.4 \leq \phi_B \leq -3.4$ ~~& $-3.4 < \phi_B < -1.4$ ~~& $\phi_B = -12.41$ (SR) ~~& $\phi_{min} \leq \phi_B \leq -12.42$ \\
~~&~~& ~~& \& $-1.4 \leq \phi_B \leq \infty$ ~~&  ~~&  ~~&  \\\\
~~&$<0$~~& 0.5 ~~&  $2.4 \leq \phi_B \leq \infty $ ~~& $-3.92 < \phi_B < 2.4$ ~~& $\phi_B = -3.92$ (NSR) ~~& None \\\\
~~&~~& 5 ~~&  $-12.39 \leq \phi_B \leq 1$ ~~& $1 < \phi_B < 2.7$ ~~& $\phi_B = -12.4$ (SR) ~~& $\phi_{min} \leq \phi_B \leq -12.41$ \\
~~&~~& ~~& \& $2.7 \leq \phi_B \leq \infty$ ~~&  ~~&  ~~&  \\\\
\hline\hline
\end{tabular}}
\label{tab:n012_alpha_dphi}
\end{center}
\end{table*}
\begin{table*}
\caption{This table corresponds to the $T-model$. We display the number of $e$-foldings $N_{inf}$ and other parameters for different choices of $\phi_B$ in two cases of $\alpha$, see  Eq.(\ref{eq:Talphac}). This table is for $\dot{\phi}_B > 0$. Due to the symmetric nature of the $T-model$, one can obtain similar results for  $\dot{\phi}_B < 0$. }
\begin{center}
\resizebox{\textwidth}{!}{%
\begin{tabular}{cccccccccc}
\hline\hline
\\
$\alpha$~~ & $\phi_B$~~~  & Inflation~~~ & $t/t_{pl}$~~~ & $\epsilon$~~ & $w$ ~~& $N_{inf}$ &~~~$r_{w}^c/r_w$&~~~${w}^B$\\\\
\hline
10 ~~ & $-5.3$~~~& begin~~~& $5.40007 \times 10^3$ ~~~& 0.99~~ & $-1/3$ ~~& ~~~& ~~~&\\
~~ & ~~~& slow-roll~~~& $4.66194 \times 10^4$ ~~~& $5.0 \times 10^{-3}$~~ & $-0.996$ ~~& 79.23 ~~~&$r_{w}^c < r_w$ ~~~& $>0$\\
~~ & ~~~& end~~~& $2.032 \times 10^6$ ~~~& 1.0~~ & $-1/3$ ~~& ~~~& ~~~& \\\\
~~ & $-4.9$~~~& begin~~~& $6.05487 \times 10^3$ ~~~& 0.99~~ & $-1/3$ ~~& ~~~& ~~~&\\
~~ & ~~~& slow-roll~~~& $5.01307 \times 10^4$ ~~~& $7.1 \times 10^{-3}$~~ & $-0.994$ ~~& 60.28 ~~~&$r_{w}^c = r_w$ ~~~& $>0$\\
~~ & ~~~& end~~~& $1.745 \times 10^6$ ~~~& 0.99~~ & $-1/3$ ~~& ~~~& ~~~& \\\\
~~ & $-4$~~~& begin~~~& $8.594 \times 10^3$ ~~~& 1.0~~ & $-1/3$ ~~& ~~~& ~~~&\\
~~ & ~~~& slow-roll~~~& $6.21983 \times 10^4$ ~~~& $1.9 \times 10^{-2}$~~ & $-0.987$ ~~& 27.57 ~~~&$r_{w}^c > r_w$ ~~~& $>0$\\
~~ & ~~~& end~~~& $1.133 \times 10^6$ ~~~& 1.0~~ & $-1/3$ ~~& ~~~& ~~~& \\\\
~~ & $1.1$~~~& begin~~~& $6.25705 \times 10^3$ ~~~& 1.0~~ & $-1/3$ ~~& ~~~& ~~~&\\
~~ & ~~~& slow-roll~~~& $2.04834 \times 10^4$ ~~~& $3.1 \times 10^{-5}$~~ & $-1$ ~~& 60.59 ~~~&$r_{w}^c = r_w$ ~~~& $>0$\\
~~ & ~~~& end~~~& $1.751 \times 10^6$ ~~~& 0.99~~ & $-1/3$ ~~& ~~~& ~~~& \\\\
~~ & $2$~~~& begin~~~& $4.98597 \times 10^3$ ~~~& 1.0~~ & $-1/3$ ~~& ~~~& ~~~&\\
~~ & ~~~& slow-roll~~~& $1.7624 \times 10^4$ ~~~& $1.5 \times 10^{-4}$~~ & $-1$ ~~& 103.17 ~~~&$r_{w}^c < r_w$ ~~~& $>0$\\
~~ & ~~~& end~~~& $2.356 \times 10^6$ ~~~& 1.0~~ & $-1/3$ ~~& ~~~& ~~~& \\\\
$10^{10}$~~ & $-5.5$~~~& begin~~~& $1.106433 \times 10^4$ ~~~& 0.99~~ & $-1/3$ ~~& ~~~& ~~~&
\\
~~ & ~~~& slow-roll~~~& $9.60811 \times 10^4$ ~~~& $7.1 \times 10^{-3}$~~ & $-0.995$ ~~& 77.91 ~~~&$r_{w}^c < r_w$ ~~~& $>0$\\
~~ & ~~~& end~~~& $4.345 \times 10^6$ ~~~& 0.99~~ & $-1/3$ ~~& ~~~& ~~~& \\\\
~~ & $-5.1$~~~& begin~~~& $1.2549 \times 10^4$ ~~~& 0.99~~ & $-1/3$ ~~& ~~~& ~~~&
\\
~~ & ~~~& slow-roll~~~& $1.04701 \times 10^5$ ~~~& $8.3 \times 10^{-3}$~~ & $-0.994$ ~~& 60.49 ~~~&$r_{w}^c = r_w$ ~~~& $>0$\\
~~ & ~~~& end~~~& $3.804 \times 10^6$ ~~~& 0.99~~ & $-1/3$ ~~& ~~~& ~~~& \\\\
~~ & $0.5$~~~& begin~~~& $1.57596 \times 10^4$ ~~~& 0.99~~ & $-1/3$ ~~& ~~~& ~~~&
\\
~~ & ~~~& slow-roll~~~& $4.82807 \times 10^4$ ~~~& $1.2 \times 10^{-4}$~~ & $-1$ ~~& 41.88 ~~~&$r_{w}^c > r_w$ ~~~& $>0$\\
~~ & ~~~& end~~~& $3.13 \times 10^6$ ~~~& 0.99~~ & $-1/3$ ~~& ~~~& ~~~& \\\\
~~ & $1.05$~~~& begin~~~& $1.30481 \times 10^4$ ~~~& 0.99~~ & $-1/3$ ~~& ~~~& ~~~&
\\
~~ & ~~~& slow-roll~~~& $4.19299 \times 10^4$ ~~~& $1.3 \times 10^{-5}$~~ & $-1$ ~~& 60.32 ~~~&$r_{w}^c = r_w$ ~~~& $>0$\\
~~ & ~~~& end~~~& $3.8 \times 10^6$ ~~~& 0.99~~ & $-1/3$ ~~& ~~~& ~~~& \\\\
~~ & $1.5$~~~& begin~~~& $1.14255 \times 10^4$ ~~~& 0.99~~ & $-1/3$ ~~& ~~~& ~~~&
\\
~~ & ~~~& slow-roll~~~& $3.79398 \times 10^4$ ~~~& $3.4 \times 10^{-5}$~~ & $-1$ ~~& 78.09 ~~~&$r_{w}^c < r_w$ ~~~& $>0$\\
~~ & ~~~& end~~~& $4.354 \times 10^6$ ~~~& 1.00~~ & $-1/3$ ~~& ~~~& ~~~& \\\\
~~ & $2 \times 10^{5}$~~~& begin~~~& $0.01$ ~~~& 1.0~~ & $-0.492$ ~~& ~~~& ~~~&
\\
~~ & ~~~& slow-roll~~~& $1.0$ ~~~& $0.047$~~ & $-0.984$ ~~& 713.98 ~~~&$r_{w}^c < r_w$ ~~~& $<0$\\
~~ & ~~~& end~~~& $888$ ~~~& 0.1~~ & $-1/3$ ~~& ~~~& ~~~& \\\\
\hline\hline
\end{tabular}}
\label{tab:n0_dphip}
\end{center}
\end{table*}

\section{Background evolution}
\label{sec:EOM}

In the framework of LQC, the modified Friedmann equation in a spatially flat Friedmann-Lemaitre-Robertson-Walker (FLRW) background is written as \cite{ashtekar2006}
\begin{eqnarray}
H^2=\frac{8 \pi}{3 m_{Pl}^2}~\rho \Big{(}1-\frac{\rho}{\rho_c}\Big{)}, 
\label{eq:Hub}
\end{eqnarray}
where $H=\dot{a}/a$ is the Hubble parameter, $\rho=\dot{\phi}^2/2+V(\phi)$ represents  the energy density of the inflaton field, and $V(\phi)$ is the potential of the field. 
The dot denotes a derivative with respect to the cosmic time $t$, $m_{Pl}$ is the Planck mass and $\rho_c$ is the critical energy density that corresponds to the maximum value of energy density, and is given by $\rho_c \simeq 0.41 m_{pl}^4$ \cite{Meissne,Domagala}.

The conservation equation in the context of LQC remains the same as in the classical theory
\begin{eqnarray}
\dot{\rho}+3H(\rho+p)=0.
\label{eq:conser}
\end{eqnarray}
Here, $p$ denotes the pressure of the matter field.  Eq.(\ref{eq:conser}) gives the Klein-Gordon equation for a single scalar field
\begin{eqnarray}
\ddot{\phi}+3H \dot{\phi}+ \frac{dV(\phi)}{d\phi}=0.
\label{eq:ddphi}
\end{eqnarray}
 Eq.(\ref{eq:Hub}) tells that at $\rho=\rho_c$, $H=0$ that means quantum bounce occurs at $\rho=\rho_c$. The numerical evolution of the background with the bouncing phase has been extensively studied in the literature. One of the main results is that a desired slow-roll inflation is achieved \cite{psingh2006,Mielczarek,zhang2007,chen2015,alam2017,Tao2017a,Tao2017b,ashtekar2011}. Keeping this in mind, we shall study ``bounce and slow-roll" with the generalized $\alpha-$attractor model (see Sec. \ref{sec:alpha}).

Let us first examine the evolution equations for a general potential $V(\phi)$. We  numerically solve  Eqs.(\ref{eq:Hub}) and (\ref{eq:ddphi}) with the initial values of $a(t)$, $\phi(t)$ and $\dot{\phi}(t)$ at a specific time. A natural option of the time is at the bounce $t=t_B$, for which we have
\begin{eqnarray}
\rho &=& \rho_c = \frac{1}{2}\dot{\phi}^2(t_B)+V(\phi(t_B)), \nonumber\\
\dot{a}(t_B)&=&0, 
\label{eq:bounce}
\end{eqnarray}
from which we find  
\begin{eqnarray}
\dot{\phi}(t_B) &=& \pm \sqrt{2 \Big{(} \rho_c - V(\phi(t_B)) \Big{)}}.
\label{eq:bounce2}
\end{eqnarray}
Without loss of the generality, we can always choose  
\begin{eqnarray}
a(t_B) &=& 1.
\label{eq:bounce3}
\end{eqnarray}
Hereafter, we shall read $\phi(t_B)$ and $\dot{\phi}(t_B)$ as $\phi_B$ and $\dot{\phi}_B$. From  Eq.(\ref{eq:bounce2}), one can clearly see that for a given potential, the initial values will be uniquely identified by $\phi_B$ only. Subsequently, we shall consider two cases: (a) PIV:~~ $\dot{\phi}_B > 0$ and (b) NIV: ~$\dot{\phi}_B < 0$.

Second, we introduce the following quantities that are essential for this paper \cite{alam2017,Tao2017a,Tao2017b}.

(1) The equation of state (EoS) $w(\phi)$ for the inflaton field is defined as
\begin{eqnarray}
w(\phi) = \frac{\dot{\phi}^2/2-V(\phi)}{\dot{\phi}^2/2+V(\phi)}.
\label{eq:w}
\end{eqnarray}
In the slow-roll inflationary phase, $w(\phi)\simeq-1$.

To differentiate the initial conditions for being dominated by the kinetic energy (KE) or potential energy (PE) at the bounce, we also introduce the quantity $w^B$, so that
\begin{equation}
w^B \equiv  w(\phi) \Big{\vert}_{\phi=\phi_B}
= \begin{cases}  > 0, \qquad \text{KE} > \text{PE} \\
  = 0, \qquad \text{KE}=\text{PE} \\
 < 0, \qquad \text{KE} < \text{PE} \end{cases}
\label{eq:wb}
\end{equation}

(2) The slow-roll parameter $\epsilon_H$, that is expressed in terms of the Hubble parameter  and its derivatives,
\begin{eqnarray}
\epsilon_H = - \frac{\dot{H}}{H^2}.
\label{eq:epsilon}
\end{eqnarray}
During the slow-roll inflation, $\epsilon_H \ll 1$.

(3) The number of $e$-folds $N_{inf}$ during the slow-roll inflation is given by
\begin{eqnarray}
N_{inf} = ln \Big{(} \frac{a_{end}}{a_i} \Big{)} =  \int_{t_i}^{t_{end}} H(t) dt \nonumber \\
 = \int_{\phi_i}^{\phi_{end}} \frac{H}{\dot{\phi}} d\phi \simeq \int_{\phi_{end}}^{\phi_i} \frac{V}{V_{\phi}} d\phi, 
\label{eq:Ninf}
\end{eqnarray}
where $a_i$ ($a_{end}$) represents the expansion factor when the inflation  starts  (ends),  i.e. $\ddot{a}(t_i) \gtrsim 0$ and  $w(\phi_{end})=-1/3$.

(4) Using  Eqs.(\ref{eq:Hub}) and (\ref{eq:ddphi}), we obtain an analytical expression  of the scale factor $a(t)$ during the bouncing phase. In this phase, if the potential is very small compared to the kinetic energy, 
 then Eqs.(\ref{eq:Hub}) and (\ref{eq:ddphi}) become
\begin{eqnarray}
&& H^2 = \frac{8 \pi}{3 m_{Pl}^2}~\frac{1}{2}\dot{\phi^2} \Big{(}1-\frac{\dot{\phi^2}}{2\rho_c}\Big{)},\nonumber\\
&& \ddot{\phi}+3H \dot{\phi}=0.
\label{eq:Hreduce}
\end{eqnarray}
We solve the above equations analytically, and find \cite{alam2017,Tao2017a,Tao2017b}
\begin{eqnarray}
\dot{\phi} &=& \pm \sqrt{2 \rho_c} \left( \frac{a_B}{a(t)} \right)^3,\nonumber \\
a(t) &=& a_B \left( 1+ \delta \frac{t^2}{t_{Pl}^2} \right)^{1/6},
\label{eq:a}
\end{eqnarray}
where $t_{Pl}$  denotes the Planck time, and $\delta = {24 \pi \rho_c}/{m_{Pl}^{4}}$ is a dimensionless parameter.

(5) We define a quantity $r_w$, that is the ratio between the kinetic and potential energies, 
\begin{eqnarray}
r_{w} &\equiv & \frac{\frac{1}{2}\dot{\phi}^2}{V(\phi)}.
\label{eq:rw}
\end{eqnarray}
Following  Eq.(\ref{eq:rw}), one can define $r_{w}^c$ that corresponds to   $N_{inf} \simeq 60$ during the slow-roll inflation,
\begin{equation}
r_{w}^c \equiv  \frac{\frac{1}{2}\dot{\phi}^2}{V(\phi)} \Big{\vert}_{N_{inf}\simeq 60}
= \begin{cases} r_{w}^c > r_{w} \qquad N_{inf} < 60 \\
 r_{w}^c = r_{w} \qquad N_{inf}\simeq 60 \\
 r_{w}^c < r_{w} \qquad N_{inf} > 60. \end{cases}
\label{eq:rwc}
\end{equation}
In the following section, we shall discuss the generalized $\alpha-$attractor model in the context of PIV and NIV at the quantum bounce.

\section{$\alpha-$attractor model}
\label{sec:alpha}

In this section, we shall study ``bounce and slow-roll" with the generalized  $\alpha-$attractor model in the framework of LQC. Let us consider the following form of the potential \cite{alam2018,linder15}:  
\begin{eqnarray}
V(\phi) &=& \alpha c^2  \frac{\left[\tanh \left( \frac{\phi }{\sqrt{6\alpha }}\right)\right] ^2}{\left[ 1+\tanh \left( \frac{\phi }{\sqrt{6\alpha }}\right) \right]^{2n}},
\label{eq:potGen}
\end{eqnarray}
where the parameters $\alpha$ and $c$ have the dimensions of $M_{Pl}^2$ and $M_{Pl}$, and $M_{Pl}=m_{Pl}/\sqrt{8 \pi}$ is the reduced Planck mass. The parameter $n$ takes the values $n = 0,1,2,3...$ For large field values ($\phi \rightarrow \infty$), the generalized $\alpha-$attractor potential becomes flatten, and for small field values ($\phi \rightarrow 0$), it behaves as a quadratic one. For different values of $n$,  Eq.(\ref{eq:potGen}) gives the following forms of the potentials. \\

For $n=0$, we have 
\begin{eqnarray}
V(\phi) &=& \alpha c^2  \left[ \tanh \left( \frac{\phi }{\sqrt{6\alpha }}\right) \right]^2.
\label{eq:Tpot}
\end{eqnarray}
In the literature,  Eq.(\ref{eq:Tpot}) is known as $T-model$ \cite{alpha,alpha2,alpha3}, and also represents GL model for $\alpha=1/9$ \cite{GL}. We find values of $\alpha$ and $c$ that are consistent with the Planck 2015 results for inflationary universe \cite{Planck2015}. Here, we write only those values which shall be used in figures and tables. However, one can also obtain other combinations (see Appendix A)
\begin{eqnarray}
\alpha &=& 10 m_{Pl}^2, \qquad~~~ c = 1.8 \times 10^{-5} m_{Pl}\nonumber \\
\alpha &=& 10^{10} m_{Pl}^2, \qquad c = 8.2 \times 10^{-6} m_{Pl}.
\label{eq:Talphac}
\end{eqnarray}

For $n=1$, we have 
\begin{eqnarray}
V(\phi) &=& \frac{\alpha c^2}{4}  \left(1-e^{-\sqrt{\frac{2}{3\alpha}}\phi} \right)^2.
\label{eq:Epot}
\end{eqnarray}
This is called $E-model$ (generalization of the Starobinsky model) \cite{alpha1}.  Eq.(\ref{eq:Epot}) corresponds to the Starobinsky model when $\alpha=1$ \cite{staro1980}. For the numerical evolution, combinations of $\alpha$ and $c$ that are in agreement with the Planck data \cite{Planck2015}, are given as (see Appendix A)
\begin{eqnarray}
\alpha &=& 0.1 m_{Pl}^2, \qquad c = 3.3 \times 10^{-4} m_{Pl},\nonumber \\
\alpha &=& 5 m_{Pl}^2, \qquad~~ c = 4.9 \times 10^{-5} m_{Pl}.
\label{eq:Ealphac}
\end{eqnarray}

For $n=2$, we have
\begin{eqnarray}
V(\phi) &=& \alpha c^2  \frac{\left[\tanh \left( \frac{\phi }{\sqrt{6\alpha }}\right)\right]^2}{\left[ 1+\tanh \left( \frac{\phi }{\sqrt{6\alpha }}\right) \right]^{4}}.
\label{eq:n2pot}
\end{eqnarray}
Values of $\alpha$ and $c$ that are compatible with the Planck data \cite{Planck2015} are given by (see Appendix A)
\begin{eqnarray}
\alpha &=& 0.5 m_{Pl}^2, \qquad c = 2.9 \times 10^{-4} m_{Pl},\nonumber \\
\alpha &=& 5 m_{Pl}^2, \qquad~~ c = 9.4 \times 10^{-5} m_{Pl}.
\label{eq:n2alphac}
\end{eqnarray}
Here, we are interested in the dynamics of the inflaton field having potentials (\ref{eq:Tpot}), (\ref{eq:Epot}) and (\ref{eq:n2pot}). Numerically, we shall solve Eqs.(\ref{eq:Hub}), (\ref{eq:ddphi}) with (\ref{eq:Tpot}), 
(\ref{eq:Epot}) and (\ref{eq:n2pot}), and examine whether following the quantum bounce, a desired slow-roll inflation exists or not.

Before proceeding, let us first consider the inflationary potentials (\ref{eq:Tpot}), (\ref{eq:Epot}) and (\ref{eq:n2pot}) that are shown in Fig. \ref{fig:pot}. The predictions of  Eq.(\ref{eq:Tpot}) with (\ref{eq:Epot}) and (\ref{eq:n2pot}) are similar but not identical as the main difference is in the shape of the potentials. The potential (\ref{eq:Tpot}) is symmetric about $\phi=0$, whereas potentials (\ref{eq:Epot}) and (\ref{eq:n2pot}) are not symmetric in nature.

Second, we present the range of $\alpha$ (depending on $\phi_B$) having inflationary/non-inflationary phase for the models under consideration in Table \ref{tab:n012_alpha}. Following this, we choose some values of $\alpha$ in each case and use them to draw the figures with different values of $\phi_B$, and its corresponding range for inflationary/non-inflationary phase with PIV and NIV are shown in Table \ref{tab:n012_alpha_dphi}.

\begin{figure*}[tbp]
\begin{center}
\begin{tabular}{ccc}
{\includegraphics[width=2.1in,height=1.65in,angle=0]{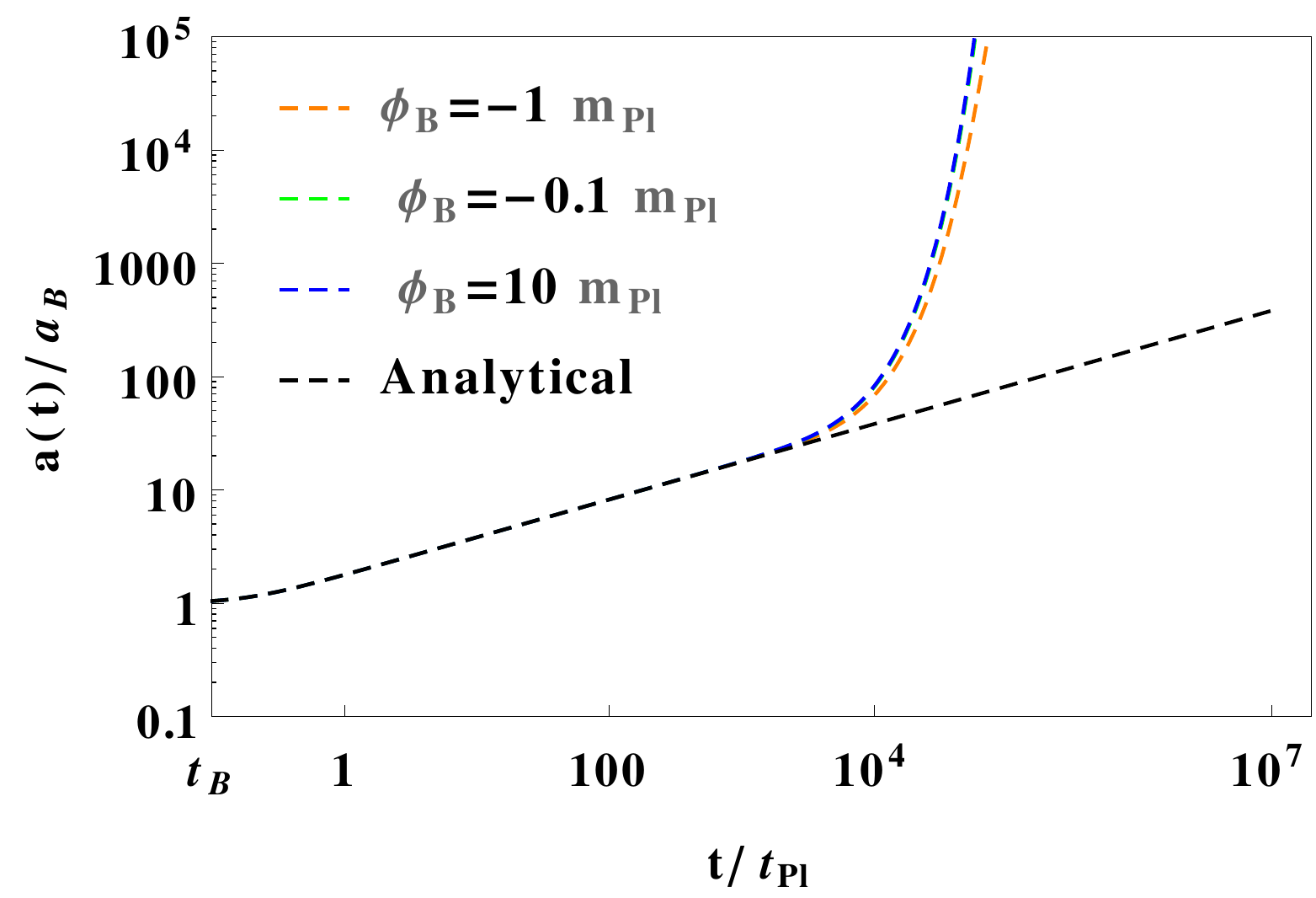}} &
{\includegraphics[width=2.1in,height=1.6in,angle=0]{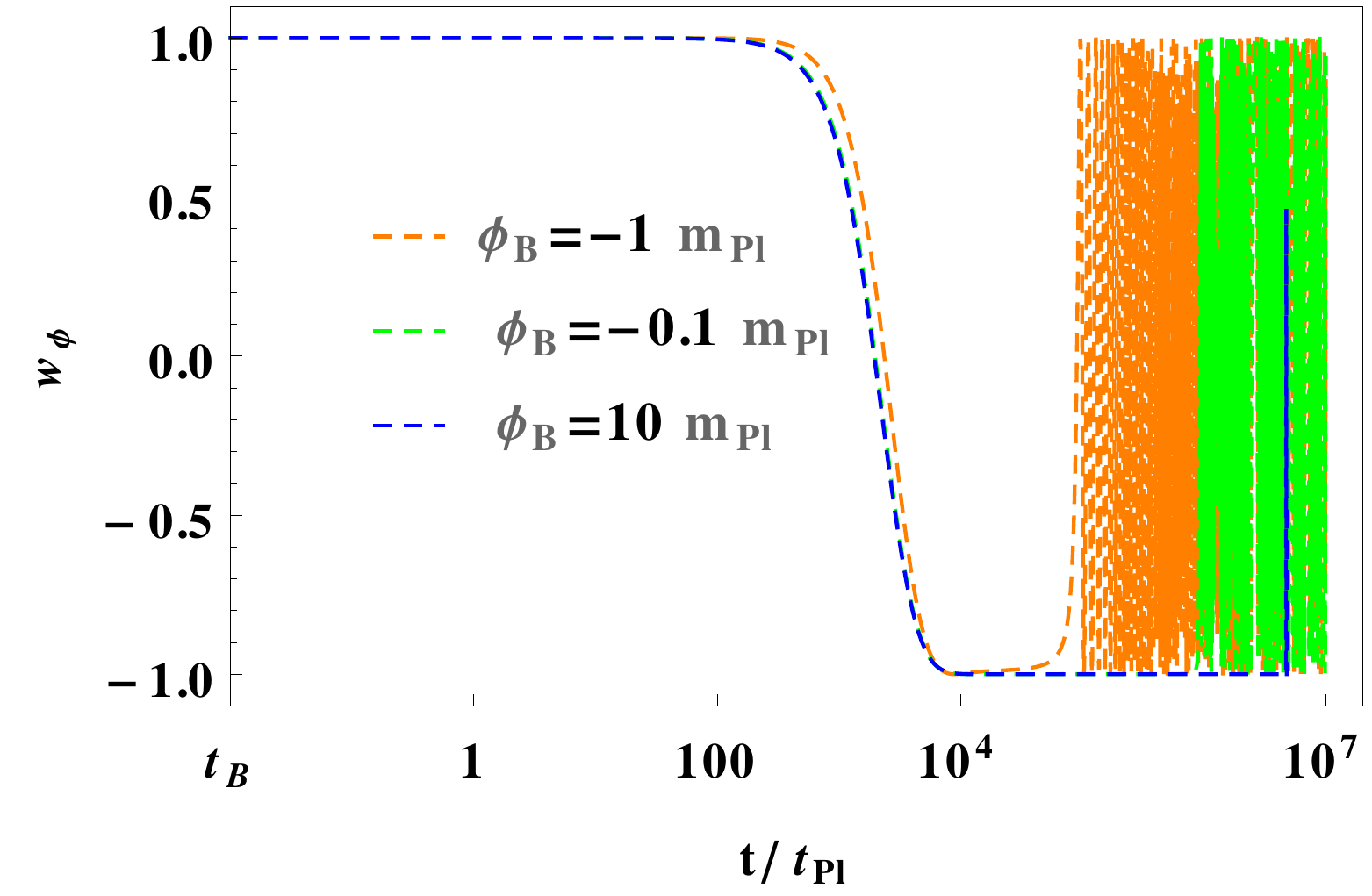}} &
{\includegraphics[width=2.0in,height=1.6in,angle=0]{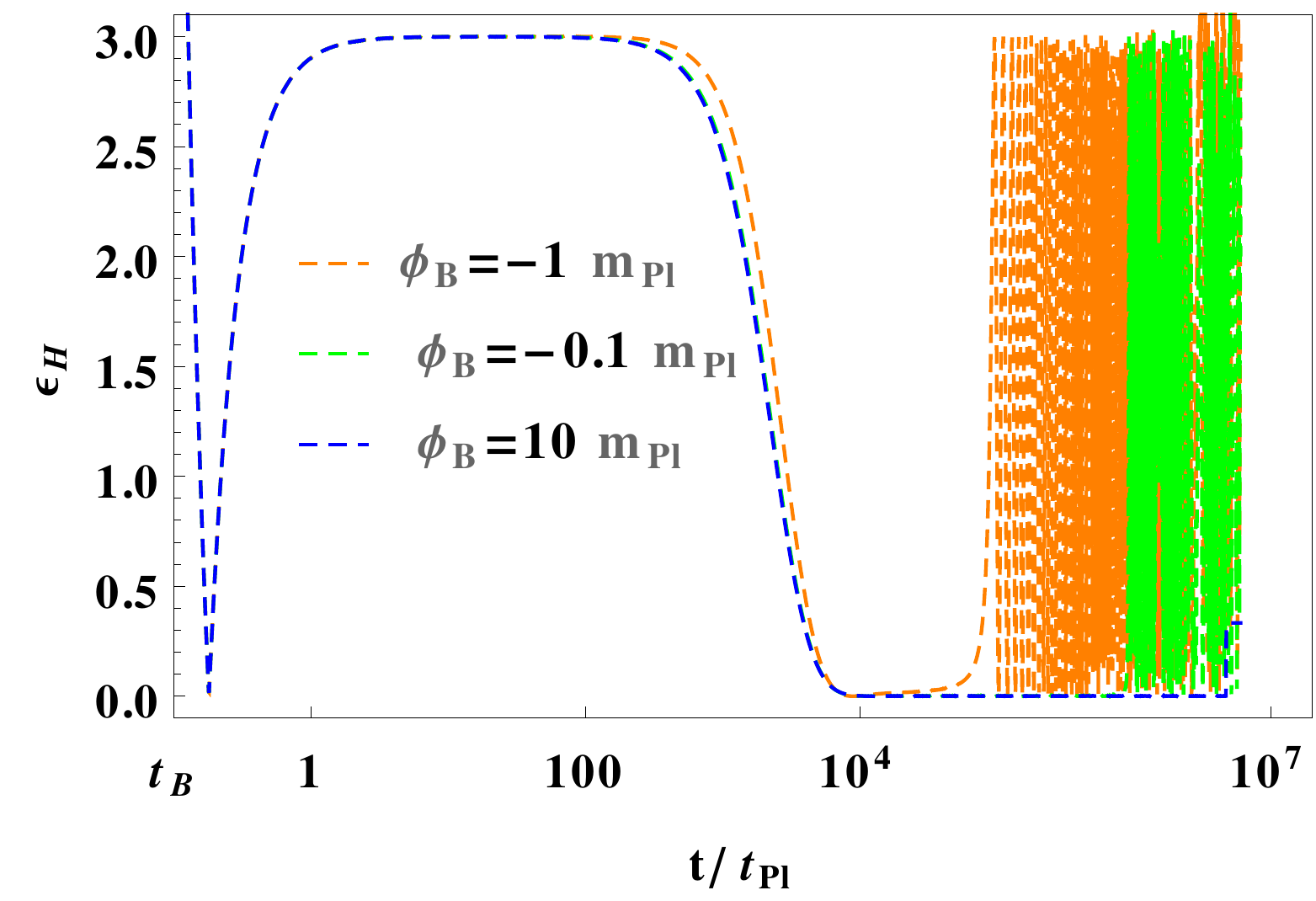}}
\\
{\includegraphics[width=2.1in,height=1.65in,angle=0]{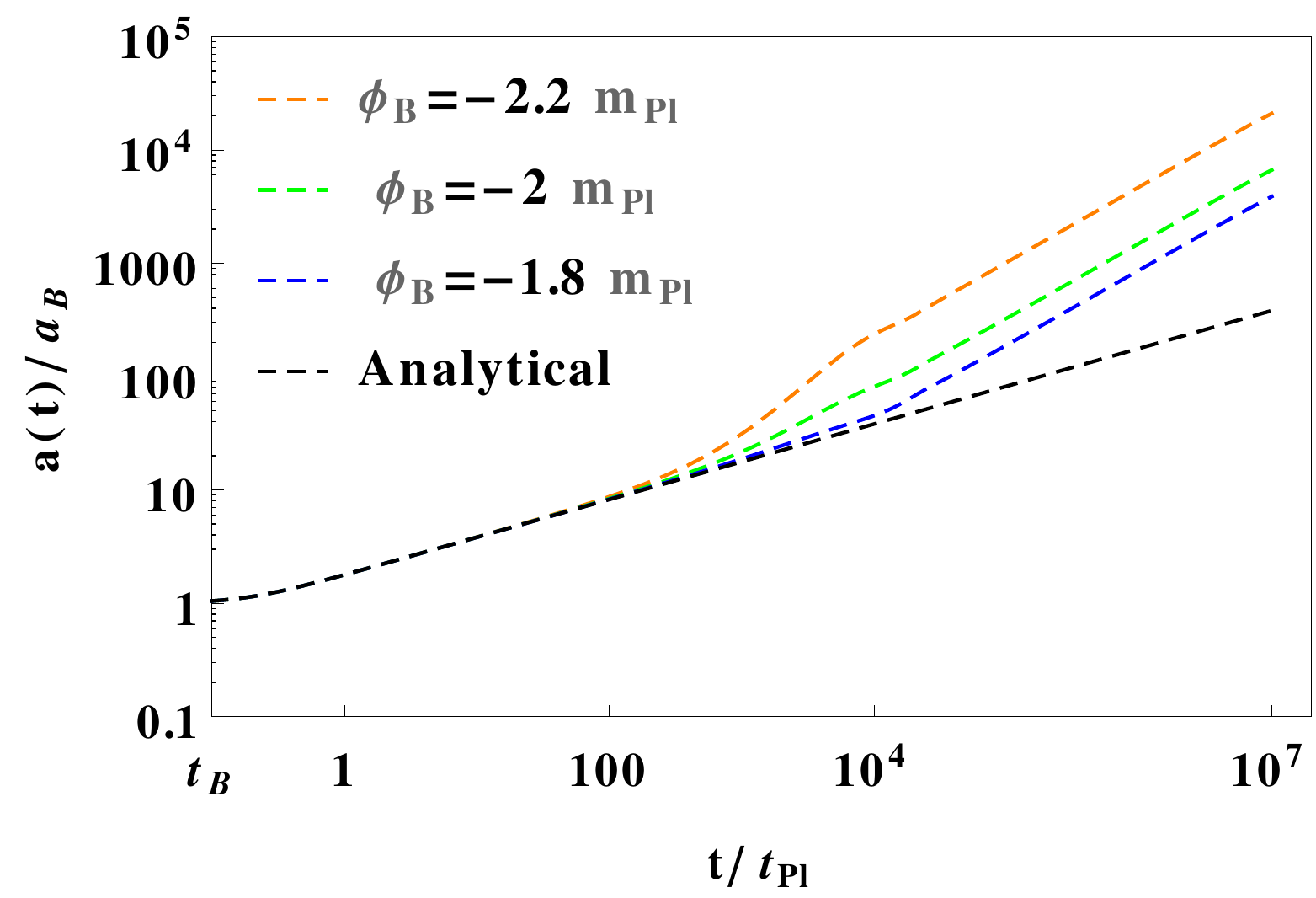}} &
{\includegraphics[width=2.1in,height=1.6in,angle=0]{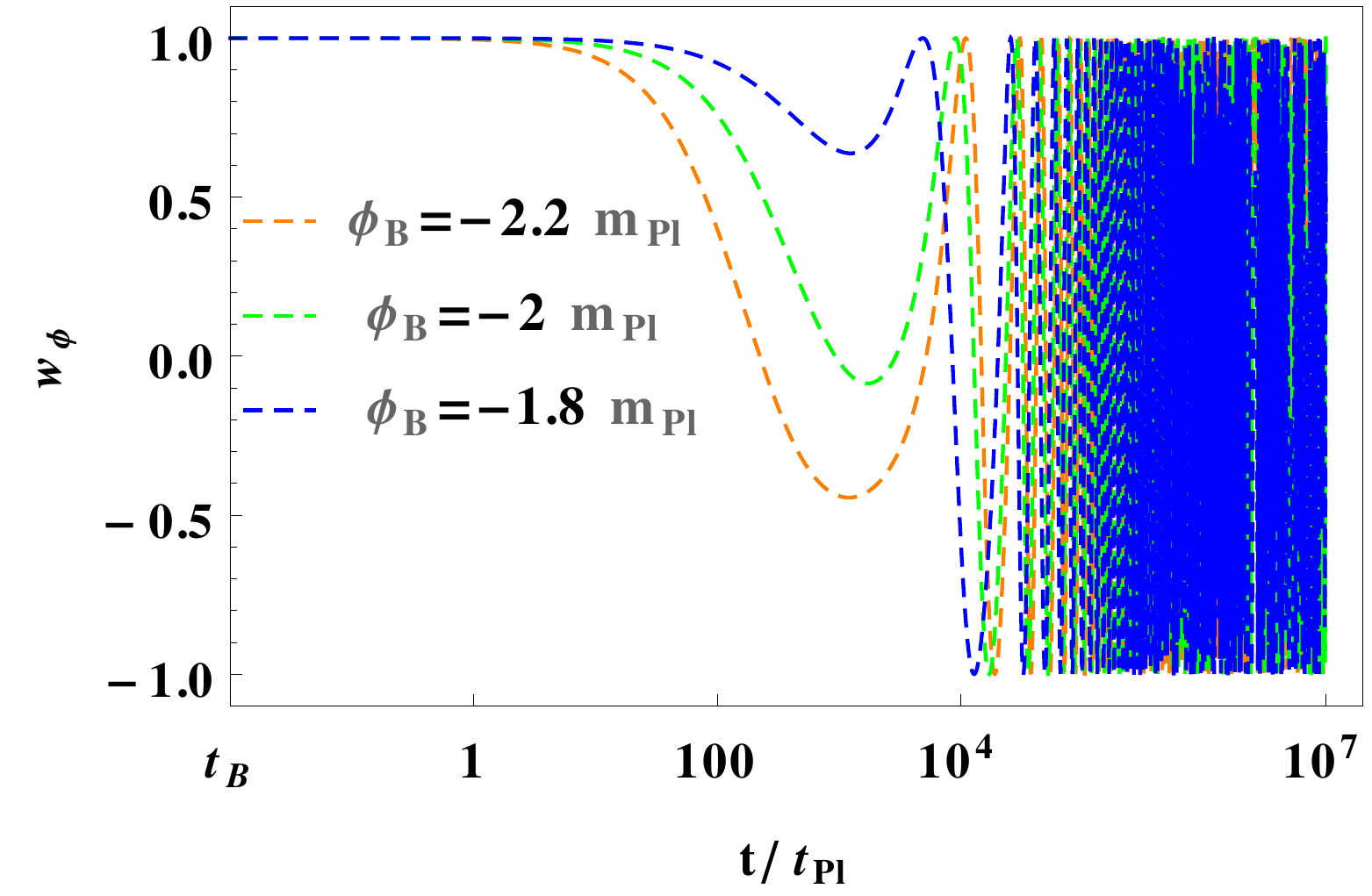}} &
{\includegraphics[width=2.0in,height=1.6in,angle=0]{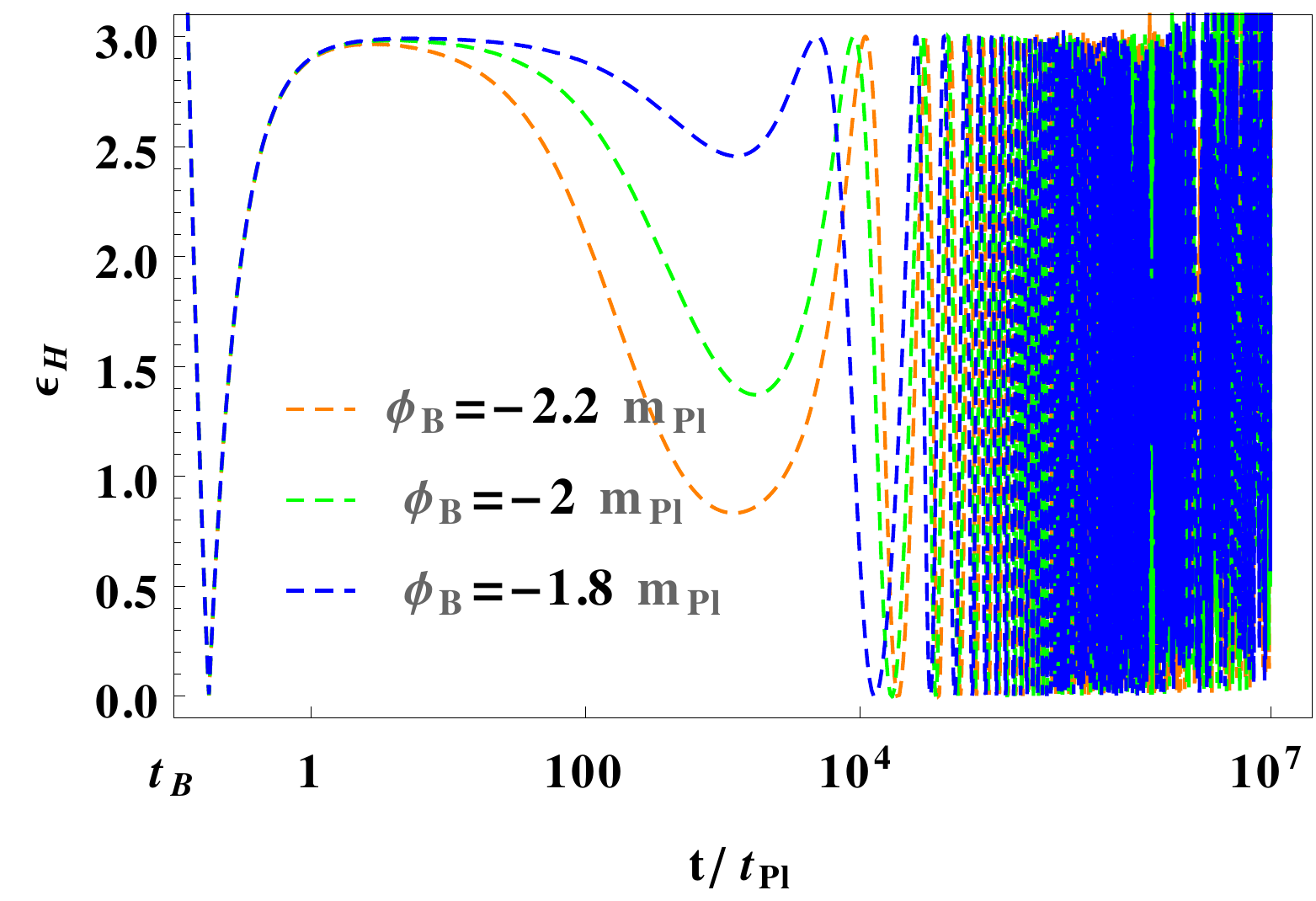}}
\\
{\includegraphics[width=2.1in,height=1.6in,angle=0]{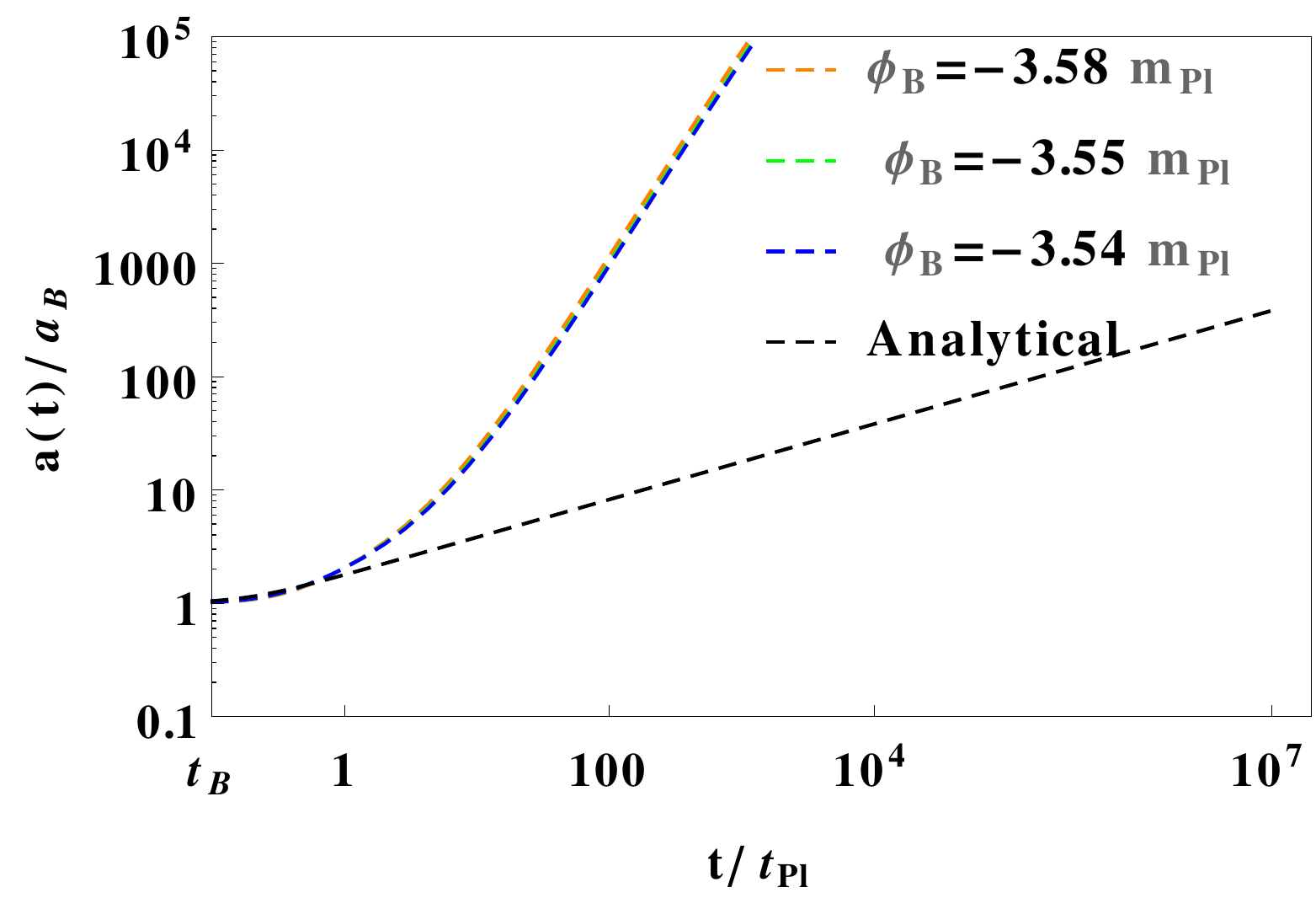}} & 
{\includegraphics[width=2.1in,height=1.6in,angle=0]{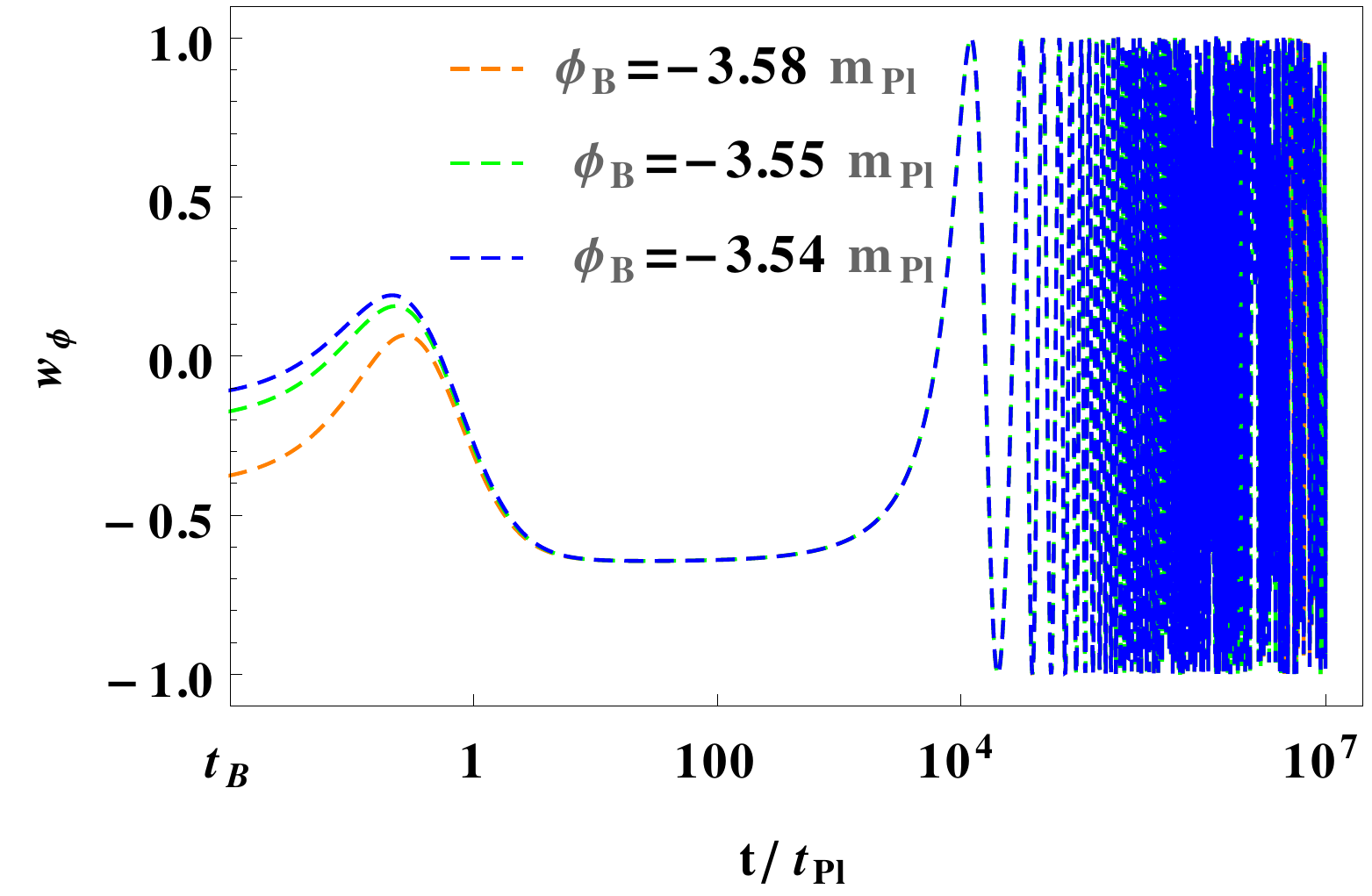}} & 
{\includegraphics[width=2.0in,height=1.6in,angle=0]{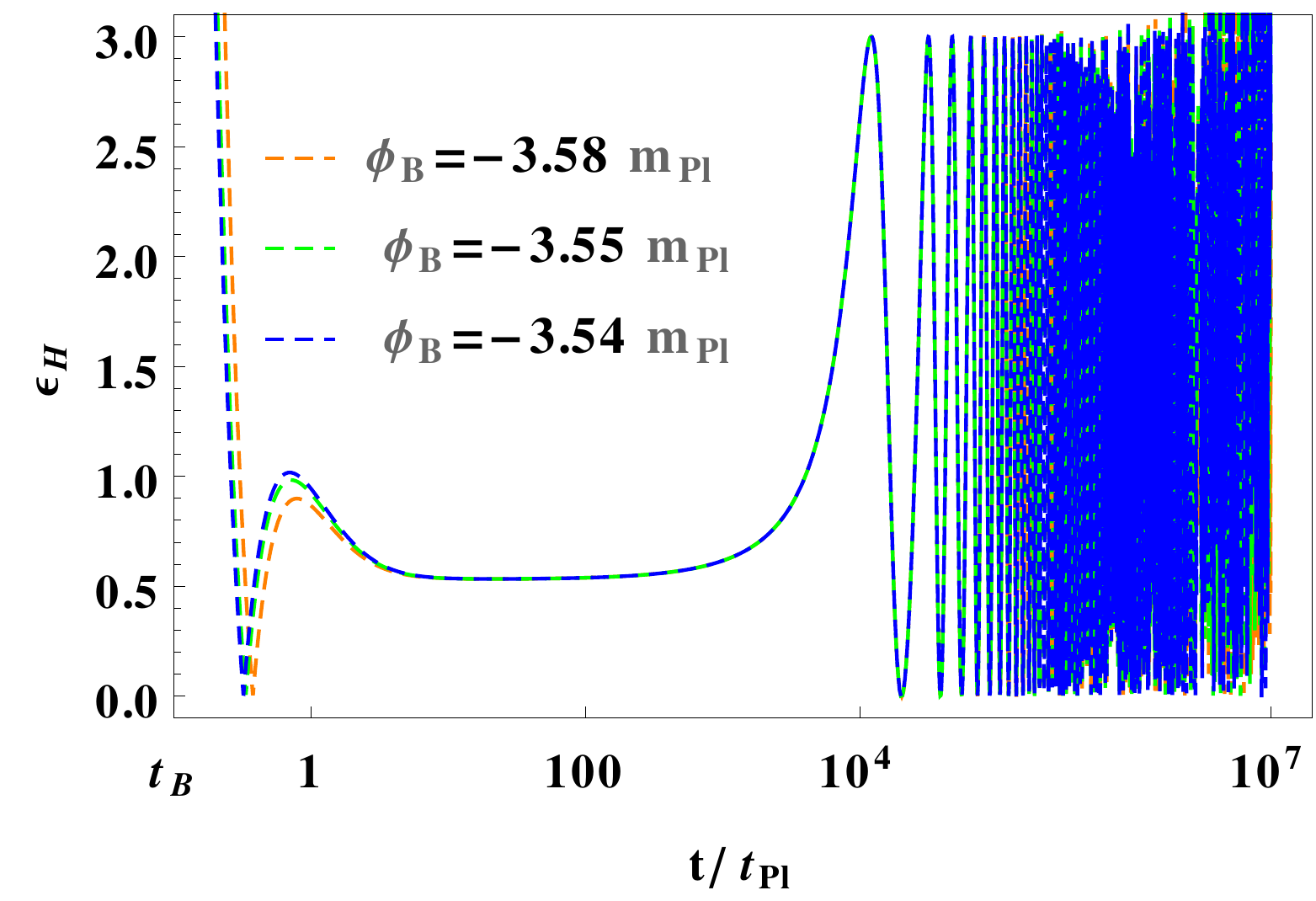}} 
\end{tabular}
\end{center}
\caption{ This figure exhibits the numerical evolution of $a(t)$, $w(\phi)$ and $\epsilon_H$ for $E-model$ (\ref{eq:Epot}) with $\dot{\phi}_B>0$. Top and middle panels are plotted for KED initial conditions whereas bottom ones are for PED initial values. Here, only KED (except a small subset) initial values lead to the slow-roll inflationary phase,  while the PED and a subset of KED initial conditions do not provide a slow-roll inflation phase. We choose $\alpha=0.1 m_{Pl}^2$, $c=3.3 \times 10^{-4}m_{Pl}$, and $m_{Pl}=1$. }
\label{fig:n1alpha01_dphp}
\end{figure*}
\begin{figure*}[tbp]
\begin{center}
\begin{tabular}{ccc}
{\includegraphics[width=2.1in,height=1.65in,angle=0]{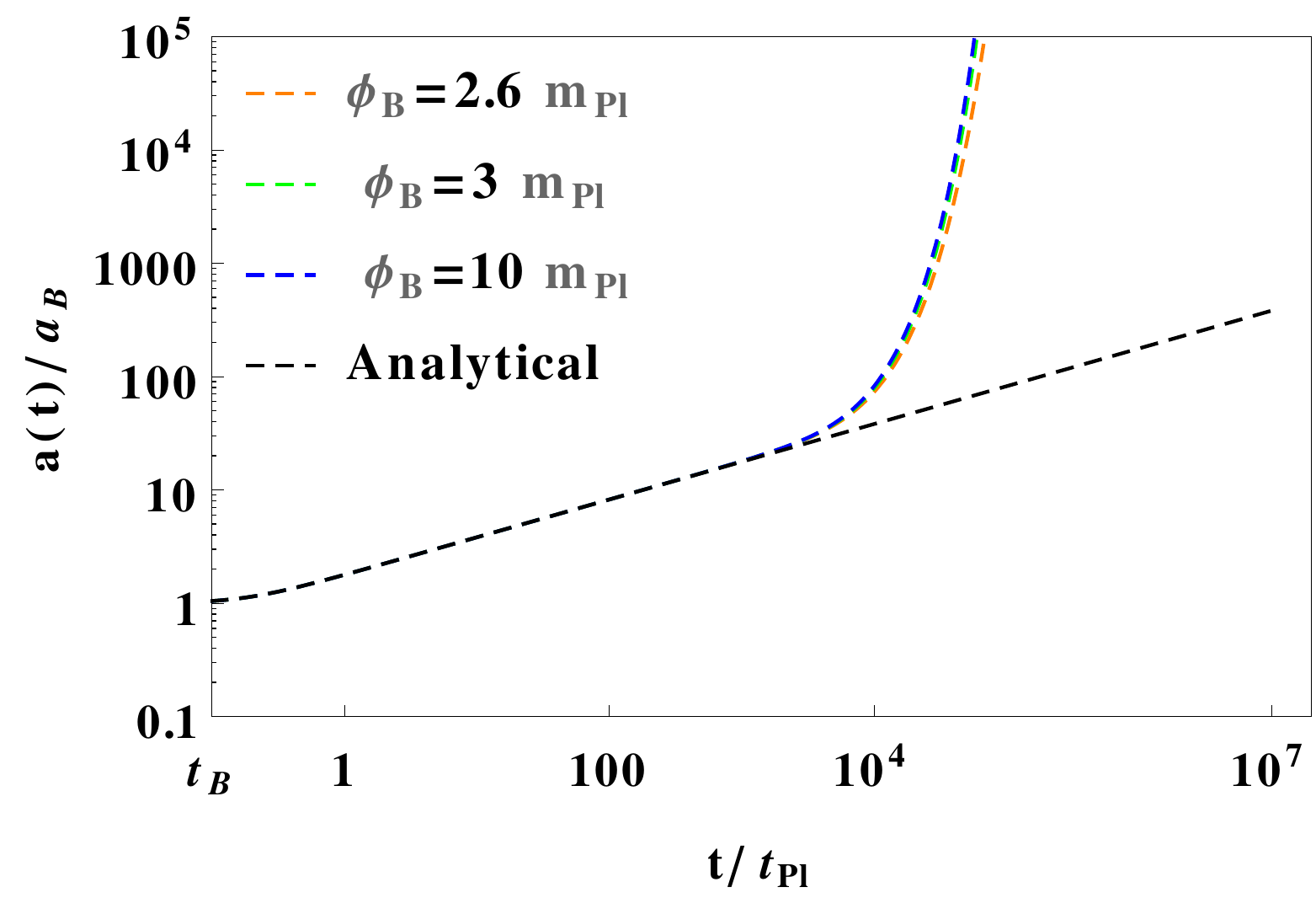}} &
{\includegraphics[width=2.1in,height=1.6in,angle=0]{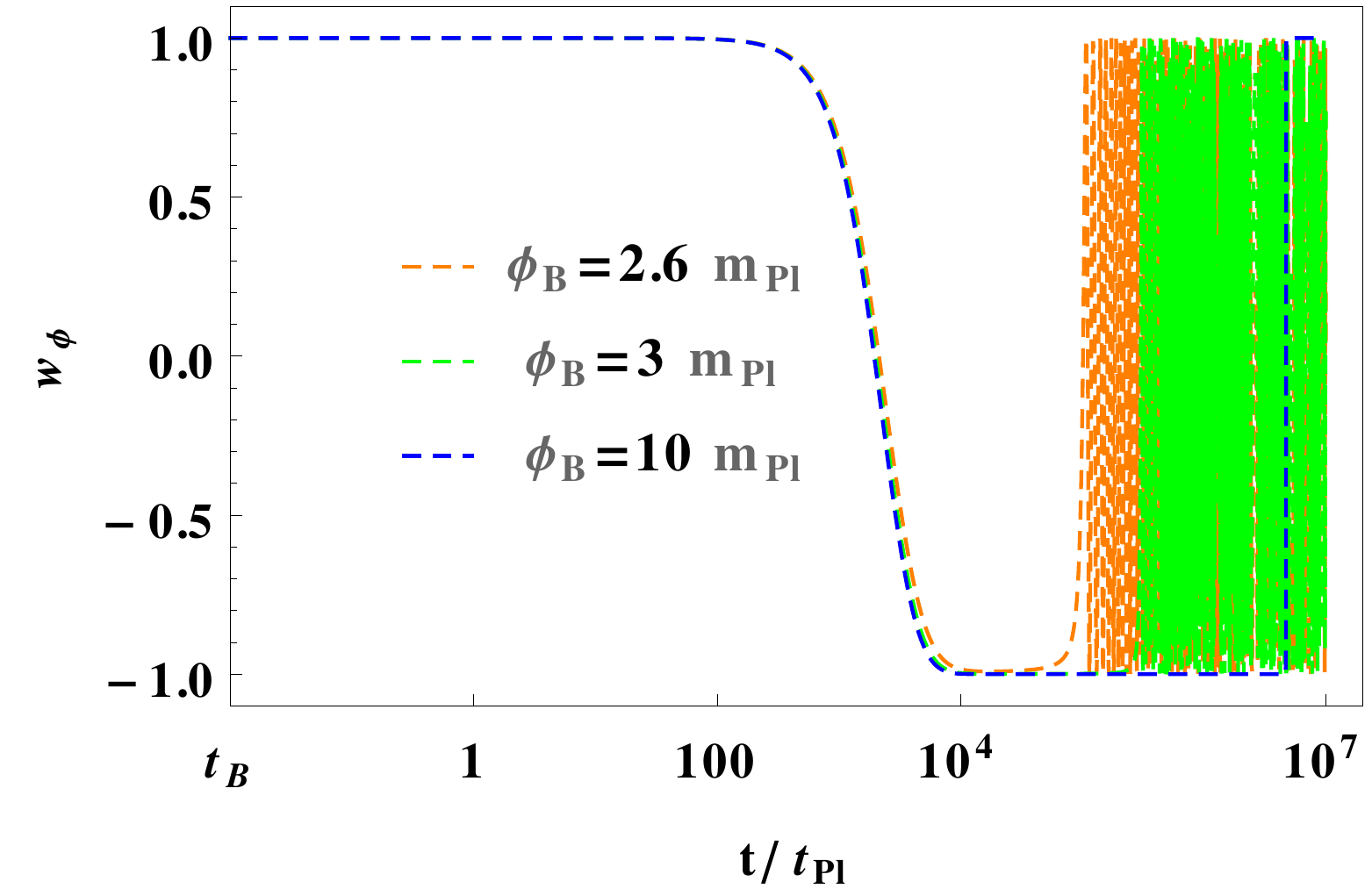}} &
{\includegraphics[width=2.0in,height=1.6in,angle=0]{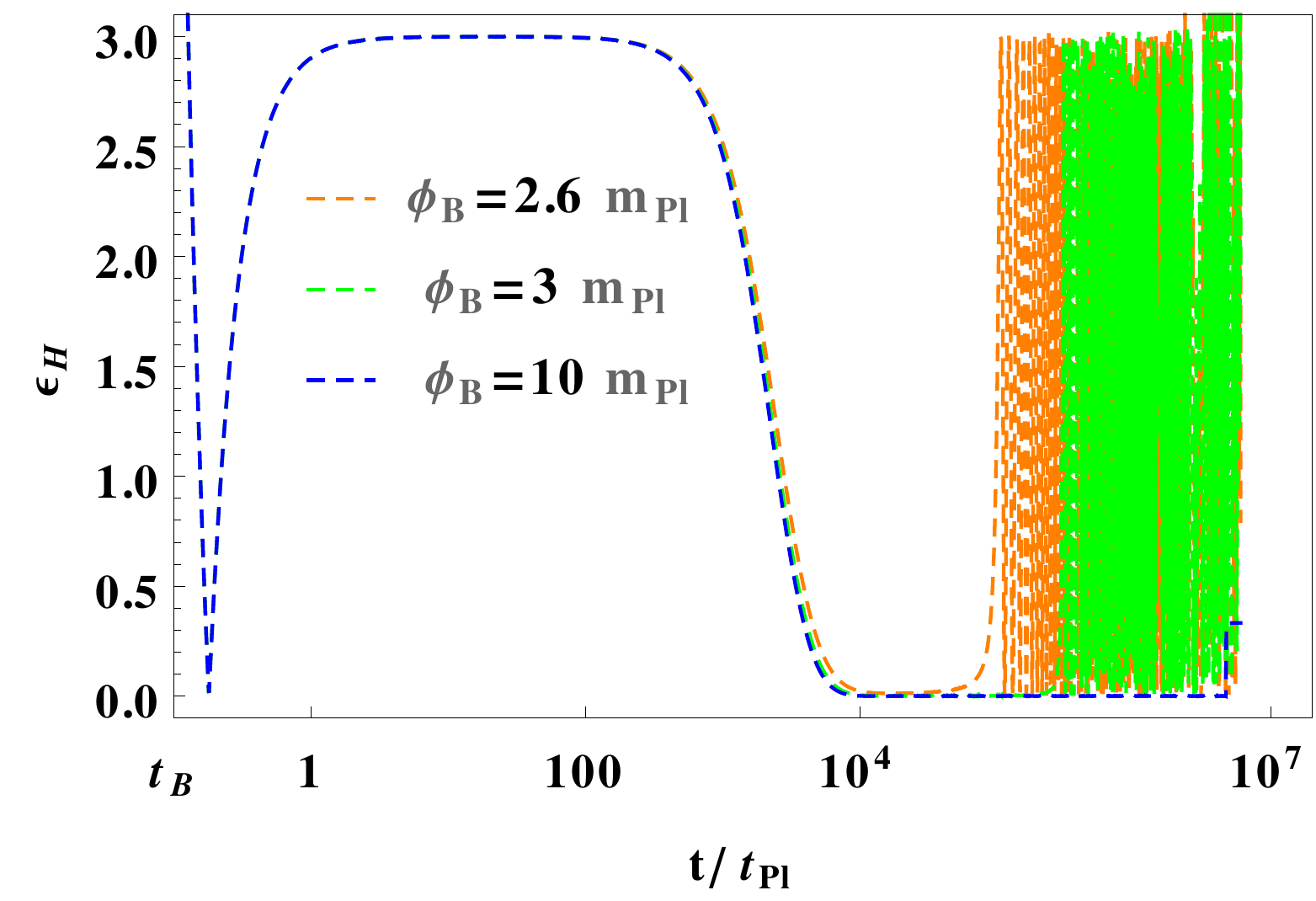}}
\\
{\includegraphics[width=2.1in,height=1.65in,angle=0]{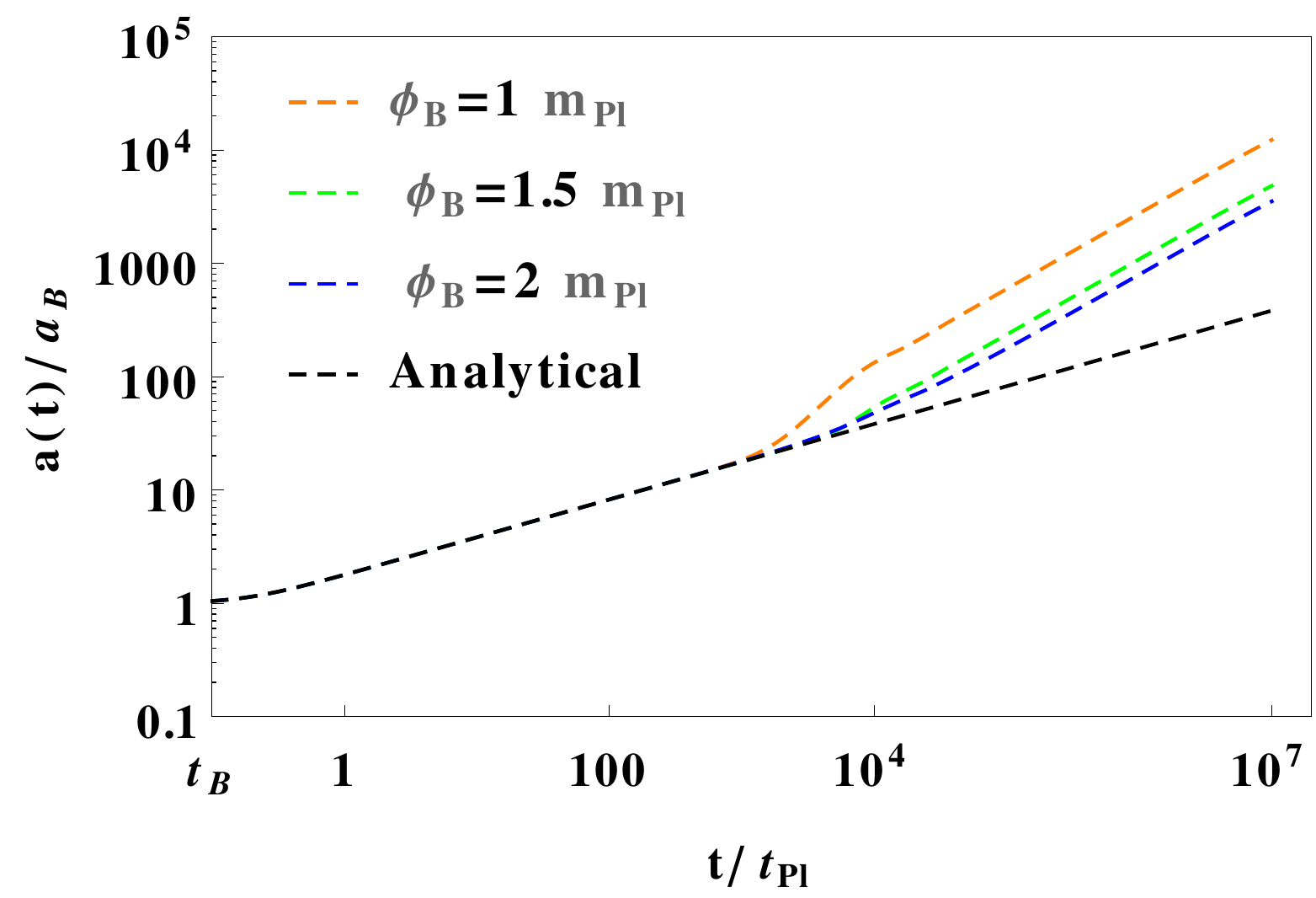}} &
{\includegraphics[width=2.1in,height=1.6in,angle=0]{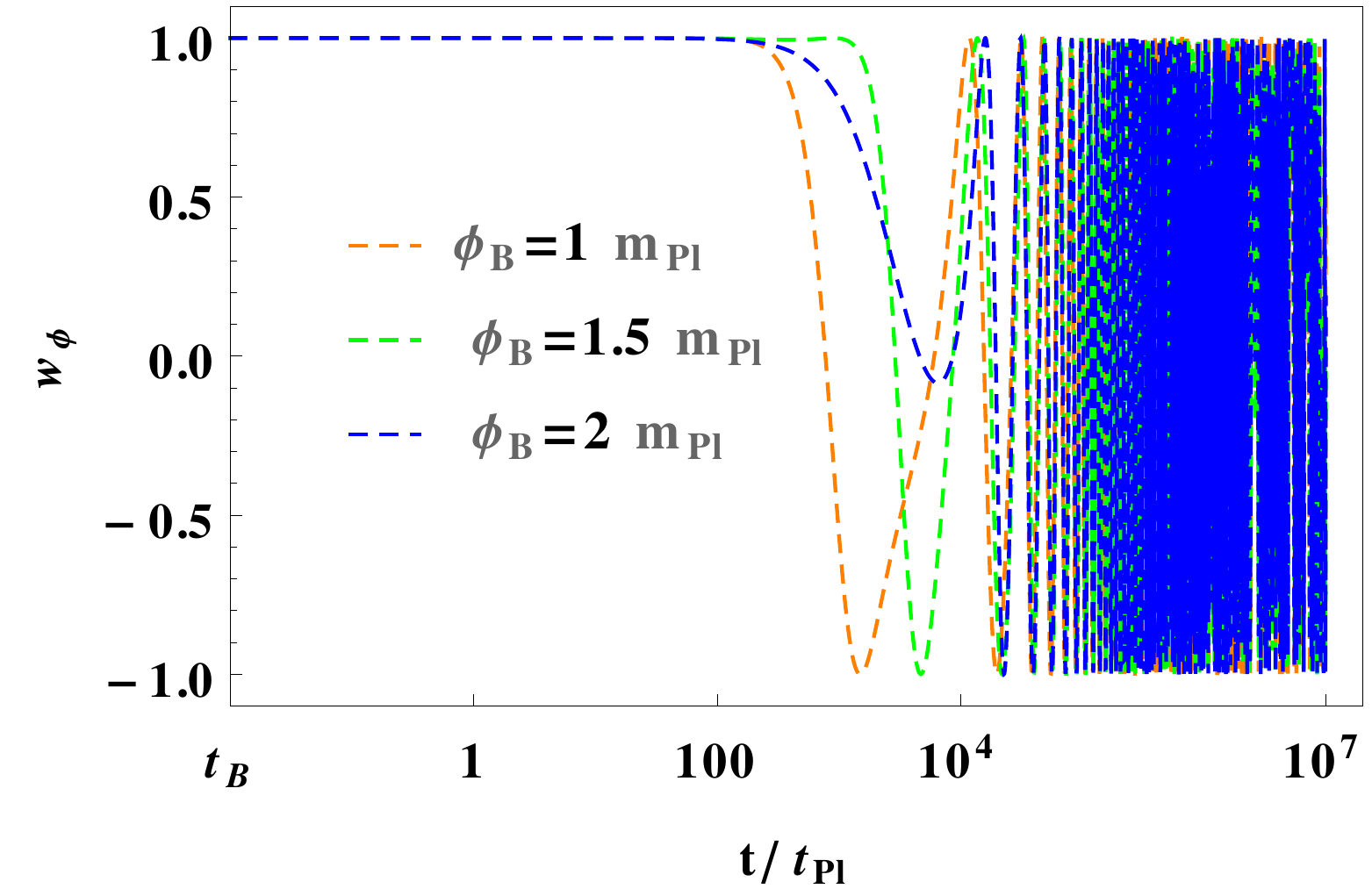}} &
{\includegraphics[width=2.0in,height=1.6in,angle=0]{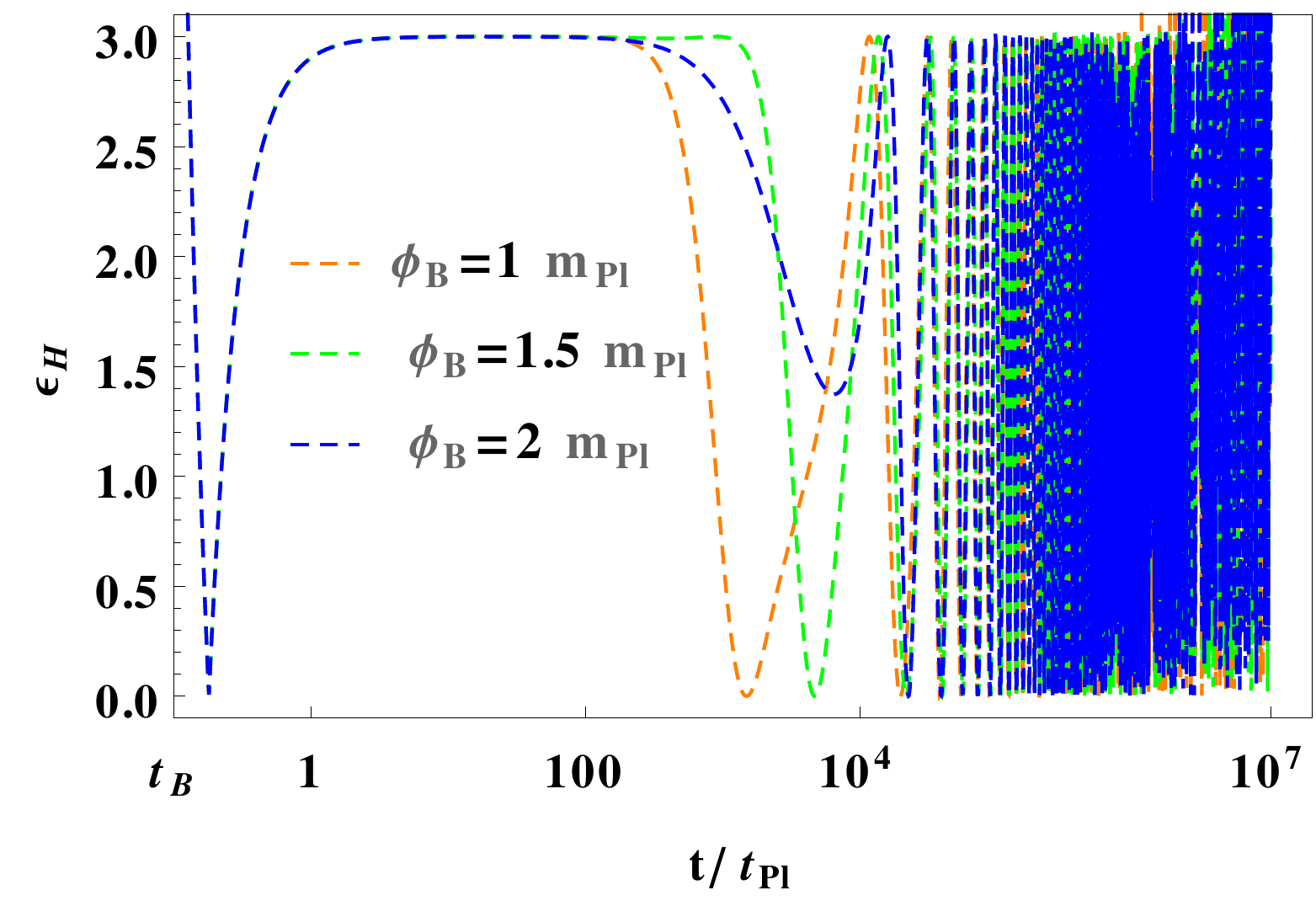}}
\\
{\includegraphics[width=2.1in,height=1.6in,angle=0]{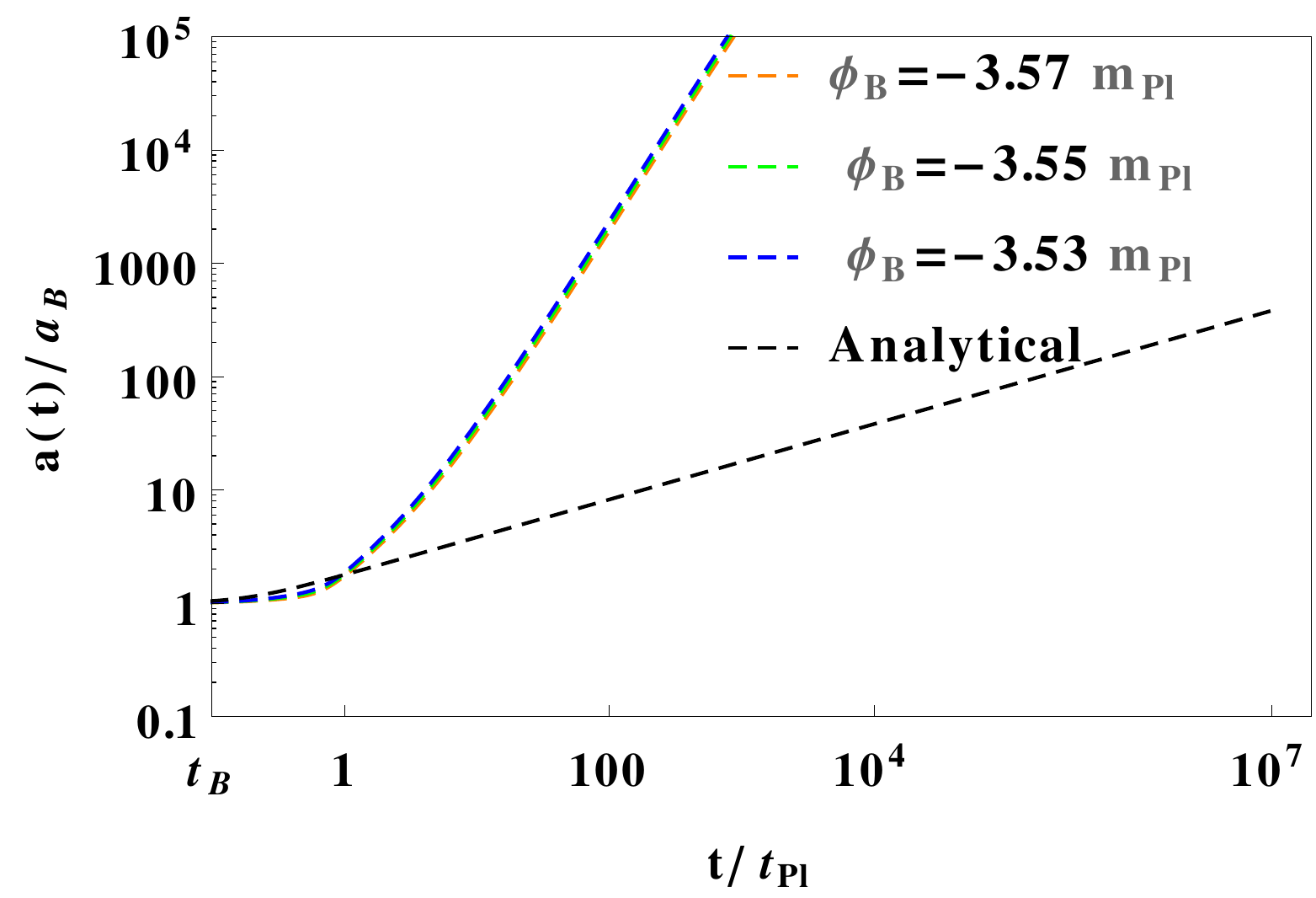}} & 
{\includegraphics[width=2.1in,height=1.6in,angle=0]{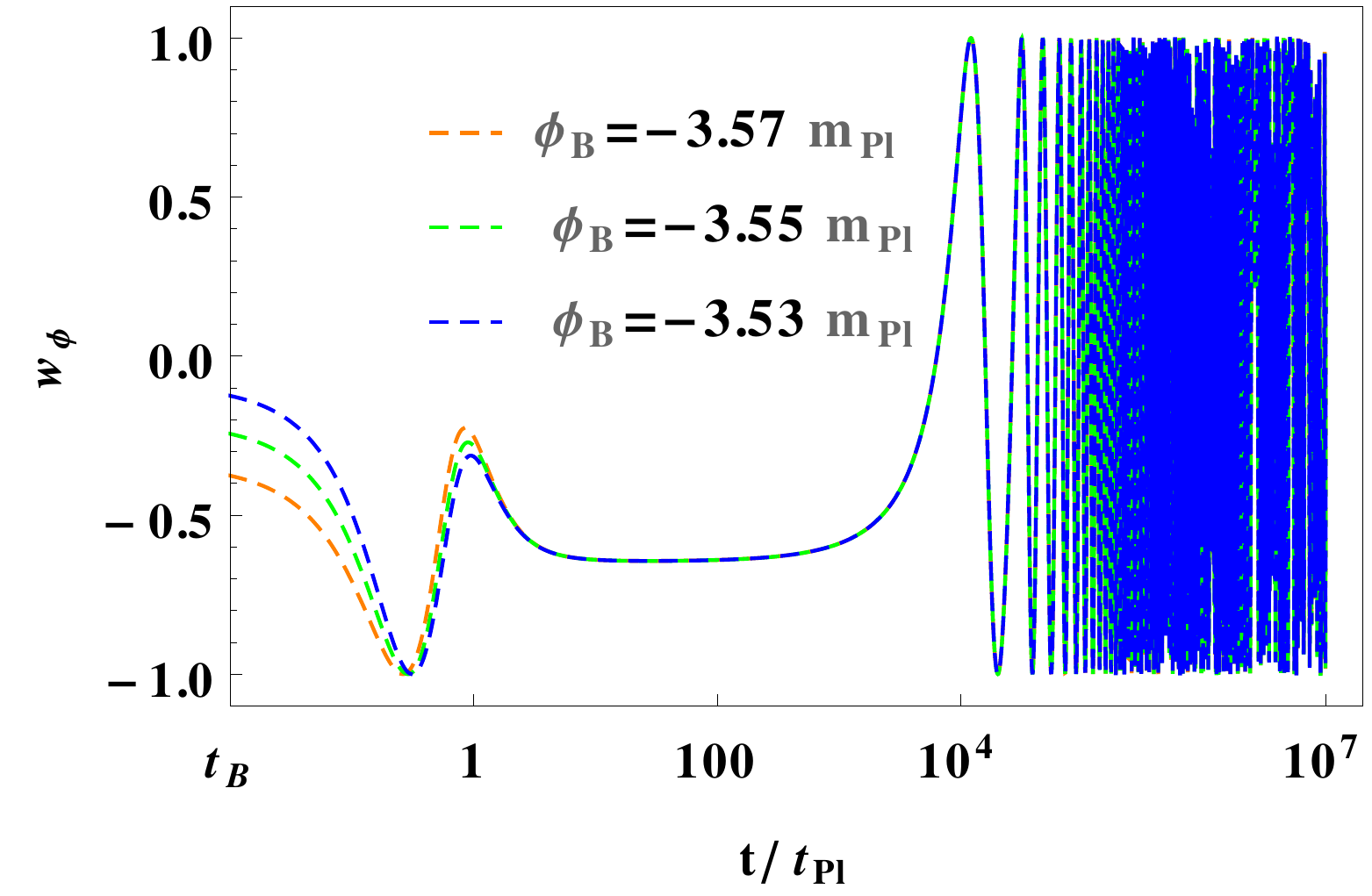}} & 
{\includegraphics[width=2.0in,height=1.6in,angle=0]{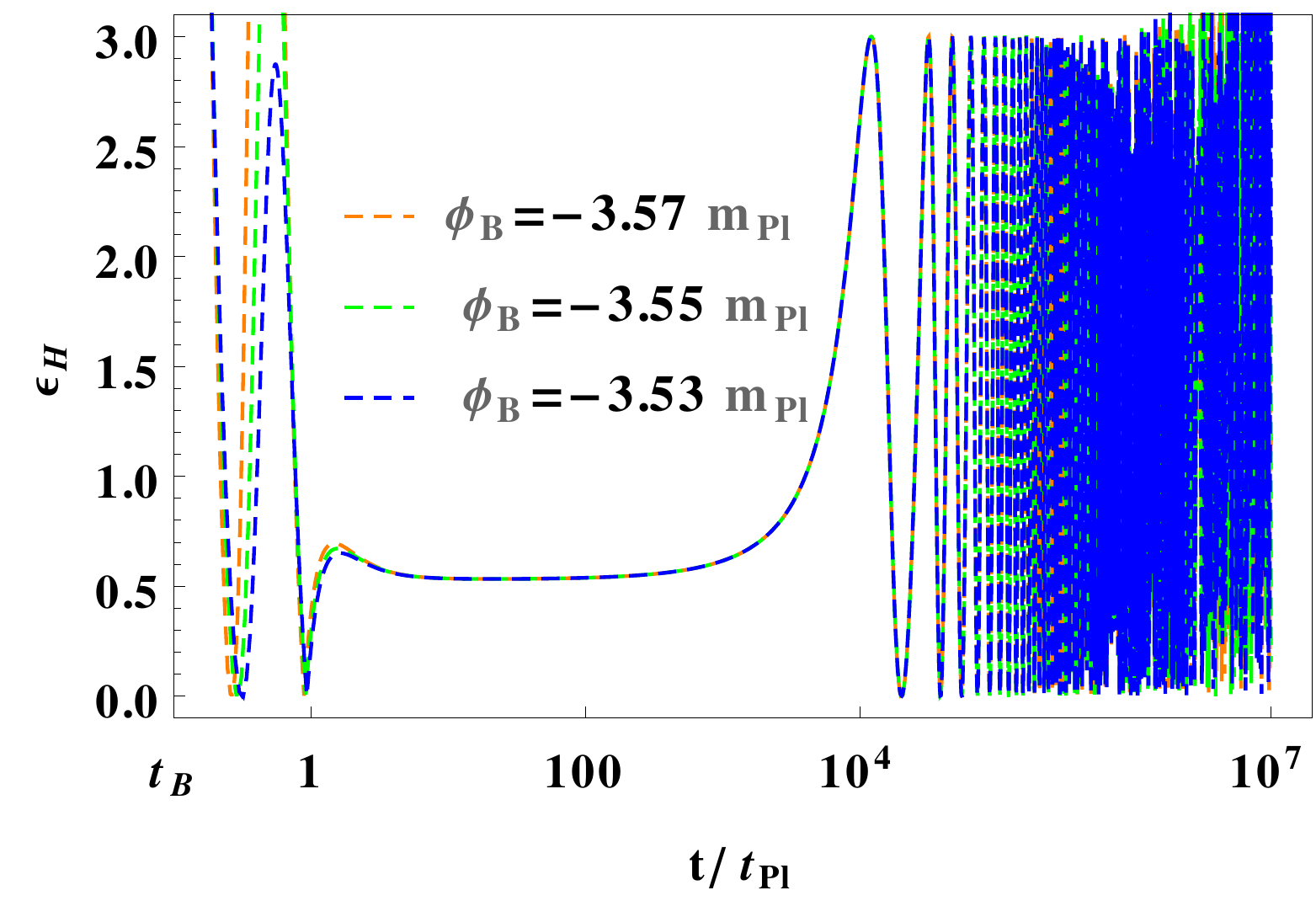}} 
\end{tabular}
\end{center}
\caption{ This figure is similar to Fig. \ref{fig:n1alpha01_dphp} but with $\dot{\phi}_B<0$.}
\label{fig:n1alpha01_dphn}
\end{figure*}
\begin{figure*}[tbp]
\begin{center}
\begin{tabular}{ccc}
{\includegraphics[width=2.1in,height=1.65in,angle=0]{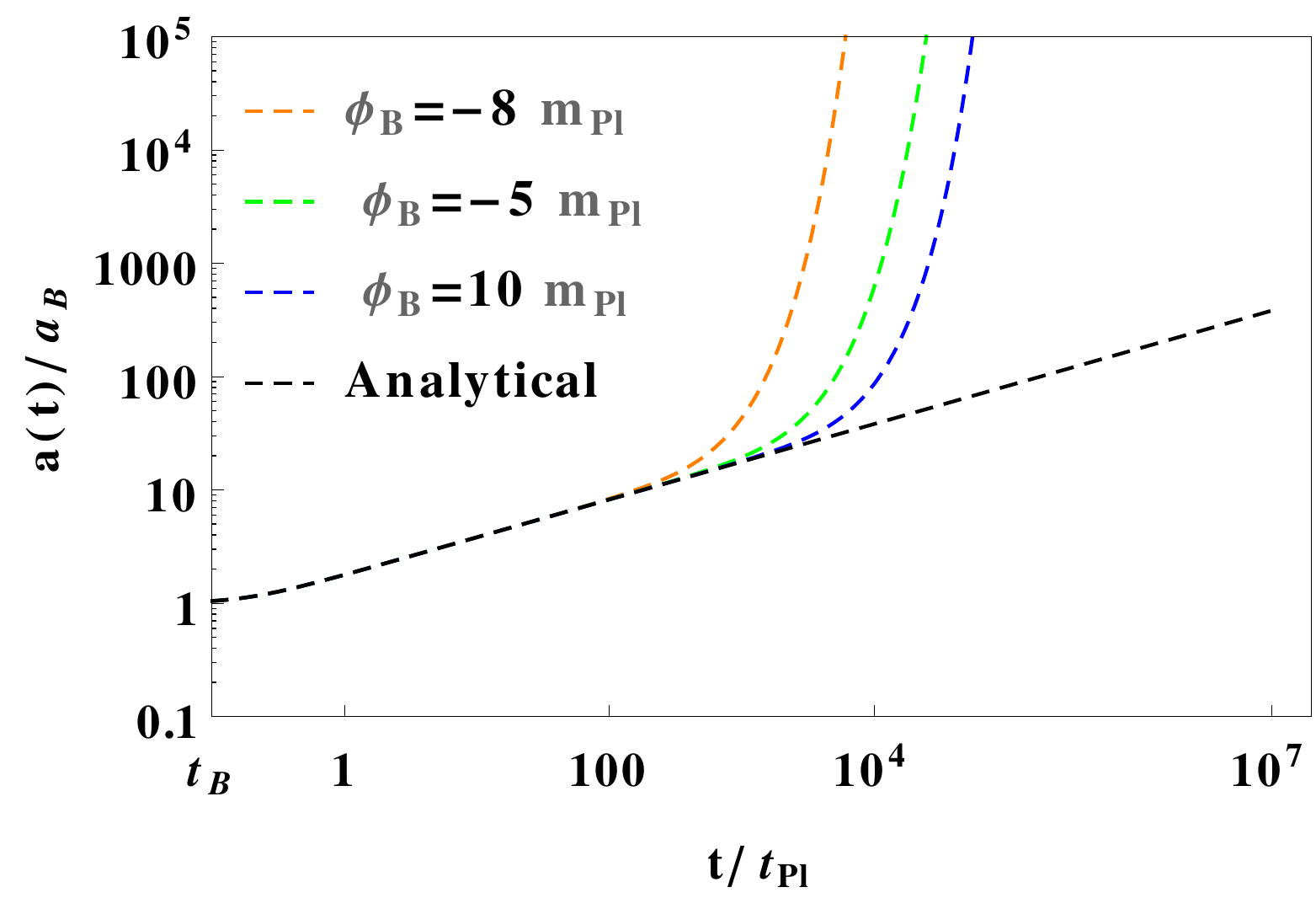}} &
{\includegraphics[width=2.1in,height=1.6in,angle=0]{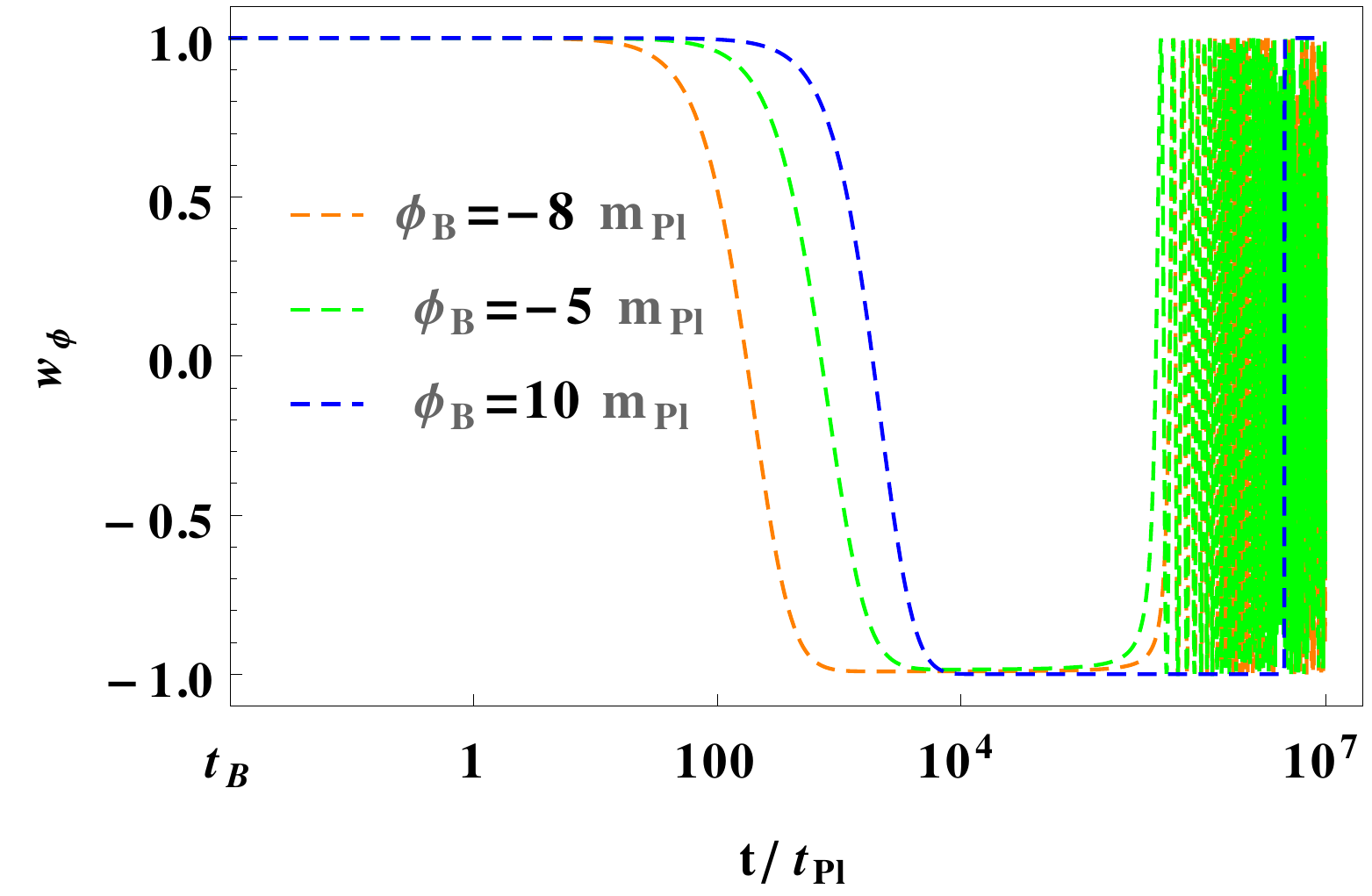}} &
{\includegraphics[width=2.0in,height=1.6in,angle=0]{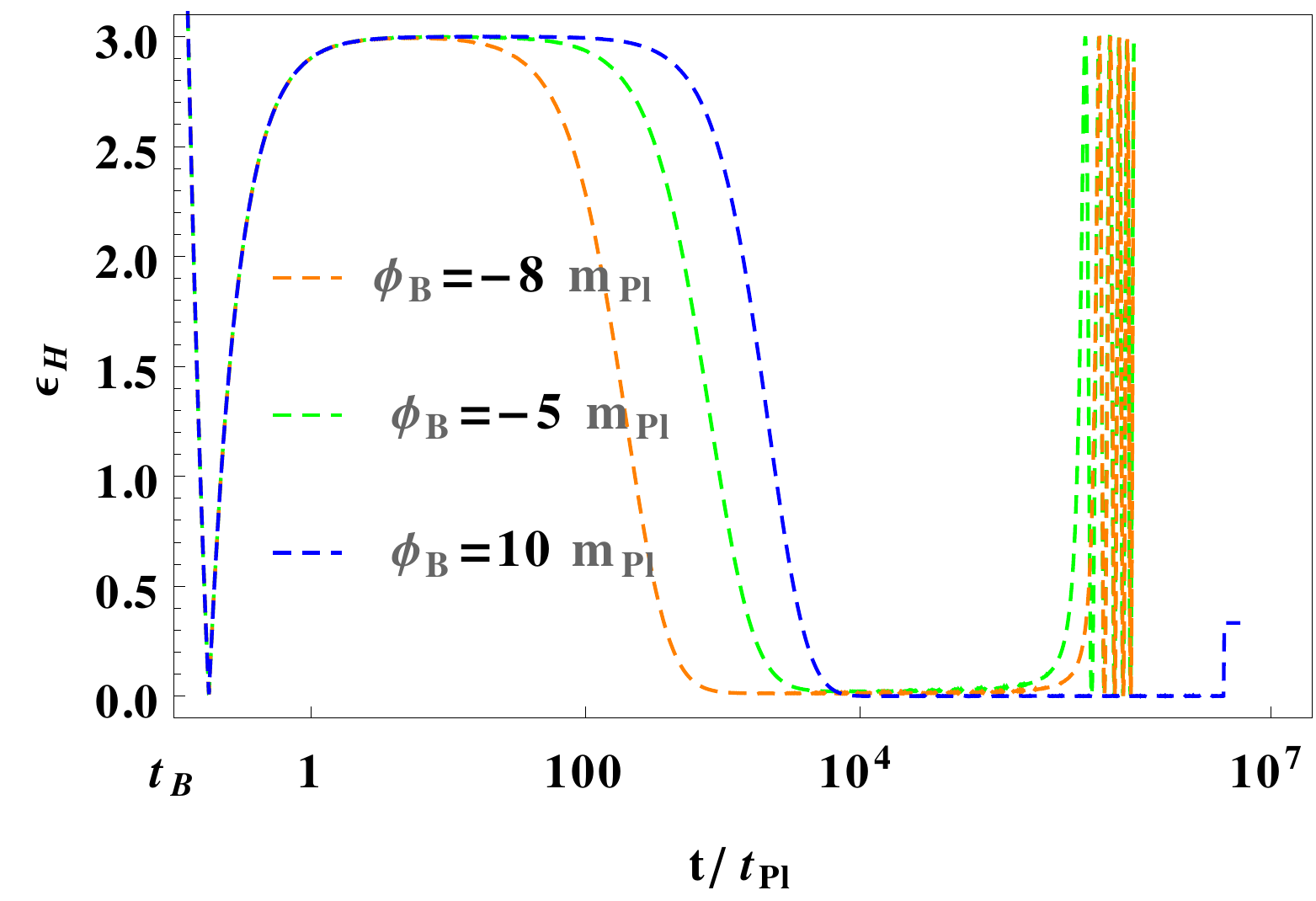}}
 \\
{\includegraphics[width=2.1in,height=1.65in,angle=0]{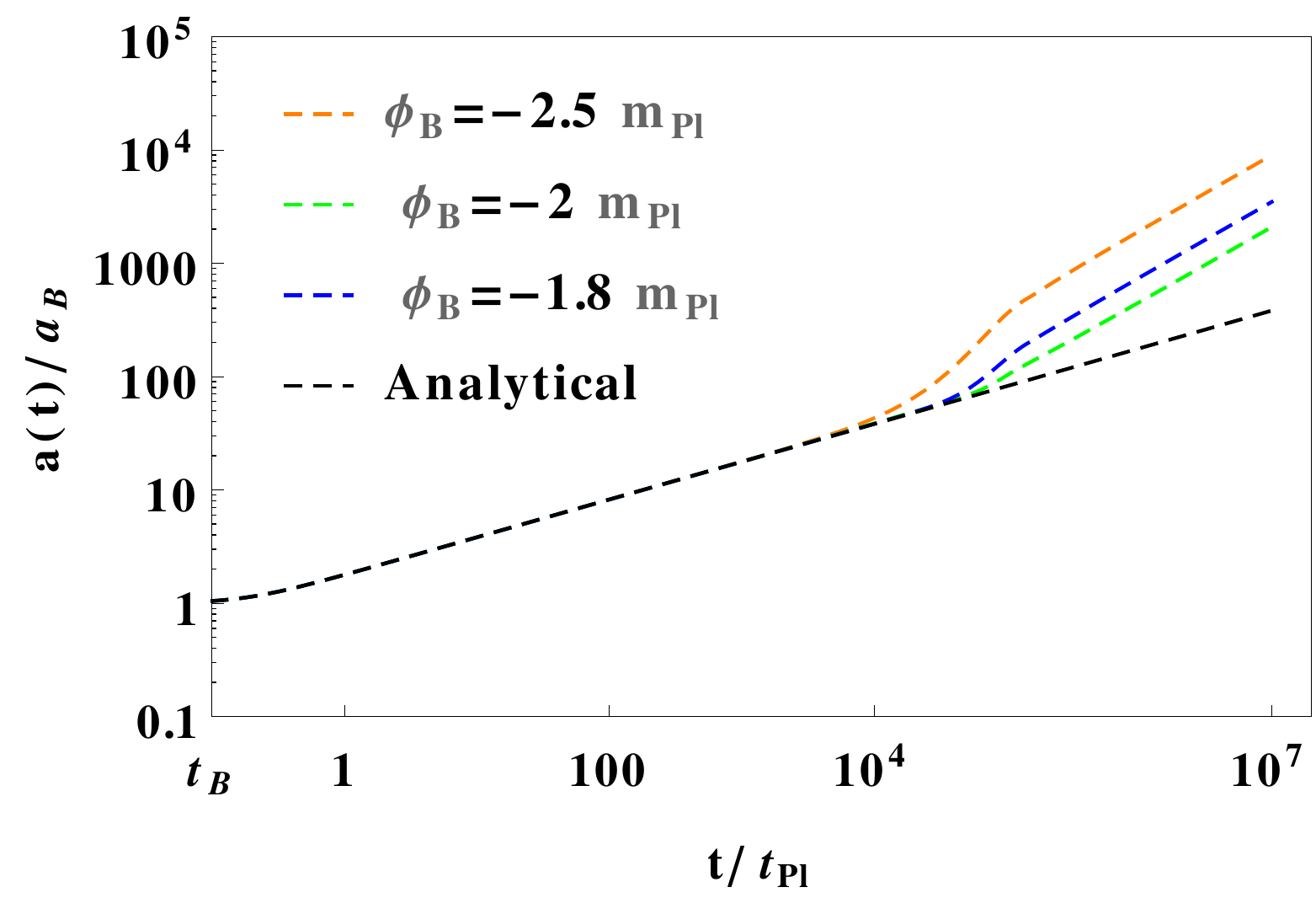}} &
{\includegraphics[width=2.1in,height=1.6in,angle=0]{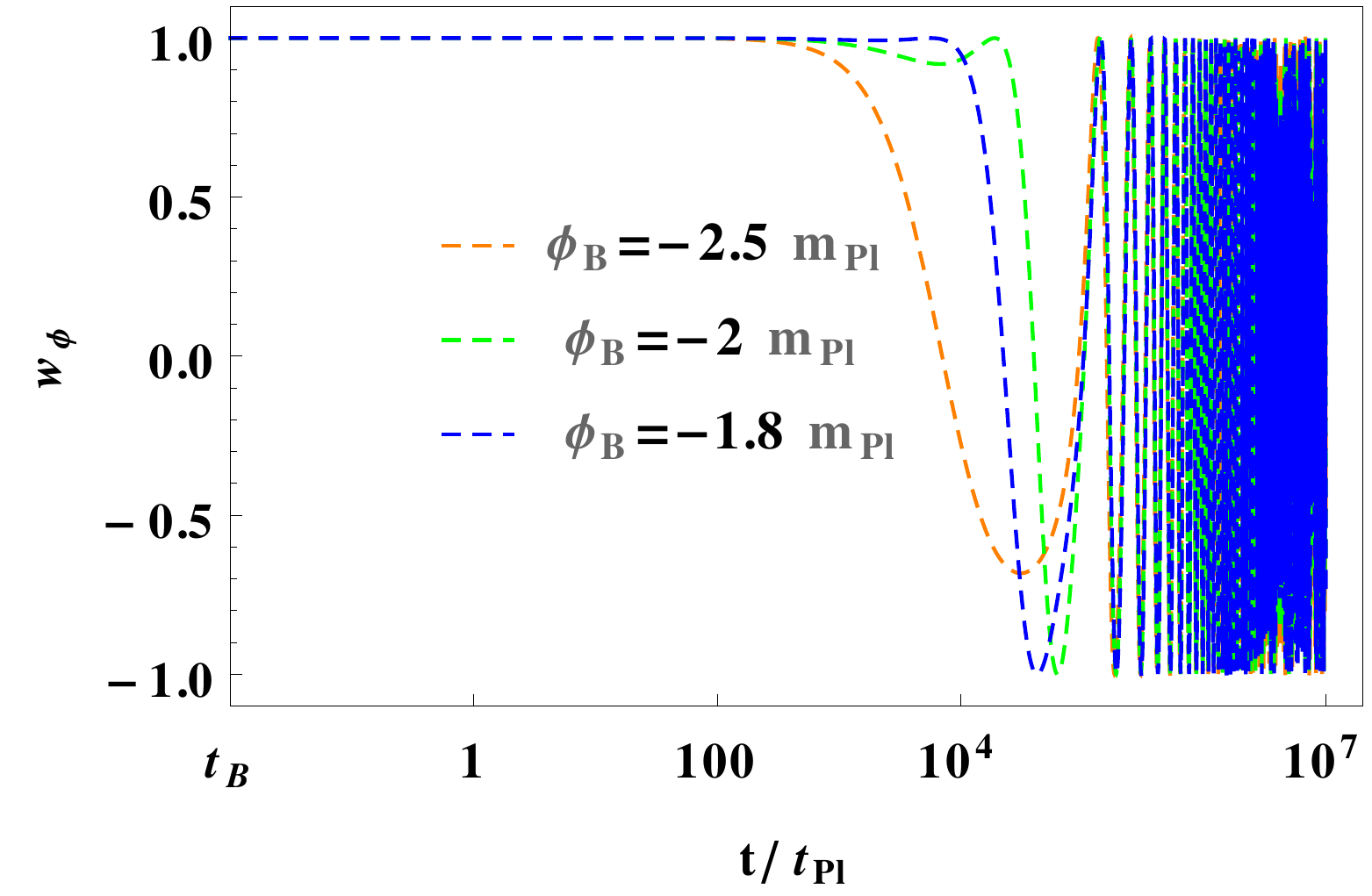}} &
{\includegraphics[width=2.0in,height=1.6in,angle=0]{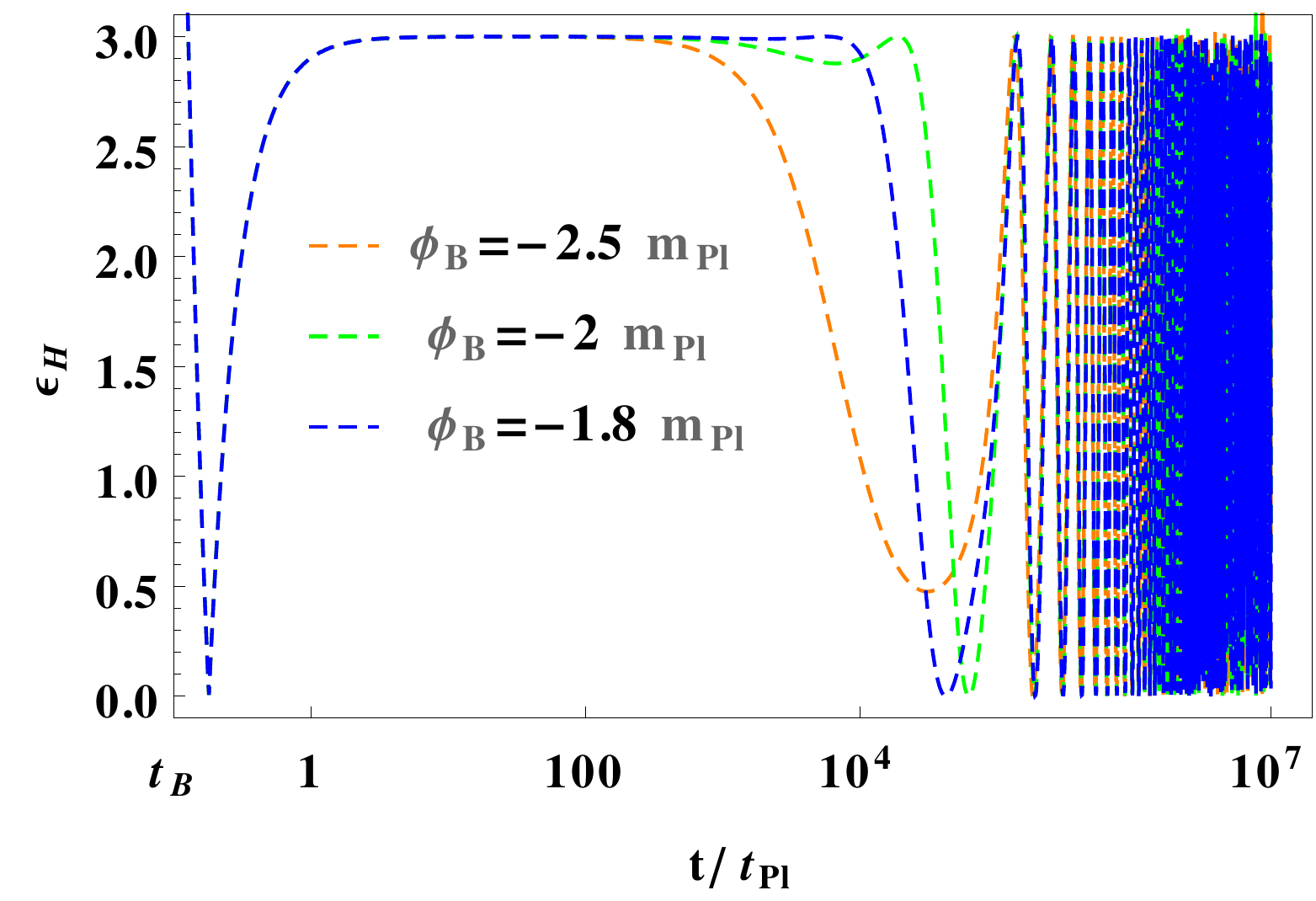}}
 \\ 
{\includegraphics[width=2.1in,height=1.6in,angle=0]{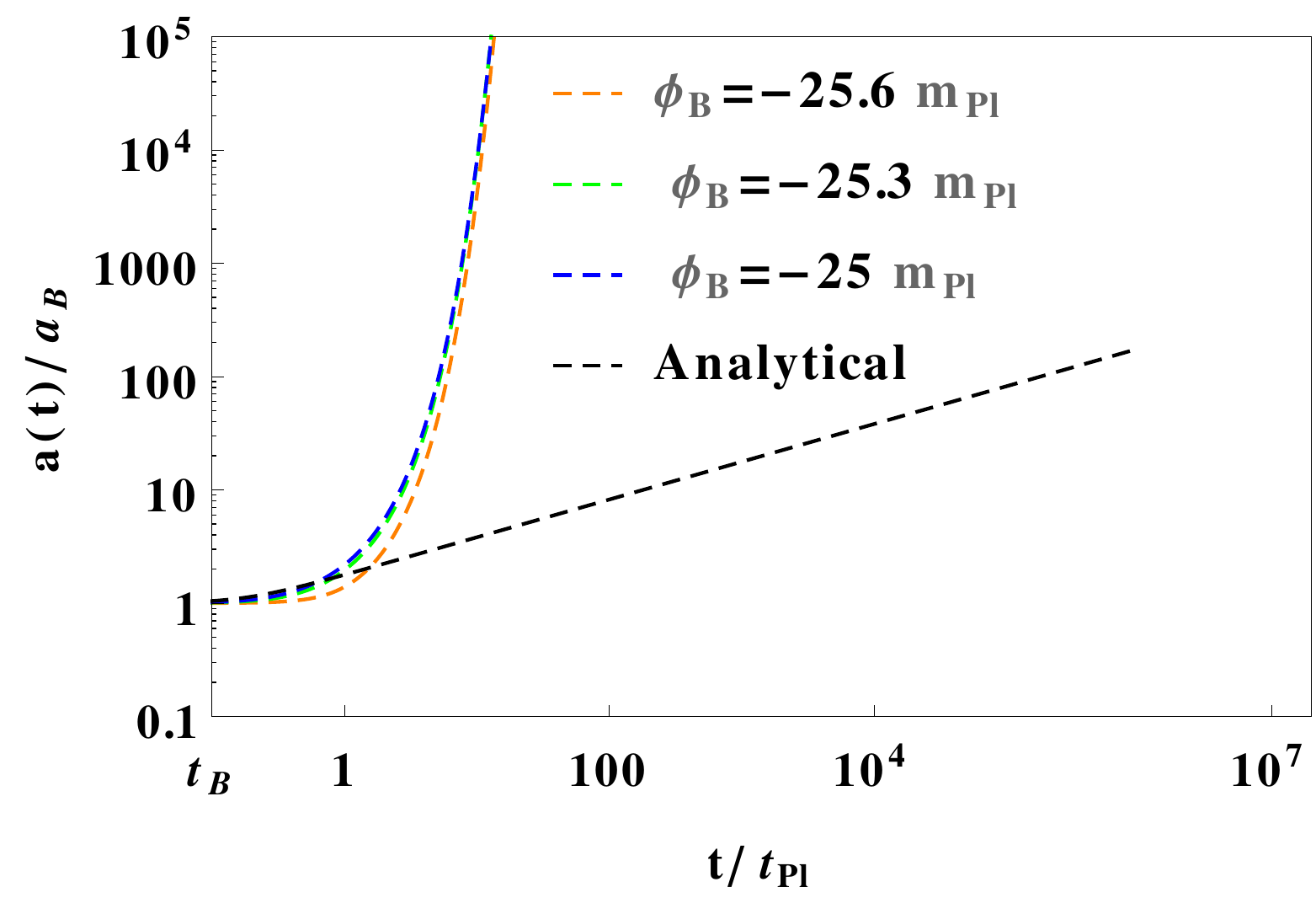}} & 
{\includegraphics[width=2.1in,height=1.6in,angle=0]{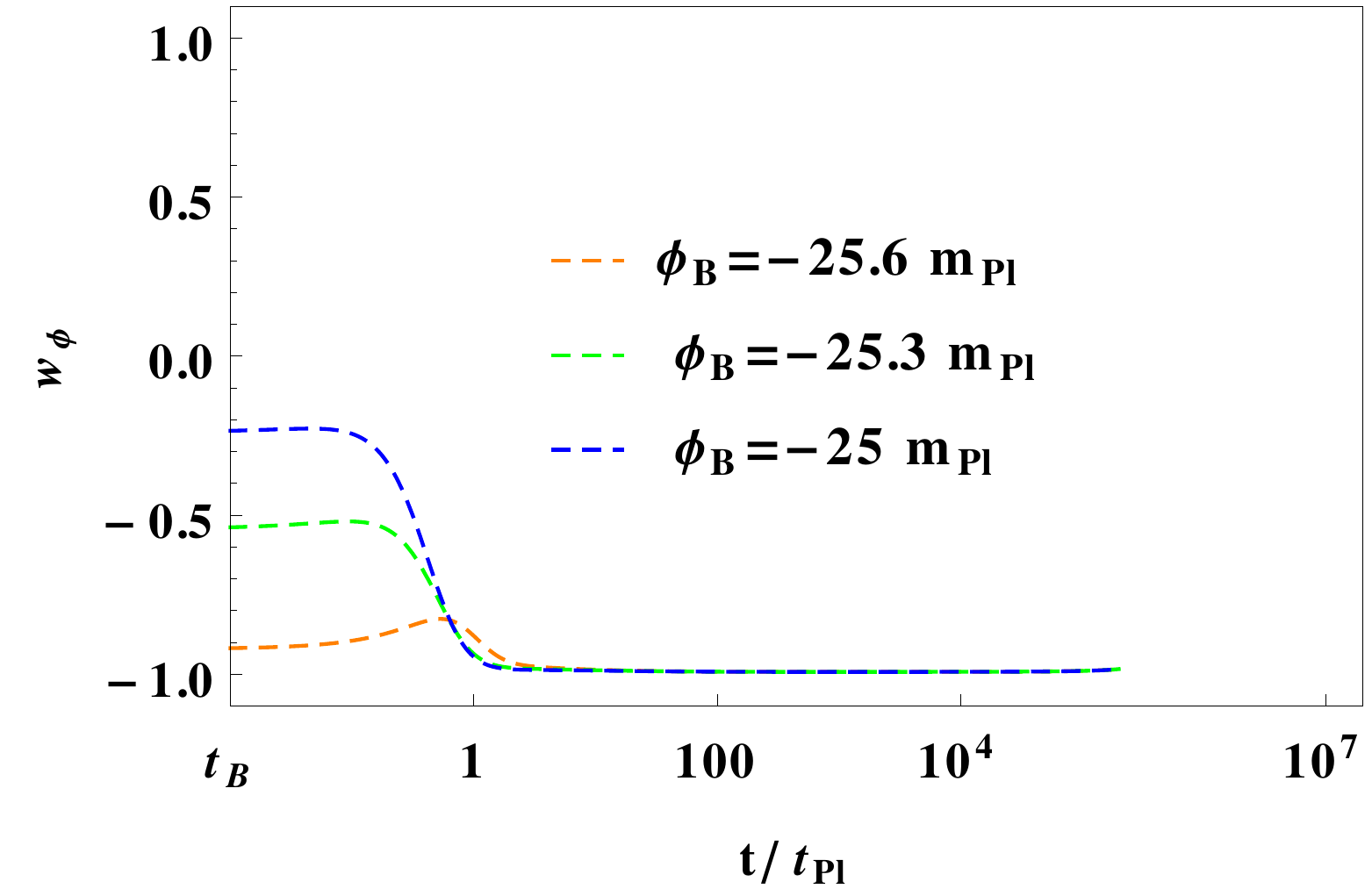}} & 
{\includegraphics[width=2.0in,height=1.6in,angle=0]{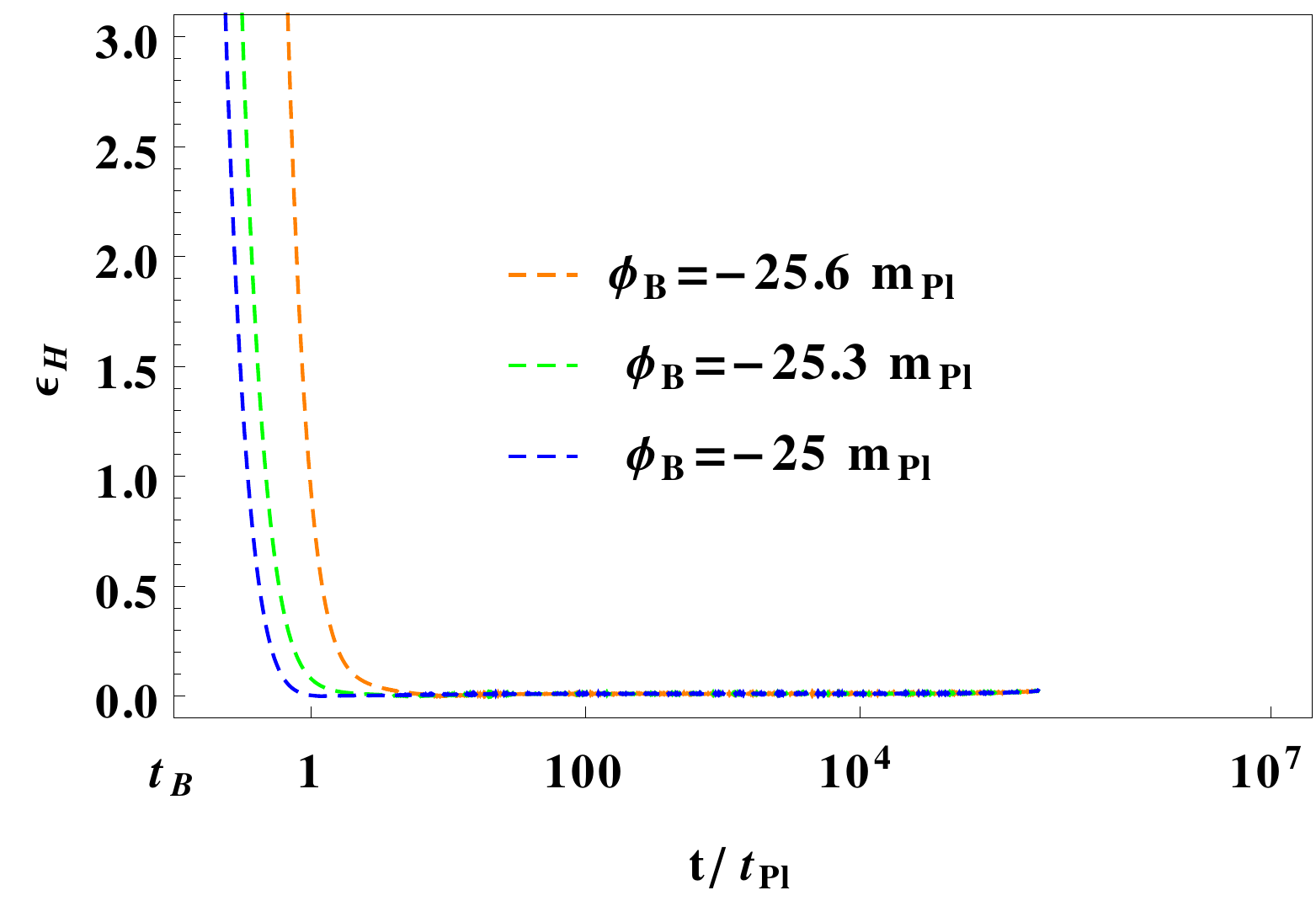}} 
\end{tabular}
\end{center}
\caption{ This figure  is for $E-model$ (\ref{eq:Epot}) with $\dot{\phi}_B>0$. Top (KED, except  a small subset) and bottom (PED) panels represent the slow-roll inflation region whereas the middle panels (subset of KED) do not. Here, we take $\alpha=5 m_{Pl}^2$, $c=4.9 \times 10^{-5}m_{Pl}$, and $m_{Pl}=1$.}
\label{fig:n1alpha5_dphp}
\end{figure*}
\begin{figure*}[tbp]
\begin{center}
\begin{tabular}{ccc}
{\includegraphics[width=2.1in,height=1.65in,angle=0]{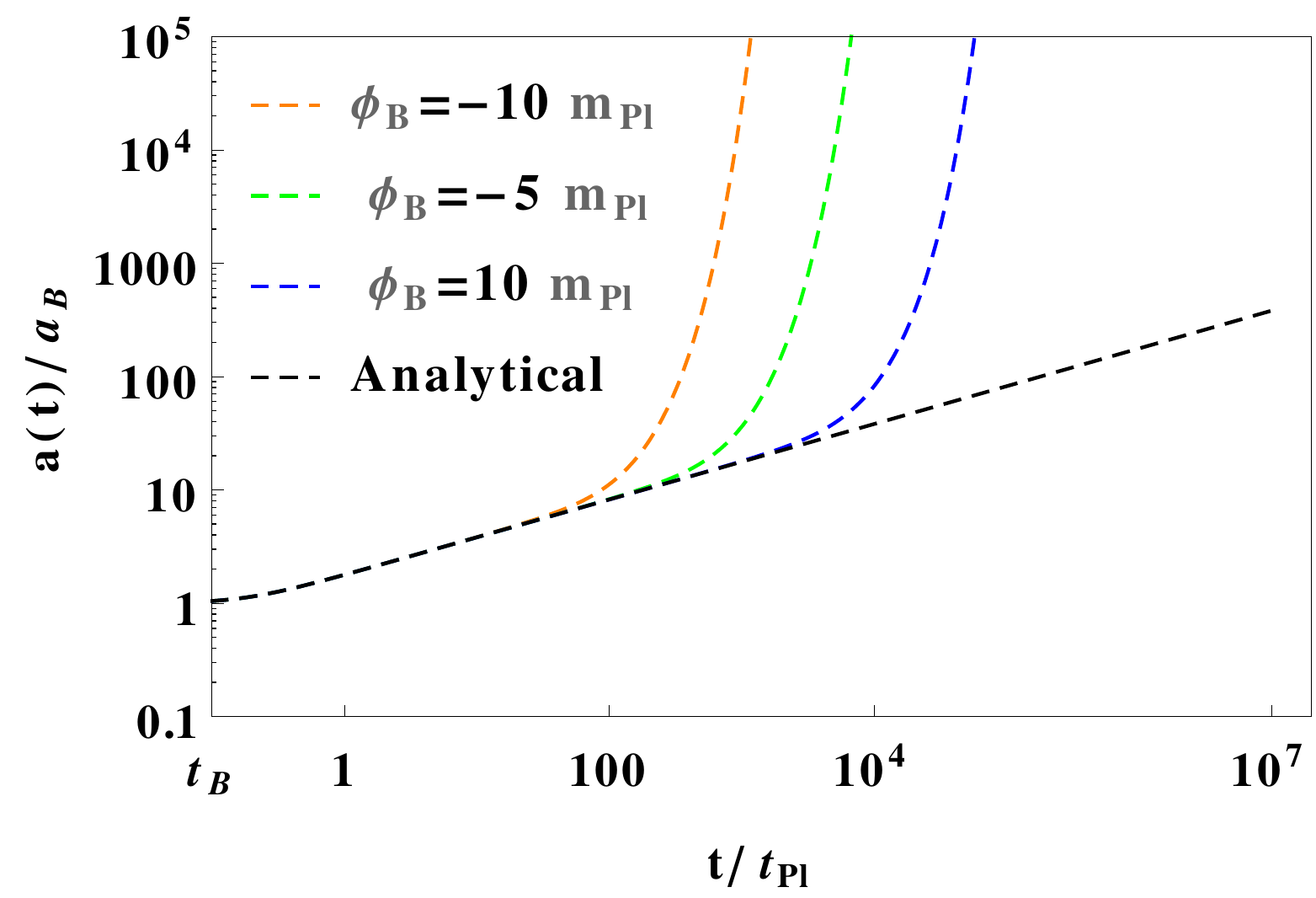}} &
{\includegraphics[width=2.1in,height=1.6in,angle=0]{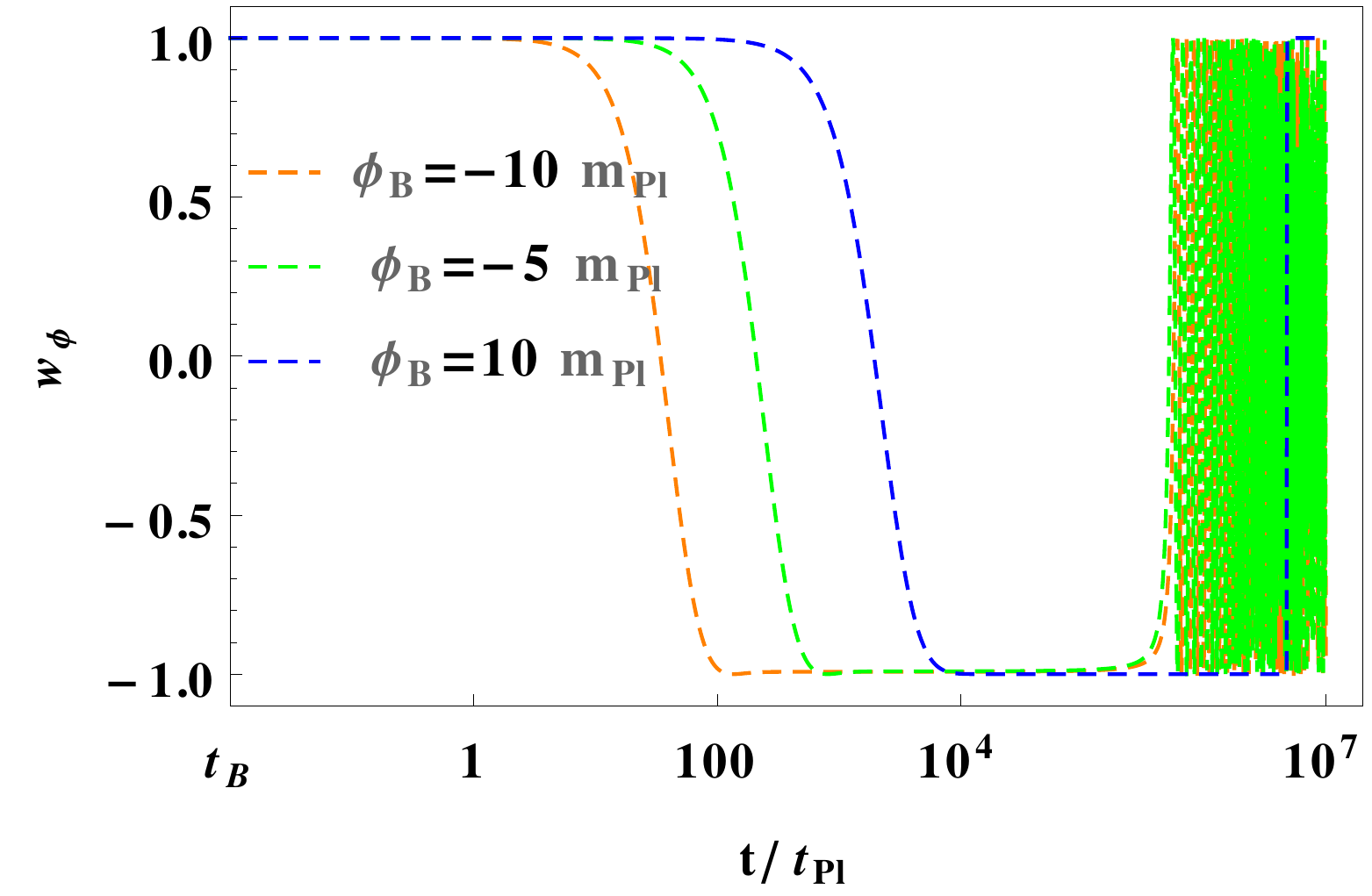}} &
{\includegraphics[width=2.0in,height=1.6in,angle=0]{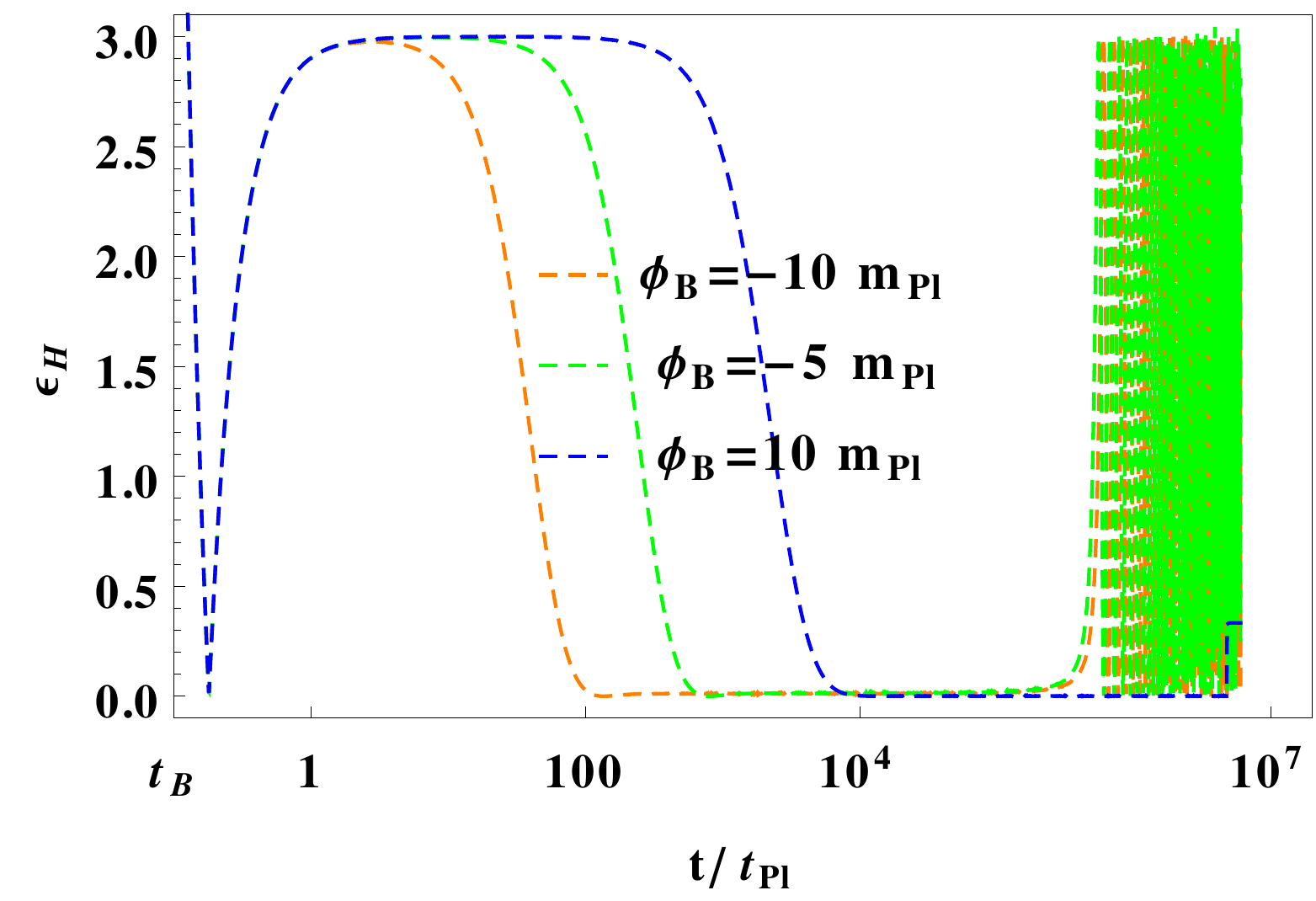}}
 \\
{\includegraphics[width=2.1in,height=1.65in,angle=0]{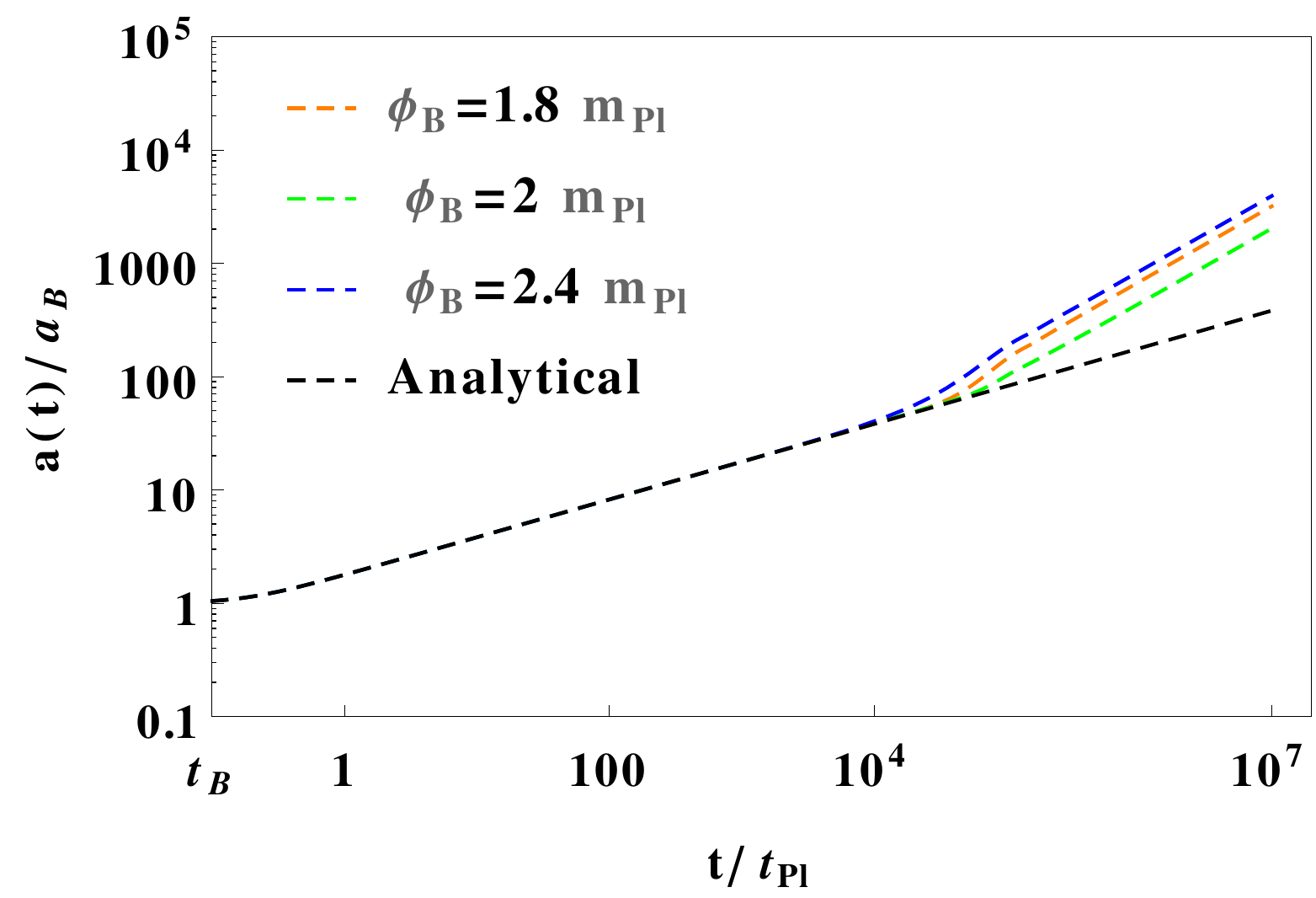}} &
{\includegraphics[width=2.1in,height=1.6in,angle=0]{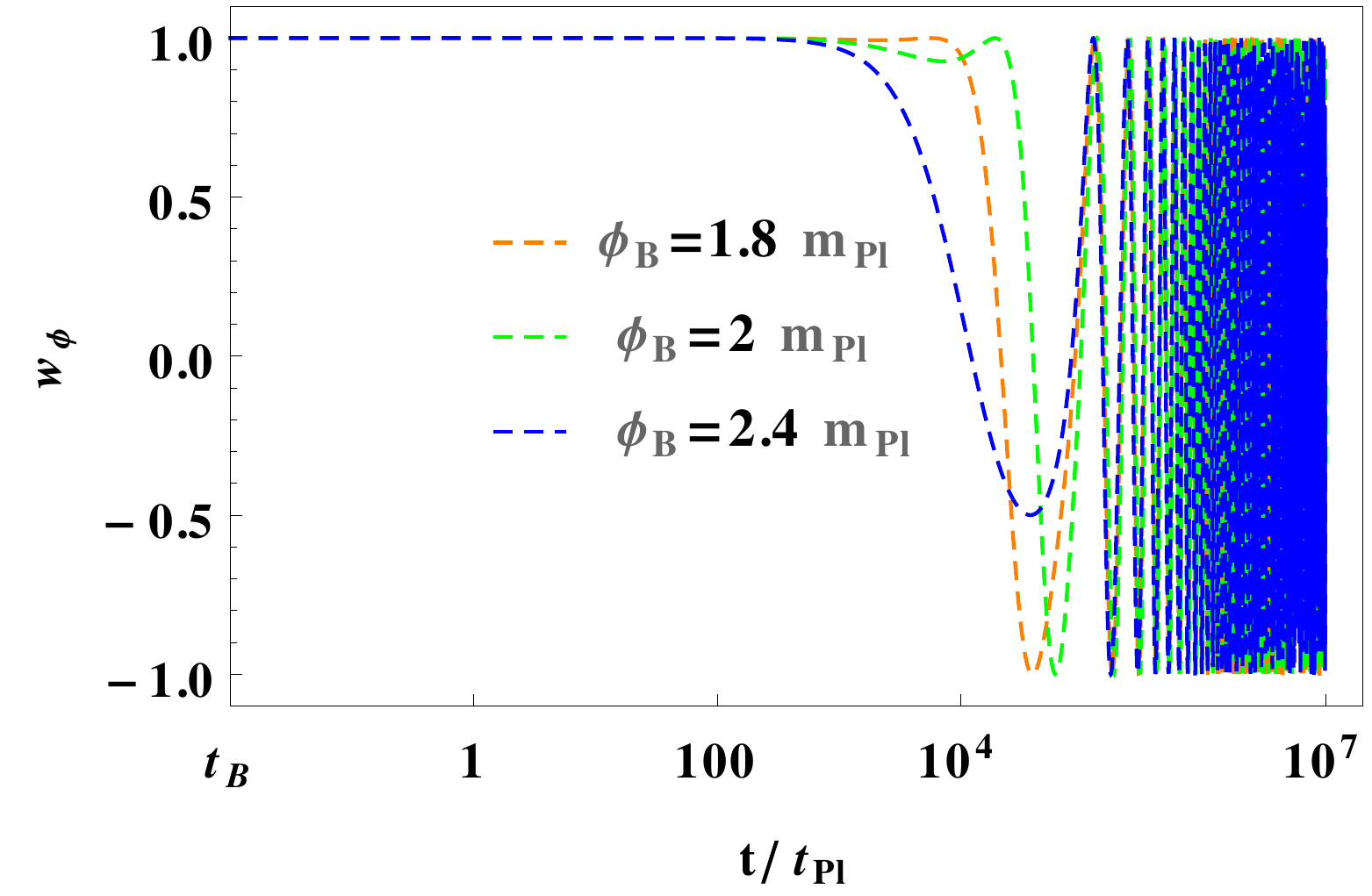}} &
{\includegraphics[width=2.0in,height=1.6in,angle=0]{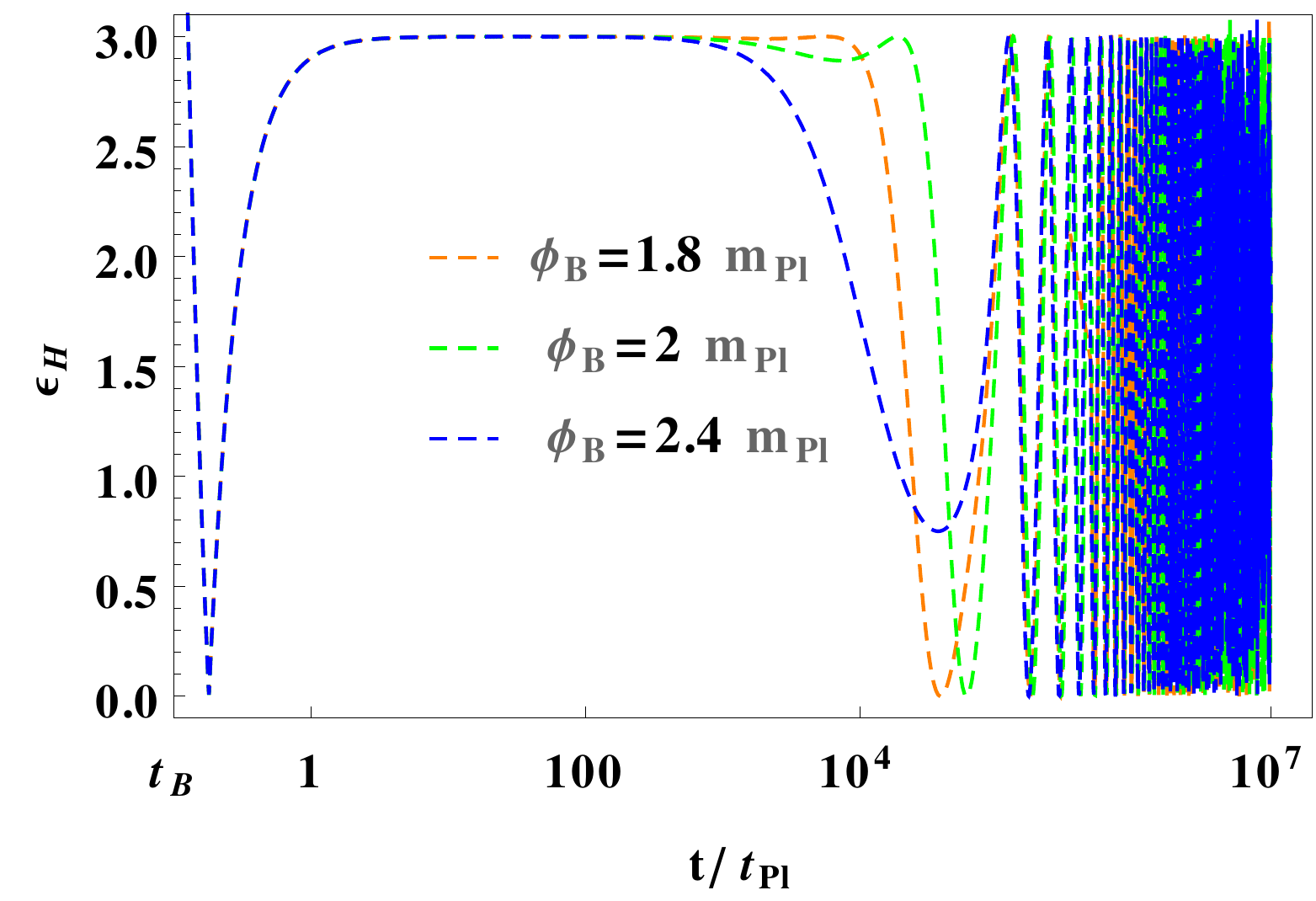}}
 \\ 
{\includegraphics[width=2.1in,height=1.6in,angle=0]{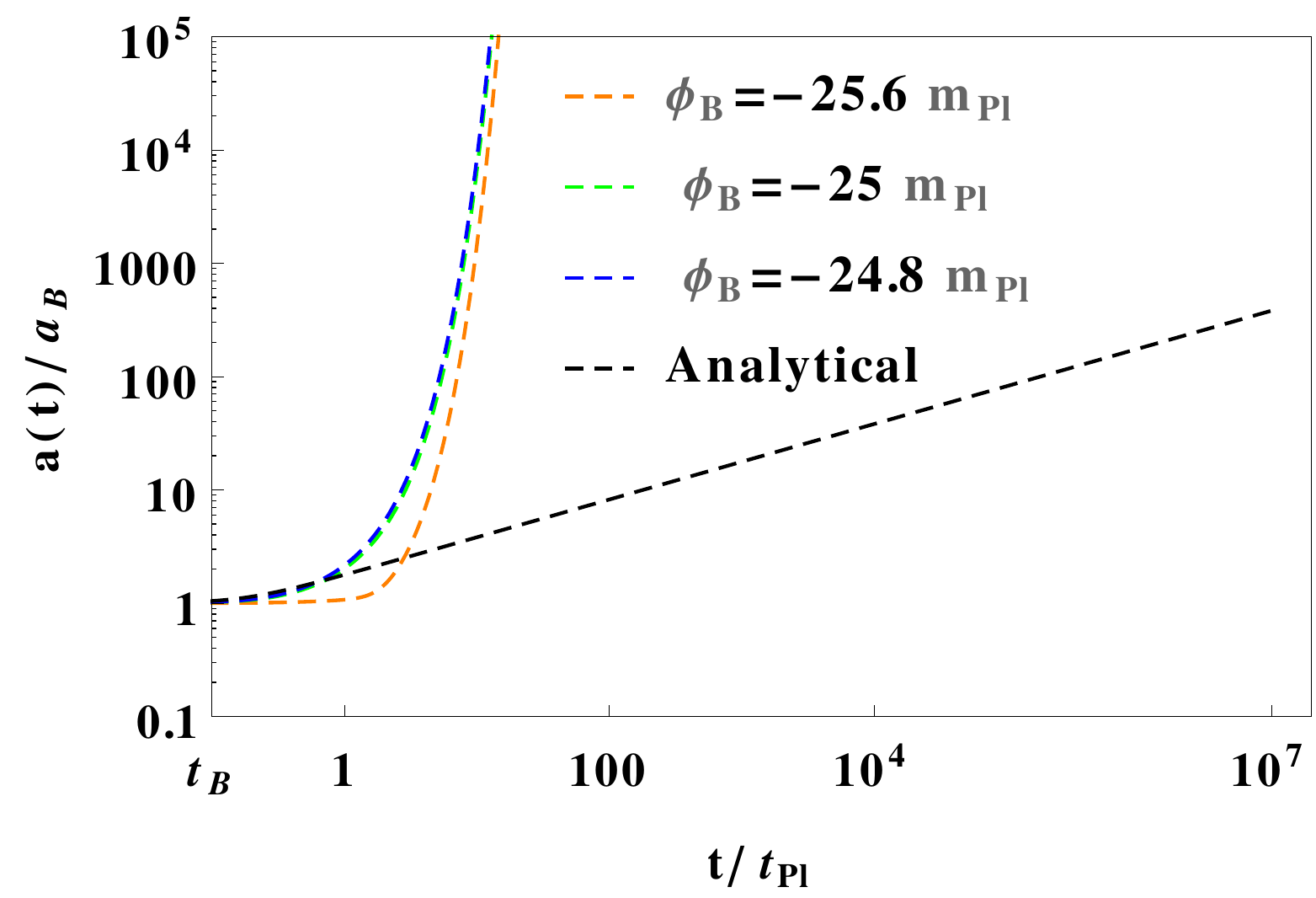}} & 
{\includegraphics[width=2.1in,height=1.6in,angle=0]{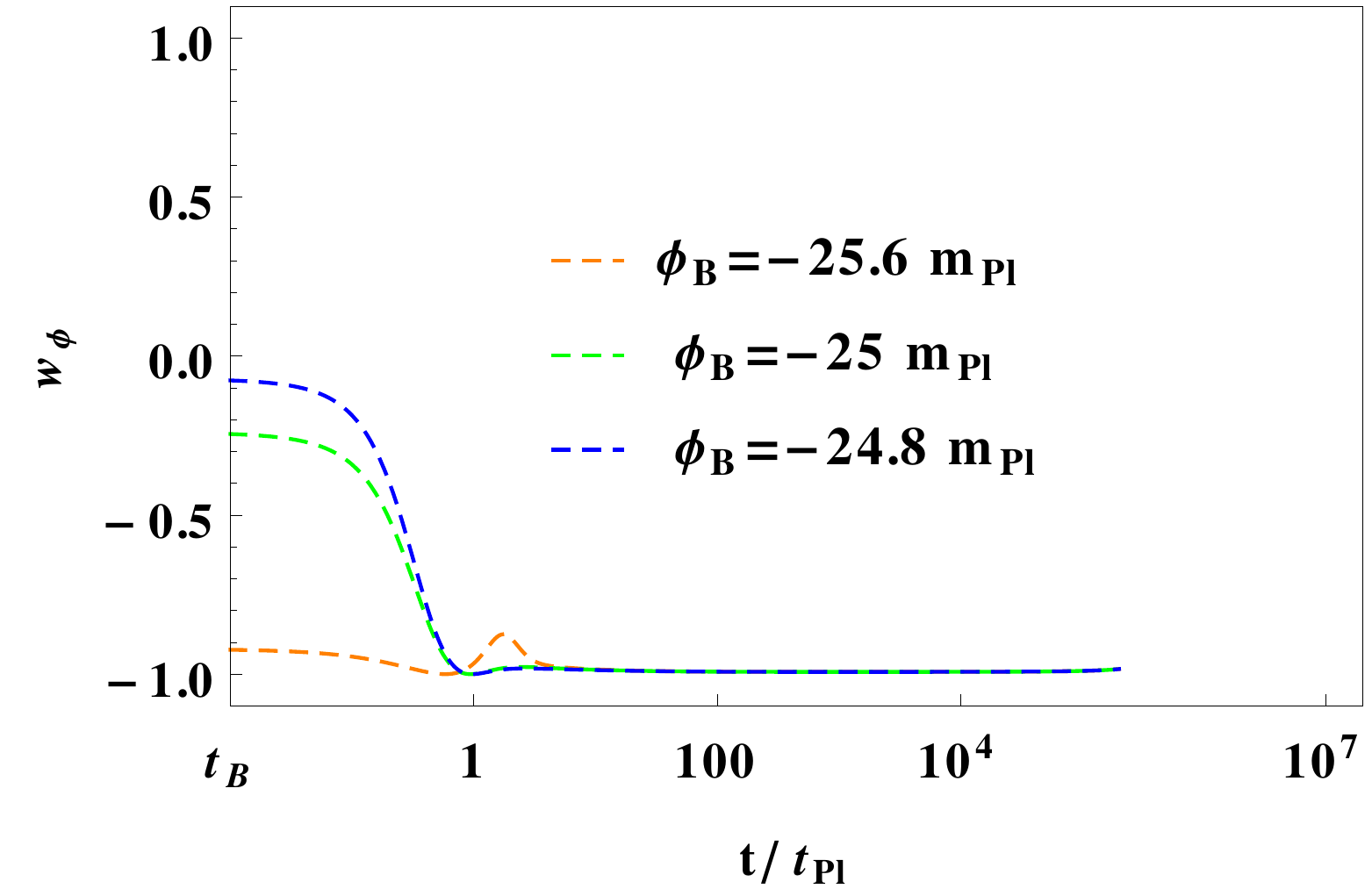}} & 
{\includegraphics[width=2.0in,height=1.6in,angle=0]{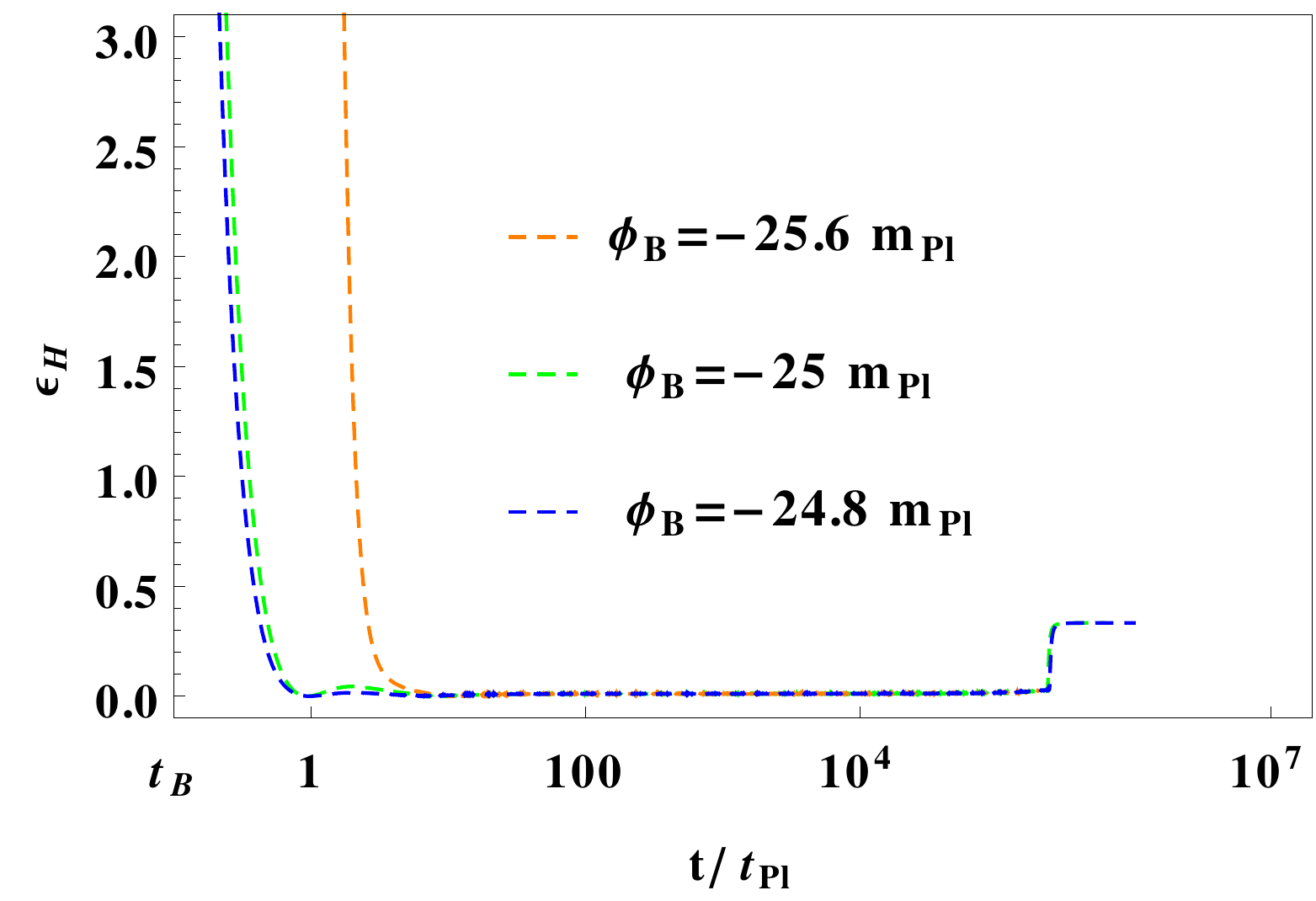}} 
\end{tabular}
\end{center}
\caption{ This figure is same as Fig. \ref{fig:n1alpha5_dphp} but with $\dot{\phi}_B<0$.
}
\label{fig:n1alpha5_dphn}
\end{figure*}
\begin{figure*}[tbp]
\begin{center}
\begin{tabular}{ccc}
{\includegraphics[width=2.1in,height=1.65in,angle=0]{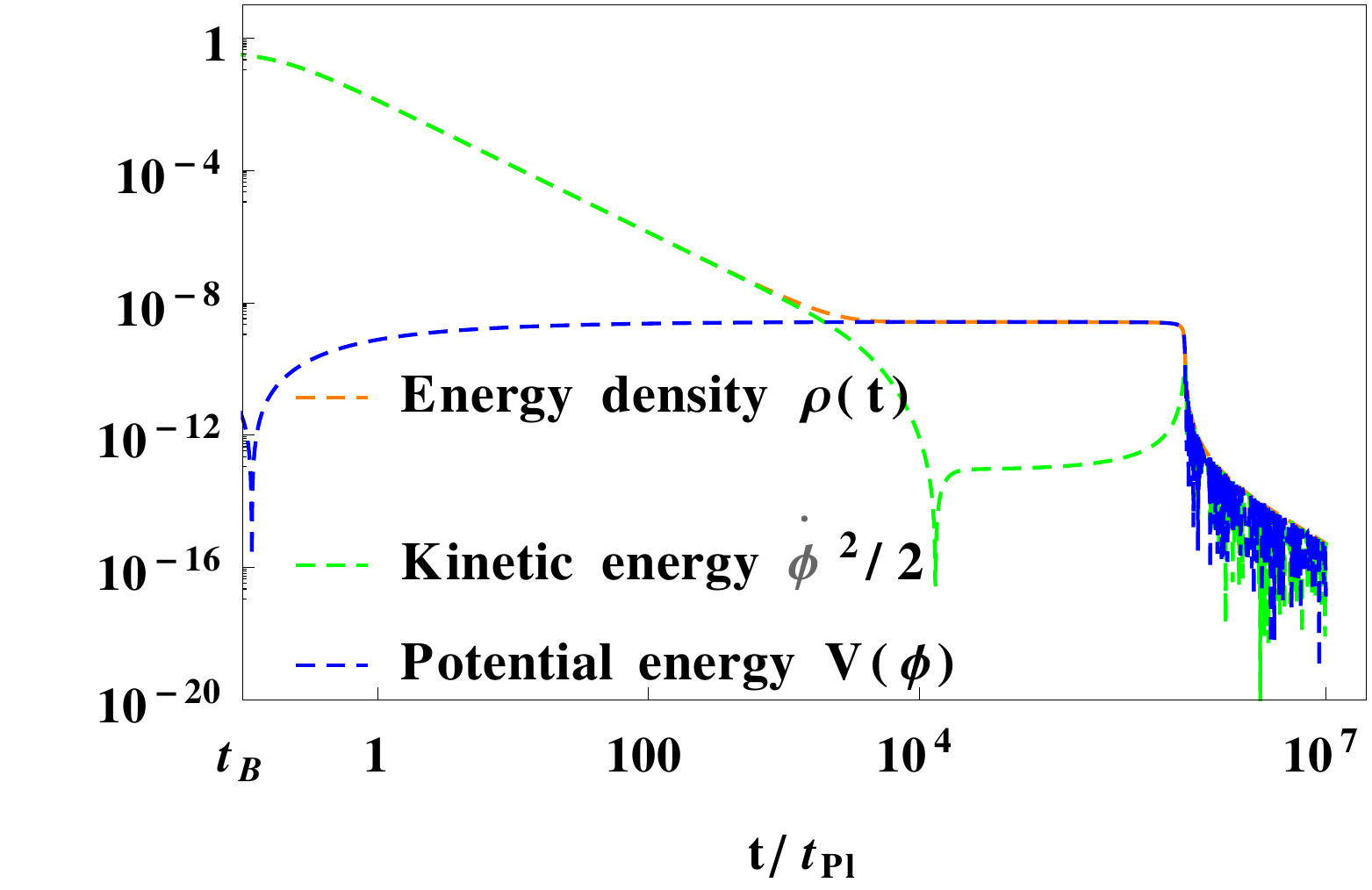}} &
{\includegraphics[width=2.1in,height=1.6in,angle=0]{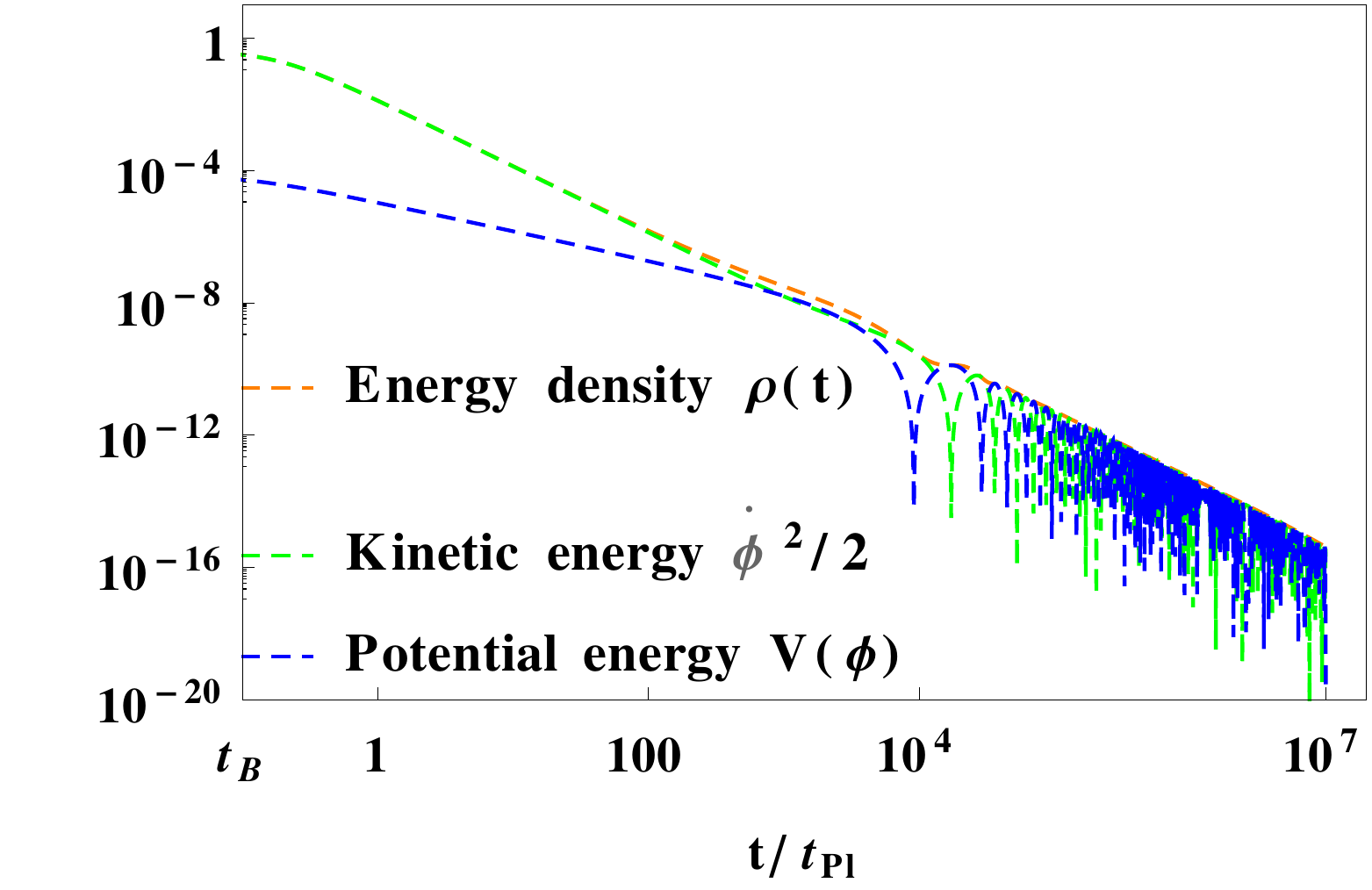}} &
{\includegraphics[width=2.0in,height=1.6in,angle=0]{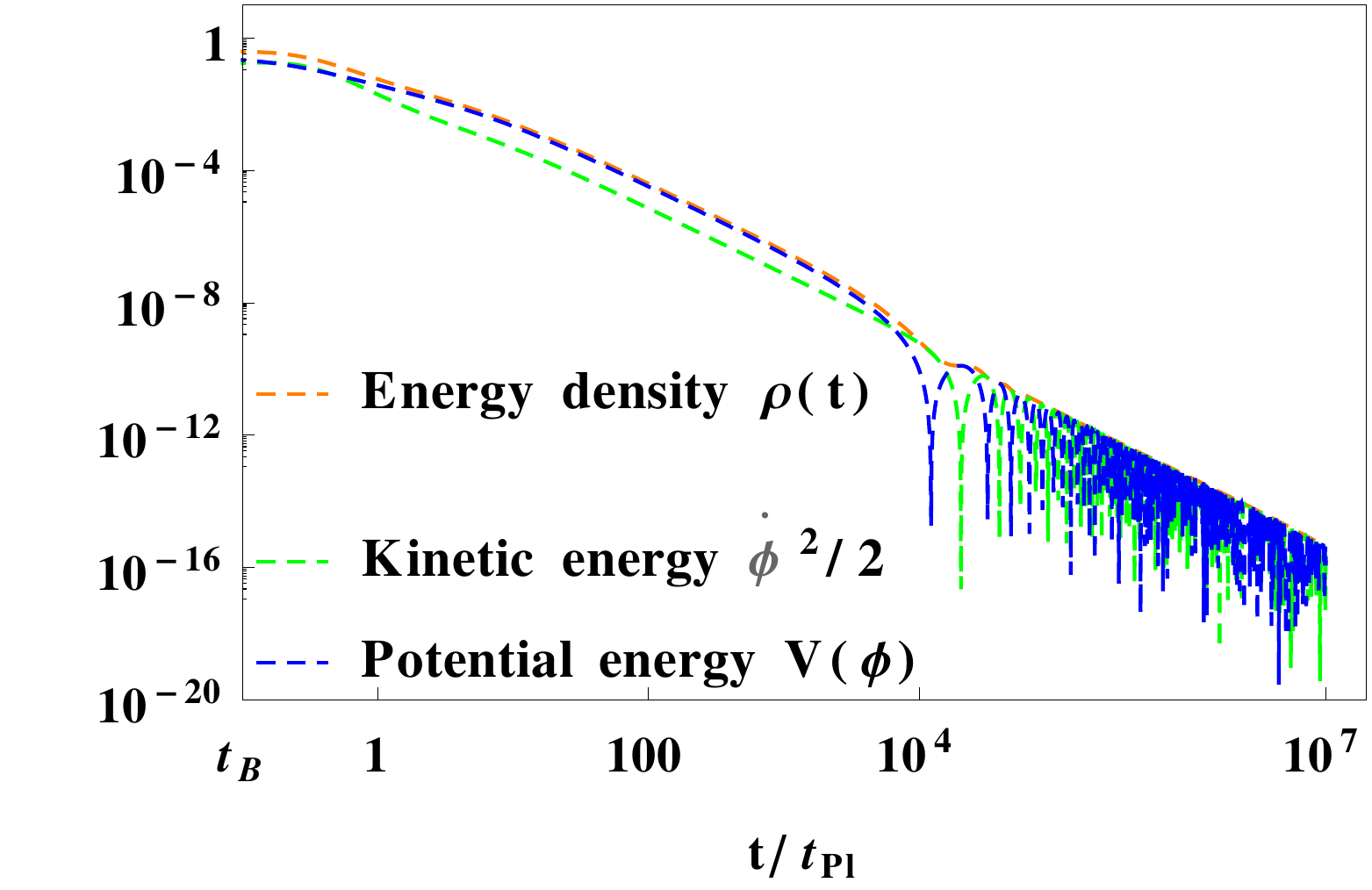}}
 \\
{\includegraphics[width=2.1in,height=1.65in,angle=0]{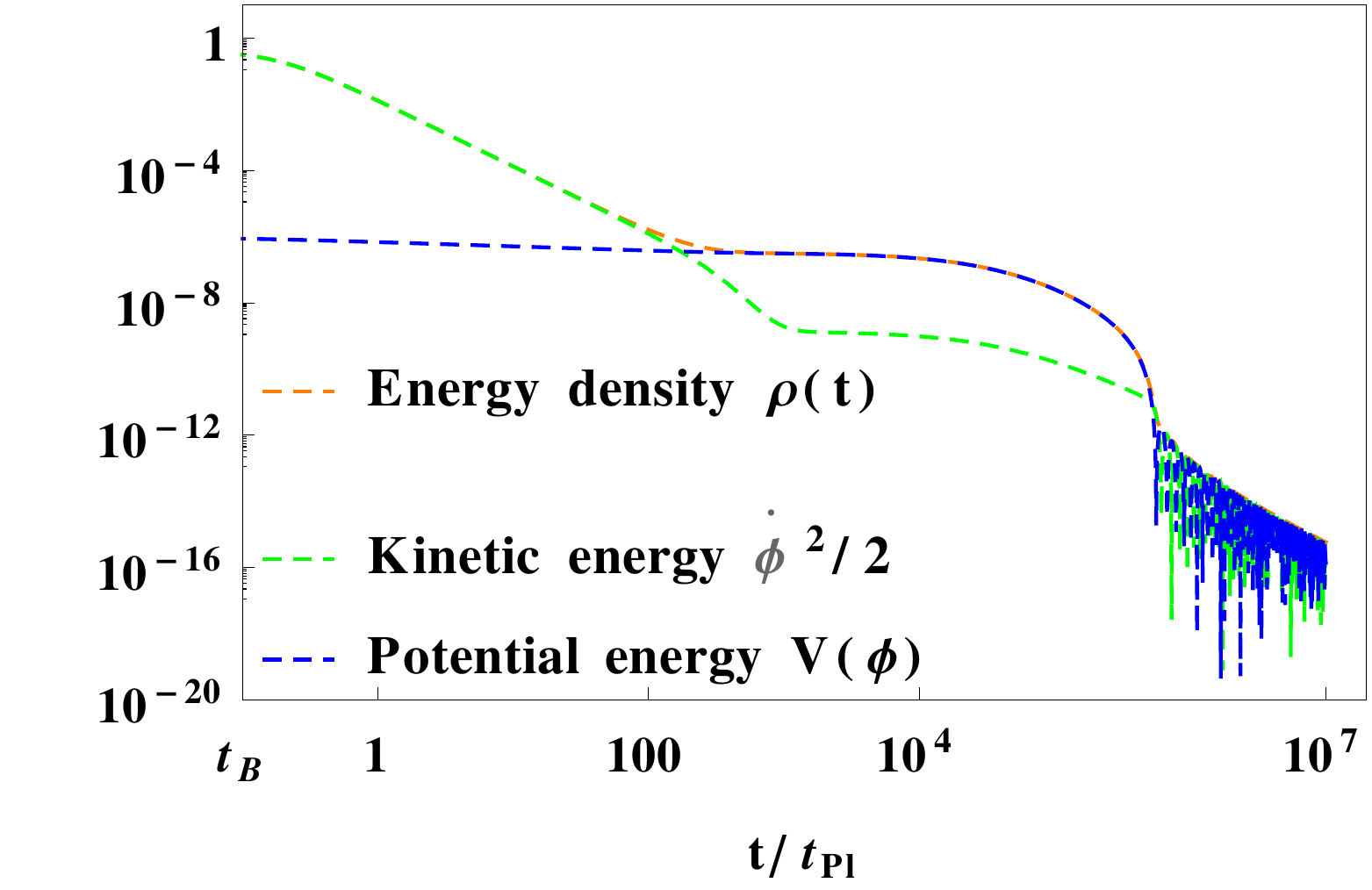}} &
{\includegraphics[width=2.1in,height=1.6in,angle=0]{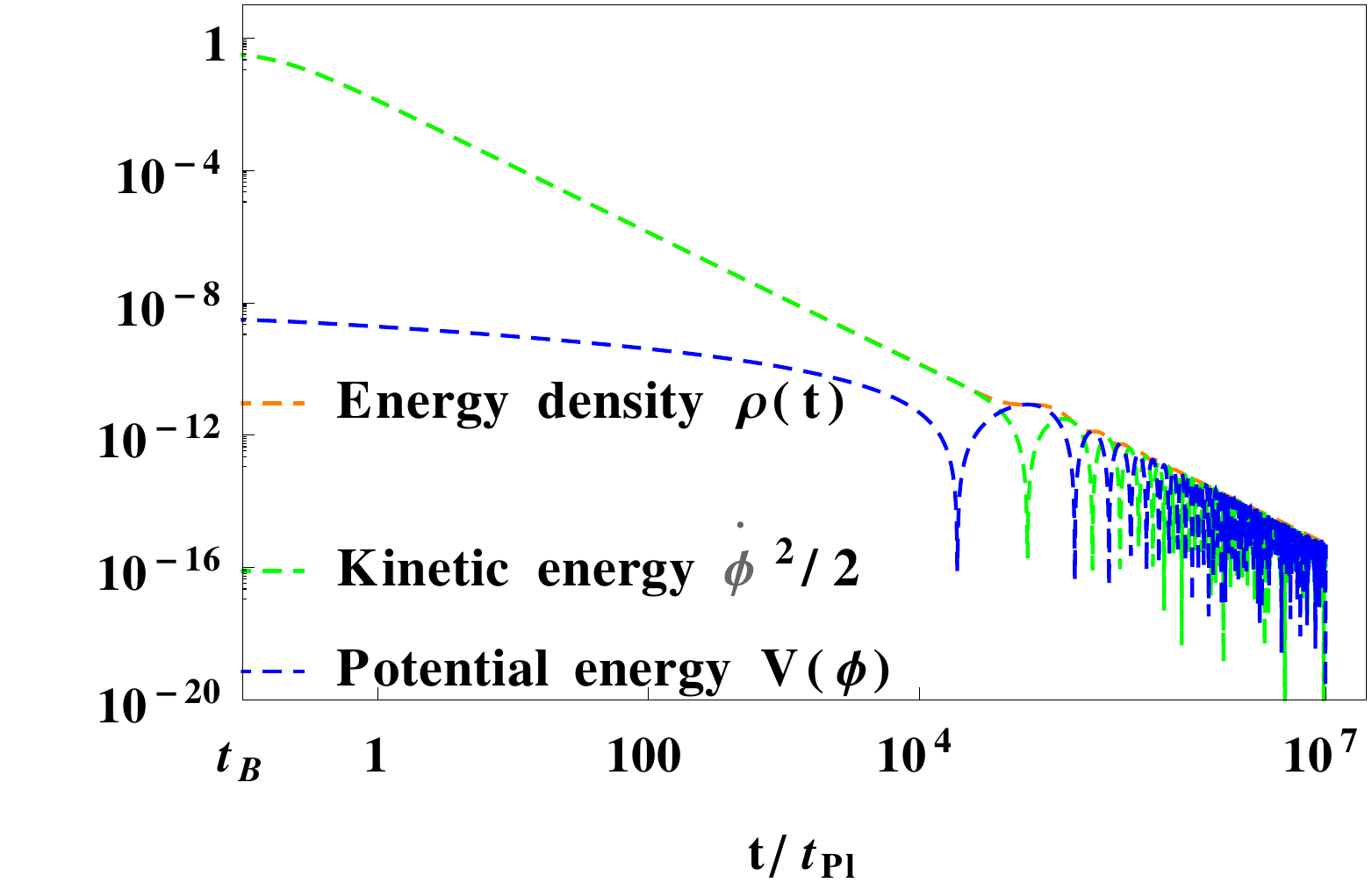}} &
{\includegraphics[width=2.0in,height=1.6in,angle=0]{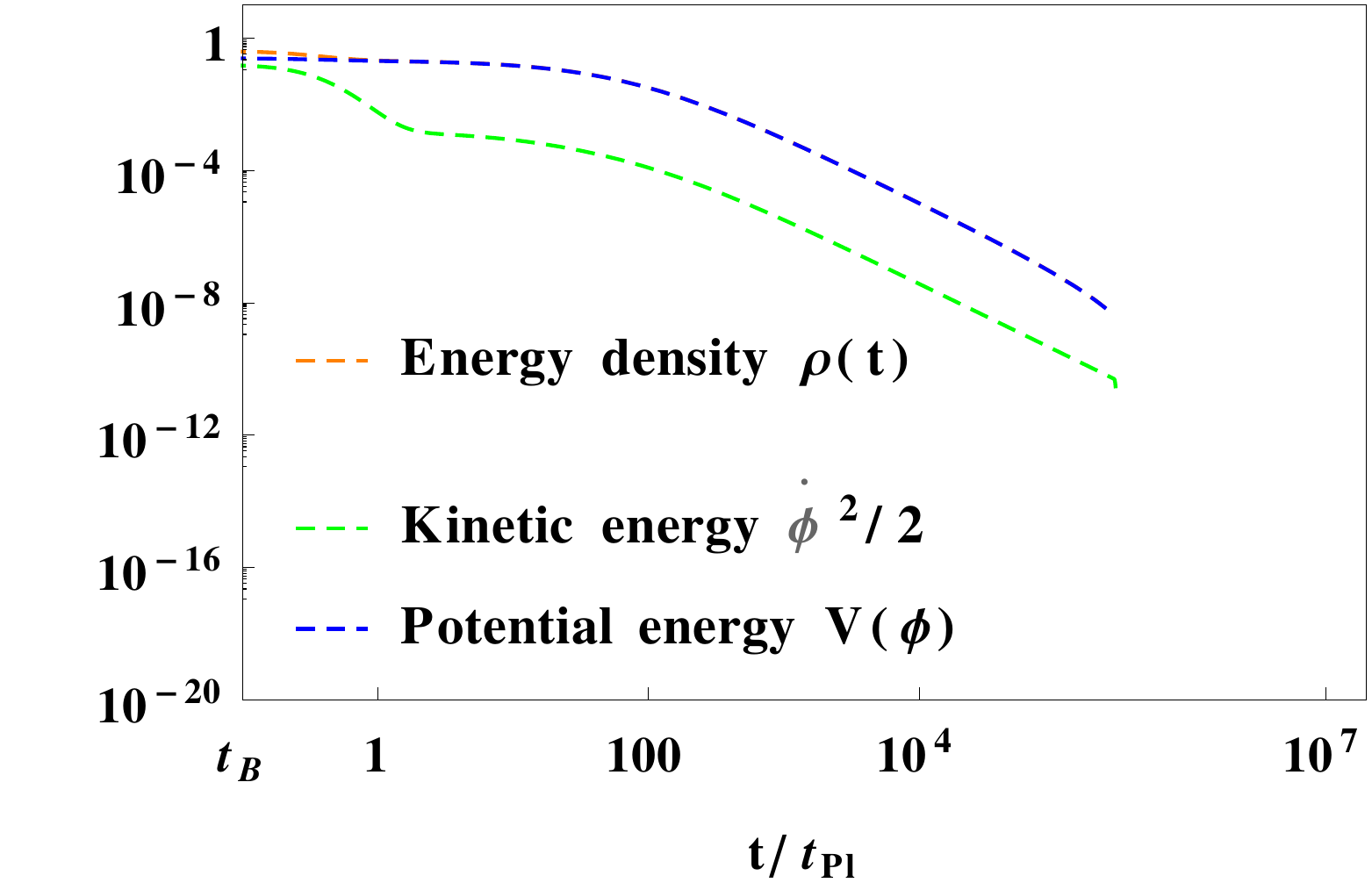}}
\end{tabular}
\end{center}
\caption{ In this figure the kinetic and potential energies are plotted  for  $E-model$ with $\alpha=0.1m_{Pl}^2$ (Top) and $\alpha=5m_{Pl}^2$ (Bottom) for $\dot{\phi}_B>0$. Top:  left and middle panels correspond to KED initial conditions at the bounce. Left panel ( $\phi_B=-0.1 m_{Pl}$) provides the slow-roll inflation as PE dominates at the later regime, while it is not possible in the middle panel ( $\phi_B=-2 m_{Pl}$) as PE never dominates. Right panel ( $\phi_B=-3.58 m_{Pl}$) represents the PED case, but does not yield the slow-roll inflation. Bottom: Left panel ( $\phi_B=-8 m_{Pl}$) gives rise to the slow-roll inflation, whereas the middle panel ( $\phi_B=-2 m_{Pl}$) does not. However, in this case, the right panel ( $\phi_B=-25 m_{Pl}$) yield the slow-roll inflation for PED initial conditions.}
\label{fig:n1rho}
\end{figure*}
\begin{figure*}[tbp]
\begin{center}
\begin{tabular}{ccc}
{\includegraphics[width=2.1in,height=1.65in,angle=0]{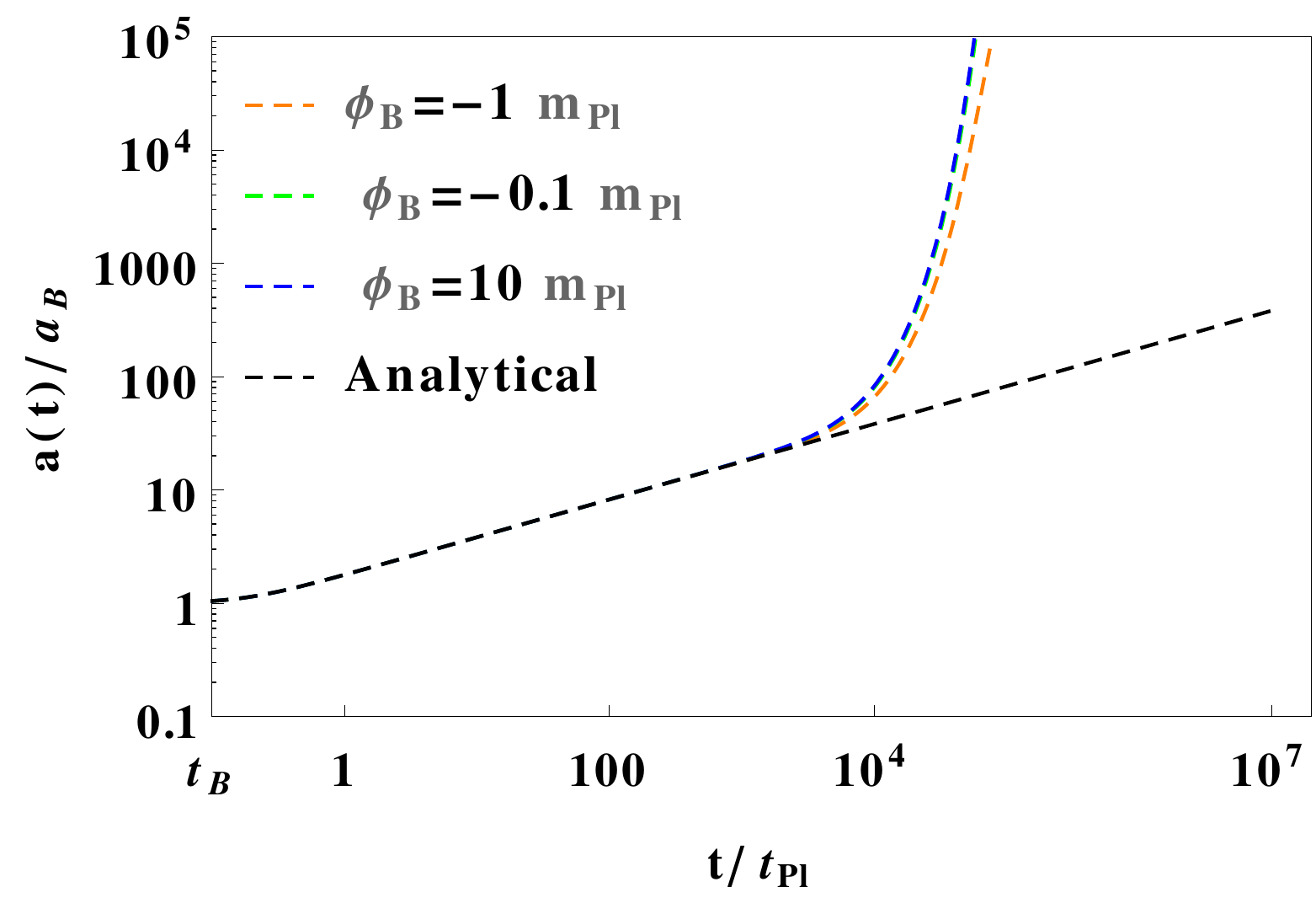}} &
{\includegraphics[width=2.1in,height=1.6in,angle=0]{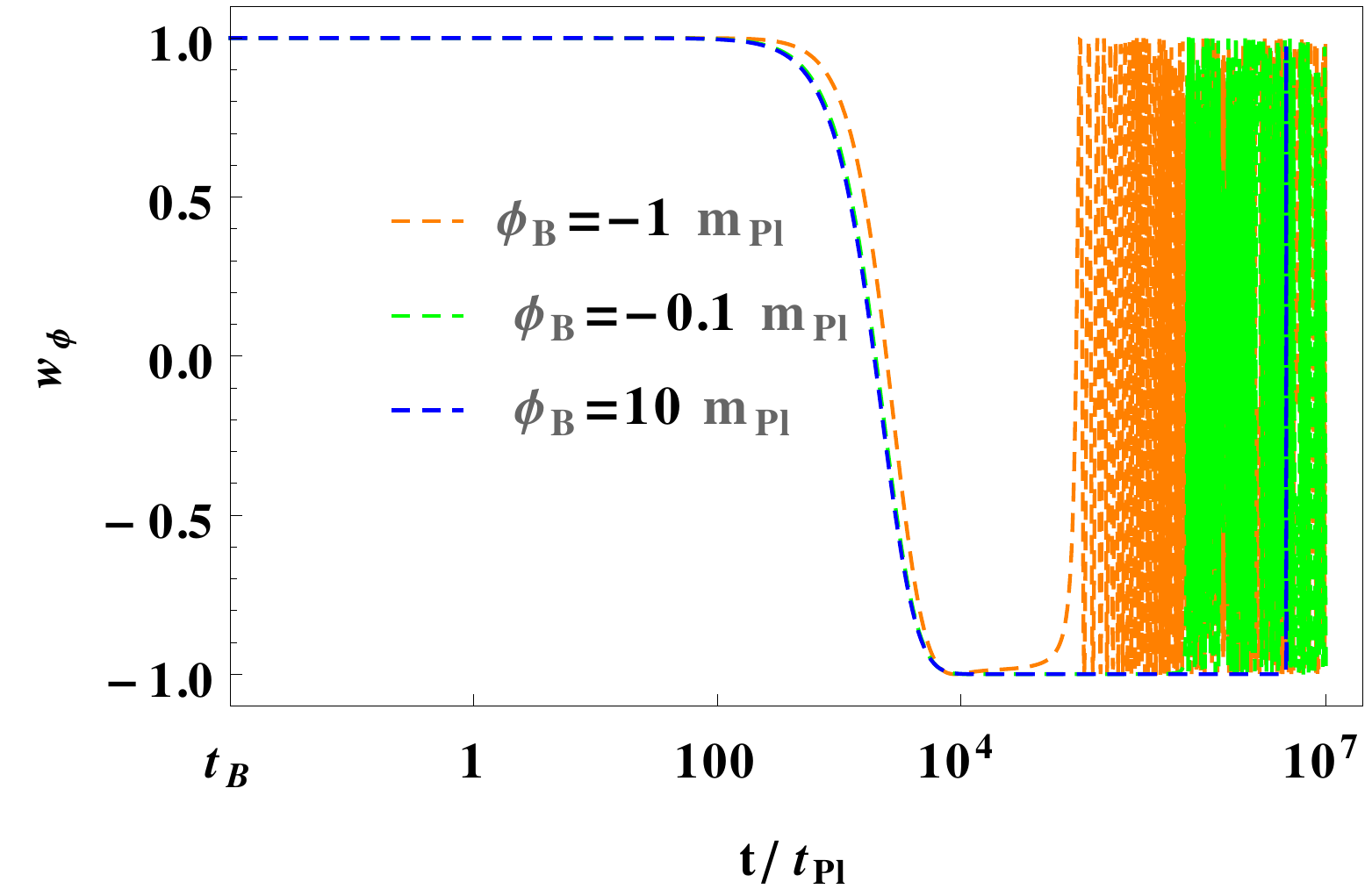}} &
{\includegraphics[width=2.0in,height=1.6in,angle=0]{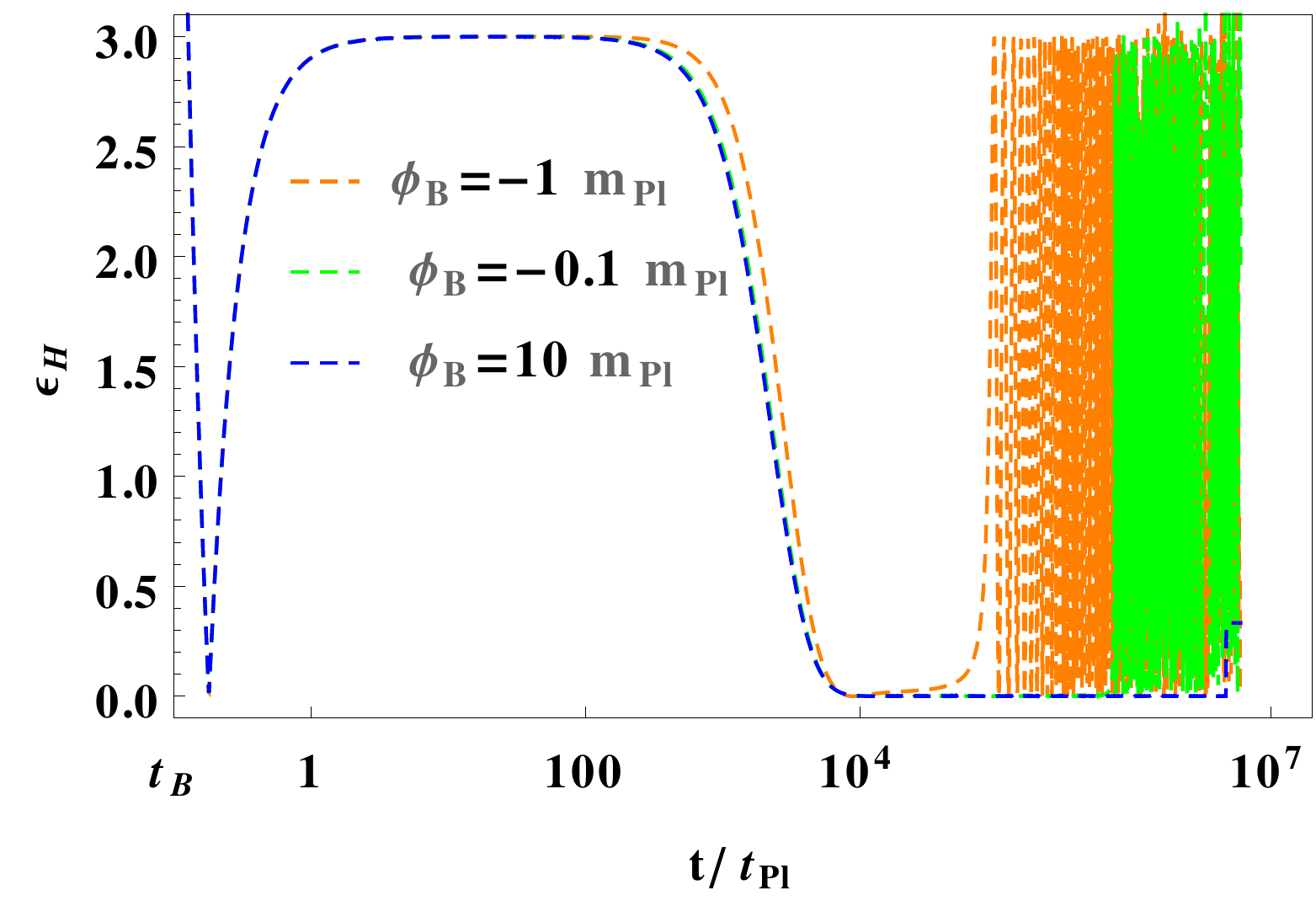}}
 \\
{\includegraphics[width=2.1in,height=1.65in,angle=0]{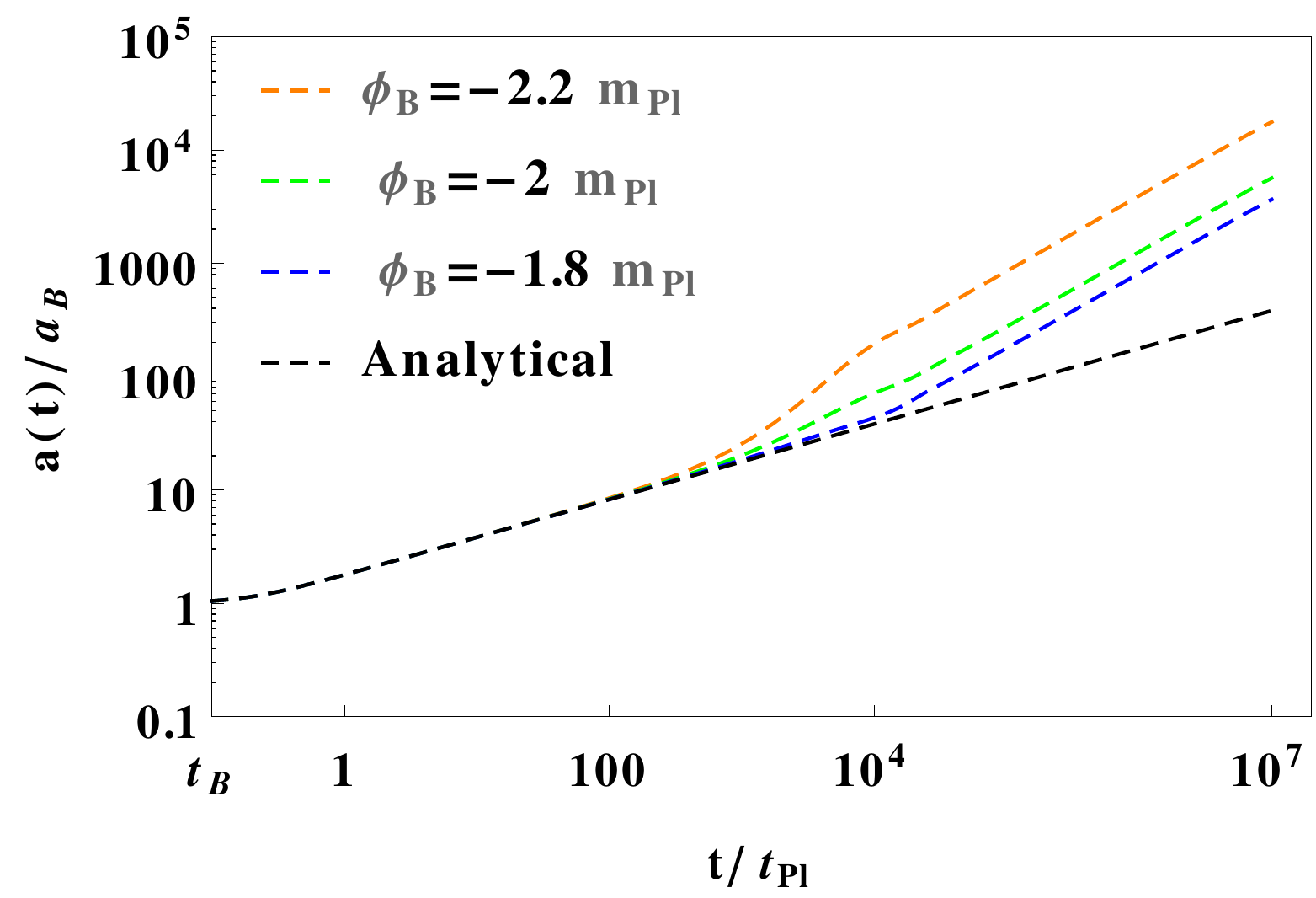}} &
{\includegraphics[width=2.1in,height=1.6in,angle=0]{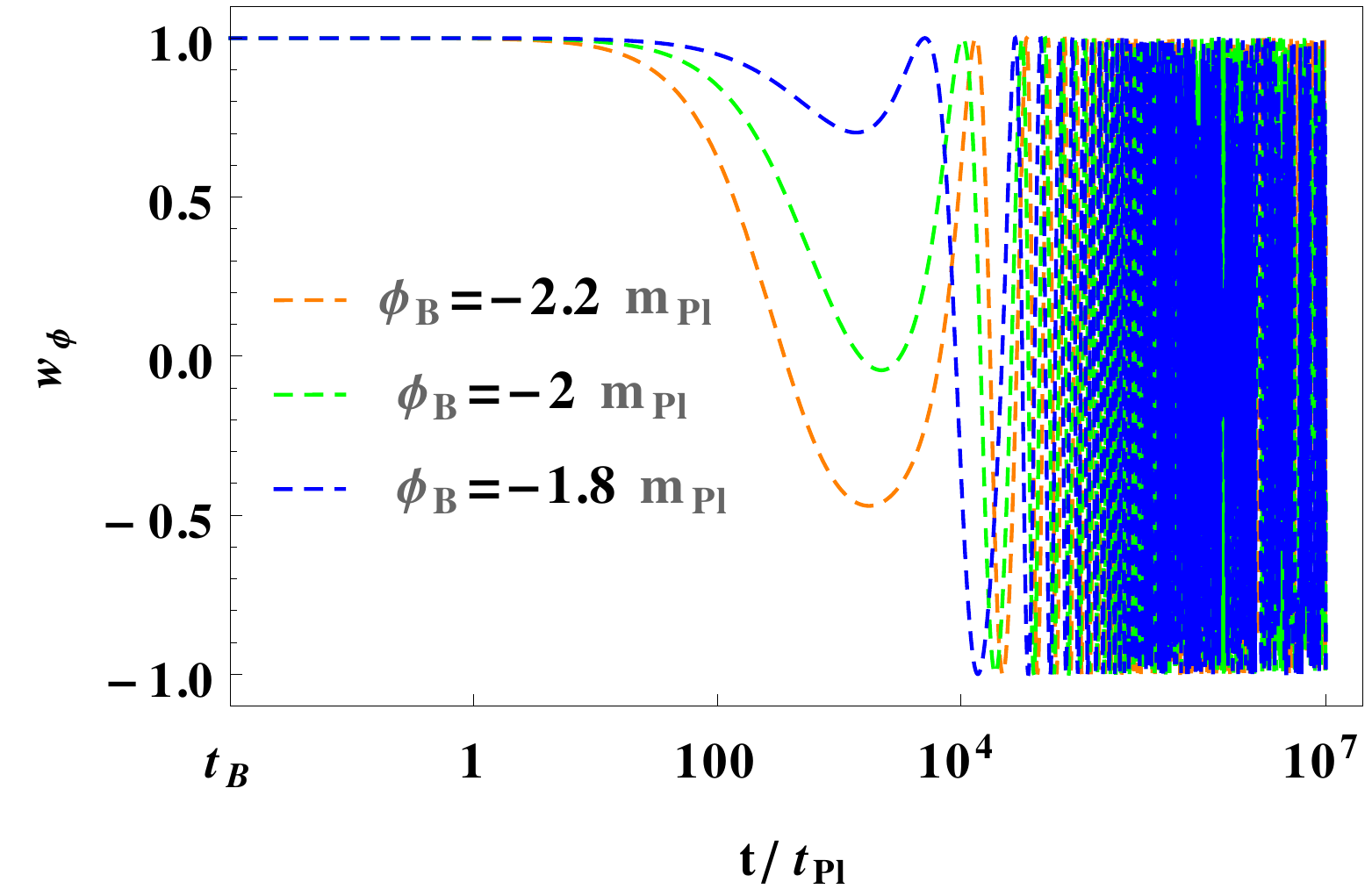}} &
{\includegraphics[width=2.0in,height=1.6in,angle=0]{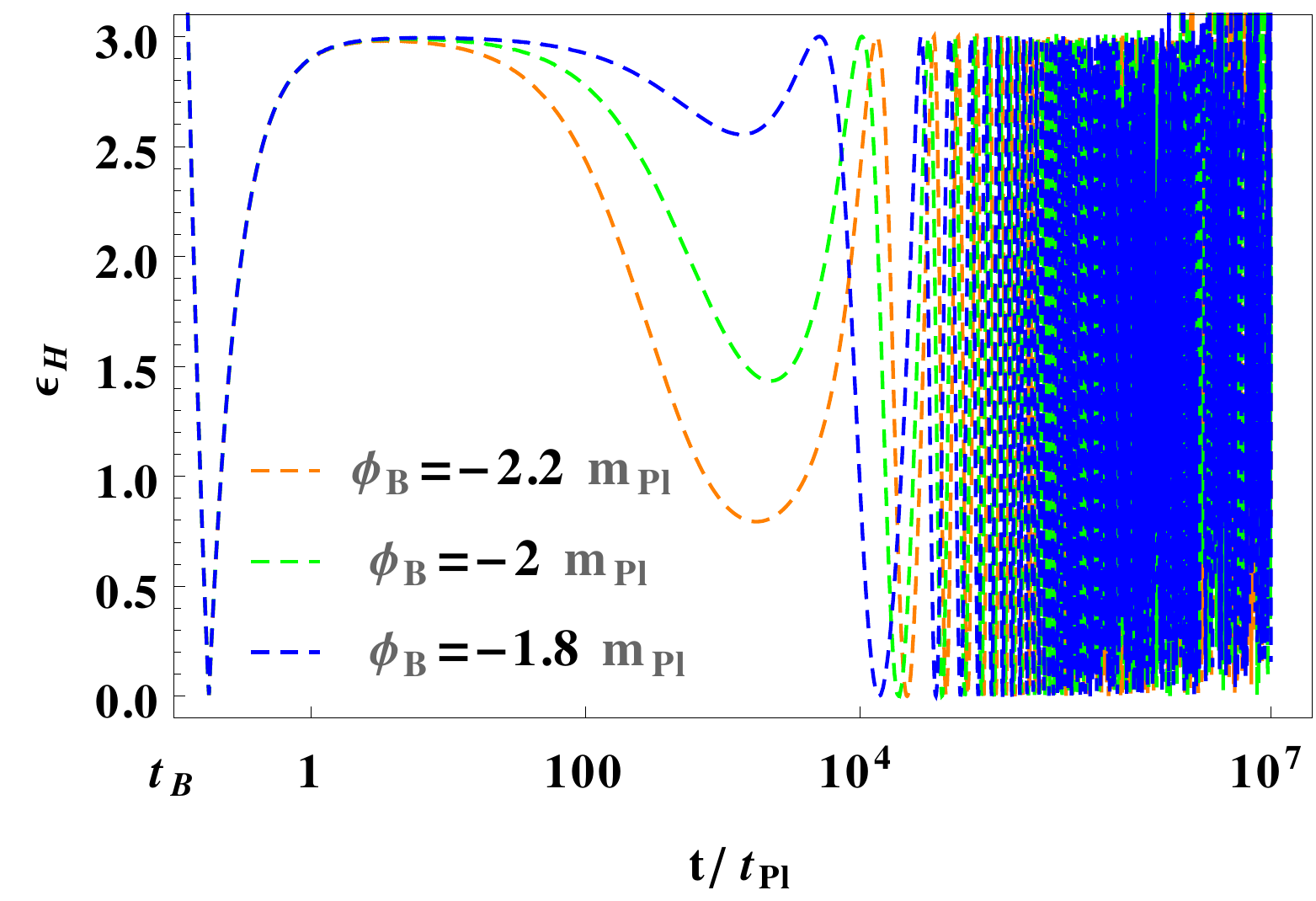}}
 \\ 
{\includegraphics[width=2.1in,height=1.6in,angle=0]{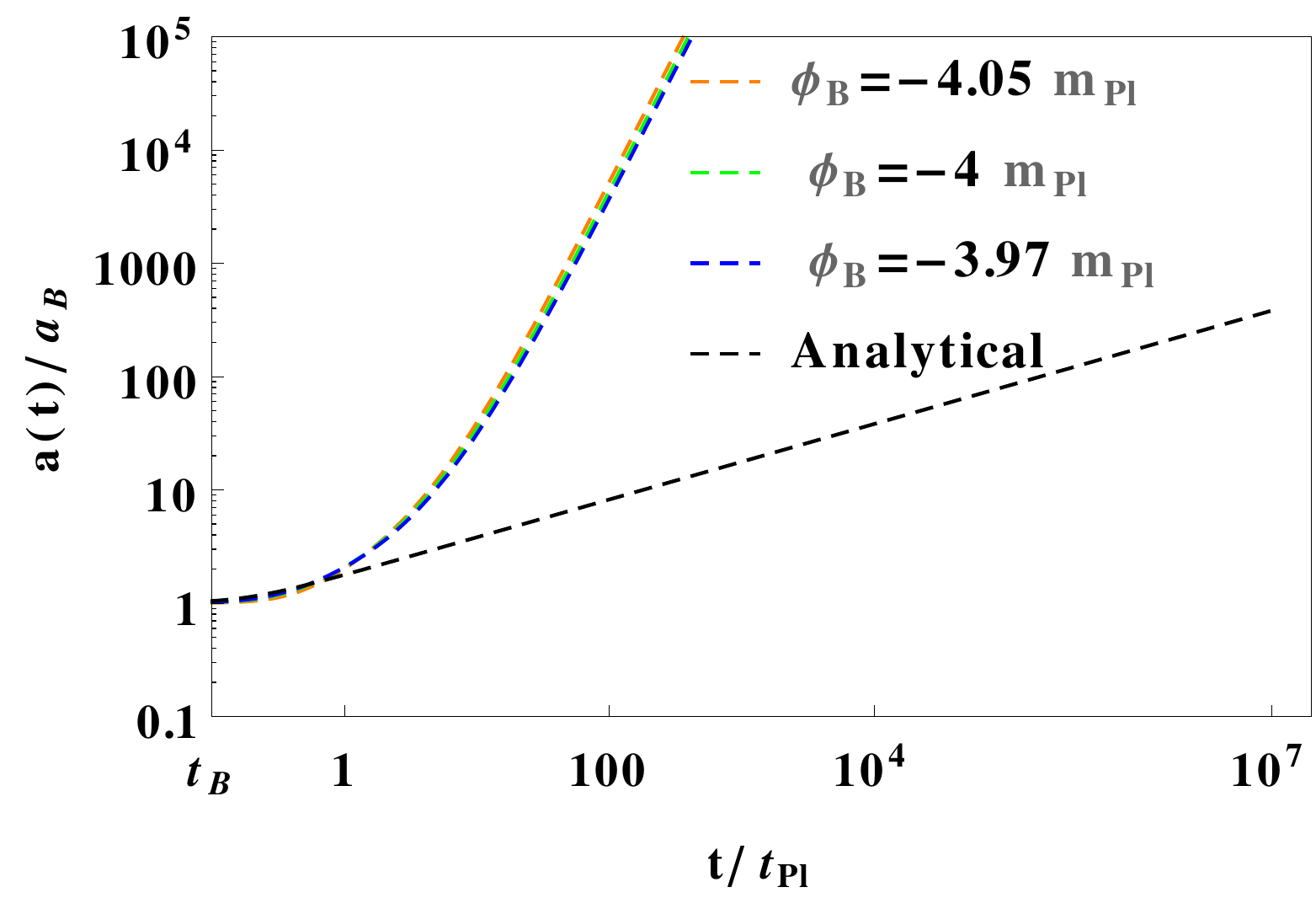}} & 
{\includegraphics[width=2.1in,height=1.6in,angle=0]{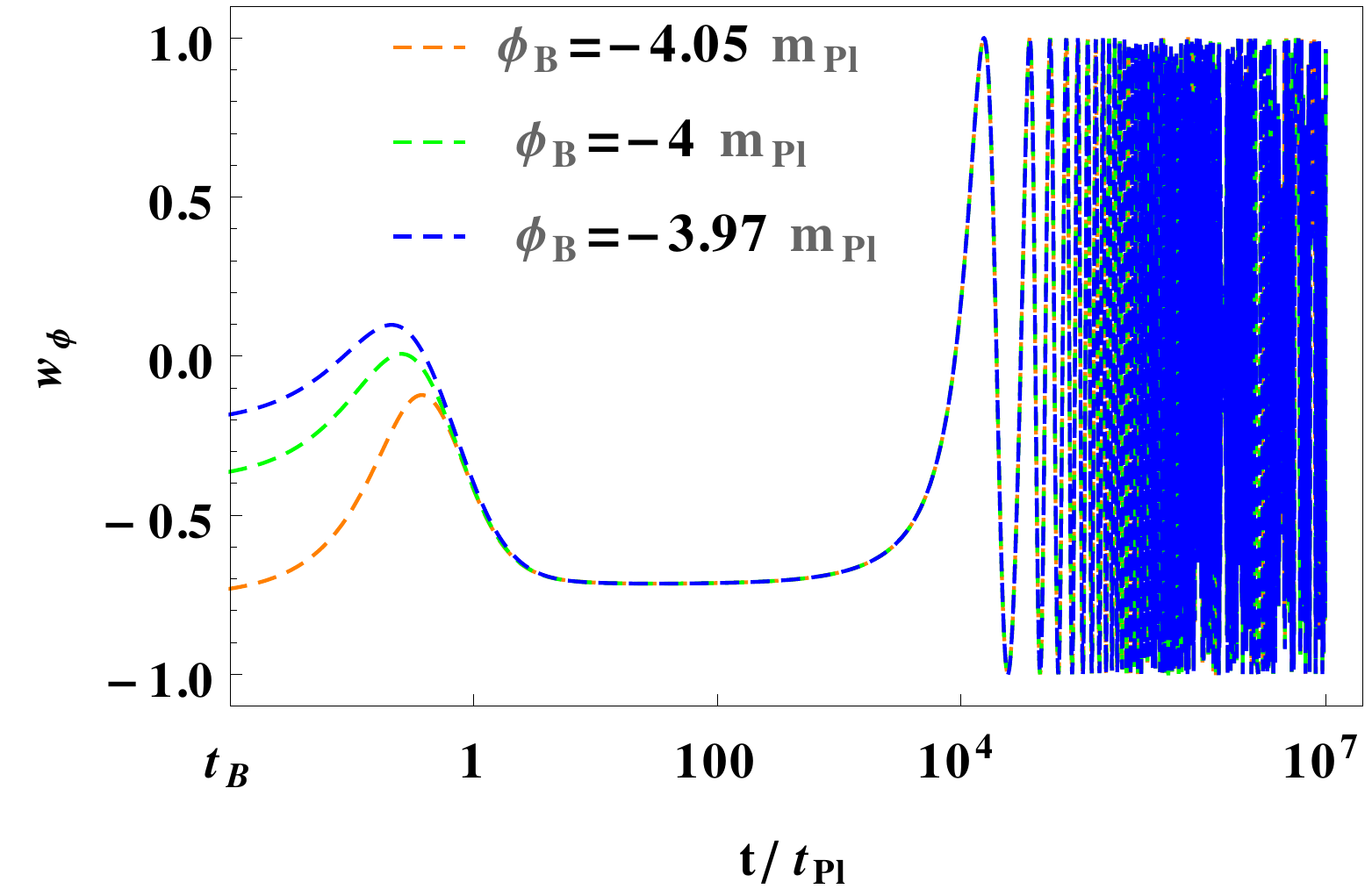}} & 
{\includegraphics[width=2.0in,height=1.6in,angle=0]{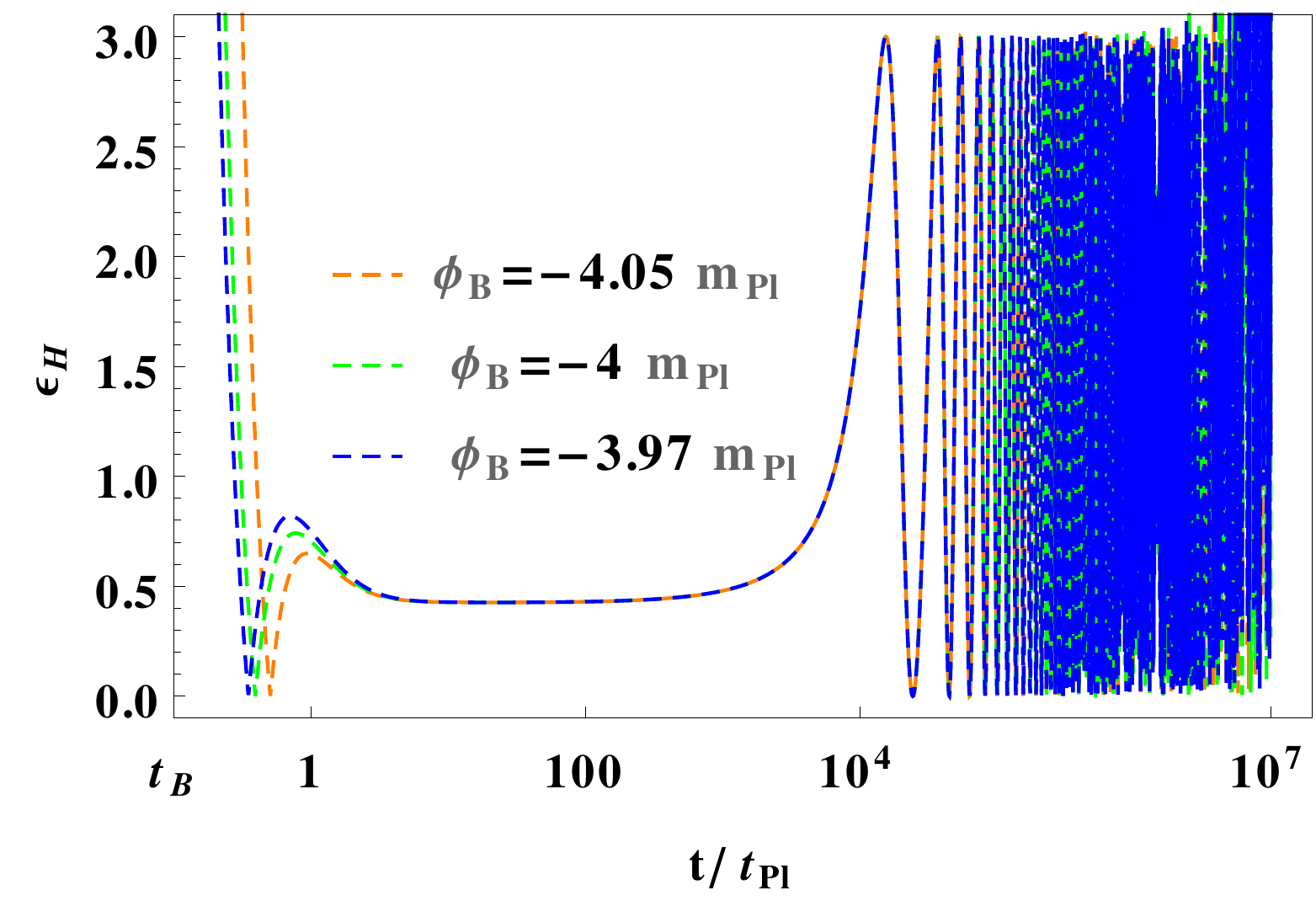}} 
\end{tabular}
\end{center}
\caption{ In this figure it is plotted for $\alpha-$model with $n=2$ (\ref{eq:n2pot}) and $\dot{\phi}_B>0$. Only top (KED, except a subset) panels  gives rise to a slow-roll inflation,
 while the middle (a subset of KED) and bottom (PED case) panels do not. We use $\alpha=0.5 m_{Pl}^2$, $c=2.9 \times 10^{-4}m_{Pl}$, and $m_{Pl}=1$.}
\label{fig:n2alpha05_dphp}
\end{figure*}
\begin{figure*}[tbp]
\begin{center}
\begin{tabular}{ccc}
{\includegraphics[width=2.1in,height=1.65in,angle=0]{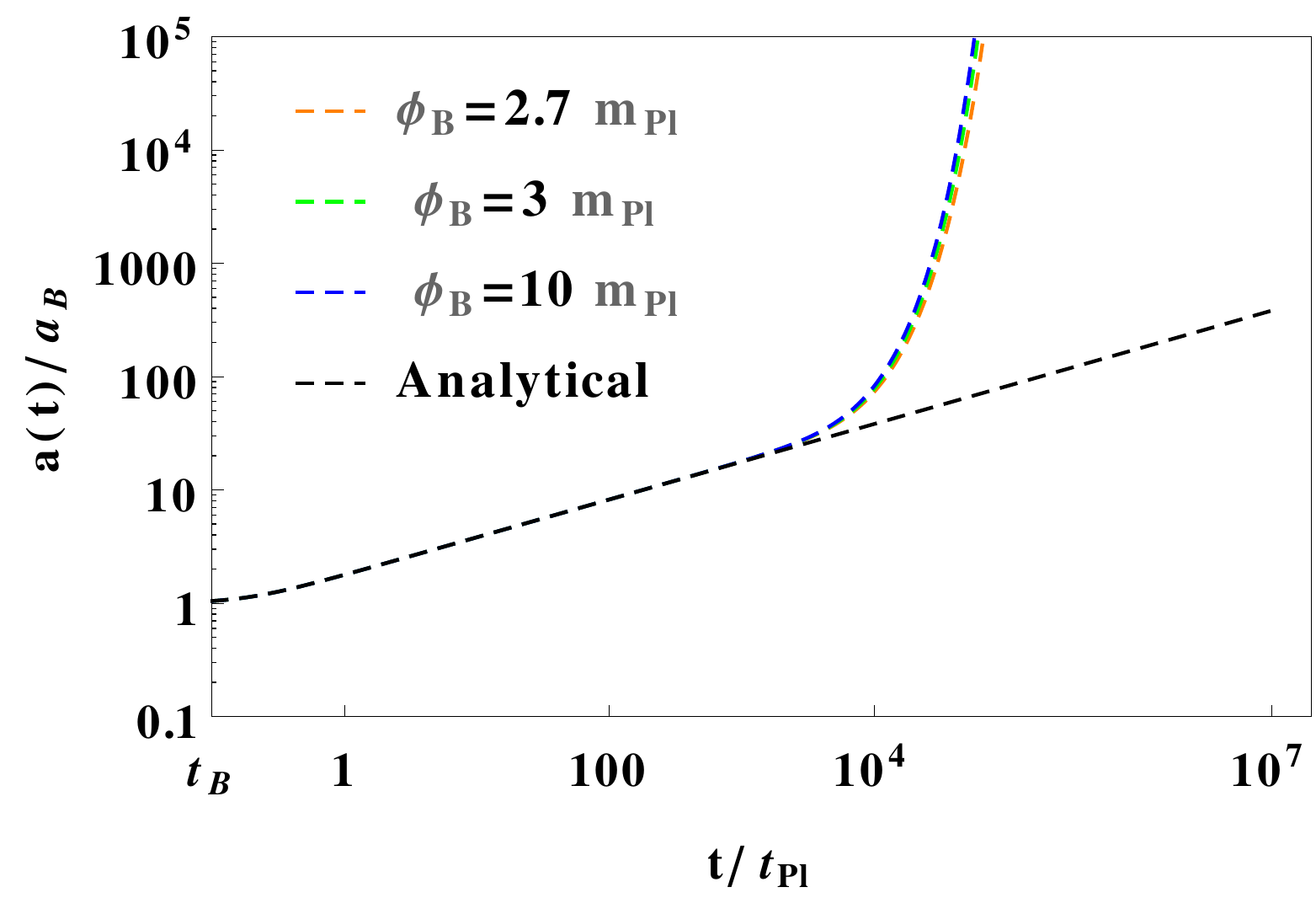}} &
{\includegraphics[width=2.1in,height=1.6in,angle=0]{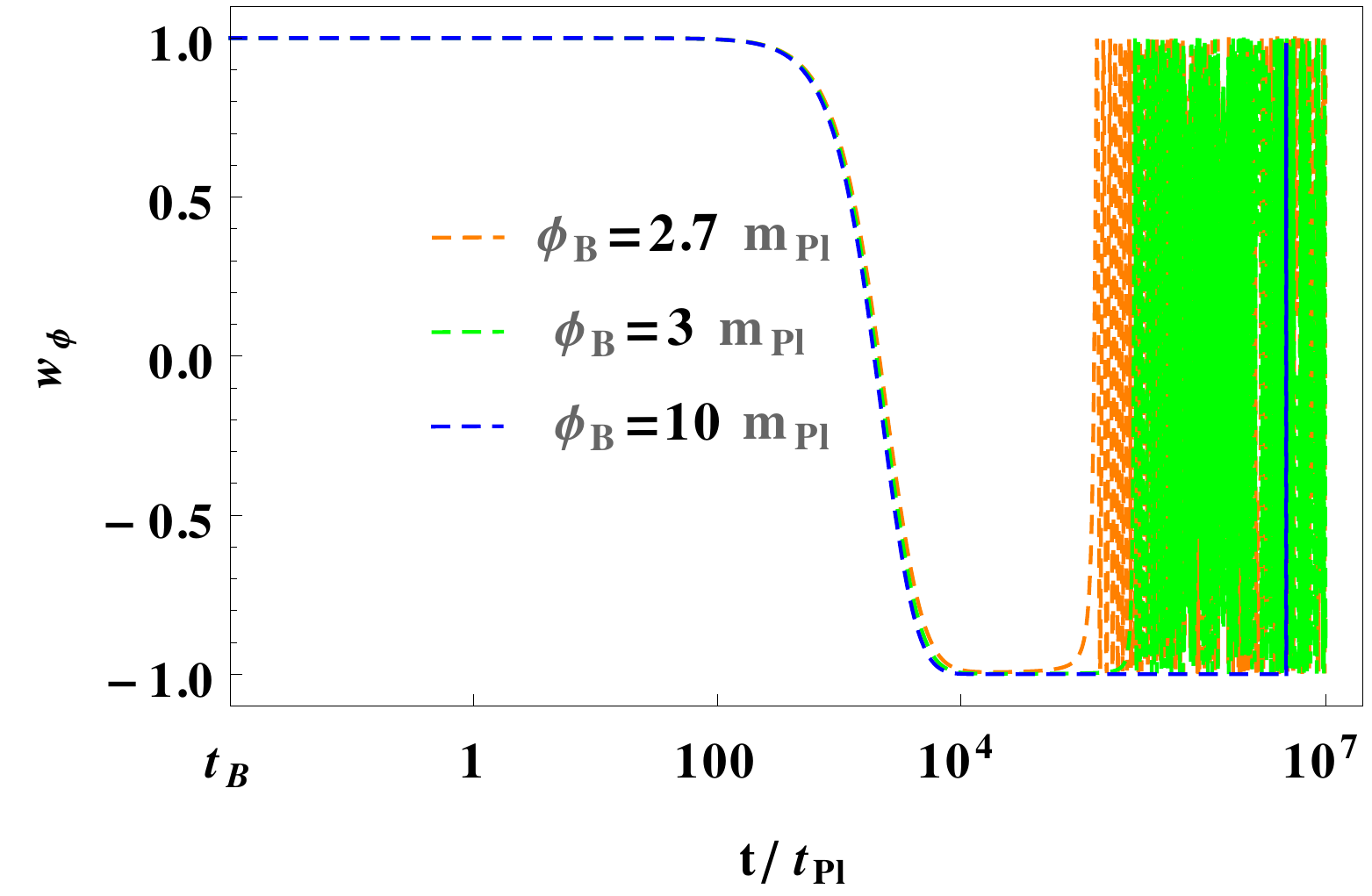}} &
{\includegraphics[width=2.0in,height=1.6in,angle=0]{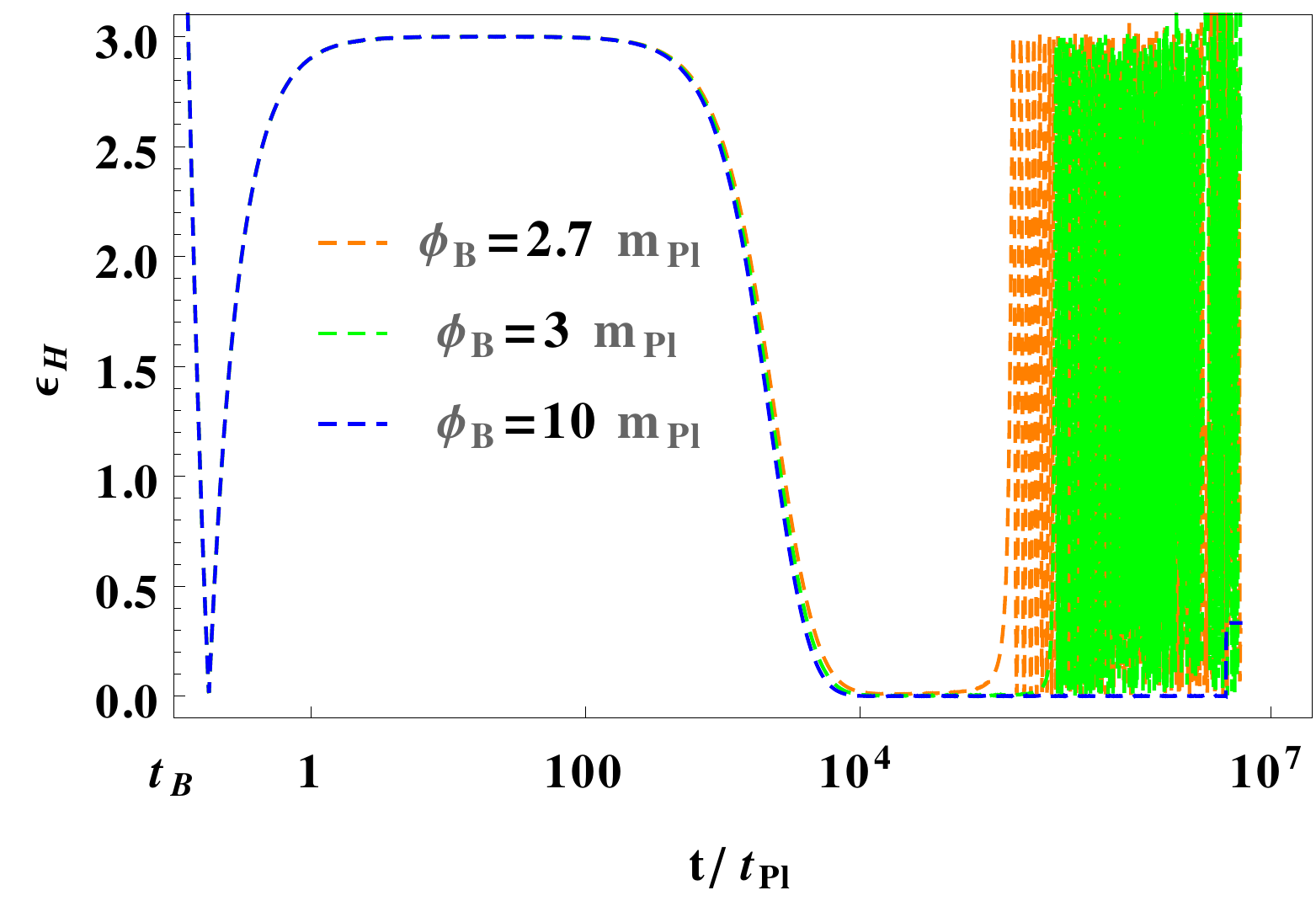}}
 \\
{\includegraphics[width=2.1in,height=1.65in,angle=0]{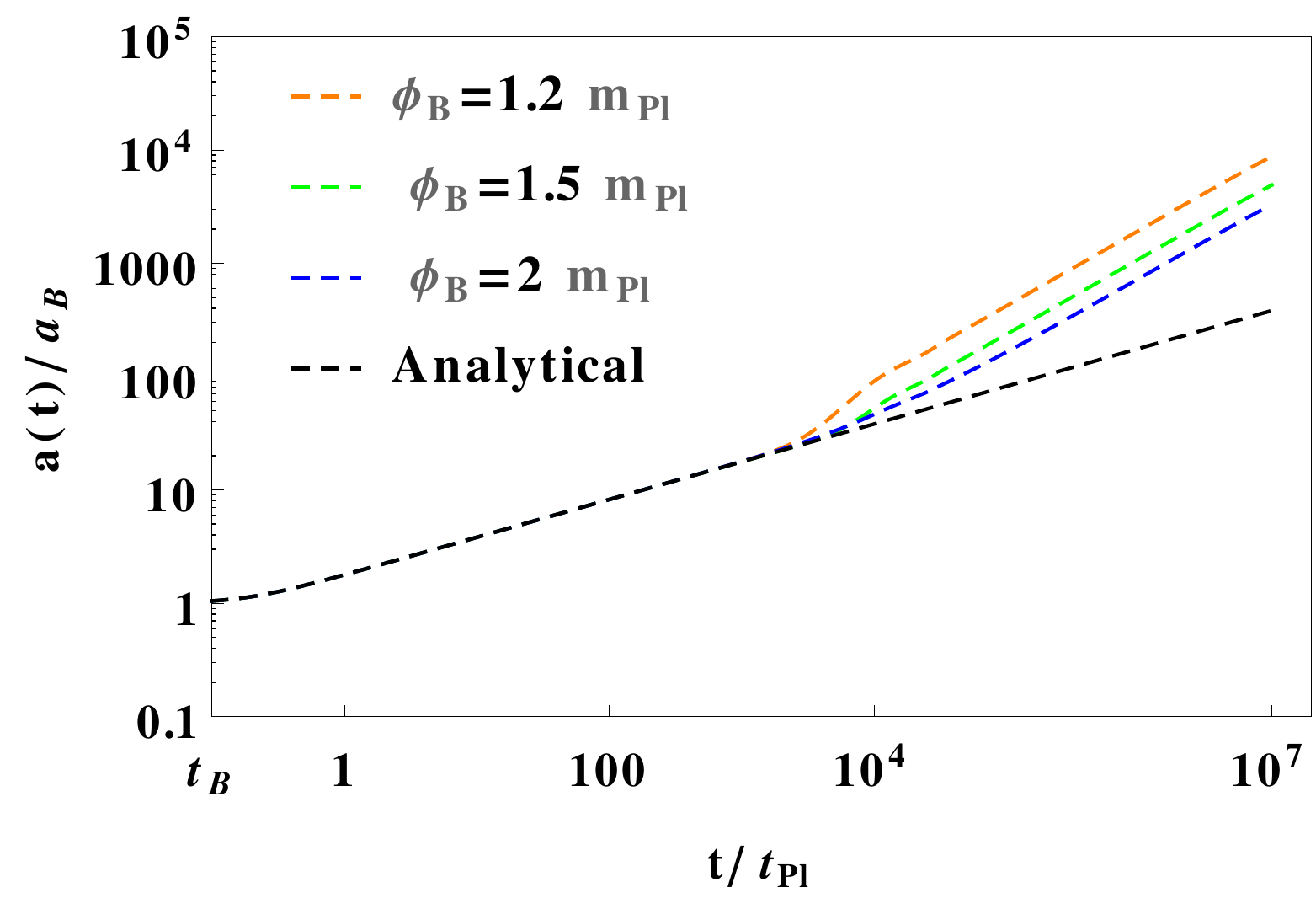}} &
{\includegraphics[width=2.1in,height=1.6in,angle=0]{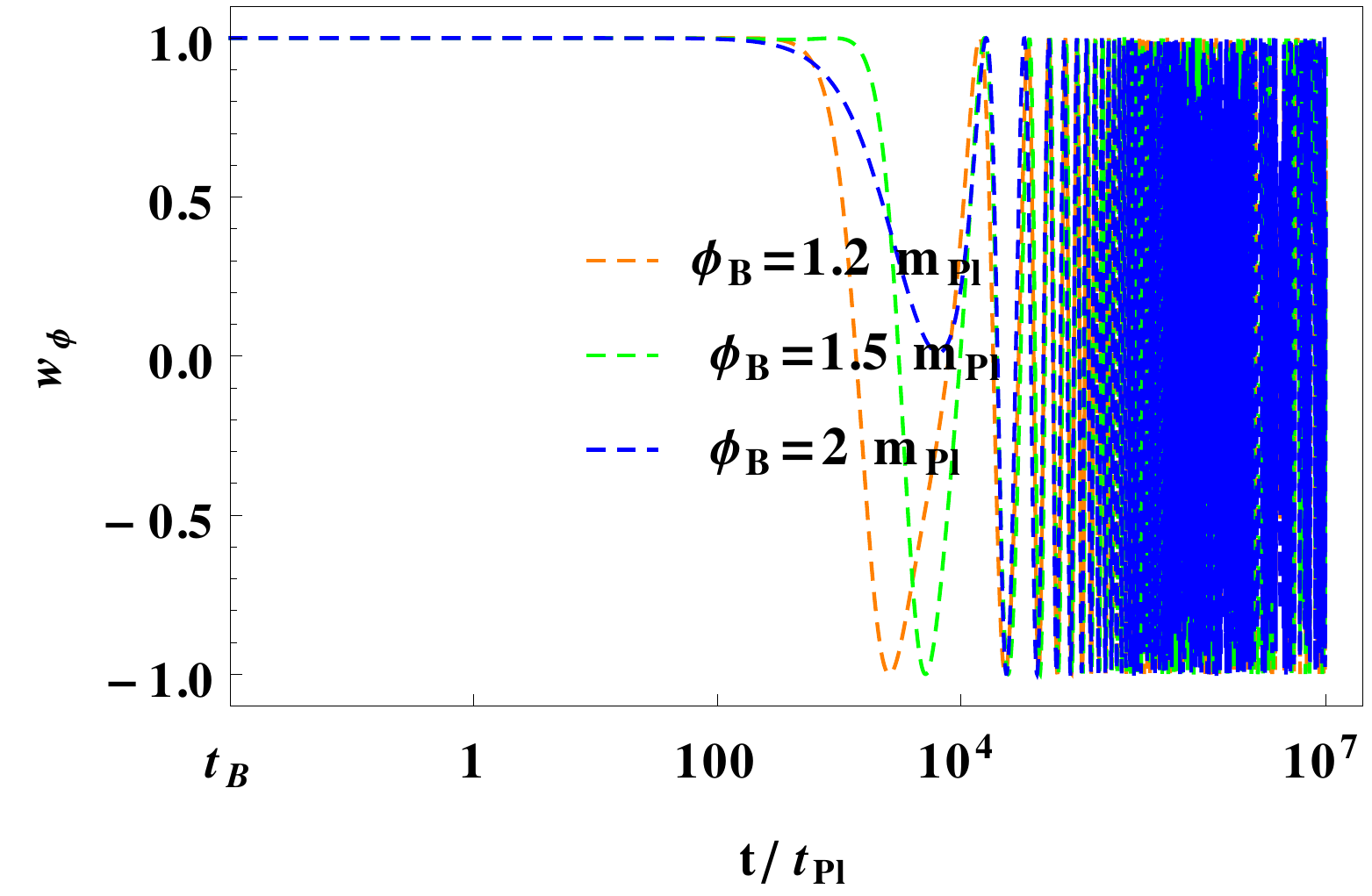}} &
{\includegraphics[width=2.0in,height=1.6in,angle=0]{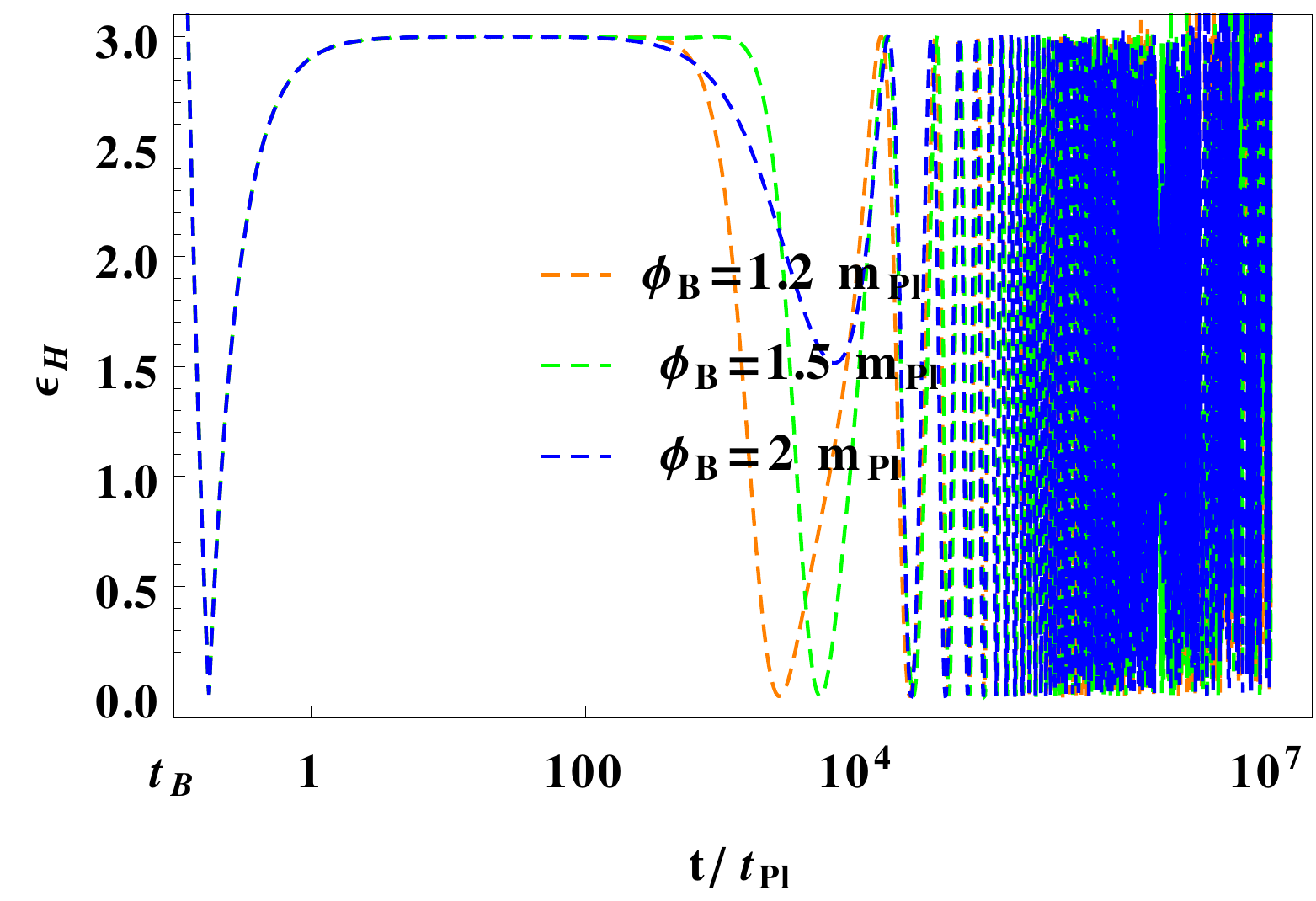}}
 \\ 
{\includegraphics[width=2.1in,height=1.6in,angle=0]{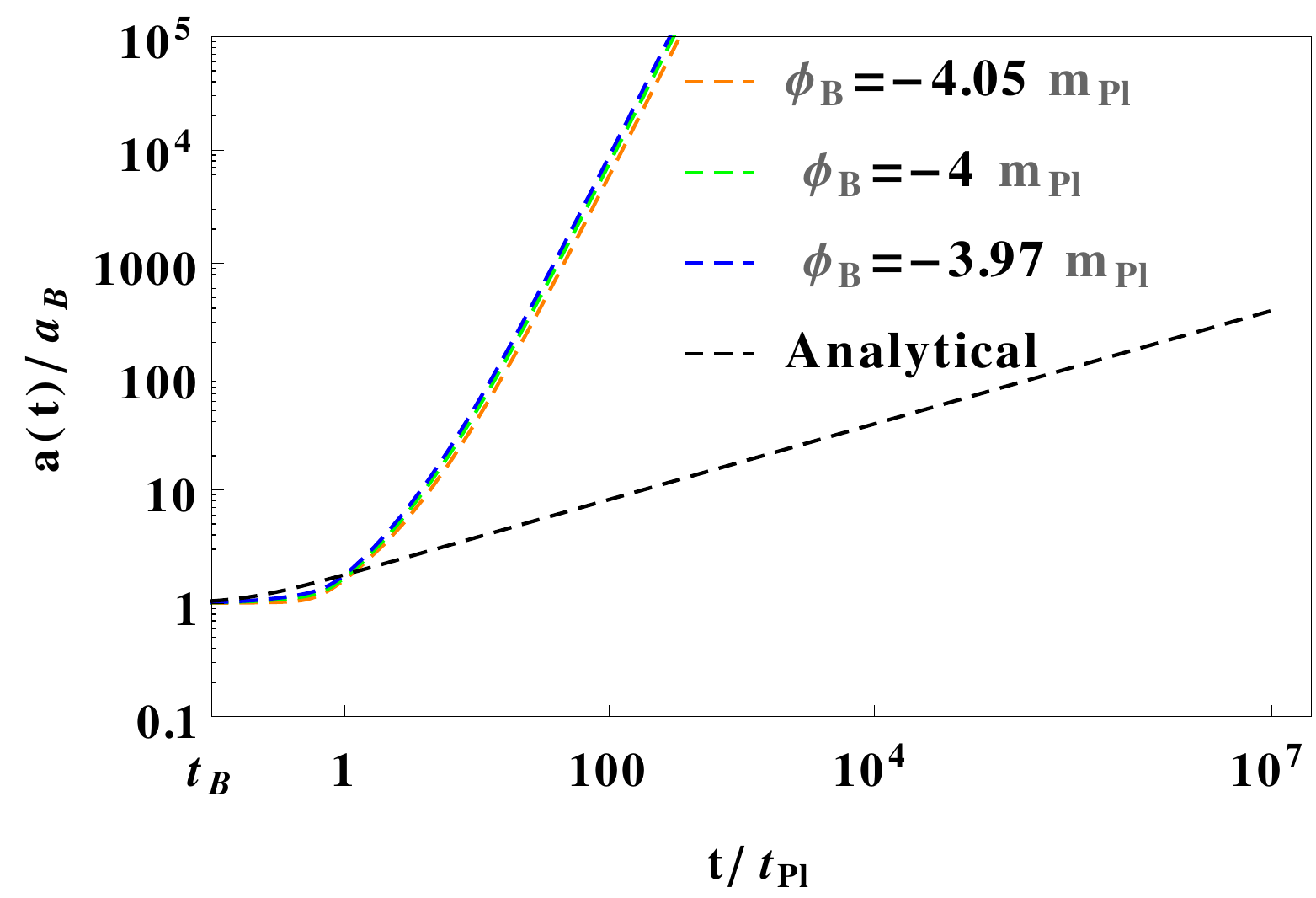}} & 
{\includegraphics[width=2.1in,height=1.6in,angle=0]{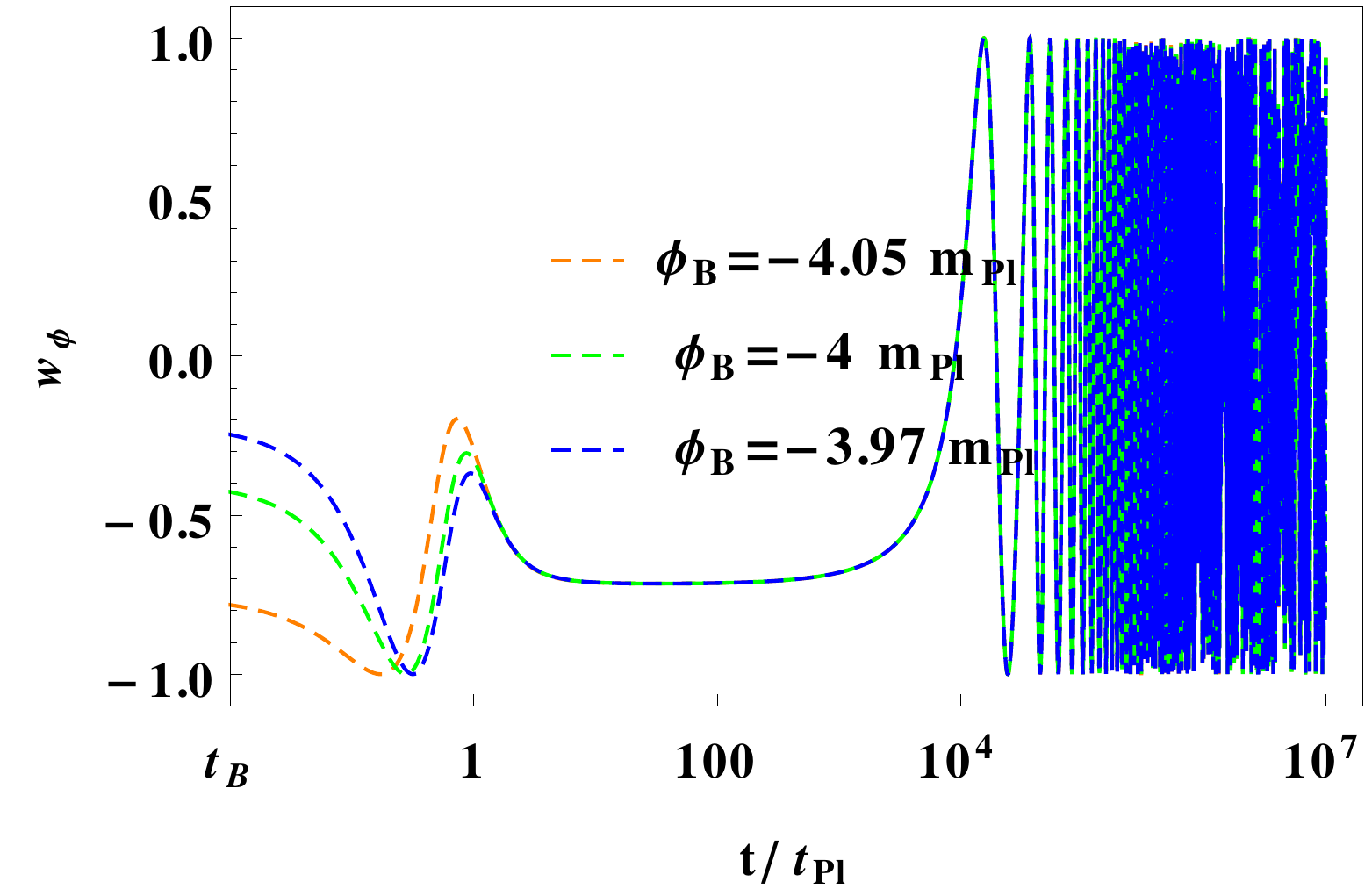}} & 
{\includegraphics[width=2.0in,height=1.6in,angle=0]{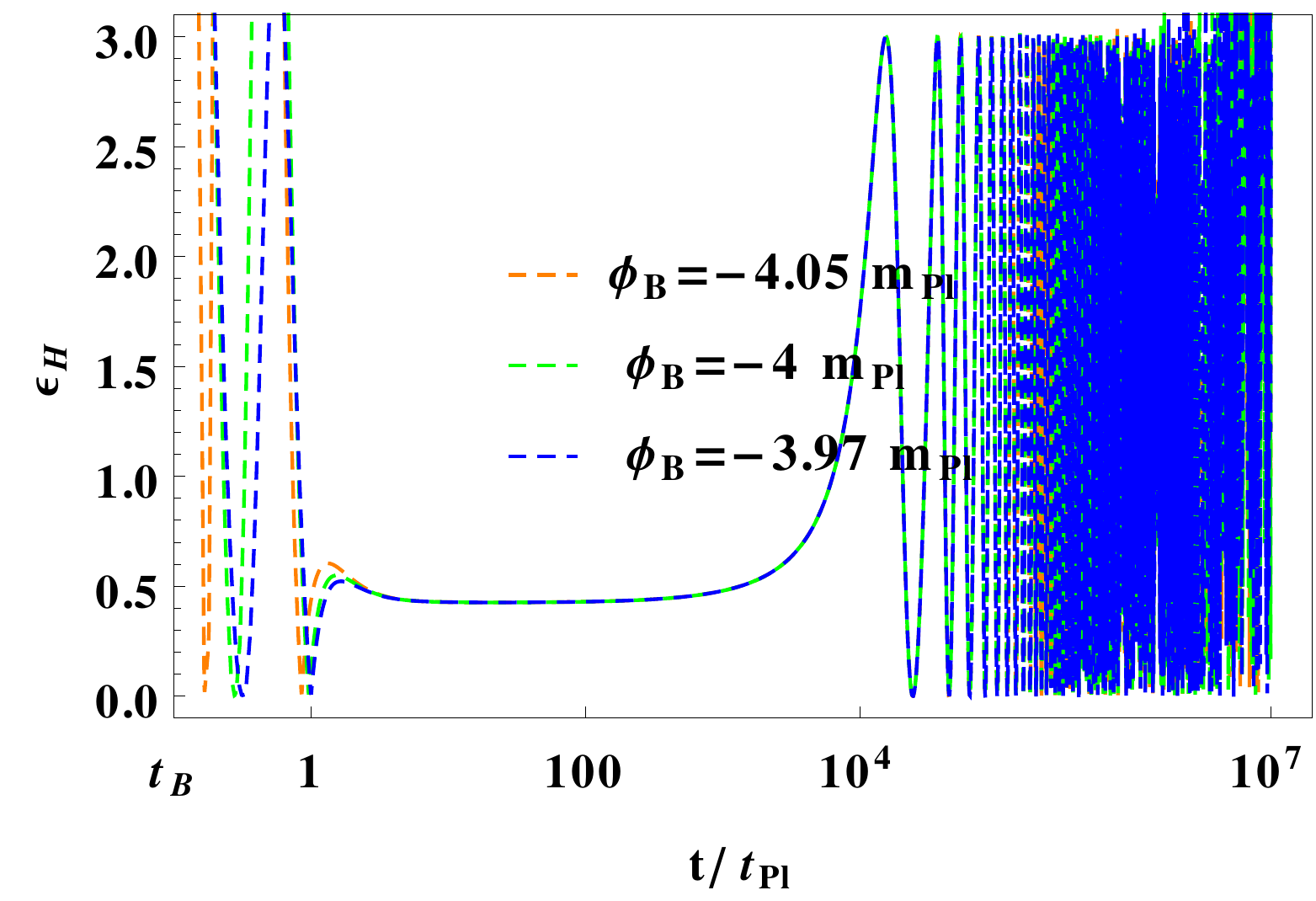}} 
\end{tabular}
\end{center}
\caption{This figure is similar to Fig. \ref{fig:n2alpha05_dphp} but with $\dot{\phi}_B<0$.}
\label{fig:n2alpha05_dphn}
\end{figure*}
\begin{figure*}[tbp]
\begin{center}
\begin{tabular}{ccc}
{\includegraphics[width=2.1in,height=1.65in,angle=0]{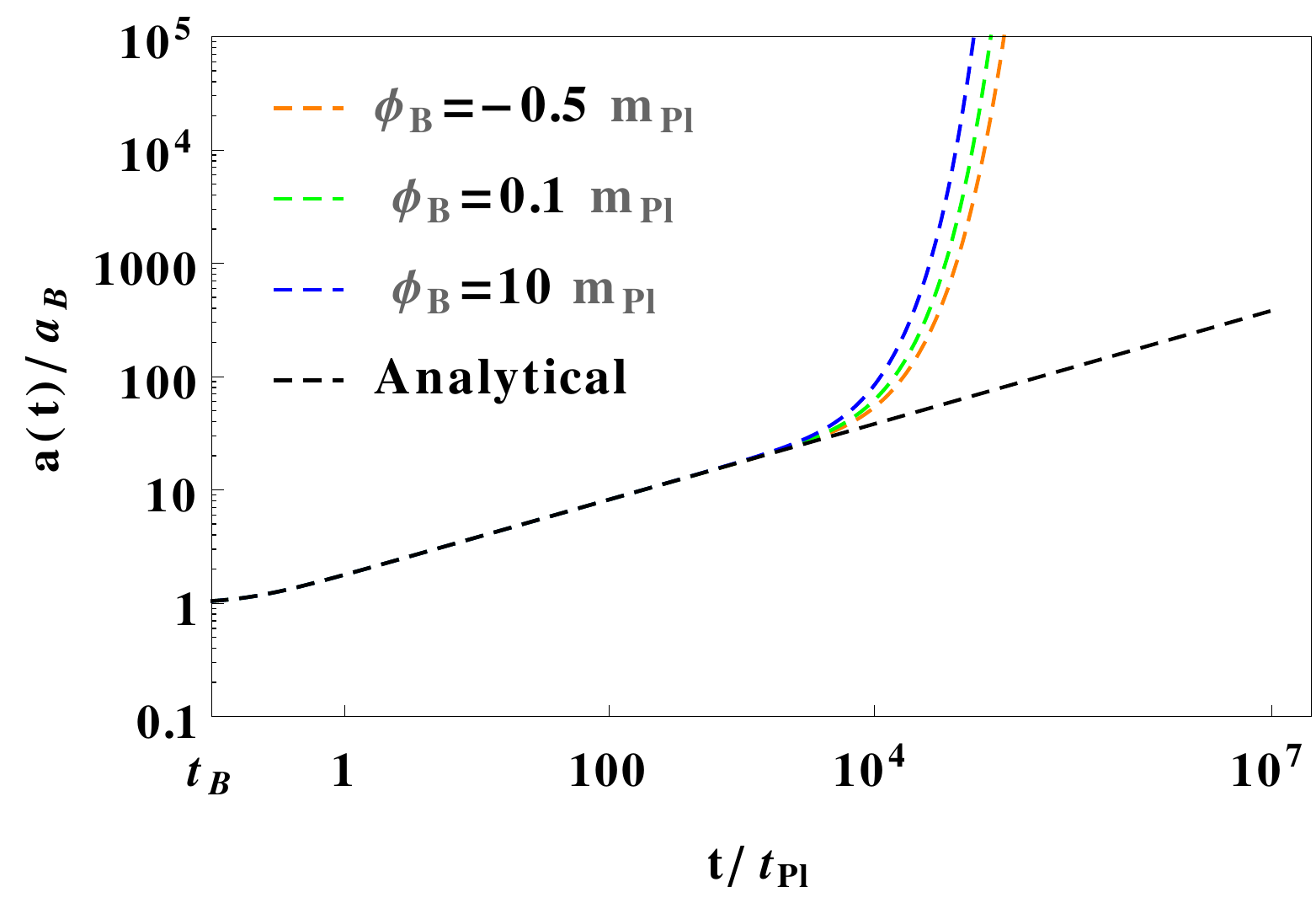}} &
{\includegraphics[width=2.1in,height=1.6in,angle=0]{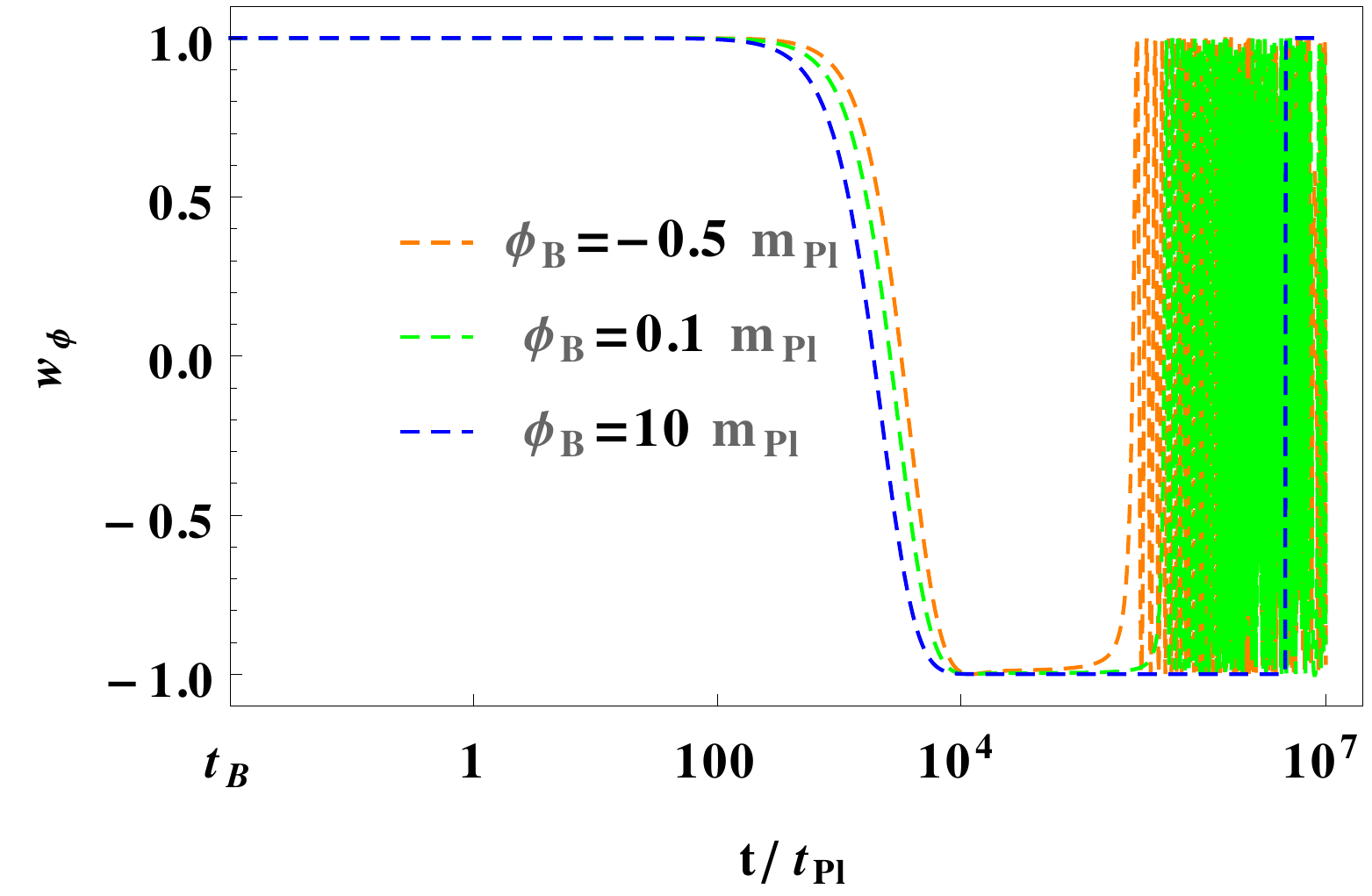}} &
{\includegraphics[width=2.0in,height=1.6in,angle=0]{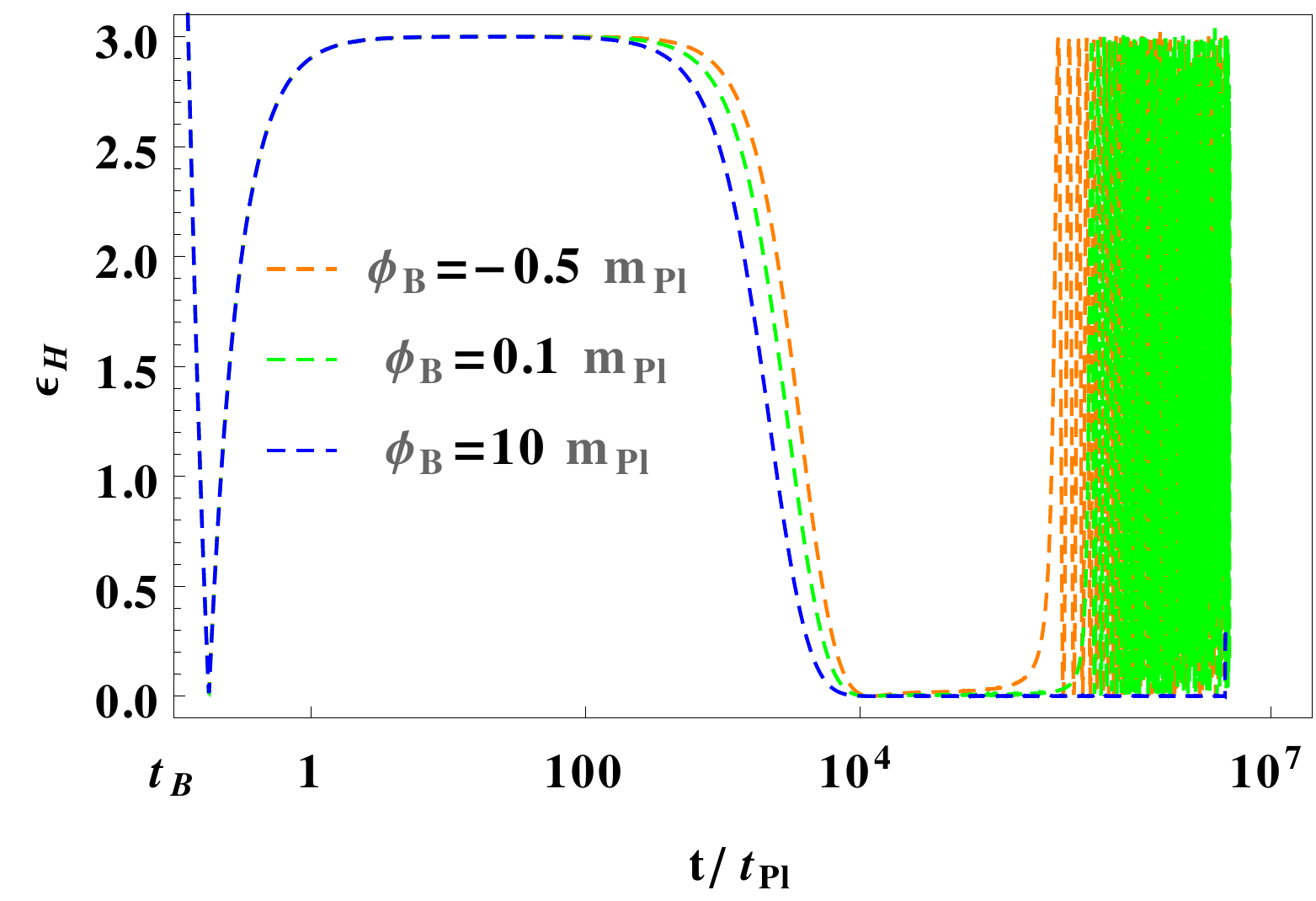}}
\\ 
{\includegraphics[width=2.1in,height=1.65in,angle=0]{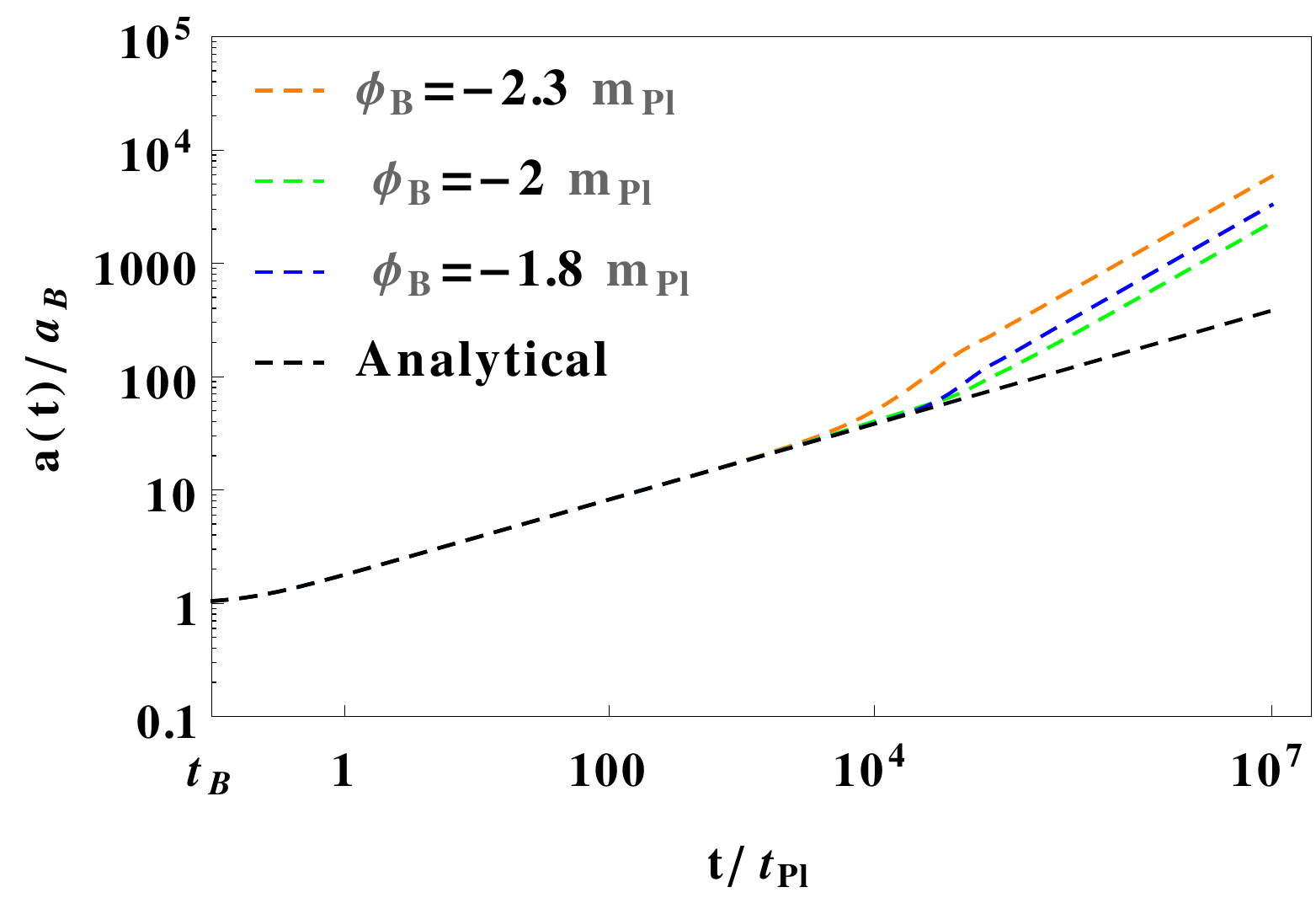}} &
{\includegraphics[width=2.1in,height=1.6in,angle=0]{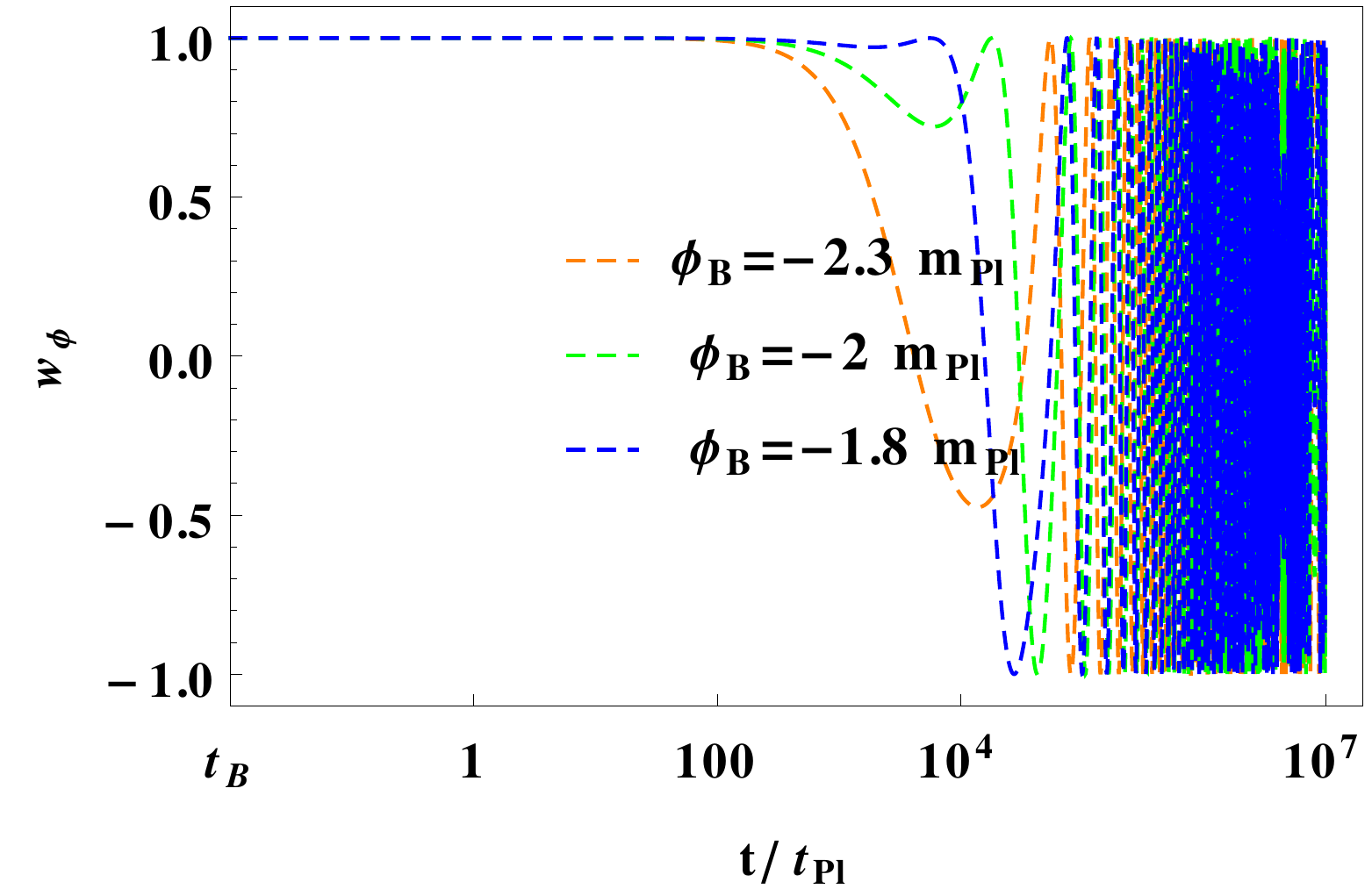}} &
{\includegraphics[width=2.0in,height=1.6in,angle=0]{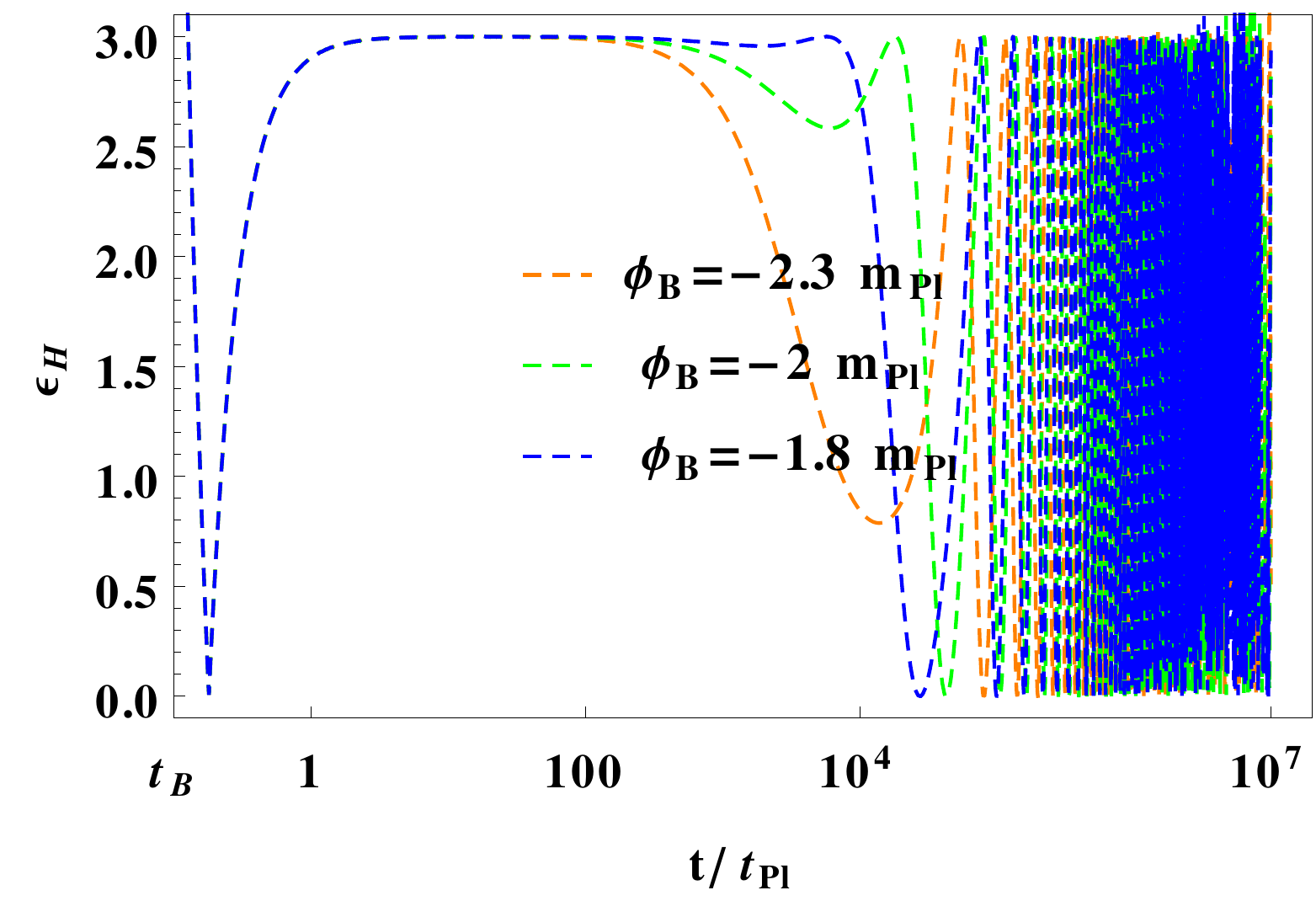}}
\\ 
{\includegraphics[width=2.1in,height=1.6in,angle=0]{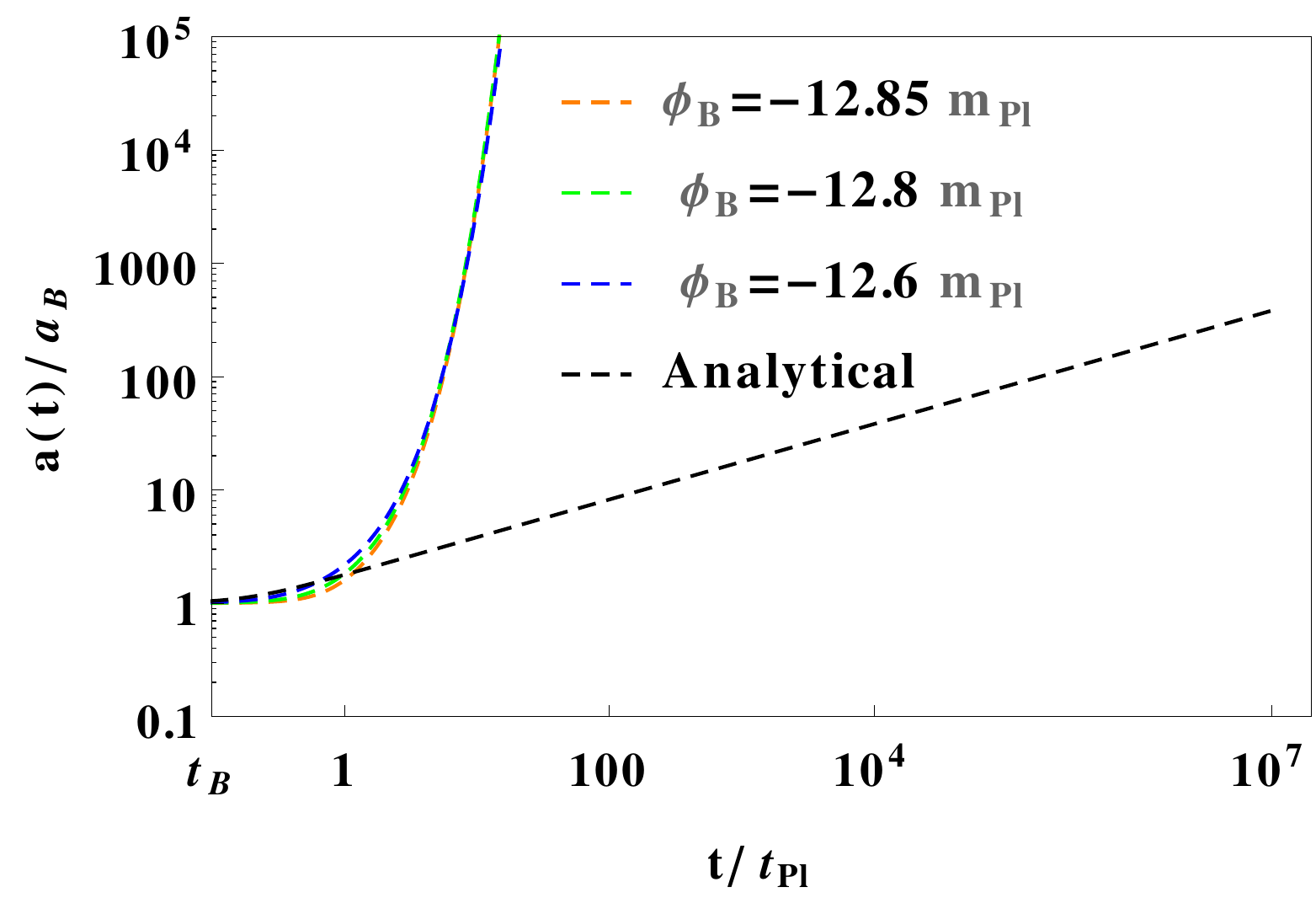}} & 
{\includegraphics[width=2.1in,height=1.6in,angle=0]{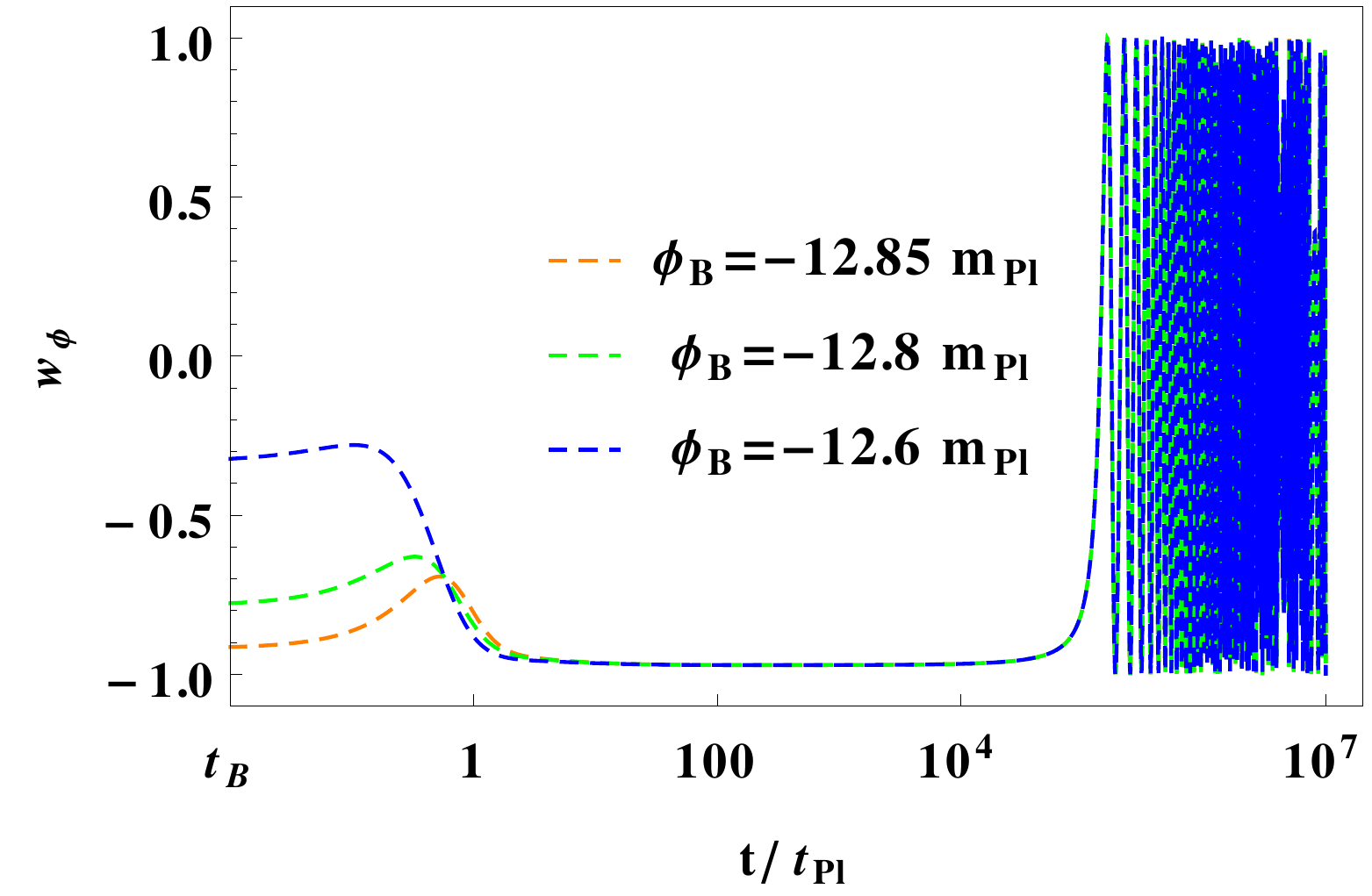}} & 
{\includegraphics[width=2.0in,height=1.6in,angle=0]{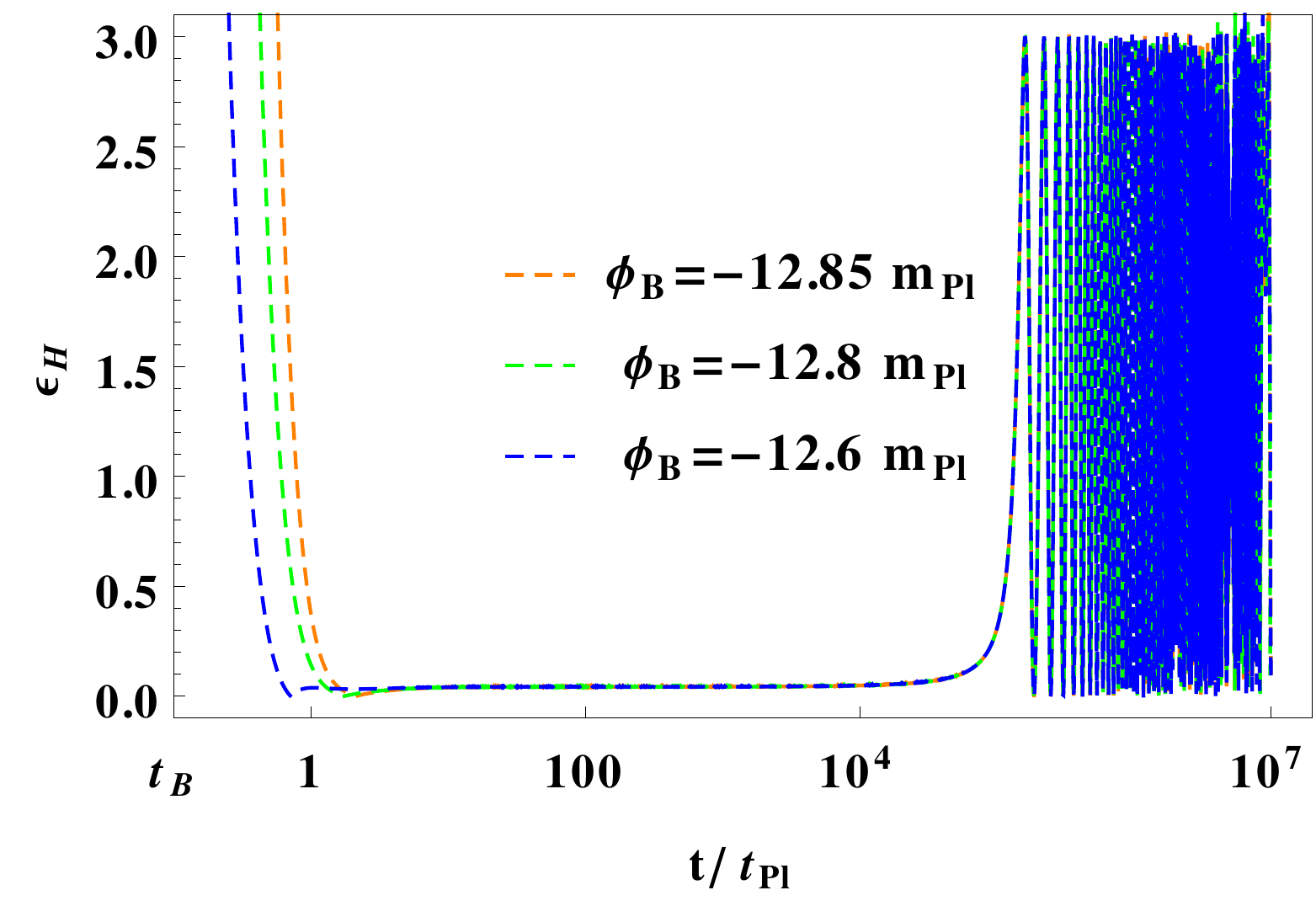}} 
\end{tabular}
\end{center}
\caption{ This figure shows the numerical evolution of $a(t)$, $w(\phi)$ and $\epsilon_H $ for $\alpha-$model with $n=2$ (\ref{eq:n2pot}) and $\dot{\phi}_B>0$. Top (slow-roll inflation) and middle (no slow-roll inflation) panels demonstrate the evolution of KED and a subset of KED initial conditions whereas bottom (slow-roll inflation) ones are for PED. Here, we take $\alpha=5 m_{Pl}^2$, $c=9.4 \times 10^{-5}m_{Pl}$, and $m_{Pl}=1$.}
\label{fig:n2alpha5_dphp}
\end{figure*}
\begin{figure*}[tbp]
\begin{center}
\begin{tabular}{ccc}
{\includegraphics[width=2.1in,height=1.65in,angle=0]{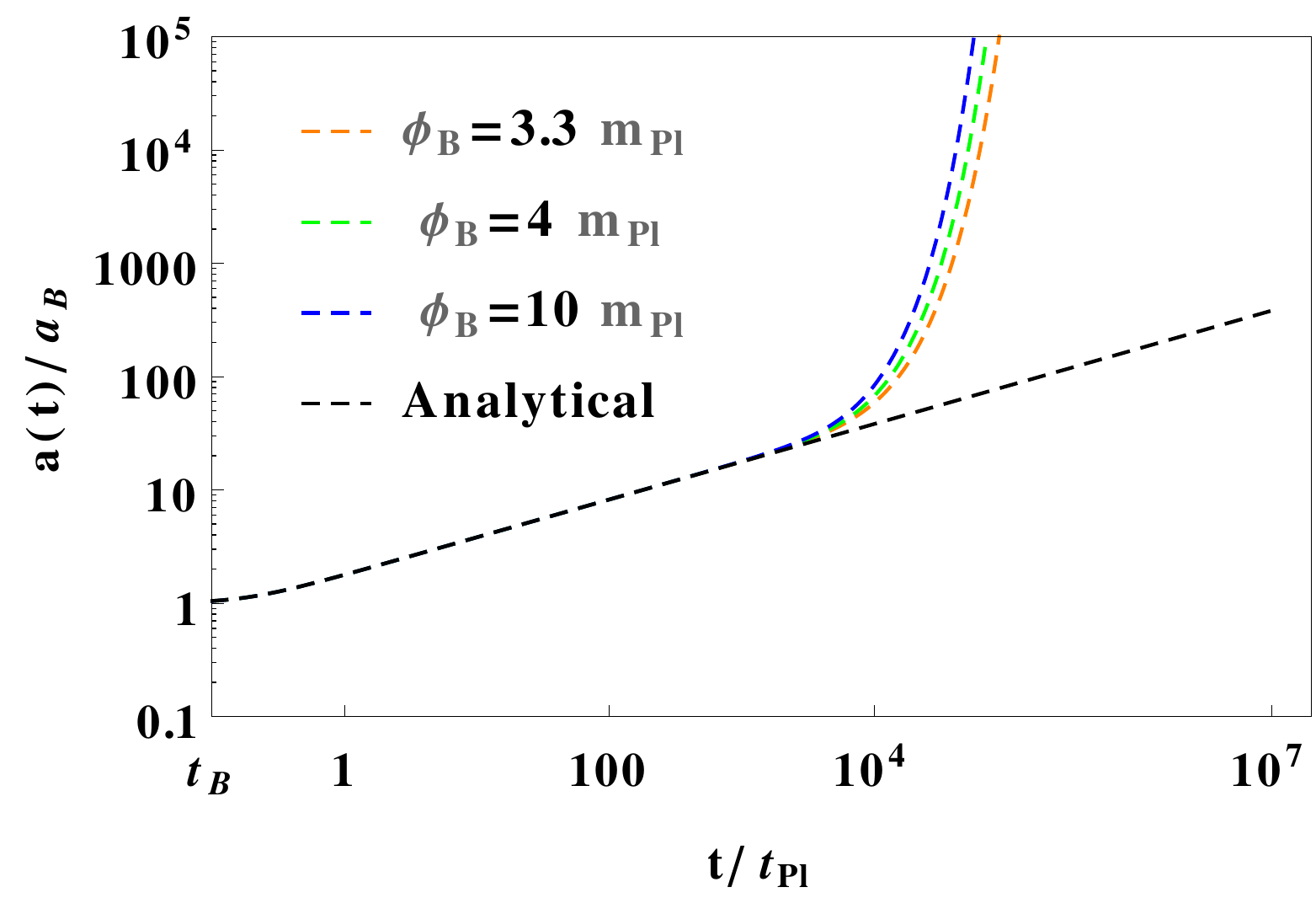}} &
{\includegraphics[width=2.1in,height=1.6in,angle=0]{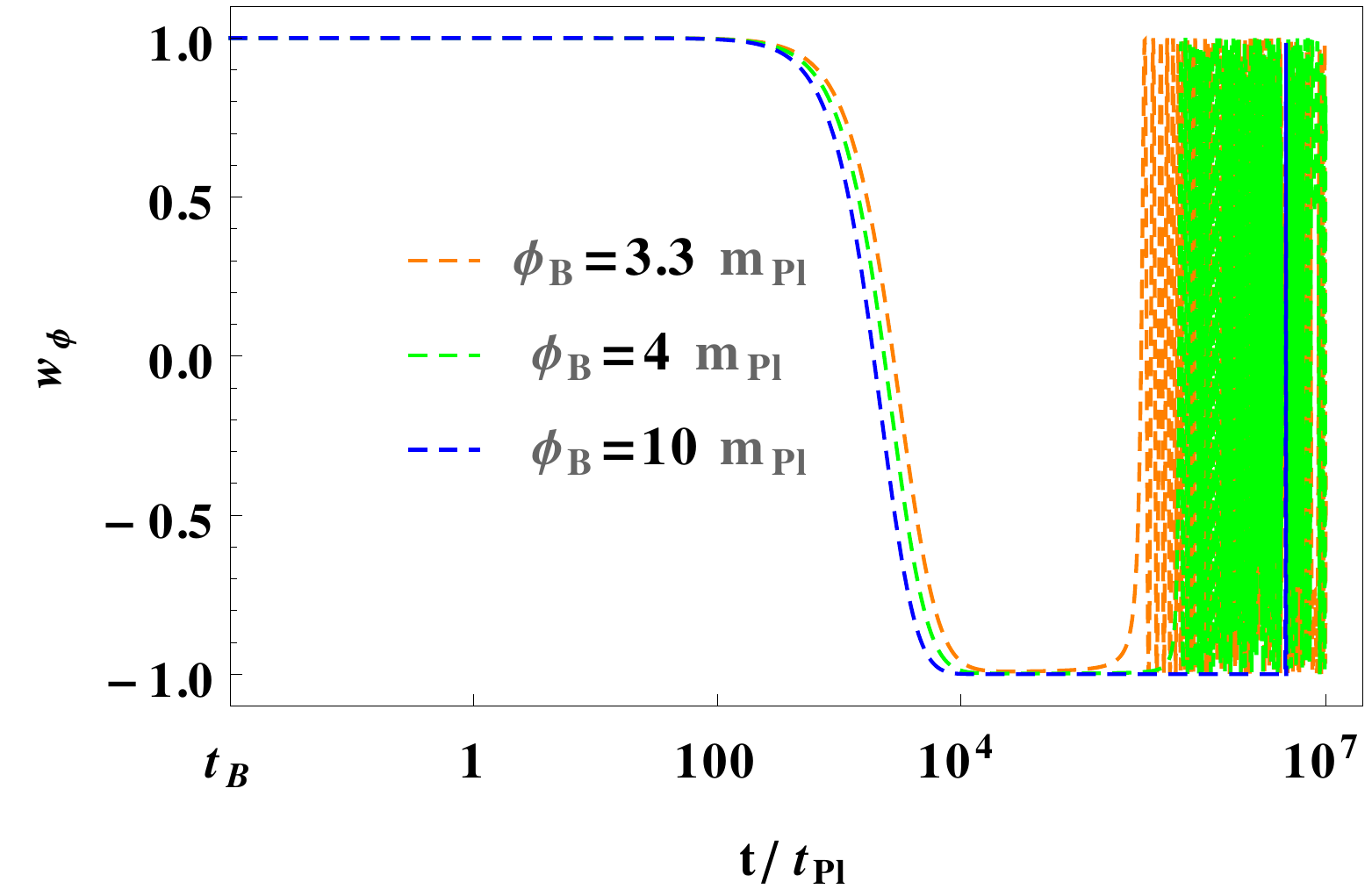}} &
{\includegraphics[width=2.0in,height=1.6in,angle=0]{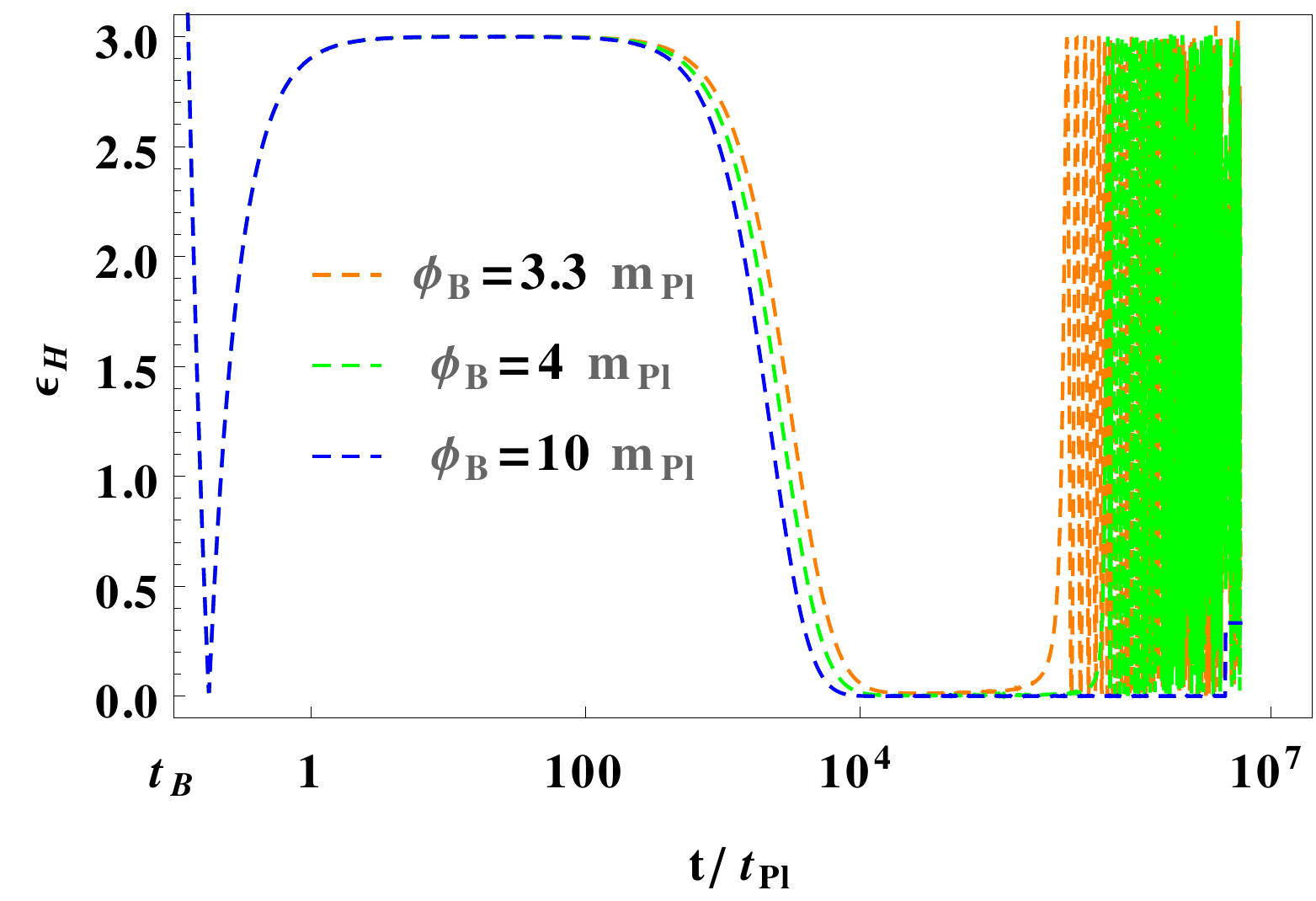}}
\\ 
{\includegraphics[width=2.1in,height=1.65in,angle=0]{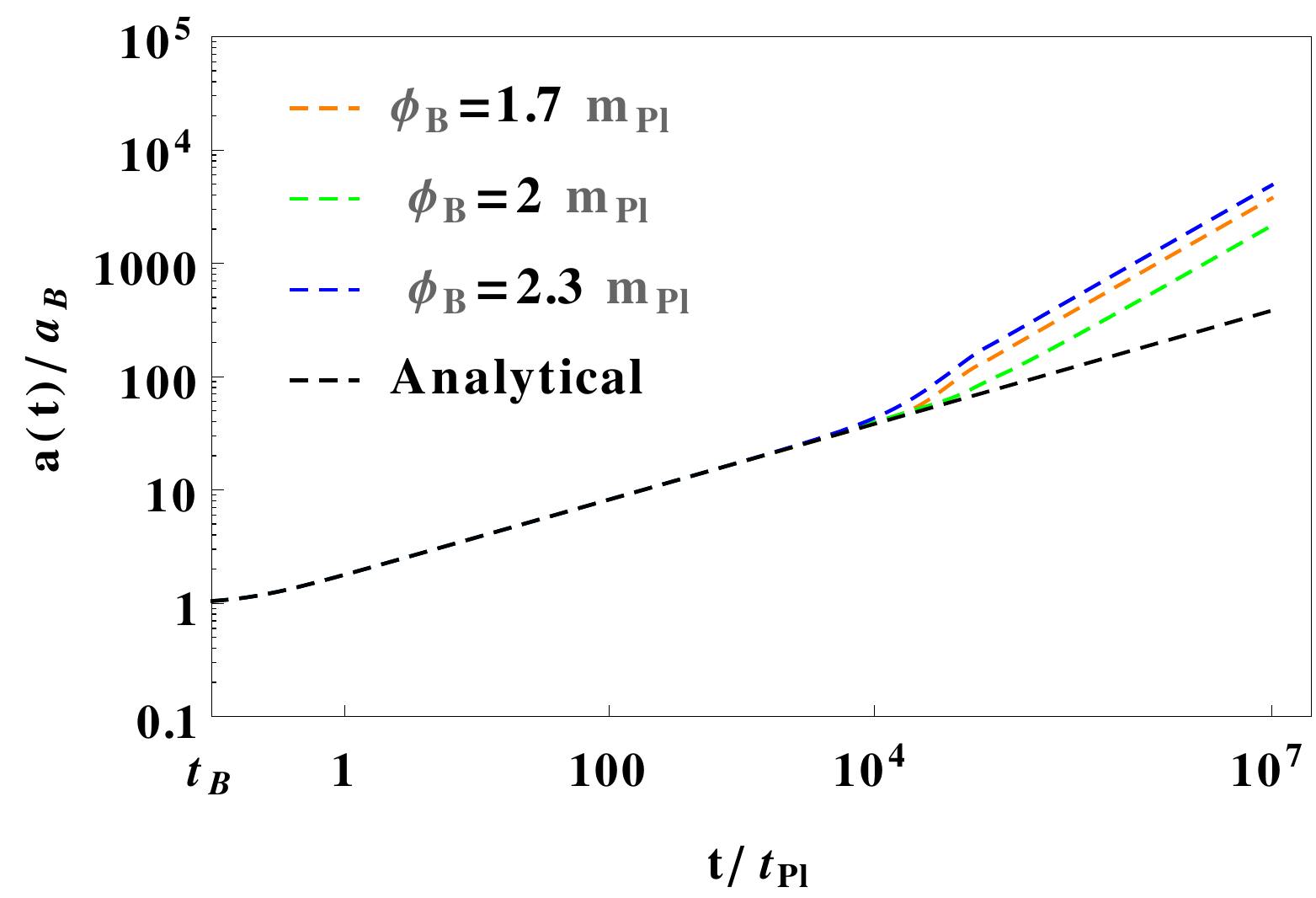}} &
{\includegraphics[width=2.1in,height=1.6in,angle=0]{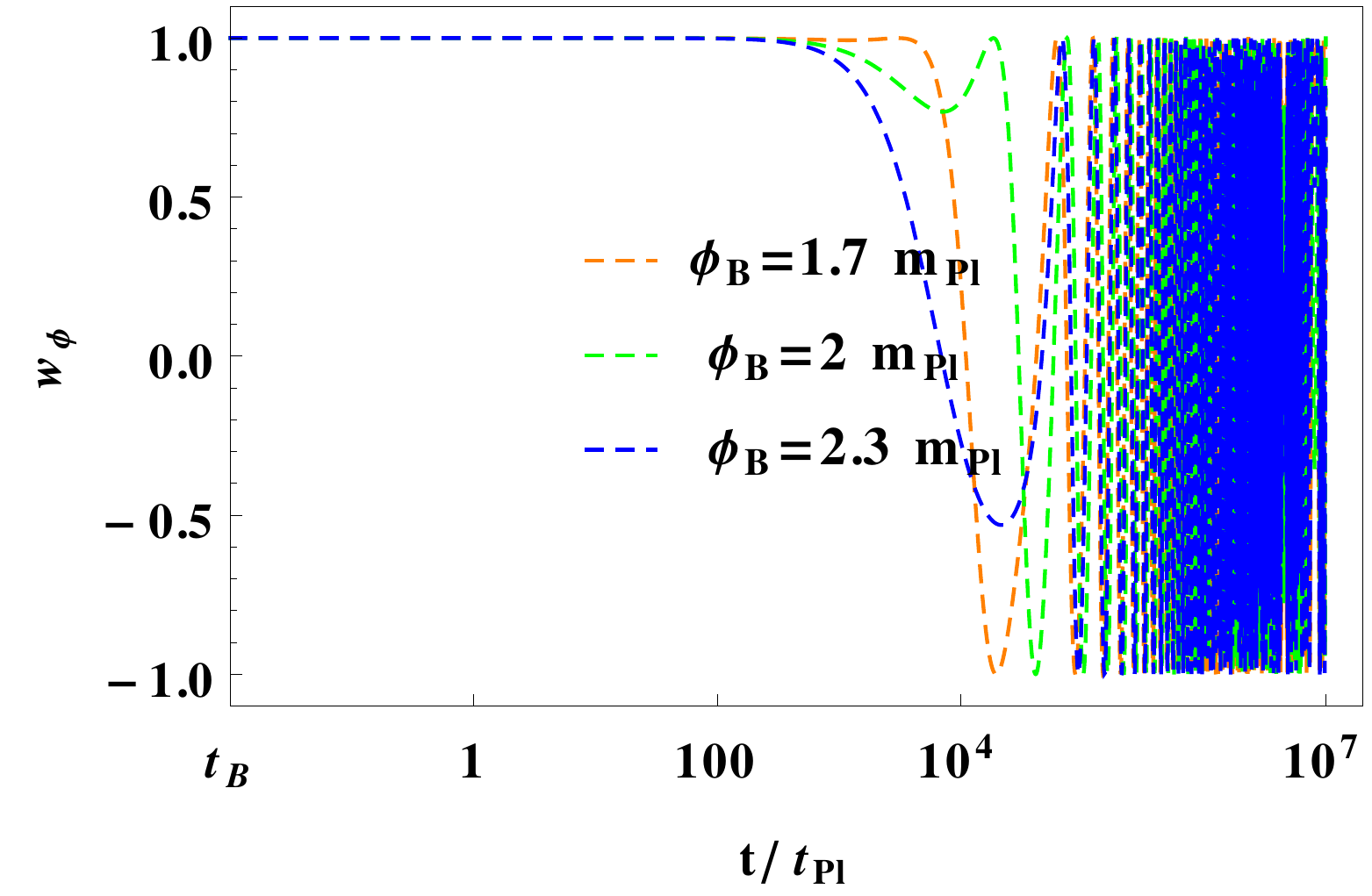}} &
{\includegraphics[width=2.0in,height=1.6in,angle=0]{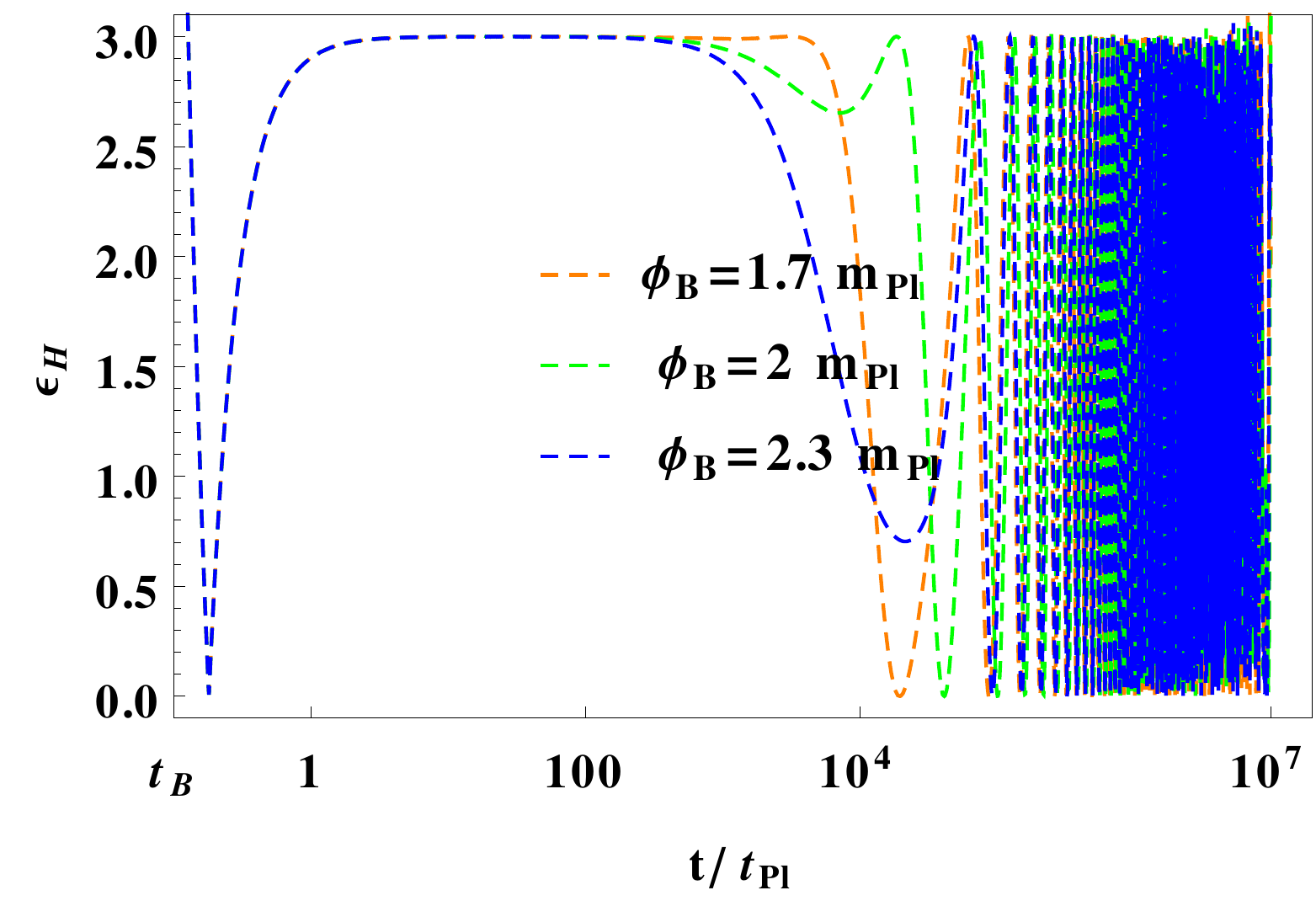}}
\\ 
{\includegraphics[width=2.1in,height=1.6in,angle=0]{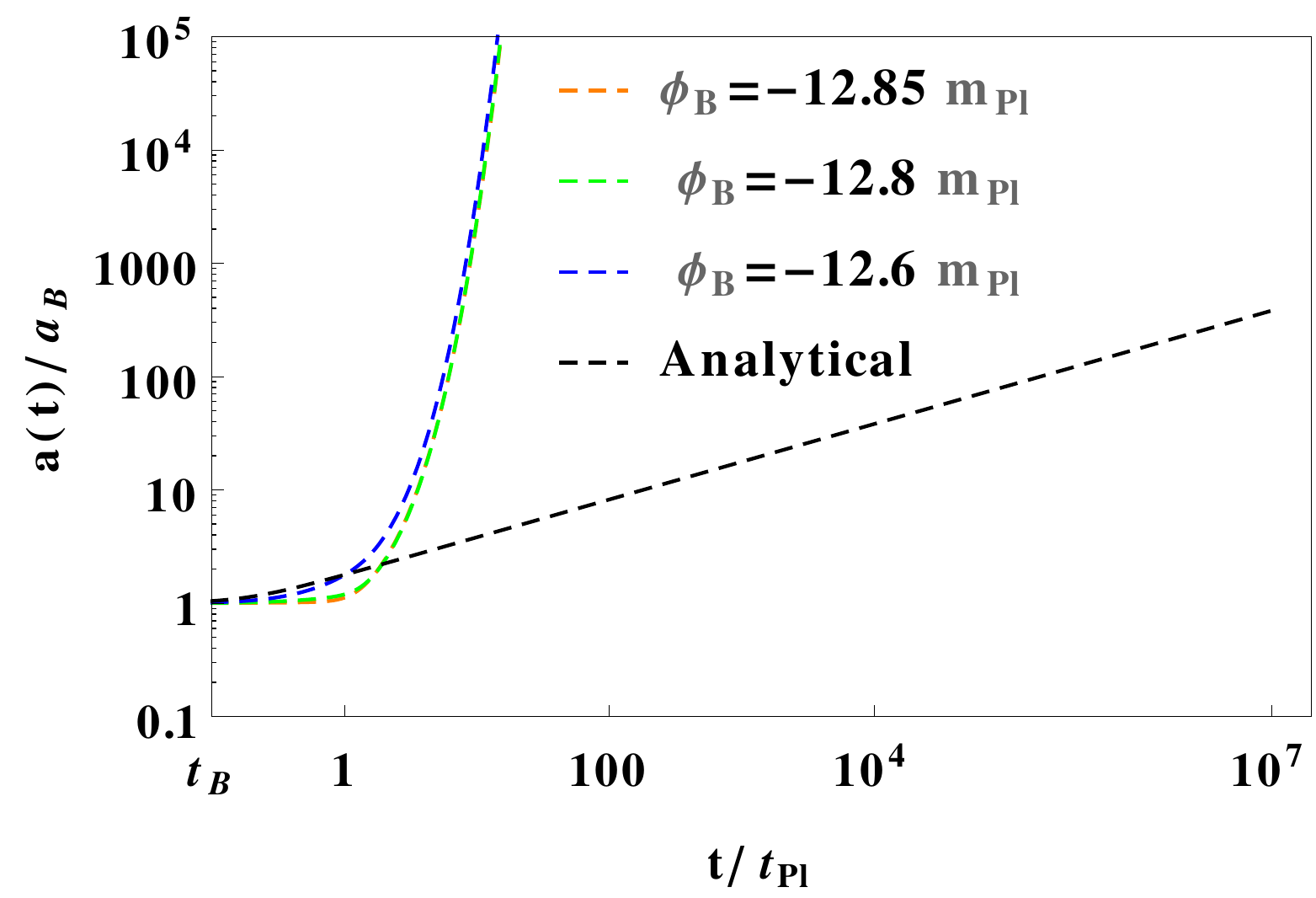}} & 
{\includegraphics[width=2.1in,height=1.6in,angle=0]{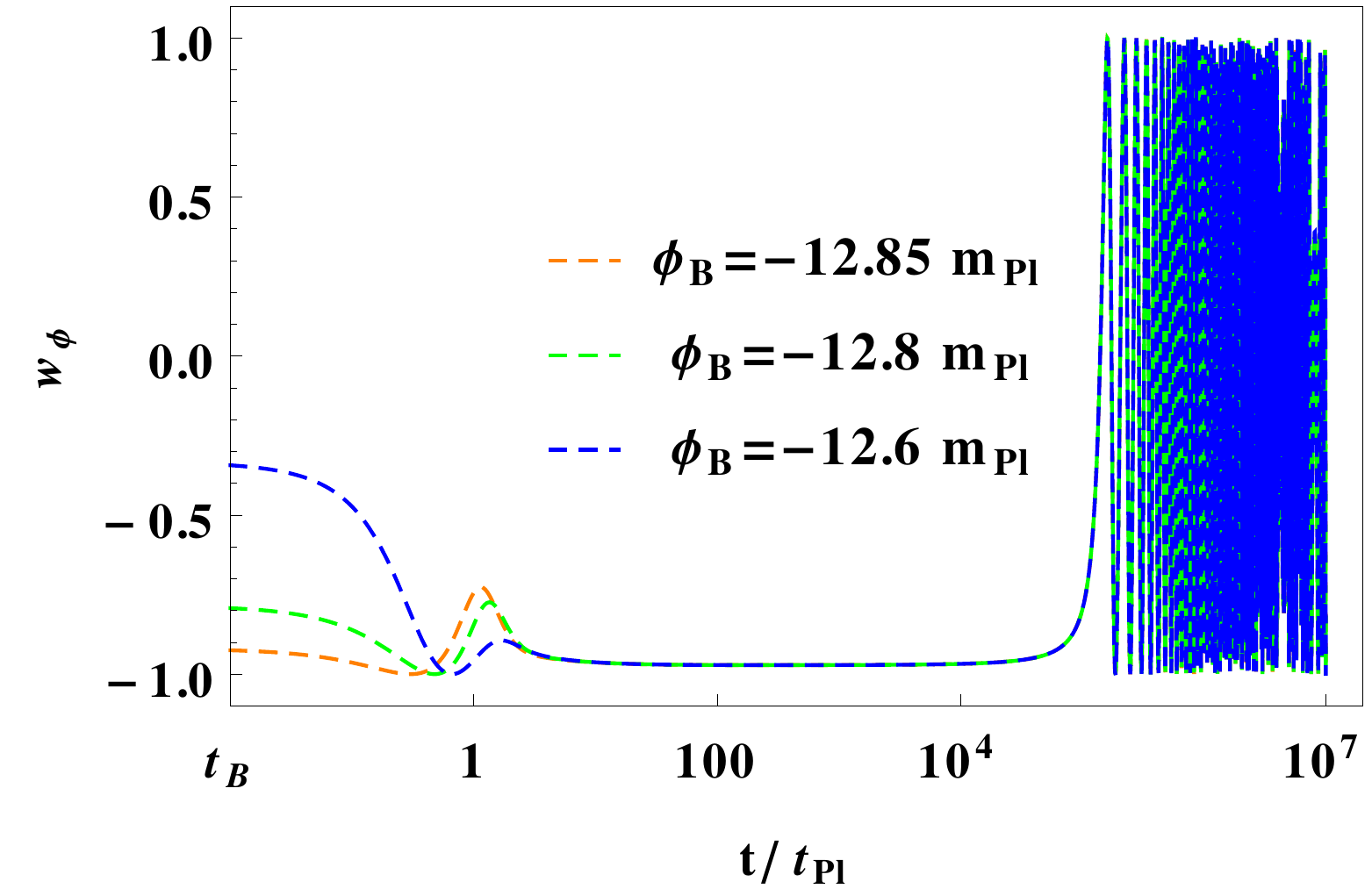}} & 
{\includegraphics[width=2.0in,height=1.6in,angle=0]{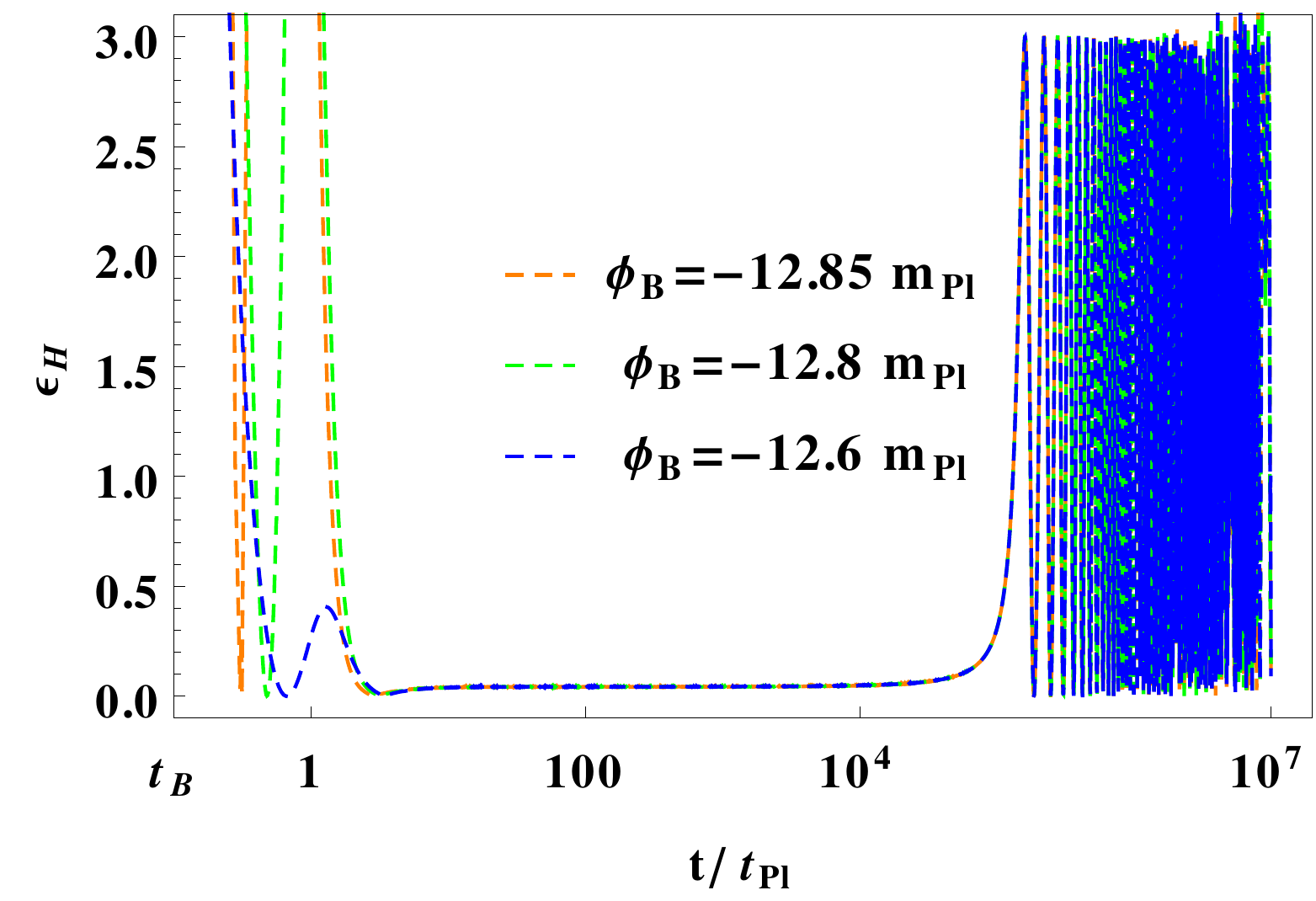}} 
\end{tabular}
\end{center}
\caption{ This figure is same as Fig. \ref{fig:n2alpha5_dphp} but for $\dot{\phi}_B<0$.}
\label{fig:n2alpha5_dphn}
\end{figure*}
\begin{figure*}[tbp]
\begin{center}
\begin{tabular}{ccc}
{\includegraphics[width=2.1in,height=1.65in,angle=0]{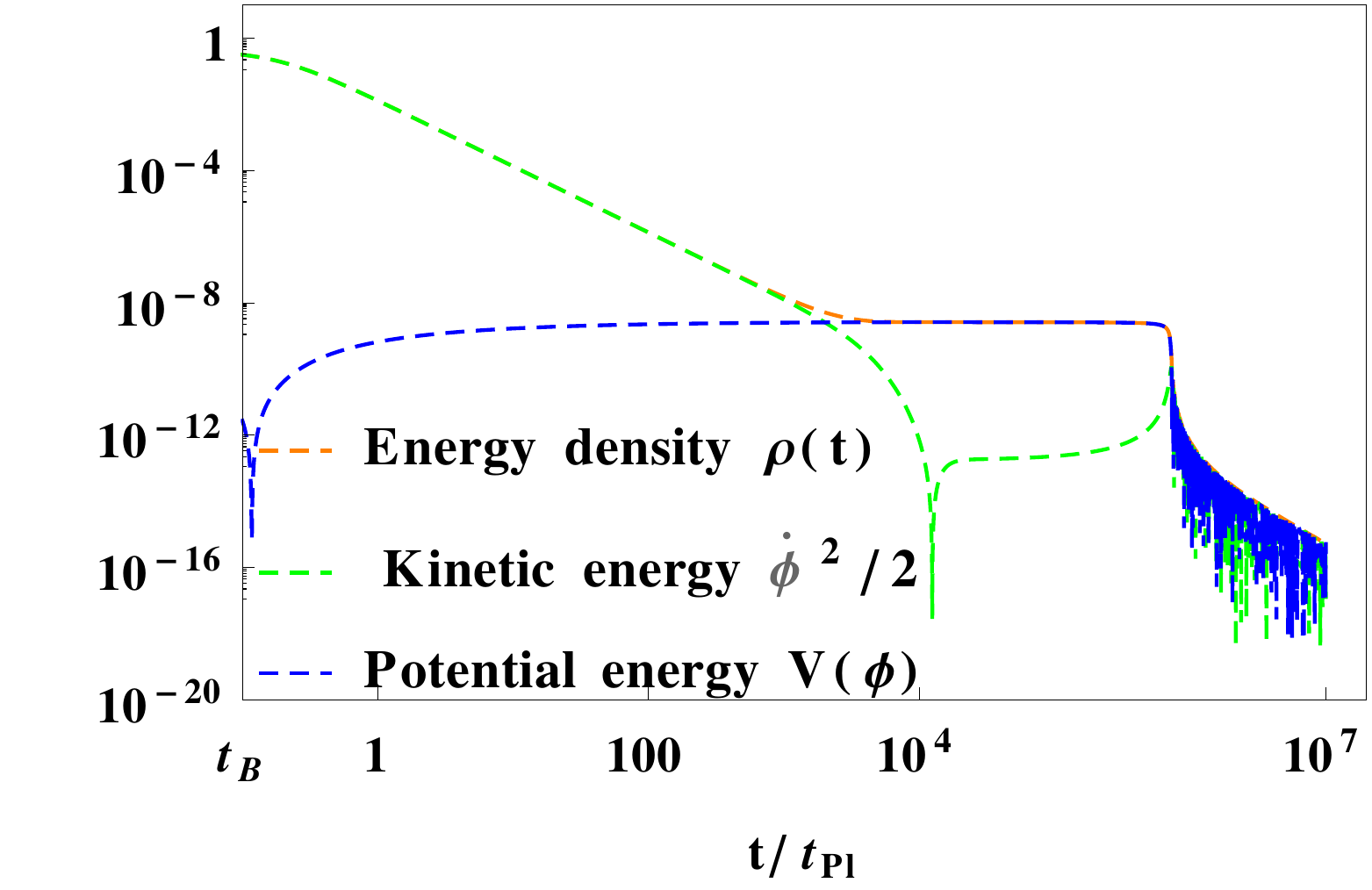}} &
{\includegraphics[width=2.1in,height=1.6in,angle=0]{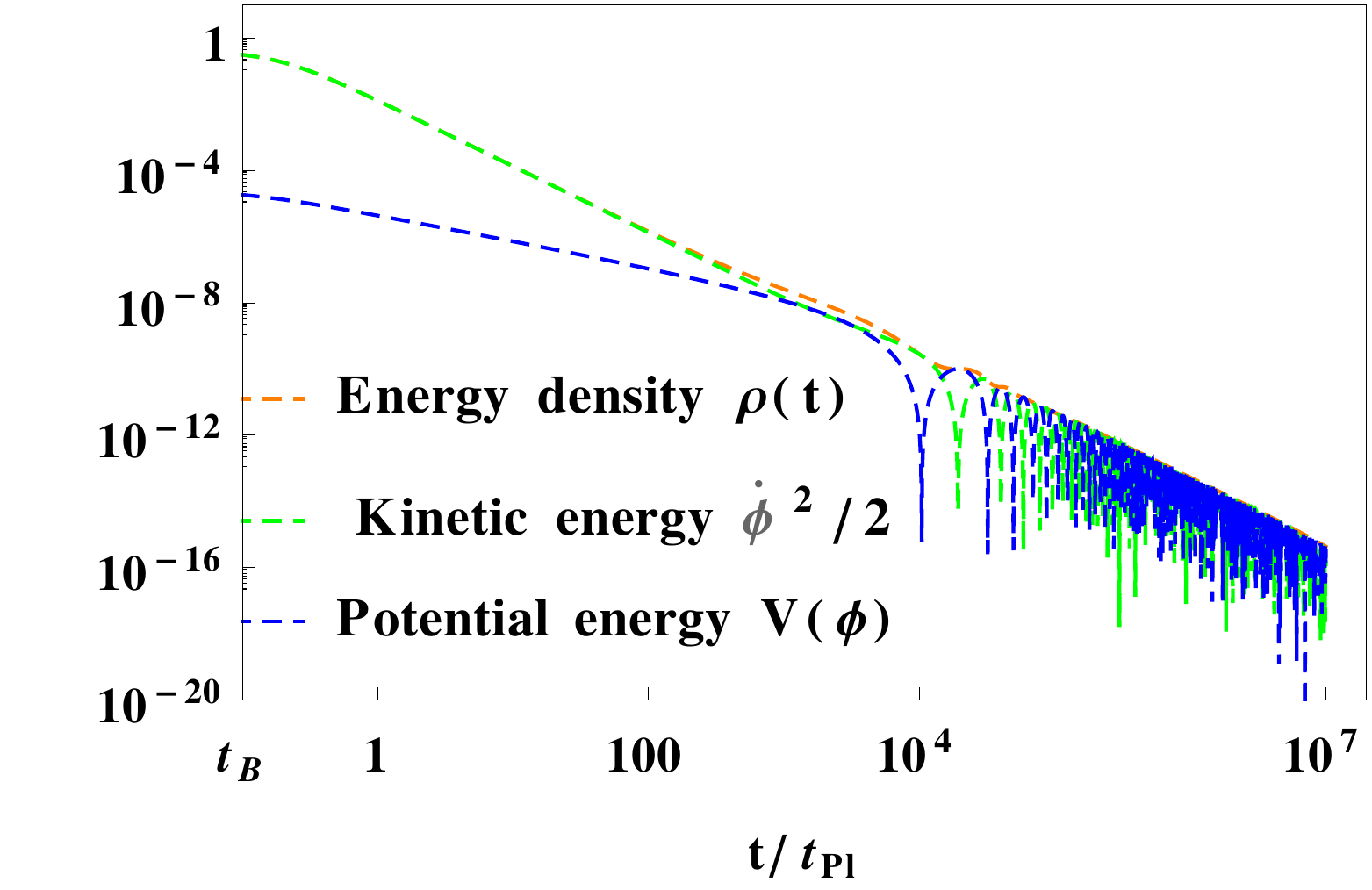}} &
{\includegraphics[width=2.0in,height=1.6in,angle=0]{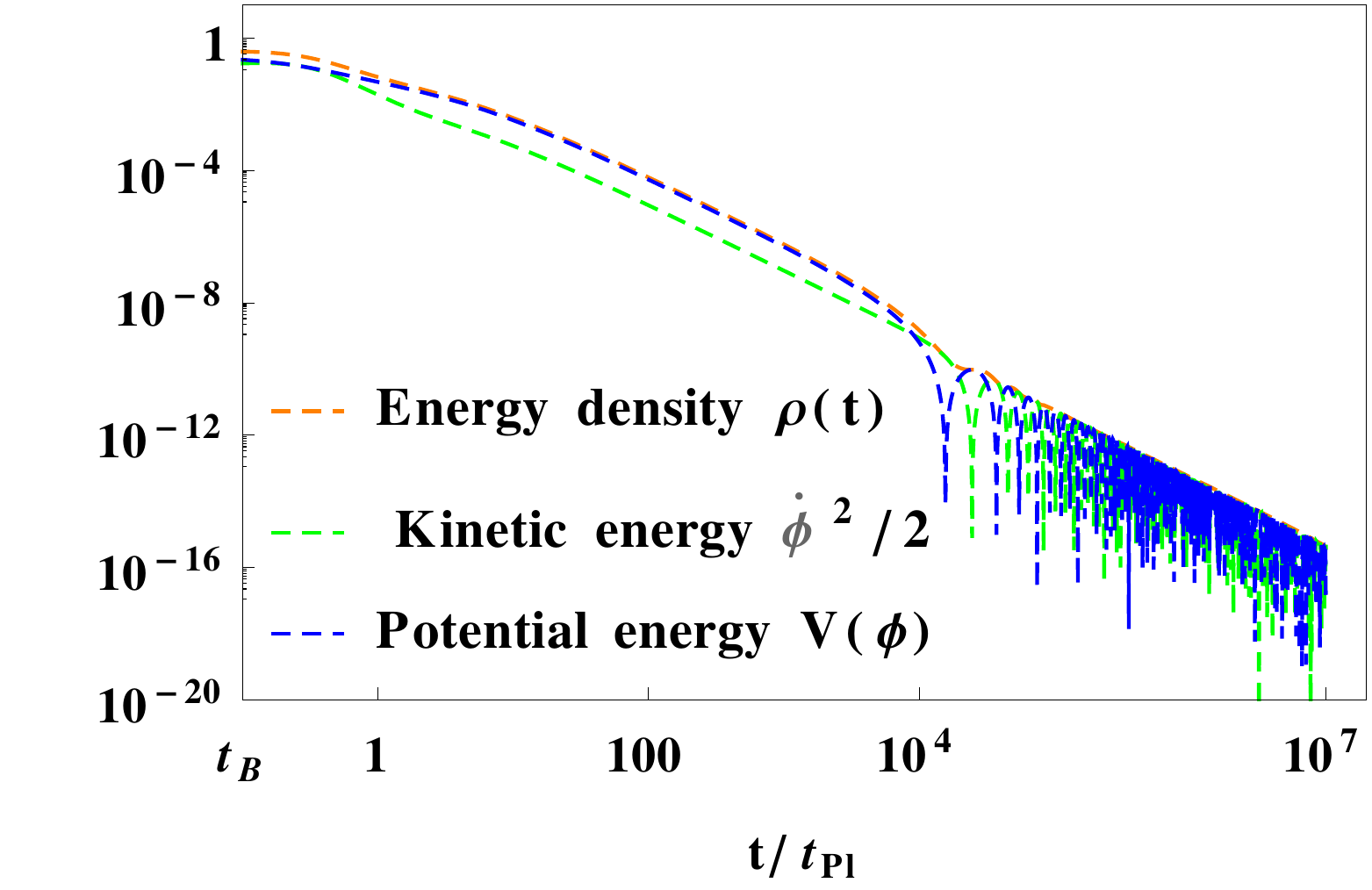}}
 \\
{\includegraphics[width=2.1in,height=1.65in,angle=0]{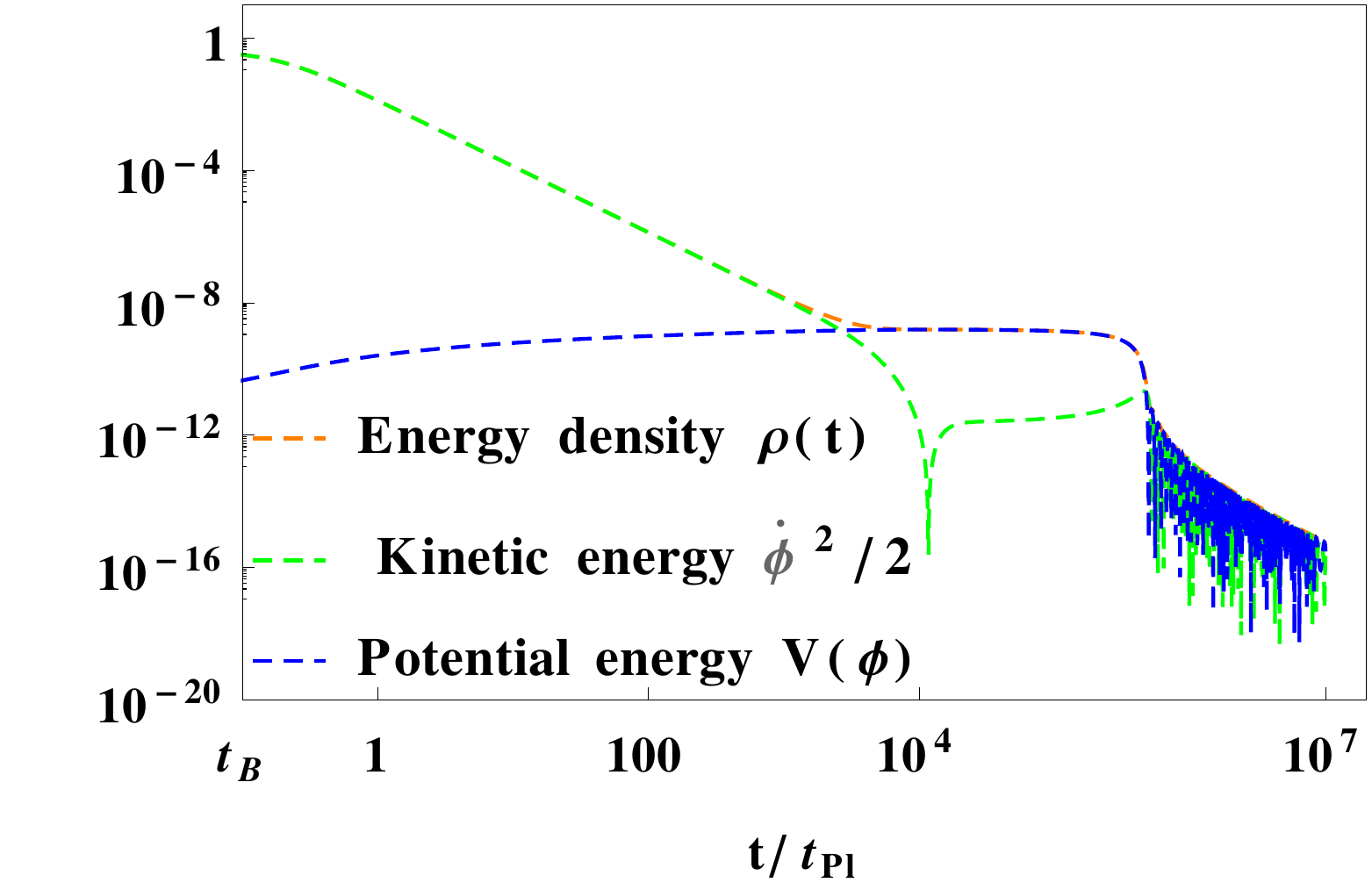}} &
{\includegraphics[width=2.1in,height=1.6in,angle=0]{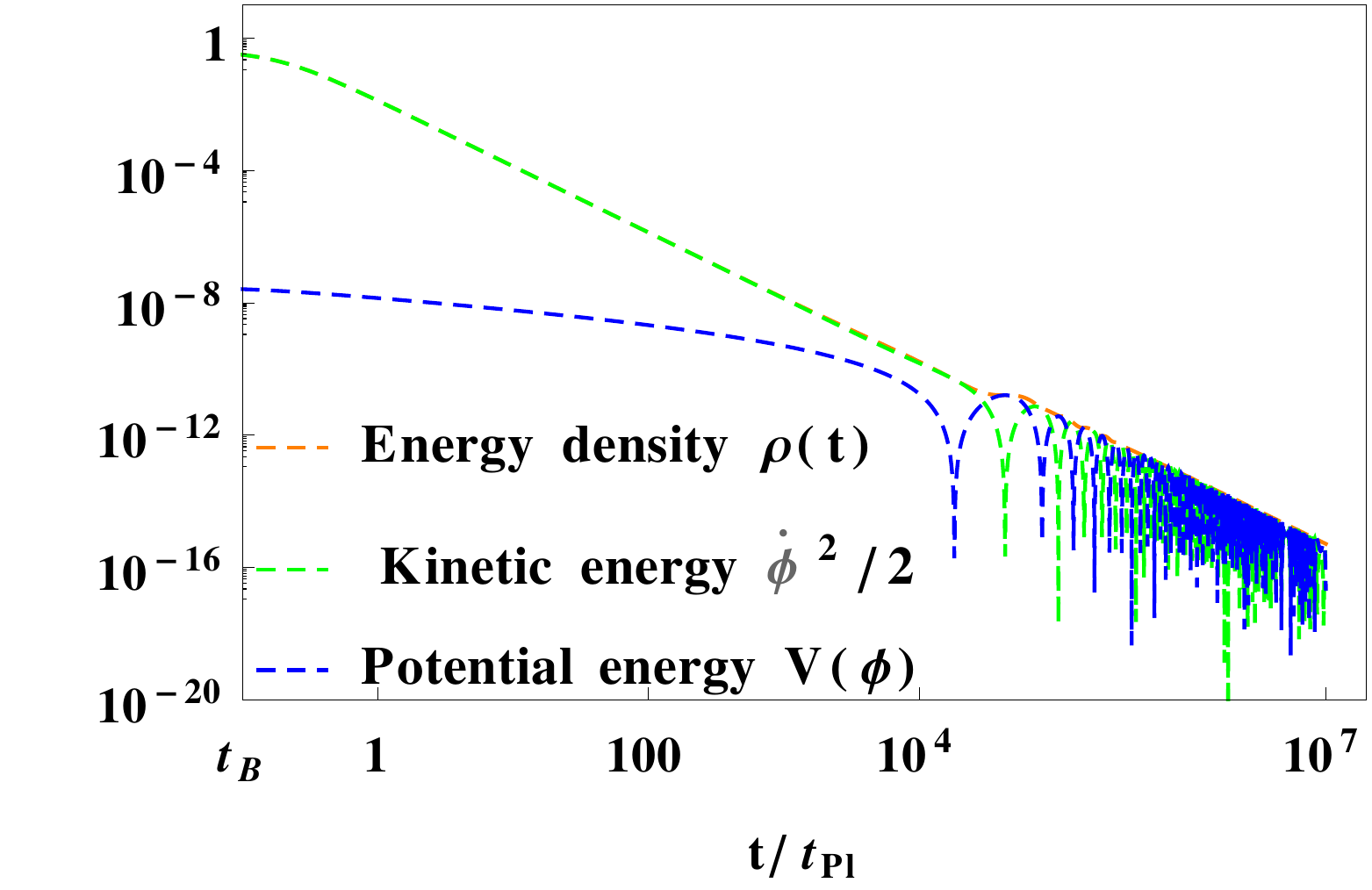}} &
{\includegraphics[width=2.0in,height=1.6in,angle=0]{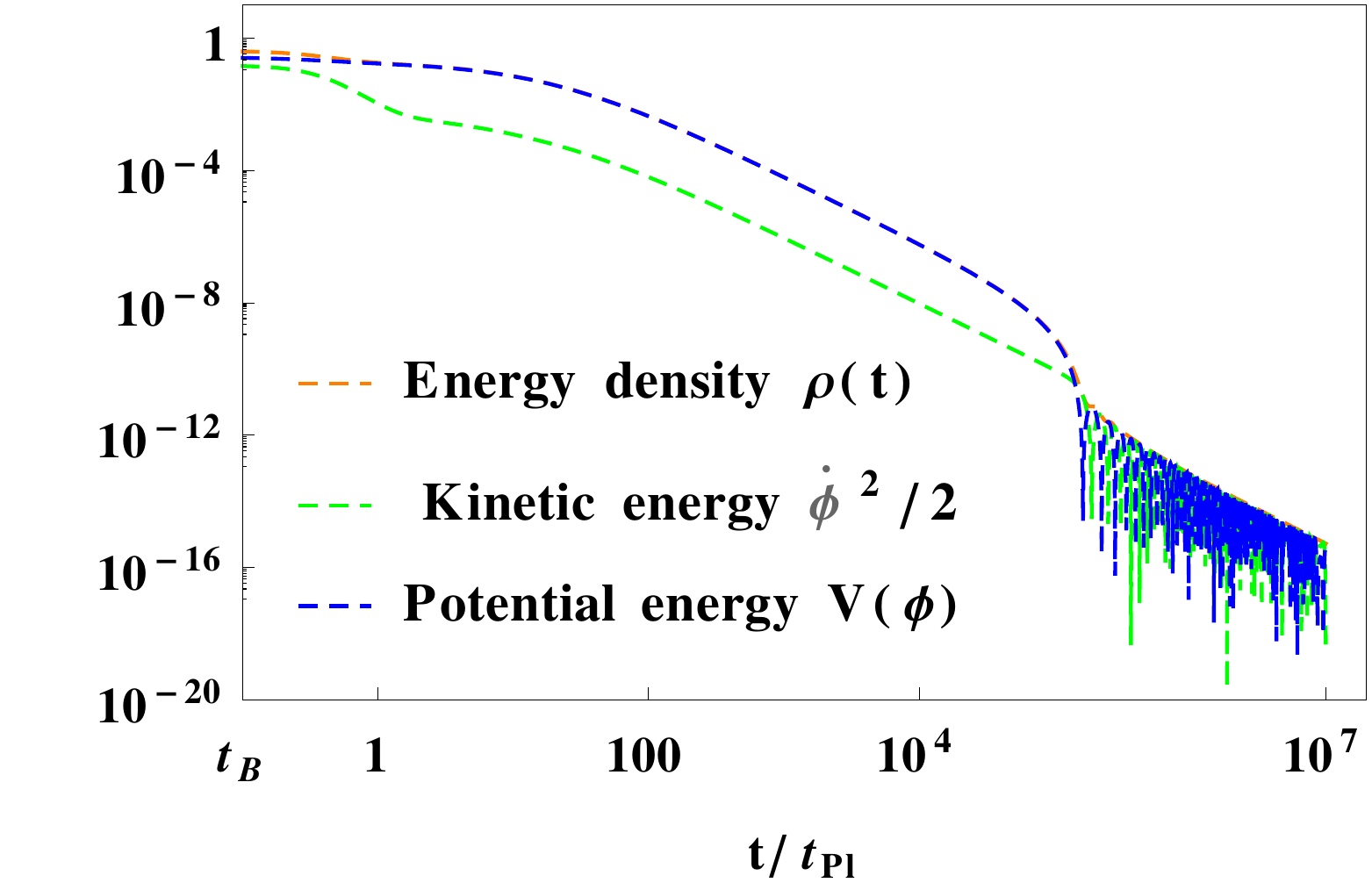}}
\end{tabular}
\end{center}
\caption{ This figure is displayed for $\alpha-$attractor with $n=2$ and $\dot{\phi}_B>0$. Top panels are plotted for $\alpha=0.5m_{Pl}^2$ with $\phi_B=-0.1 m_{Pl}$ (left), $-2$ (middle) and $-4$ (right) whereas bottom ones are for $\alpha=5m_{Pl}^2$ with $\phi_B=0.1 m_{Pl}$ (left), $-2$ (middle) and $-12.6$ (right). The rest is the  same as in Fig. \ref{fig:n1rho}. }
\label{fig:n2rho}
\end{figure*}

\subsection{T-model}
\label{sec:Tmodel}

Let us discuss some characteristics of $T-model$ [Eq.(\ref{eq:Tpot})]. We show the evolution of potential (\ref{eq:Tpot}) vs the scalar field in Fig. \ref{fig:pot}. This potential asymptotically approaches  a plateau for large field values, and as the field approaches the origin, it is oscillating. The potential is symmetric with respect to  $\phi=0$. 

We numerically solve  Eqs.(\ref{eq:Hub}) and (\ref{eq:ddphi}) with $T-model$. Here, we only consider the case $\dot{\phi}_B>0$ (PIV) because the initial conditions for $T-model$ at the bounce have symmetry $(\phi_B,\dot{\phi}_B) \rightarrow (-\phi_B,-\dot{\phi}_B)$, and the results for $\dot{\phi}_B<0$ (NIV) can be easily obtained by using the above symmetry. Further, initial conditions can be categorized into two sub-cases, namely,  KED and PED at the bounce.

For this model, we choose two values of $\alpha$, $\alpha=10m_{Pl}^2$ and $10^{10}m_{Pl}^2$, and the corresponding values  of the parameter $c$ are given by  Eq.(\ref{eq:Talphac}). In the case of $\alpha=10m_{Pl}^2$, only KED initial conditions are possible at the quantum bounce. To get both KED and PED initial conditions at the bounce, $\alpha$ should be large like $10^{10}m_{Pl}^2$ as the potential contains $c^2$ term that is very small, as can be seen from Eq.(\ref{eq:Talphac}).

First, we numerically evolve $T-model$ with the background given by Eqs.(\ref{eq:Hub}) and (\ref{eq:ddphi}) for $\alpha=10 m_{Pl}^2$. The results for a set of KED initial conditions at the bounce are presented in Fig. \ref{fig:n0alpha10_dphp}, in which the scale factor $a(t)$, the EoS $w(\phi)$ and the slow-roll parameter $\epsilon_H$ are shown for the same set of $\phi_B$. In the future evolution of $w(\phi)$ and $\epsilon_H$, we obtain inflationary and non-inflationary phases. This means, in the entire parameter space of the KED initial conditions, we also have a small subset that does not provide inflationary phase, see Fig. \ref{fig:n0alpha10_dphp} and Table \ref{tab:n012_alpha_dphi}.

From the top panels of Fig. \ref{fig:n0alpha10_dphp}, one can clearly see that the desired slow-roll inflationary phase is obtained for the chosen initial values of $\phi_B/m_{Pl}=-5, 4, 10$. In this region, $a(t)$ grows exponentially, $w(\phi) \simeq -1$ and  $\epsilon_H \ll 1$. For NIV ($\dot{\phi}_B<0$), one can obtain the same results with the replacement of $\phi_B$ by $-\phi_B$ [i.e. $\phi_B/m_{Pl}=5, -4,-10$].

From the curves of $w(\phi)$ (top panel of Fig. \ref{fig:n0alpha10_dphp}), we notice that the evolution of the universe before reheating can be split up into three different phases, namely bouncing, transition and slow-roll \cite{alam2017,Tao2017a,Tao2017b}. During the bouncing phase, the kinetic energy remains dominant, and $w(\phi) \simeq +1$. In the transition phase, $w(\phi)$ decreases drastically from $+1$ $(t/t_{Pl} \simeq 10^3)$ to $-1$ $(t/t_{Pl} \simeq 10^4)$. This transition phase is slightly short in comparison with the other two phases. In the slow-roll phase, $w(\phi)$ is close to $-1$, and remains so until the end of the slow-roll inflation. During the bouncing phase, it is remarkable to note that the evolution of $a(t)$ (top panel of Fig. \ref{fig:n0alpha10_dphp}) is independent for a wide range of initial values of $\phi_B$, and exhibits the compatible behavior with the analytical solution (\ref{eq:a}).

The entire range of KED initial conditions is from $-\infty$ to $+\infty$. In this range most of initial values provide inflationary phase. However, there is a small subset that does not give inflationary phase, see Table \ref{tab:n012_alpha_dphi}. Total number of $e$-folds $N_{inf}$ during the inflationary phase can be obtained for different values of $\phi_B$, and the range for $\dot{\phi_B}>0$ is given as (See Table \ref{tab:n012_alpha_dphi})
\begin{eqnarray}
&& \frac{\phi_B}{m_{Pl}}  \in  (-\infty, -3.1) \cup (-1.61, +\infty)\nonumber\\
&& ~~~~~~~~~~~~~ \rightarrow \text{slow-roll}~(N_{inf}>0), \nonumber\\
&& -3.1 < \phi_B \leq -1.6 \rightarrow \text{no slow-roll inflation}.
\label{eq:TNphiB}
\end{eqnarray}
To be consistent with the Planck data \cite{Planck2015}, at least 60 $e$-folds are needed during the slow-roll inflation, and to obtain it one has to require (see Table \ref{tab:n0_dphip})
\begin{eqnarray}
 \frac{\phi_B}{m_{Pl}} & \in & (-\infty, -4.9) \cup (1.1, +\infty).
\label{eq:TN60phiB}
\end{eqnarray}
In the case of initial conditions with $\dot{\phi}_B<0$, we use the symmetry $(\phi_B,\dot{\phi}_B) \rightarrow (-\phi_B,-\dot{\phi}_B)$, then the constraints are
\begin{eqnarray}
 \frac{\phi_B}{m_{Pl}}  & \in & (-\infty, 1.61) \cup (3.1, +\infty) \rightarrow \text{slow-roll}~(N_{inf}>0)\nonumber\\
&& 1.6 \leq \phi_B < 3.1 \rightarrow \text{no slow-roll}, \nonumber\\
 \frac{\phi_B}{m_{Pl}}  & \in & (-\infty, -1.1) \cup (4.9, +\infty) \rightarrow N_{inf} \gtrsim 60.
\label{eq:TNphiBsym}
\end{eqnarray}
From Table \ref{tab:n0_dphip}, one  notices that the number of $e$-folds $N_{inf}$ grows as the absolute values of $\phi_B$ increase, which implies that an absolute large value of $\phi_B$ can produce more number of $e$-folds. The similar results for power-law potentials were obtained in \cite{alam2017}. 

Next, we study $T-model$ with $\alpha=10^{10}m_{Pl}^2$. The results are displayed in Fig. \ref{fig:n0alpha10p10_dphp} for $\dot{\phi}_B>0$. Here, we use a large value of $\alpha$ to get both KED and PED initial conditions at the quantum bounce. In Fig. \ref{fig:n0alpha10p10_dphp}, we show the evolution of the scale factor $a(t)$, EoS $w(\phi)$ and slow-roll parameter $\epsilon_H$, and show the inflationary and non-inflationary phases of the universe. Top, middle and bottom panels of Fig. \ref{fig:n0alpha10p10_dphp} are obtained for the different sets of initial conditions of $\phi_B$ that correspond to KED with slow-roll (Top), without slow-roll (Middle) and PED with slow-roll (Bottom). From the  top and middle panels of Fig. \ref{fig:n0alpha10p10_dphp}, we conclude that the KED initial conditions have a subset that does not provide slow-roll inflation phases. The range of the above subset is given in Table \ref{tab:n012_alpha_dphi}.

Let us compare Top and bottom panels of Fig. \ref{fig:n0alpha10p10_dphp} that are obtained for KED and PED initial conditions. In top panels, the evolution of $a(t)$ exhibits the universal feature which is consistent with the analytical solution (\ref{eq:a}). The evolution of $w(\phi)$ shows three different phases, namely bouncing, transition and  slow-roll. In bottom panels, the universal feature of $a(t)$ is lost, and the bouncing phase no longer exists, though the slow-roll inflation $w(\phi)\simeq -1$ can still be achieved. 

The range of $\phi_B$ that provides inflationary and non-inflationary phases is given by (see Table \ref{tab:n012_alpha_dphi}):
\begin{eqnarray}
&& \frac{\phi_B}{m_{Pl}}  \in  (-\phi_{max}, -3.21) \cup (-1.8, \phi_{max}), \nonumber\\
&&~~~~~~~~~~~~~ \rightarrow \text{slow-roll}~(N_{inf}>0), \nonumber\\
&& -3.2 \leq \phi_B < -1.8 \rightarrow \text{no slow-roll inflation,}
\label{eq:TNphiB2}
\end{eqnarray}
where
\begin{eqnarray}
\phi_{max} \simeq \sqrt{6 \alpha} \arctan \text{h} \left( \sqrt{\frac{\rho_c}{\alpha c^2}} \right) \simeq 2.56 \times 10^5 m_{Pl}.
\label{eq:TNphimax2}
\end{eqnarray}
To obtain at least 60 $e$-folds    during the slow-roll inflationary phase, one has to require (see Table \ref{tab:n0_dphip}):
\begin{eqnarray}
\frac{\phi_B}{m_{Pl}}  \in  (-\phi_{max}, -5.1) \cup (1.05, \phi_{max}).
\label{eq:TN60phiB2}
\end{eqnarray}

For $\dot{\phi}_B<0$, the same results can be obtained with the symmetry $(\phi_B,\dot{\phi}_B) \rightarrow (-\phi_B,-\dot{\phi}_B)$, and are given by
\begin{eqnarray}
&& \frac{\phi_B}{m_{Pl}}  \in  (-\phi_{max}, 1.8) \cup (3.21, \phi_{max}), \nonumber\\
&&~~~~~~~~~~~~ \rightarrow \text{slow-roll} ~(N_{inf}>0)\nonumber\\
&& 1.8 < \phi_B \leq 3.2 \rightarrow \text{no slow-roll}, \nonumber\\
&& \frac{\phi_B}{m_{Pl}}  \in  (-\phi_{max}, -1.05) \cup (5.1, \phi_{max})\nonumber\\
&&~~~~~~~~~~ \rightarrow N_{inf} \gtrsim 60.
\label{eq:TNphiB2sym}
\end{eqnarray}
As mentioned in the case of $\alpha=10m_{Pl}^2$, here also one can get more e-folds    for the large absolute  values of $\phi_B$, see Table \ref{tab:n0_dphip}.
\begin{table*}
\caption{This table represents the $E-model$. We show the number of $e$-foldings $N_{inf}$ and other parameters with $\dot{\phi}_B>0$. For each choice of $\alpha$, the corresponding values of $c$ are given by  Eq.(\ref{eq:Ealphac}). }
\begin{center}
\resizebox{\textwidth}{!}{%
\begin{tabular}{ccccccccc}
\hline\hline
\\
$\alpha$~~ & $\phi_B$~~~  & Inflation~~~ & $t/t_{pl}$~~~ & $\epsilon$~~ & $w$ ~~& $N_{inf}$ &~~~$r_{w}^c/r_w$&~~~${w}^B$\\\\
\hline\hline
0.1 ~~ & $-0.5$~~~& begin~~~& $2.65082 \times 10^3$ ~~~& 0.99~~ & $-1/3$ ~~& ~~~& ~~~&\\
~~ & ~~~& slow-roll~~~& $1.10225 \times 10^4$ ~~~& $1.0 \times 10^{-4}$~~ & $-1$ ~~& 42.77 ~~~&$r_{w}^c > r_w$ ~~~& $>0$\\
~~ & ~~~& end~~~& $3.243 \times 10^5$ ~~~& 0.99~~ & $-1/3$ ~~& ~~~& ~~~& \\\\
~~ & $-0.38$~~~& begin~~~& $2.62043 \times 10^3$ ~~~& 0.99~~ & $-1/3$ ~~& ~~~& ~~~&\\
~~ & ~~~& slow-roll~~~& $1.16211 \times 10^4$ ~~~& $1.8 \times 10^{-4}$~~ & $-1$ ~~& 60.0 ~~~&$r_{w}^c = r_w$ ~~~& $>0$\\
~~ & ~~~& end~~~& $4.425 \times 10^5$ ~~~& 0.99~~ & $-1/3$ ~~& ~~~& ~~~& \\\\
~~ & $-0.3$~~~& begin~~~& $2.60392 \times 10^3$ ~~~& 0.99~~ & $-1/3$ ~~& ~~~& ~~~&\\
~~ & ~~~& slow-roll~~~& $1.20318 \times 10^4$ ~~~& $2.4 \times 10^{-4}$~~ & $-1$ ~~& 74.92 ~~~&$r_{w}^c < r_w$ ~~~& $>0$\\
~~ & ~~~& end~~~& $5.441 \times 10^6$ ~~~& 0.99~~ & $-1/3$ ~~& ~~~& ~~~& \\\\
5~~ & $-25$~~~& begin~~~& $0.3$ ~~~& 1.5~~ & $-0.478$ ~~& ~~~& ~~~&\\
~~ & ~~~& slow-roll~~~& $1.5$ ~~~& $1.3 \times 10^{-3}$~~ & $-0.978$ ~~& 713.22 ~~~&$r_{w}^c < r_w$ ~~~& $<0$\\
~~ & ~~~& end~~~& $2.963 \times 10^5$ ~~~& 0.31~~ & $-1/3$ ~~& ~~~& ~~~& \\\\
~~ & $-6$~~~& begin~~~& $5.6861 \times 10^2$ ~~~& 0.99~~ & $-1/3$ ~~& ~~~& ~~~&\\
~~ & ~~~& slow-roll~~~& $3.0 \times 10^3$ ~~~& $1.8 \times 10^{-2}$~~ & $-0.987$ ~~& 79.14 ~~~&$r_{w}^c < r_w$ ~~~& $>0$\\
~~ & ~~~& end~~~& $4.4 \times 10^5$ ~~~& 0.99~~ & $-1/3$ ~~& ~~~& ~~~& \\\\
~~ & $-5.35$~~~& begin~~~& $7.8972 \times 10^2$ ~~~& 0.99~~ & $-1/3$ ~~& ~~~& ~~~&\\
~~ & ~~~& slow-roll~~~& $6.78678 \times 10^3$ ~~~& $2.0 \times 10^{-2}$~~ & $-0.987$ ~~& 60.28 ~~~&$r_{w}^c = r_w$ ~~~& $>0$\\
~~ & ~~~& end~~~& $4.07 \times 10^5$ ~~~& 1.0~~ & $-1/3$ ~~& ~~~& ~~~& \\\\
~~ & $-5$~~~& begin~~~& $9.5418 \times 10^2$ ~~~& 0.99~~ & $-1/3$ ~~& ~~~& ~~~&\\
~~ & ~~~& slow-roll~~~& $9.55025 \times 10^3$ ~~~& $2.1 \times 10^{-2}$~~ & $-0.985$ ~~& 50.69 ~~~&$r_{w}^c > r_w$ ~~~& $>0$\\
~~ & ~~~& end~~~& $3.85 \times 10^5$ ~~~& 1.0~~ & $-1/3$ ~~& ~~~& ~~~& \\\\
~~ & $0.5$~~~& begin~~~& $4.26416 \times 10^3$ ~~~& 0.99~~ & $-1/3$ ~~& ~~~& ~~~&\\
~~ & ~~~& slow-roll~~~& $1.43245 \times 10^4$ ~~~& $2.7 \times 10^{-4}$~~ & $-1$ ~~& 47.83 ~~~&$r_{w}^c > r_w$ ~~~& $>0$\\
~~ & ~~~& end~~~& $7.85 \times 10^5$ ~~~& 0.99~~ & $-1/3$ ~~& ~~~& ~~~& \\\\
~~ & $0.76$~~~& begin~~~& $3.99721 \times 10^3$ ~~~& 1.0~~ & $-1/3$ ~~& ~~~& ~~~&\\
~~ & ~~~& slow-roll~~~& $1.39402 \times 10^4$ ~~~& $1.0 \times 10^{-4}$~~ & $-1$ ~~& 60.45 ~~~&$r_{w}^c = r_w$ ~~~& $>0$\\
~~ & ~~~& end~~~& $9.19 \times 10^5$ ~~~& 0.99~~ & $-1/3$ ~~& ~~~& ~~~& \\\\
~~ & $1$~~~& begin~~~& $3.79512 \times 10^3$ ~~~& 0.99~~ & $-1/3$ ~~& ~~~& ~~~&\\
~~ & ~~~& slow-roll~~~& $1.36658 \times 10^4$ ~~~& $3.6 \times 10^{-4}$~~ & $-1$ ~~& 73.96 ~~~&$r_{w}^c < r_w$ ~~~& $>0$\\
~~ & ~~~& end~~~& $1.054 \times 10^6$ ~~~& 0.99~~ & $-1/3$ ~~& ~~~& ~~~& \\\\
\hline\hline
\end{tabular}}
\label{tab:n1_dphip}
\end{center}
\end{table*}
\begin{table*}
\caption{This table is for the $E-model$. We exhibit different parameters with $\dot{\phi}_B<0$. }
\begin{center}
\resizebox{\textwidth}{!}{%
\begin{tabular}{ccccccccc}
\hline\hline
\\
$\alpha$~~ & $\phi_B$~~~  & Inflation~~~ & $t/t_{pl}$~~~ & $\epsilon$~~ & $w$ ~~& $N_{inf}$ &~~~$r_{w}^c/r_w$&~~~${w}^B$\\\\
\hline\hline
0.1 ~~ & $3$~~~& begin~~~& $2.62892 \times 10^3$ ~~~& 0.99~~ & $-1/3$ ~~& ~~~& ~~~&\\
~~ & ~~~& slow-roll~~~& $2.56475 \times 10^4$ ~~~& $1.6 \times 10^{-3}$~~ & $-0.999$ ~~& 39.16 ~~~&$r_{w}^c > r_w$ ~~~& $>0$\\
~~ & ~~~& end~~~& $2.992 \times 10^5$ ~~~& 0.99~~ & $-1/3$ ~~& ~~~& ~~~& \\\\
~~ & 3.15~~~& begin~~~& $2.59873 \times 10^3$ ~~~& 0.99~~ & $-1/3$ ~~& ~~~& ~~~&\\
~~ & ~~~& slow-roll~~~& $2.50846 \times 10^4$ ~~~& $3.4 \times 10^{-5}$~~ & $-0.999$ ~~& 60.48 ~~~&$r_{w}^c = r_w$ ~~~& $>0$\\
~~ & ~~~& end~~~& $4.457 \times 10^5$ ~~~& 0.99~~ & $-1/3$ ~~& ~~~& ~~~& \\\\
~~ & 3.3~~~& begin~~~& $2.57866 \times 10^3$ ~~~& 0.99~~ & $-1/3$ ~~& ~~~& ~~~&\\
~~ & ~~~& slow-roll~~~& $2.50846 \times 10^4$ ~~~& $1.1 \times 10^{-3}$~~ & $-0.999$ ~~& 92.19 ~~~&$r_{w}^c < r_w$ ~~~& $>0$\\
~~ & ~~~& end~~~& $6.611 \times 10^5$ ~~~& 0.99~~ & $-1/3$ ~~& ~~~& ~~~& \\\\
5~~ & $-25$~~~& begin~~~& $0.2$ ~~~& 4.94~~ & $-0.507$ ~~& ~~~& ~~~&\\
~~ & ~~~& slow-roll~~~& $0.8$ ~~~& $1.1 \times 10^{-2}$~~ & $-0.995$ ~~& 712.15 ~~~&$r_{w}^c < r_w$ ~~~& $<0$\\
~~ & ~~~& end~~~& $2.47527 \times 10^5$ ~~~& 0.28~~ & $-1/3$ ~~& ~~~& ~~~& \\\\
~~ & $-3$~~~& begin~~~& $5.88967 \times 10^2$ ~~~& 0.99~~ & $-1/3$ ~~& ~~~& ~~~&\\
~~ & ~~~& slow-roll~~~& $1.74169 \times 10^3$ ~~~& $6.6 \times 10^{-5}$~~ & $-1$ ~~& 80.19 ~~~&$r_{w}^c < r_w$ ~~~& $>0$\\
~~ & ~~~& end~~~& $4.42 \times 10^5$ ~~~& 1.0~~ & $-1/3$ ~~& ~~~& ~~~& \\\\
~~ & $-2.2$~~~& begin~~~& $8.361 \times 10^2$ ~~~& 0.99~~ & $-1/3$ ~~& ~~~& ~~~&\\
~~ & ~~~& slow-roll~~~& $2.42178 \times 10^3$ ~~~& $9.5 \times 10^{-5}$~~ & $-1$ ~~& 60.23 ~~~&$r_{w}^c = r_w$ ~~~& $>0$\\
~~ & ~~~& end~~~& $4.07 \times 10^5$ ~~~& 1.0~~ & $-1/3$ ~~& ~~~& ~~~& \\\\
~~ & $-2$~~~& begin~~~& $9.1693 \times 10^2$ ~~~& 0.99~~ & $-1/3$ ~~& ~~~& ~~~&\\
~~ & ~~~& slow-roll~~~& $2.6382 \times 10^3$ ~~~& $3.3 \times 10^{-6}$~~ & $-1$ ~~& 55.51 ~~~&$r_{w}^c > r_w$ ~~~& $>0$\\
~~ & ~~~& end~~~& $3.966 \times 10^5$ ~~~& 1.0~~ & $-1/3$ ~~& ~~~& ~~~& \\\\
~~ & 4.2~~~& begin~~~& $4.11346 \times 10^3$ ~~~& 0.99~~ & $-1/3$ ~~& ~~~& ~~~&\\
~~ & ~~~& slow-roll~~~& $3.77015 \times 10^4$ ~~~& $6.2 \times 10^{-3}$~~ & $-0.995$ ~~& 48.70 ~~~&$r_{w}^c > r_w$ ~~~& $>0$\\
~~ & ~~~& end~~~& $7.944 \times 10^5$ ~~~& 0.99~~ & $-1/3$ ~~& ~~~& ~~~& \\\\
~~ & $4.42$~~~& begin~~~& $3.89349 \times 10^3$ ~~~& 0.99~~ & $-1/3$ ~~& ~~~& ~~~&\\
~~ & ~~~& slow-roll~~~& $5.83076 \times 10^4$ ~~~& $3.8 \times 10^{-3}$~~ & $-0.996$ ~~& 60.33 ~~~&$r_{w}^c = r_w$ ~~~& $>0$\\
~~ & ~~~& end~~~& $9.173 \times 10^5$ ~~~& 0.99~~ & $-1/3$ ~~& ~~~& ~~~& \\\\
~~ & 5~~~& begin~~~& $3.47442 \times 10^3$ ~~~& 0.99~~ & $-1/3$ ~~& ~~~& ~~~&\\
~~ & ~~~& slow-roll~~~& $3.27722 \times 10^4$ ~~~& $2.3 \times 10^{-3}$~~ & $-0.998$ ~~& 99.13 ~~~&$r_{w}^c < r_w$ ~~~& $>0$\\
~~ & ~~~& end~~~& $1.292 \times 10^6$ ~~~& 0.99~~ & $-1/3$ ~~& ~~~& ~~~& \\\\
\hline\hline
\end{tabular}}
\label{tab:n1_dphin}
\end{center}
\end{table*}

Finally, we consider  the evolutions of the kinetic and potential energies  in Fig. \ref{fig:n0rho}, and pay particular attention on the case in which a slow-roll inflationary phase is absent, although the kinetic energy of the inflaton 
still dominates the evolution of the universe at the bounce.    Left panel is plotted for KED initial condition with $\phi_B=5 m_{Pl}$. In the bouncing phase, KE dominates the evolution whereas PE remains sub-dominant. As time increases, KE decreases 
until the transition phase in which KE falls below the PE, and thereafter, PE dominates and remains so for most of the time of the evolution, during which   the slow-roll inflation is resulted. Middle panel is shown for a value ($\phi_B=-2.4 m_{Pl}$) of a subset of KED initial conditions. In this case, PE is sub-dominant initially and remains so during the entire evolution. It never overtakes the KE. As a result,   a slow-roll inflationary phase is absent. Right panel exhibits the PED case where PE dominates generically during the whole process, and gives rise to  a slow-roll inflationary phase for a long period.

\subsection{E-model}
\label{sec:Emodel}

In this subsection, we study the features of $E-model$ [Eq.(\ref{eq:Epot})]. The   $E-model$ potential is displayed in Fig. \ref{fig:pot}. This kind of potentials is bounded below by zero i.e. $V(\phi) \geq 0$. On the positive side ($\phi \rightarrow \infty$),  the potential (\ref{eq:Epot}) achieves  a finite value $V(\phi) \rightarrow \alpha c^2/4$, whereas in the negative side ($\phi \rightarrow -\infty$) it diverges. Hence, this potential is asymmetric. In LQC, the total energy density can not exceed the value of $\rho_c$. Therefore, the critical energy density constrains the initial values of $\phi_B$ as $(\phi_{min}, \infty)$,
where
\begin{eqnarray}
\phi_{min} &\simeq & \sqrt{6 \alpha} \arctan \text{h} \left( \frac{\sqrt{\rho_c}}{\sqrt{\alpha c^2}-\sqrt{\rho_c}} \right)\nonumber\\
 &\simeq & -3.64 m_{Pl}~ \text{for}~ \alpha=0.1 m_{Pl}^2,\nonumber\\
 &\simeq & -25.65 m_{Pl}~ \text{for}~ \alpha=5 m_{Pl}^2.
\label{eq:Ephimin}
\end{eqnarray}
The $E-model$ reduces to the Starobinsky model for $\alpha=1$. Here, we shall not discuss the Starobinsky model as the  evolutions and the phase space analysis have been already studied in detail in  \cite{Tao2017a,Tao2017b,Bonga2016}. Hence, in this sub-section, we shall investigate $E-model$ with different values of $\alpha$ ($\alpha\not=1$).

From  Eq.(\ref{eq:Ephimin}), one can obtain $\phi_{min}$ for the given value of $\alpha$ and $c$. First,  let us work with $\alpha=0.1 m_{Pl}^2$ and $c=3.3 \times 10^{-4} m_{Pl}$ [Eq.(\ref{eq:Ealphac})]. In this case, we have $\phi_{min} \simeq -3.64 m_{Pl}$. Numerically, we examine the whole range of the inflaton field in order to identify the initial conditions that can give rise to the slow-roll inflation. We find the KED (PED) evolution exists in a very long (narrow) range and given by  (see, Table \ref{tab:n012_alpha_dphi}):

$\bullet$ For $\dot{\phi}_B>0$, we have
\begin{eqnarray}
&& \frac{\phi_B}{m_{Pl}}  \in  (-1.51, +\infty) \rightarrow \text{KED (SR)}, \nonumber\\
&& -3.52 < \phi_B \leq -1.5 \rightarrow \text{subset of KED (NSR)}, \nonumber\\
 && \frac{\phi_B}{m_{Pl}} = -3.52 \rightarrow \text{KED=PED (NSR)}, \nonumber\\
&&  \frac{\phi_B}{m_{Pl}}  \in  (\phi_{min}, -3.53) \rightarrow \text{PED (NSR)},
\label{eq:EphiBP}
\end{eqnarray}
where $\phi_{min}$ is given by  Eq.(\ref{eq:Ephimin}). 

$\bullet$ For $\dot{\phi}_B<0$, we obtain 
\begin{eqnarray}
&&  \frac{\phi_B}{m_{Pl}}  \in  (2.4, +\infty) \rightarrow \text{KED (SR)}, \nonumber\\
&& -3.51 < \phi_B < 2.4 \rightarrow \text{subset of KED (NSR)}, \nonumber\\
 && \frac{\phi_B}{m_{Pl}} =  -3.51 \rightarrow \text{KED=PED (NSR)}, \nonumber\\
&&  \frac{\phi_B}{m_{Pl}}  \in  (\phi_{min}, -3.52) \rightarrow \text{PED (NSR)}.
\label{eq:EphiBN}
\end{eqnarray}

 The results of the background evolution for KED and PED initial conditions are shown in Figs. \ref{fig:n1alpha01_dphp} and \ref{fig:n1alpha01_dphn}, with $\dot{\phi}_B>0$ and $\dot{\phi}_B<0$, respectively. In both figures, the evolutions of $a(t)$, $w(\phi)$ and $\epsilon_H$ are obtained numerically for the same set of the initial values of $\phi_B$. In the case of KED initial conditions, the evolution of $a(t)$ is universal during the bouncing phase as it does not depends on the form of the potentials nor on the initial values of $\phi_B$, and can be well approximated by the analytical solution (\ref{eq:a}). This is mainly due to the fact that the potential remains very small in comparison with the kinetic one during the whole bouncing phase, and its effects on the evolution of the background is almost negligible. From the evolution of $w(\phi)$, one can see that the background evolution is divided into three different phases: bouncing, transition and slow-roll. The period of transition phase is very small in comparison with the other two. In the bouncing phase,  $w(\phi) \simeq +1$, while in the transition phase it suddenly decreases from $+1~  (t/t_{Pl} \approx 10^3)$ to $-1 ~ (t/t_{Pl} \approx 10^4)$. In the slow-roll inflationary phase, it is very close to $-1$ until the end of the slow-roll inflation. In the KED case, we also have a subset that does not provide the slow-roll inflation,  which is shown clearly  in the middle panels of Figs. \ref{fig:n1alpha01_dphp} and \ref{fig:n1alpha01_dphn}. The range of this subset is presented in Table \ref{tab:n012_alpha_dphi}. In the case of PED initial conditions, the universality of the scale factor $a(t)$ is lost, and the bouncing phase does not exist any more, and the slow-roll inflationary phase can not be obtained. See the bottom panels of Figs. \ref{fig:n1alpha01_dphp} and \ref{fig:n1alpha01_dphn}.

In Tables \ref{tab:n1_dphip} and \ref{tab:n1_dphin}, we display the initial values of $\phi_B$ that provide the desired slow-roll inflation, from which one can see that, for the successful inflation at least 60 $e$-folds are needed and to obtain this, the values of $\phi_B$ should   be in the range of
\begin{eqnarray}
&&  \frac{\phi_B}{m_{Pl}}  \in  (-0.38, +\infty) \rightarrow  N_{inf} \gtrsim 60 ~\text{for}~ \dot{\phi}_B>0, \nonumber\\
&&  \frac{\phi_B}{m_{Pl}}   \in  (3.15, +\infty) \rightarrow  N_{inf} \gtrsim 60 ~\text{for}~ \dot{\phi}_B<0,
\label{eq:ENphiB}
\end{eqnarray}
within which, Tables \ref{tab:n1_dphip} and \ref{tab:n1_dphin} exhibit that $N_{inf}$ grows as $\phi_B$ increases.

Next, we work with $\alpha=5 m_{Pl}^2$ and $c=4.9 \times 10^{-5} m_{Pl}$ [Eq.(\ref{eq:Ealphac})]. In this case, $\phi_{min} \simeq -25.65 m_{Pl}$. We numerically search the entire range of $\phi_B$, and find the initial values that can lead to the slow-roll inflation. Here, KED (PED) evolution has large (small) range and given by (see, Table \ref{tab:n012_alpha_dphi}),
\begin{eqnarray}
&&  \frac{\phi_B}{m_{Pl}}   \in  (-24.70, -3.11) \cup (-1.4, +\infty) \rightarrow \text{KED (SR)}, \nonumber\\
&& -3.1 \leq \phi_B < -1.4 \rightarrow \text{subset of KED (NSR)}, \nonumber\\
&&  \frac{\phi_B}{m_{Pl}}  = -24.71 \rightarrow \text{KED=PED (SR)}, \nonumber\\
&&  \frac{\phi_B}{m_{Pl}}   \in  (\phi_{min}, -24.72) \rightarrow \text{PED (SR)}, 
\label{eq:E5phiBP}
\end{eqnarray}
for $\dot{\phi}_B>0$, and 
\begin{eqnarray}
&&  \frac{\phi_B}{m_{Pl}}   \in  (-24.69, 1.4) \cup (3, +\infty) \rightarrow \text{KED (SR)}, \nonumber\\
&& 1.4 < \phi_B < 3 \rightarrow \text{subset of KED (NSR)}, \nonumber\\
&&  \frac{\phi_B}{m_{Pl}}  =  -24.7 \rightarrow \text{KED=PED (SR)}, \nonumber\\
&&  \frac{\phi_B}{m_{Pl}}  \in  (\phi_{min}, -24.71) \rightarrow \text{PED (SR)}
\label{eq:E5phiBN}
\end{eqnarray}
for $\dot{\phi}_B<0$.

We show the results of the background evolution for KED and PED initial conditions in Figs. \ref{fig:n1alpha5_dphp} and \ref{fig:n1alpha5_dphn}, with $\dot{\phi}_B>0$ and $\dot{\phi}_B<0$, respectively. In both figures, we show the evolutions of $a(t)$, $w(\phi)$ and $\epsilon_H$ for the same set of initial values of $\phi_B$. In the bouncing phase, the numerical evolution of $a(t)$ is compatible with the analytical solution (\ref{eq:a}) in the case of KED initial conditions whereas such a universality   disappears in the PED case. From the evolution of $w(\phi)$, we obtain three different phases, namely bouncing, transition and  slow-roll inflation in the KED case,  while in the PED case the bouncing and transition phases no longer exist, though the slow-roll inflation can still be achieved.

In this case, the entire range of $\phi_B$ (except for a small subset of KED) lead to the slow-roll inflation as shown in Table \ref{tab:n012_alpha_dphi}. However, this is not possible in  the case of $\alpha=0.1 m_{Pl}^2$ and the Starobinsky model \cite{Tao2017a,Tao2017b,Bonga2016} where the PED and a subset of KED initial conditions do not provide the slow-roll inflation.

In Tables \ref{tab:n1_dphip} and \ref{tab:n1_dphin}, we present the initial values of $\phi_B$ that lead to the slow-roll inflation, from which one can clearly see that, in order to get at least 60 $e$-folds  during the slow-roll inflation, one has to require
\begin{eqnarray}
&&  \frac{\phi_B}{m_{Pl}}   \in  (\phi_{min},-5.35) \cup (0.76, +\infty), \nonumber\\
&& ~~~~~~~~~~~~ \rightarrow  N_{inf} \gtrsim 60,
\label{eq:E5NphiBP}
\end{eqnarray}
for $\dot{\phi}_B>0$, and 
\begin{eqnarray}
&&  \frac{\phi_B}{m_{Pl}}  \in  (\phi_{min},-2.2) \cup (4.42, +\infty), \nonumber\\
&& ~~~~~~~~~~~~~~ \rightarrow  N_{inf} \gtrsim 60
\label{eq:E5NphiBN}
\end{eqnarray}
for $\dot{\phi}_B<0$.
In the above ranges, $N_{inf}$ grows as  $\phi_B$ increases.

Finally, we compare the numerical evolutions of the KE and the PE, which are shown in Fig. \ref{fig:n1rho}. Left panels (top and bottom) correspond to KED case at the bounce. Initially KE dominates and PE sub-dominates. As the evolution arrives  in the transition phase both the energies become comparable. Soon  PE becomes dominant, whereby  a slow-roll inflation is resulted. Middle panels (top and bottom) are displayed for a subset of KED initial conditions where the slow-roll inflation cannot be obtained as the PE remains sub-dominant. Right panels (top and bottom) is for the PED case. In top right ($\alpha=0.1 m_{Pl}^2$), the slow-roll inflation is not possible as the PE remains sub-dominant throughout the whole evolution,  while the bottom right ($\alpha=5 m_{Pl}^2$) provides the slow-roll inflation.

It is remarkable to note that the $E-model$ with small values of $\alpha$ (like $ \alpha=0.1 m_{Pl}^2$ etc.) does not provide a slow-roll inflation for the entire range of $\phi_B$. More preciously, PED and a subset of KED initial conditions do not lead to the slow-roll inflation. Though, a large range of KED initial values give rise to the slow-roll inflation. Such results are consistent with the Starobinsky model \cite{Tao2017a,Tao2017b,Bonga2016}. However, when the $E-model$ has large values of $\alpha$ (like $ \alpha=5 m_{Pl}^2$), the whole range of initial values of $\phi_B$ (except a subset of KED initial conditions) produces the slow-roll inflation phase.

\subsection{$\alpha-$attractor model with $n=2$}
\label{sec:n=2}

We now turn to   consider the $\alpha$-attractor model with $n=2$ [Eq.(\ref{eq:n2pot})]. The evolution of potential (\ref{eq:n2pot}) is shown in Fig. \ref{fig:pot}. Similar to $E-model$, the potential (\ref{eq:n2pot}) is bounded below by zero ($V(\phi) \geq 0$), and gives finite value $V(\phi)\simeq 2.8 \times 10^{-9} m_{Pl}^4$ (for $\alpha=5m_{Pl}^2 $) at $\phi \rightarrow +\infty$, whereas it shows divergence at $\phi \rightarrow -\infty$. Therefore, this is an asymmetric potential. In LQC, the maximum energy density is $\rho_c$ that constrains the initial values of $\phi_B$ as $(\phi_{min}, \infty)$, and $\phi_{min}$ is given as
\begin{eqnarray}
\phi_{min} &\simeq & \sqrt{6 \alpha} \arctan \text{h} \left[\sqrt{\mu^2-1}-\mu  \right]\nonumber\\
 &\simeq & -4.08 m_{Pl}~ \text{for}~ \alpha=0.5 m_{Pl}^2, \nonumber\\
 &\simeq & -12.88 m_{Pl}~ \text{for}~ \alpha=5 m_{Pl}^2,
\label{eq:n2phimin}
\end{eqnarray}
where
\begin{eqnarray}
\mu = 1-\sqrt{\frac{\alpha c^2}{4\rho_c}}.
\end{eqnarray}
Let us solve the background  equations (\ref{eq:Hub}) and (\ref{eq:ddphi}) with (\ref{eq:n2pot}) numerically for $\alpha=0.5 m_{Pl}^2$ and $c=2.9 \times 10^{-4} m_{Pl}$. In this case,
$\phi_{min} \simeq -4.08 m_{Pl}$. Similar to the $T-$ and $E-models$, the initial conditions are divided into two sub-classes; KED and PED, see Table \ref{tab:n012_alpha_dphi}, and is given by
\begin{eqnarray}
&&  \frac{\phi_B}{m_{Pl}}  \in  (-1.39, +\infty) \rightarrow \text{KED (SR)}, \nonumber\\
&& -3.93 < \phi_B \leq -1.4 \rightarrow \text{subset of KED (NSR)}, \nonumber\\
&&  \frac{\phi_B}{m_{Pl}} = -3.93 \rightarrow \text{KED=PED (NSR)}, \nonumber\\
&&  \frac{\phi_B}{m_{Pl}} \in  (\phi_{min}, -3.94) \rightarrow \text{PED (NSR)},
\label{eq:n2phiBP}
\end{eqnarray}
for $\dot{\phi}_B>0$, and 
\begin{eqnarray}
&&  \frac{\phi_B}{m_{Pl}}  \in  (2.4, +\infty) \rightarrow \text{KED (SR)}, \nonumber\\
&& -3.92 < \phi_B < 2.4 \rightarrow \text{subset of KED (NSR)}, \nonumber\\
&&  \frac{\phi_B}{m_{Pl}} =  -3.92 \rightarrow \text{KED=PED (NSR)}, \nonumber\\
&&  \frac{\phi_B}{m_{Pl}} \in  (\phi_{min}, -3.93) \rightarrow \text{PED (NSR)}, 
\label{eq:n2phiBN}
\end{eqnarray}
for $\dot{\phi}_B<0$, 
where $\phi_{min}$ is given by  Eq.(\ref{eq:n2phimin}). The numerical results are illustrated in Figs. \ref{fig:n2alpha05_dphp} and \ref{fig:n2alpha05_dphn} for a set of KED and PED initial conditions with $\dot{\phi}_B>0$ and $\dot{\phi}_B<0$, respectively. 
The explanation of these figures is quite similar to  the case of $E-model$. Therefore, we shall not repeat again. Here, we shall discuss the rest of results for the model (\ref{eq:n2pot}). In Tables \ref{tab:n2_dphip} and \ref{tab:n2_dphin}, we demonstrate the different inflationary parameters. Looking at both tables, the range of $\phi_B$ that is restricted to produce enough $e-$folds for the desired slow-roll inflation, is given by
\begin{eqnarray}
&&  \frac{\phi_B}{m_{Pl}}  \in  (-0.3, +\infty) \rightarrow  N_{inf} \gtrsim 60 ~\text{for}~ \dot{\phi}_B>0, \nonumber\\
&&  \frac{\phi_B}{m_{Pl}} \in  (3.23, +\infty) \rightarrow  N_{inf} \gtrsim 60 ~\text{for}~ \dot{\phi}_B<0,
\label{eq:n2N60phiB}
\end{eqnarray}
within which,   one can infer that the number of $e-$folds grows as the values of $\phi_B$ increase as shown in Tables \ref{tab:n2_dphip} and \ref{tab:n2_dphin}.
   
In the case of  $\alpha=5 m_{Pl}^2$ and $c=9.4 \times 10^{-5} m_{Pl}$, the numerical results are displayed in Figs. \ref{fig:n2alpha5_dphp} and \ref{fig:n2alpha5_dphn}. The range of $\phi_B$ is given as follows (see Table \ref{tab:n012_alpha_dphi}):
\begin{eqnarray}
&&  \frac{\phi_B}{m_{Pl}}  \in  (-12.4, -3.4) \cup (-1.4, +\infty) \rightarrow \text{KED (SR)}, \nonumber\\
&& -3.4 < \phi_B < -1.4 \rightarrow \text{subset of KED (NSR)}, \nonumber\\
&&  \frac{\phi_B}{m_{Pl}}  =  -12.41\rightarrow \text{KED=PED (SR)}, \nonumber\\
&&  \frac{\phi_B}{m_{Pl}}   \in   (\phi_{min}, -12.42) \rightarrow \text{PED (SR)}, 
\label{eq:n2phiB2P}
\end{eqnarray}
for $\dot{\phi}_B>0$, and 
\begin{eqnarray}
&&  \frac{\phi_B}{m_{Pl}}  \in  (-12.39, 1) \cup (2.7, +\infty) \rightarrow \text{KED (SR)}, \nonumber\\
&& 1 < \phi_B < 2.7 \rightarrow \text{subset of KED (NSR)}, \nonumber\\
&&  \frac{\phi_B}{m_{Pl}}  =  -12.4\rightarrow \text{KED=PED (SR)}, \nonumber\\
&&  \frac{\phi_B}{m_{Pl}}  \in   (\phi_{min}, -12.41) \rightarrow \text{PED (SR)},
\label{eq:n2phiB2N}
\end{eqnarray}
for $\dot{\phi}_B<0$, 
where $\phi_{min}$ is given by  Eq.(\ref{eq:n2phimin}). To obtain enough $e-$folds for the desired slow-roll inflarion, the range of $\phi_B$ requires as (see Tables  \ref{tab:n2_dphip} and \ref{tab:n2_dphin}):
\begin{eqnarray}
&&  \frac{\phi_B}{m_{Pl}}  \in  (\phi_{min}, -6) \cup (0.45, +\infty), \nonumber\\
&& ~~~~~~~~~~~~ \rightarrow  N_{inf} \gtrsim 60 ~\text{for}~ \dot{\phi}_B>0, \nonumber\\
&&  \frac{\phi_B}{m_{Pl}}  \in  (\phi_{min}, -3.6) \cup (4.04, +\infty),  \nonumber\\
&& ~~~~~~~~~~~~ \rightarrow  N_{inf} \gtrsim 60 ~\text{for}~ \dot{\phi}_B<0.
\label{eq:n2N60phiB2}
\end{eqnarray}
In the above range, $N_{inf}$ grows as the absolute value of $\phi_B$ increases that are displayed in Tables  \ref{tab:n2_dphip} and \ref{tab:n2_dphin}.
\begin{table*}
\caption{Table for $\alpha-$attractor model with $n=2$ [Eq.(\ref{eq:n2pot})]. Different inflationary parameters are shown for $\dot{\phi}_B>0$. }
\begin{center}
\resizebox{\textwidth}{!}{%
\begin{tabular}{ccccccccc}
\hline\hline
\\
$\alpha$~~ & $\phi_B$~~~  & Inflation~~~ & $t/t_{pl}$~~~ & $\epsilon$~~ & $w$ ~~& $N_{inf}$ &~~~$r_{w}^c/r_w$&~~~${w}^B$\\\\
\hline\hline
0.5 ~~ & $-0.4$~~~& begin~~~& $2.66349 \times 10^3$ ~~~& 0.99~~ & $-1/3$ ~~& ~~~& ~~~&\\
~~ & ~~~& slow-roll~~~& $1.10693 \times 10^4$ ~~~& $1.4 \times 10^{-4}$~~ & $-1$ ~~& 46.84 ~~~&$r_{w}^c > r_w$ ~~~& $>0$\\
~~ & ~~~& end~~~& $3.592 \times 10^5$ ~~~& 1.0~~ & $-1/3$ ~~& ~~~& ~~~& \\\\
~~ & $-0.3$~~~& begin~~~& $2.63635 \times 10^3$ ~~~& 0.99~~ & $-1/3$ ~~& ~~~& ~~~&\\
~~ & ~~~& slow-roll~~~& $1.15044 \times 10^4$ ~~~& $1.9 \times 10^{-4}$~~ & $-1$ ~~& 60.52 ~~~&$r_{w}^c = r_w$ ~~~& $>0$\\
~~ & ~~~& end~~~& $4.535 \times 10^5$ ~~~& 0.99~~ & $-1/3$ ~~& ~~~& ~~~& \\\\
~~ & $-0.1$~~~& begin~~~& $2.59876 \times 10^3$ ~~~& 0.99~~ & $-1/3$ ~~& ~~~& ~~~&\\
~~ & ~~~& slow-roll~~~& $1.24129 \times 10^4$ ~~~& $2.0 \times 10^{-4}$~~ & $-1$ ~~& 99.85 ~~~&$r_{w}^c < r_w$ ~~~& $>0$\\
~~ & ~~~& end~~~& $7.217 \times 10^5$ ~~~& 0.99~~ & $-1/3$ ~~& ~~~& ~~~& \\\\
~~ & $0.1$~~~& begin~~~& $2.57597 \times 10^3$ ~~~& 0.99~~ & $-1/3$ ~~& ~~~& ~~~&\\
~~ & ~~~& slow-roll~~~& $1.33583 \times 10^4$ ~~~& $2.1 \times 10^{-4}$~~ & $-1$ ~~& 163.90 ~~~&$r_{w}^c < r_w$ ~~~& $>0$\\
~~ & ~~~& end~~~& $1.20567 \times 10^6$ ~~~& 0.99~~ & $-1/3$ ~~& ~~~& ~~~& \\\\
5~~ & $-12.6$~~~& begin~~~& $0.22$ ~~~& 4.41~~ & $-1/3$ ~~& ~~~& ~~~&\\
~~ & ~~~& slow-roll~~~& $2.72126 \times 10^2$ ~~~& $4.1 \times 10^{-2}$~~ & $-0.971$ ~~& 183.29 ~~~&$r_{w}^c < r_w$ ~~~& $<0$\\
~~ & ~~~& end~~~& $1.31 \times 10^5$ ~~~& 0.99~~ & $-1/3$ ~~& ~~~& ~~~& \\\\
~~ & $-6.5$~~~& begin~~~& $48.135 $ ~~~& 0.99~~ & $-1/3$ ~~& ~~~& ~~~&\\
~~ & ~~~& slow-roll~~~& $4.86753 \times 10^2$ ~~~& $4.4 \times 10^{-2}$~~ & $-0.970$ ~~& 70.13 ~~~&$r_{w}^c < r_w$ ~~~& $>0$\\
~~ & ~~~& end~~~& $1.284 \times 10^5$ ~~~& 1.0~~ & $-1/3$ ~~& ~~~& ~~~& \\\\
~~ & $-6$~~~& begin~~~& $73.68$ ~~~& 0.99~~ & $-1/3$ ~~& ~~~& ~~~&\\
~~ & ~~~& slow-roll~~~& $7.11715 \times 10^2$ ~~~& $4.5 \times 10^{-2}$~~ & $-0.969$ ~~& 60.56 ~~~&$r_{w}^c = r_w$ ~~~& $>0$\\
~~ & ~~~& end~~~& $1.268 \times 10^5$ ~~~& 0.99~~ & $-1/3$ ~~& ~~~& ~~~& \\\\
~~ & $-5.5$~~~& begin~~~& $1.13474 \times 10^2$ ~~~& 0.99~~ & $-1/3$ ~~& ~~~& ~~~&\\
~~ & ~~~& slow-roll~~~& $1.04352 \times 10^3$ ~~~& $4.7 \times 10^{-2}$~~ & $-0.968$ ~~& 51.10 ~~~&$r_{w}^c > r_w$ ~~~& $>0$\\
~~ & ~~~& end~~~& $1.245 \times 10^5$ ~~~& 0.99~~ & $-1/3$ ~~& ~~~& ~~~& \\\\
~~ & $0.1$~~~& begin~~~& $3.3795 \times 10^3$ ~~~& 0.99~~ & $-1/3$ ~~& ~~~& ~~~&\\
~~ & ~~~& slow-roll~~~& $1.16382 \times 10^4$ ~~~& $2.3 \times 10^{-4}$~~ & $-1$ ~~& 39.39 ~~~&$r_{w}^c > r_w$ ~~~& $>0$\\
~~ & ~~~& end~~~& $4.615 \times 10^5$ ~~~& 0.99~~ & $-1/3$ ~~& ~~~& ~~~& \\\\
~~ & $0.45$~~~& begin~~~& $3.13922 \times 10^3$ ~~~& 1.0~~ & $-1/3$ ~~& ~~~& ~~~&\\
~~ & ~~~& slow-roll~~~& $1.16529 \times 10^4$ ~~~& $1.5 \times 10^{-4}$~~ & $-1$ ~~& 60.23 ~~~&$r_{w}^c = r_w$ ~~~& $>0$\\
~~ & ~~~& end~~~& $6.36 \times 10^5$ ~~~& 1.0~~ & $-1/3$ ~~& ~~~& ~~~& \\\\
~~ & $0.7$~~~& begin~~~& $3.0145 \times 10^3$ ~~~& 0.99~~ & $-1/3$ ~~& ~~~& ~~~&\\
~~ & ~~~& slow-roll~~~& $1.17477 \times 10^4$ ~~~& $7.9 \times 10^{-6}$~~ & $-1$ ~~& 79.72 ~~~&$r_{w}^c < r_w$ ~~~& $>0$\\
~~ & ~~~& end~~~& $7.91 \times 10^5$ ~~~& 1.0~~ & $-1/3$ ~~& ~~~& ~~~& \\\\
\hline\hline
\end{tabular}}
\label{tab:n2_dphip}
\end{center}
\end{table*}
\begin{table*}
\caption{This table designates the $\alpha-$attractor model with $n=2$ [Eq.(\ref{eq:n2pot})] for $\dot{\phi}_B<0$. }
\begin{center}
\resizebox{\textwidth}{!}{%
\begin{tabular}{ccccccccc}
\hline\hline
\\
$\alpha$~~ & $\phi_B$~~~  & Inflation~~~ & $t/t_{pl}$~~~ & $\epsilon$~~ & $w$ ~~& $N_{inf}$ &~~~$r_{w}^c/r_w$&~~~${w}^B$\\\\
\hline\hline
0.5 ~~ & $3.1$~~~& begin~~~& $2.64113 \times 10^3$ ~~~& 0.99~~ & $-1/3$ ~~& ~~~& ~~~&\\
~~ & ~~~& slow-roll~~~& $2.49628 \times 10^4$ ~~~& $1.4 \times 10^{-3}$~~ & $-0.999$ ~~& 43.04 ~~~&$r_{w}^c > r_w$ ~~~& $>0$\\
~~ & ~~~& end~~~& $3.327 \times 10^5$ ~~~& 1.0~~ & $-1/3$ ~~& ~~~& ~~~& \\\\
~~ & $3.23$~~~& begin~~~& $2.61291 \times 10^3$ ~~~& 0.99~~ & $-1/3$ ~~& ~~~& ~~~&\\
~~ & ~~~& slow-roll~~~& $2.51853 \times 10^4$ ~~~& $7.6 \times 10^{-5}$~~ & $-0.999$ ~~& 60.66 ~~~&$r_{w}^c = r_w$ ~~~& $>0$\\
~~ & ~~~& end~~~& $4.544 \times 10^5$ ~~~& 0.99~~ & $-1/3$ ~~& ~~~& ~~~& \\\\
~~ & $3.3$~~~& begin~~~& $2.60143 \times 10^3$ ~~~& 0.99~~ & $-1/3$ ~~& ~~~& ~~~&\\
~~ & ~~~& slow-roll~~~& $4.06202 \times 10^4$ ~~~& $1.8 \times 10^{-3}$~~ & $-0.999$ ~~& 72.67 ~~~&$r_{w}^c < r_w$ ~~~& $>0$\\
~~ & ~~~& end~~~& $5.367 \times 10^5$ ~~~& 1.0~~ & $-1/3$ ~~& ~~~& ~~~& \\\\
~~ & $3.5$~~~& begin~~~& $2.57631 \times 10^3$ ~~~& 0.99~~ & $-1/3$ ~~& ~~~& ~~~&\\
~~ & ~~~& slow-roll~~~& $3.62962 \times 10^4$ ~~~& $1.5 \times 10^{-3}$~~ & $-0.999$ ~~& 120.25 ~~~&$r_{w}^c < r_w$ ~~~& $>0$\\
~~ & ~~~& end~~~& $8.598 \times 10^5$ ~~~& 1.0~~ & $-1/3$ ~~& ~~~& ~~~& \\\\
5~~ & $-12.6$~~~& begin~~~& $0.3$ ~~~& 1.9~~ & $-0.804$ ~~& ~~~& ~~~&\\
~~ & ~~~& slow-roll~~~& $0.666$ ~~~& $1.0 \times 10^{-4}$~~ & $-1$ ~~& 186.30
~~~&$r_{w}^c < r_w$ ~~~& $<0$\\
~~ & ~~~& end~~~& $1.312 \times 10^5$ ~~~& 0.99~~ & $-1/3$ ~~& ~~~& ~~~& \\\\
~~ & $-5$~~~& begin~~~& $32.08 $ ~~~& 0.99~~ & $-1/3$ ~~& ~~~& ~~~&\\
~~ & ~~~& slow-roll~~~& $82.95$ ~~~& $1.2 \times 10^{-5}$~~ & $-1$ ~~& 81.34 ~~~&$r_{w}^c < r_w$ ~~~& $>0$\\
~~ & ~~~& end~~~& $1.294 \times 10^5$ ~~~& 0.99~~ & $-1/3$ ~~& ~~~& ~~~& \\\\
~~ & $-3.6$~~~& begin~~~& $81.36$ ~~~& 0.99~~ & $-1/3$ ~~& ~~~& ~~~&\\
~~ & ~~~& slow-roll~~~& $2.0934 \times 10^2$ ~~~& $4.3 \times 10^{-6}$~~ & $-1$ ~~& 60.28 ~~~&$r_{w}^c = r_w$ ~~~& $>0$\\
~~ & ~~~& end~~~& $1.266 \times 10^5$ ~~~& 0.99~~ & $-1/3$ ~~& ~~~& ~~~& \\\\
~~ & $-3$~~~& begin~~~& $1.2204 \times 10^2$ ~~~& 0.99~~ & $-1/3$ ~~& ~~~& ~~~&\\
~~ & ~~~& slow-roll~~~& $3.12689 \times 10^2$ ~~~& $1.0 \times 10^{-5}$~~ & $-1$ ~~& 51.42 ~~~&$r_{w}^c > r_w$ ~~~& $>0$\\
~~ & ~~~& end~~~& $1.246 \times 10^5$ ~~~& 0.99~~ & $-1/3$ ~~& ~~~& ~~~& \\\\
~~ & $3.8$~~~& begin~~~& $3.218 \times 10^3$ ~~~& 0.99~~ & $-1/3$ ~~& ~~~& ~~~&\\
~~ & ~~~& slow-roll~~~& $2.58609 \times 10^4$ ~~~& $4.0 \times 10^{-3}$~~ & $-0.997$ ~~& 44.70 ~~~&$r_{w}^c > r_w$ ~~~& $>0$\\
~~ & ~~~& end~~~& $5.068 \times 10^5$ ~~~& 0.99~~ & $-1/3$ ~~& ~~~& ~~~& \\\\
~~ & $4.04$~~~& begin~~~& $3.07306 \times 10^3$ ~~~& 1.0~~ & $-1/3$ ~~& ~~~& ~~~&\\
~~ & ~~~& slow-roll~~~& $2.80442 \times 10^4$ ~~~& $2.3 \times 10^{-3}$~~ & $-0.998$ ~~& 60.65 ~~~&$r_{w}^c = r_w$ ~~~& $>0$\\
~~ & ~~~& end~~~& $6.39 \times 10^5$ ~~~& 0.99~~ & $-1/3$ ~~& ~~~& ~~~& \\\\
~~ & $4.2$~~~& begin~~~& $2.99612 \times 10^3$ ~~~& 0.99~~ & $-1/3$ ~~& ~~~& ~~~&\\
~~ & ~~~& slow-roll~~~& $3.12867 \times 10^4$ ~~~& $1.3 \times 10^{-3}$~~ & $-0.998$ ~~& 73.38 ~~~&$r_{w}^c < r_w$ ~~~& $>0$\\
~~ & ~~~& end~~~& $7.41 \times 10^5$ ~~~& 1.0~~ & $-1/3$ ~~& ~~~& ~~~& \\\\
\hline\hline
\end{tabular}}
\label{tab:n2_dphin}
\end{center}
\end{table*}

\begin{figure*}[tbp]
\begin{center}
\begin{tabular}{cc}
{\includegraphics[width=2.8in,height=2.5in,angle=0]{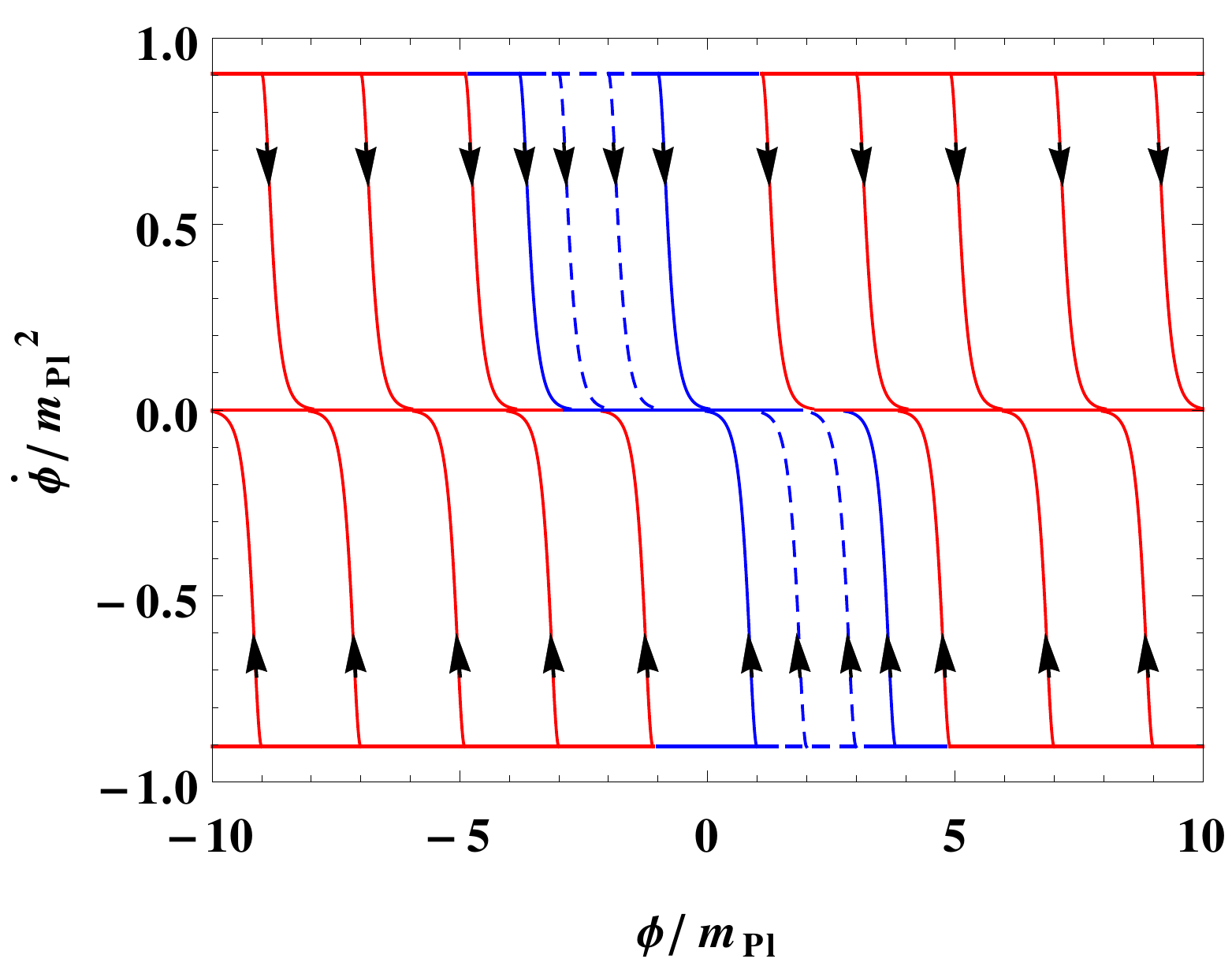}} &
{\includegraphics[width=2.8in,height=2.5in,angle=0]{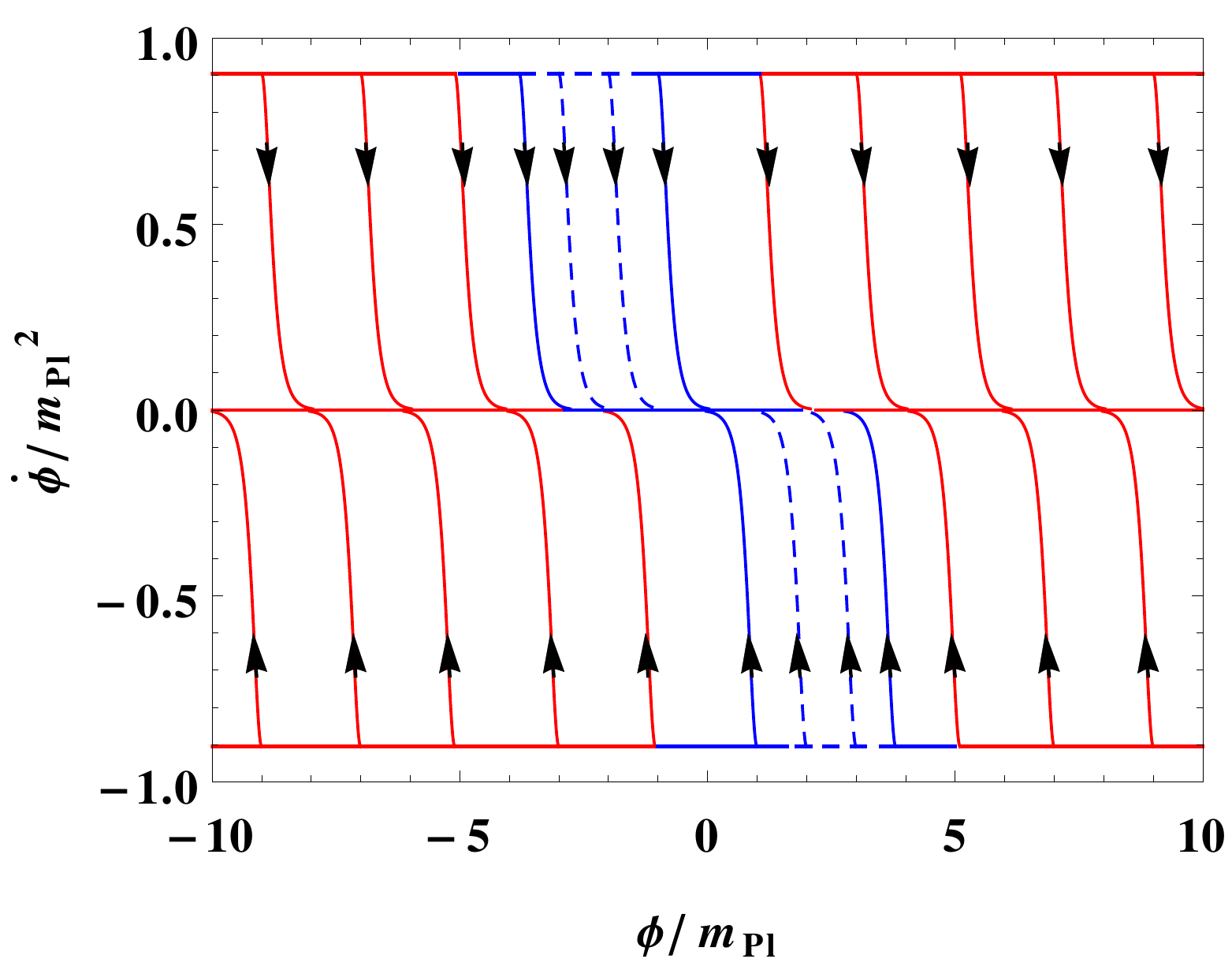}} 
\end{tabular}
\end{center}
\caption{ This figure represents the phase portraits of $T-model$ in $(\phi/m_{Pl}, \dot{\phi}/m_{Pl}^2)$ plane. All trajectories (with arrowheads) begin at the bounce at which we have    $\rho=\rho_c$ (boundary curve without arrowheads). This  surface extends as $\phi \rightarrow \pm \infty$ (left, $\alpha=10m_{Pl}^2$) and  $\phi \rightarrow \pm \phi_{max}$ (right, $\alpha=10^{10}m_{Pl}^2$, see  Eq.(\ref{eq:TNphimax2})), but here we exhibit only a small portion of it. The solid curves (red and blue) provide the slow-roll inflation, which are consistent with observations only by the red ones, and the dashed (blue) curves denotes the non-inflationary phase.}
\label{fig:Tport}
\end{figure*}
\begin{figure*}[tbp]
\begin{center}
\begin{tabular}{cc}
{\includegraphics[width=2.8in,height=2.5in,angle=0]{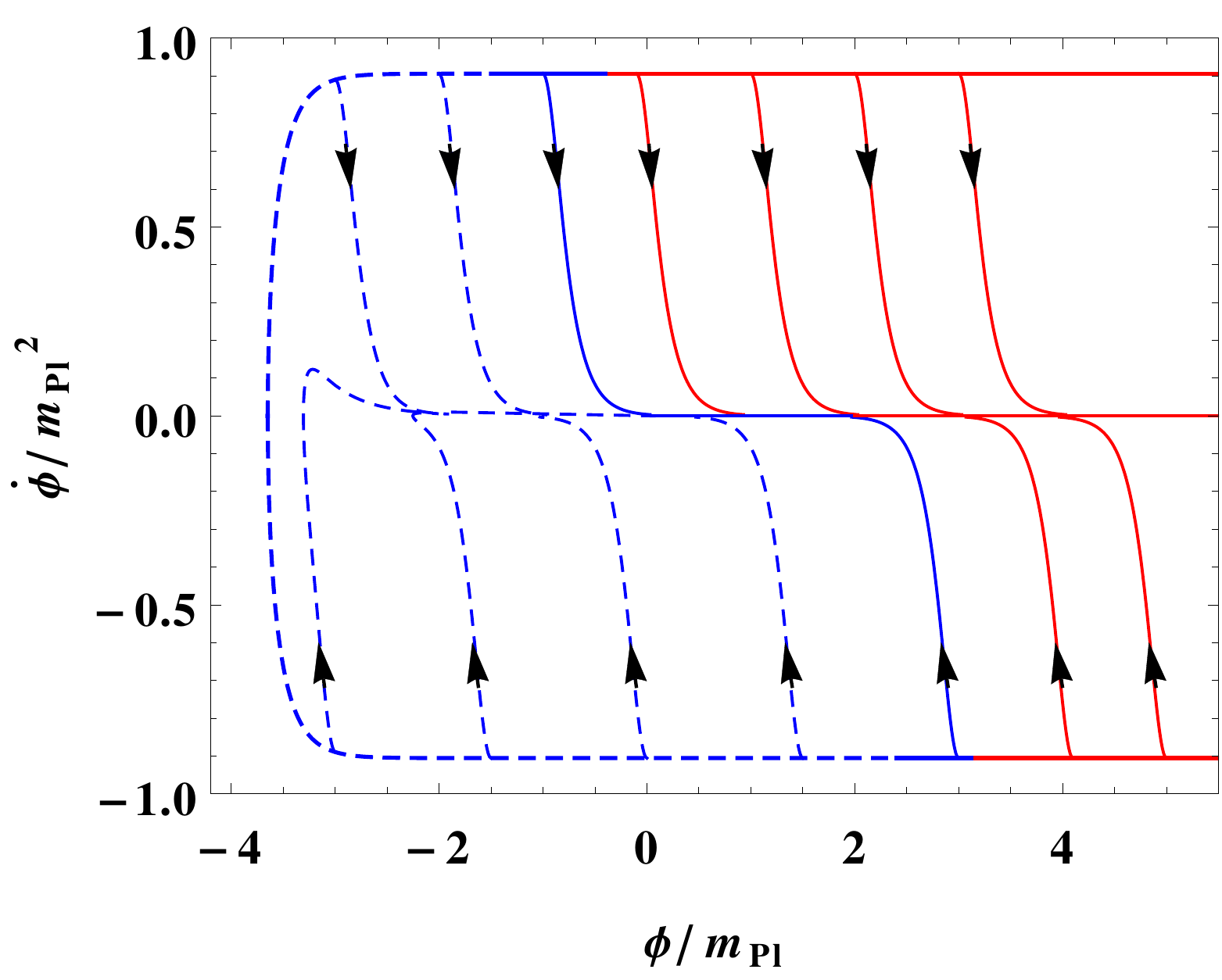}} &
{\includegraphics[width=2.8in,height=2.5in,angle=0]{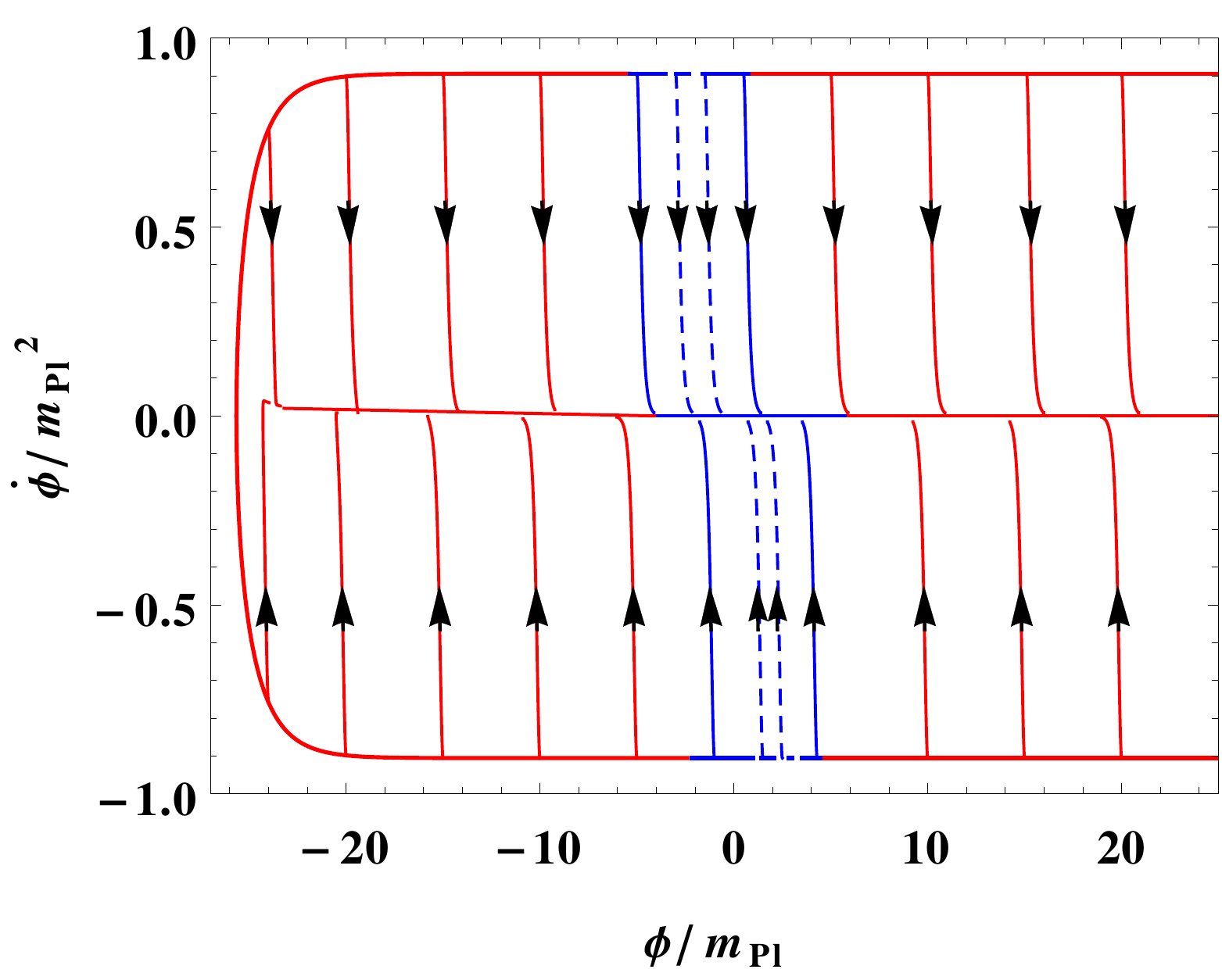}} 
\end{tabular}
\end{center}
\caption{ This figure shows the phase space trajectories of $E-model$ in $(\phi/m_{Pl}, \dot{\phi}/m_{Pl}^2)$ plane. All trajectories (with arrowheads) start at the bounce that corresponds to $\rho=\rho_c$ (boundary curve without arrowheads). This initial  surface extends from $\phi = - \phi_{min}$ to $\infty$ (left, $\alpha=0.1m_{Pl}^2$) and (right, $\alpha=5m_{Pl}^2$, see  Eq.(\ref{eq:Ephimin})), but here we display only a  part of it. The red curves provide the desired slow-roll inflation, whereas the blue (solid) ones do not. The dashed (blue) curves represent the case without  inflation. Since the figure extends from all the way to $\infty$, the fractions of blue (dashed and solid) curves are extremely small in comparison with the red ones. Hence, one  concludes that the existence of the slow-roll inflation is almost unavoidable.}
\label{fig:Eport}
\end{figure*}
\begin{figure*}[tbp]
\begin{center}
\begin{tabular}{cc}
{\includegraphics[width=2.8in,height=2.5in,angle=0]{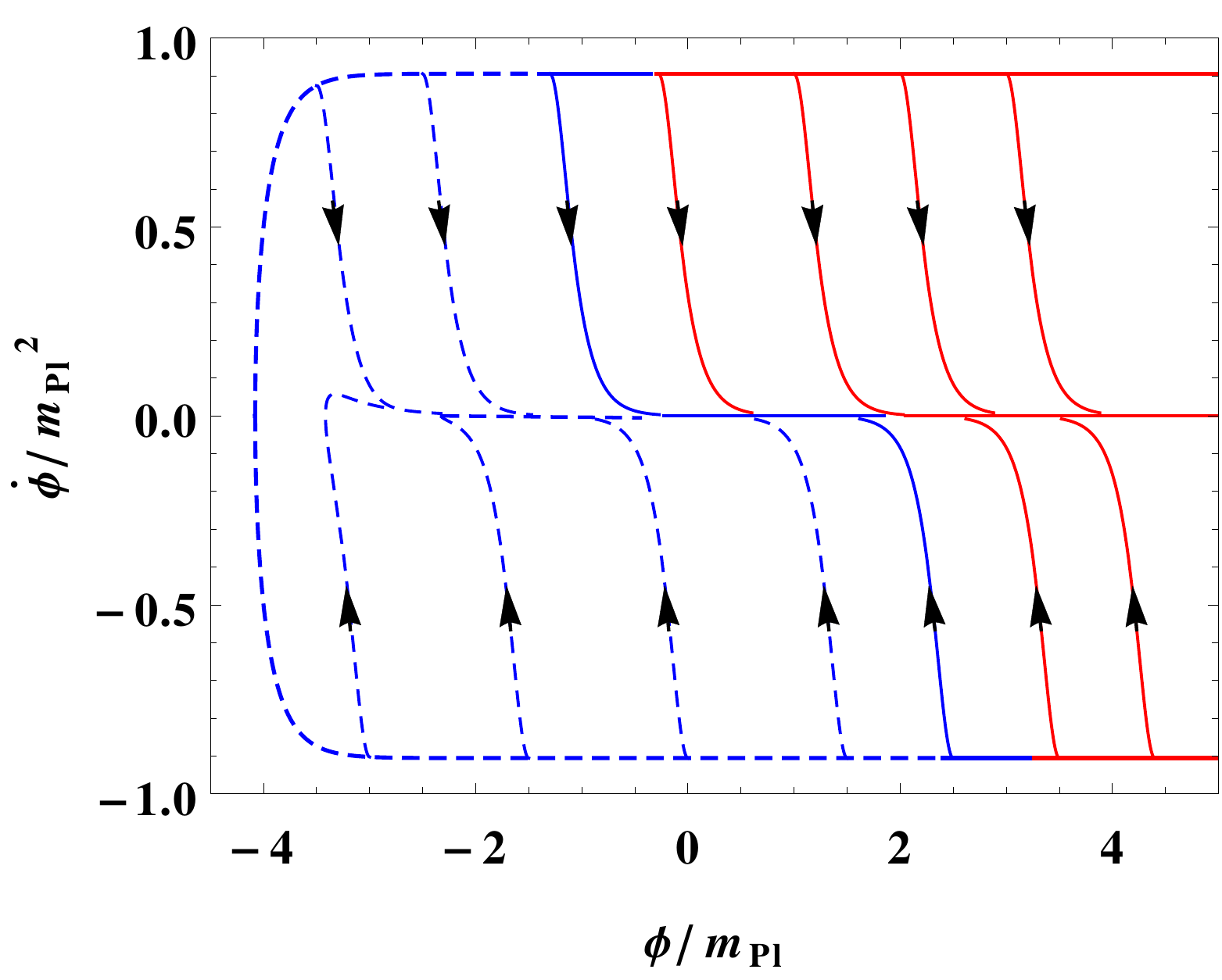}} &
{\includegraphics[width=2.8in,height=2.5in,angle=0]{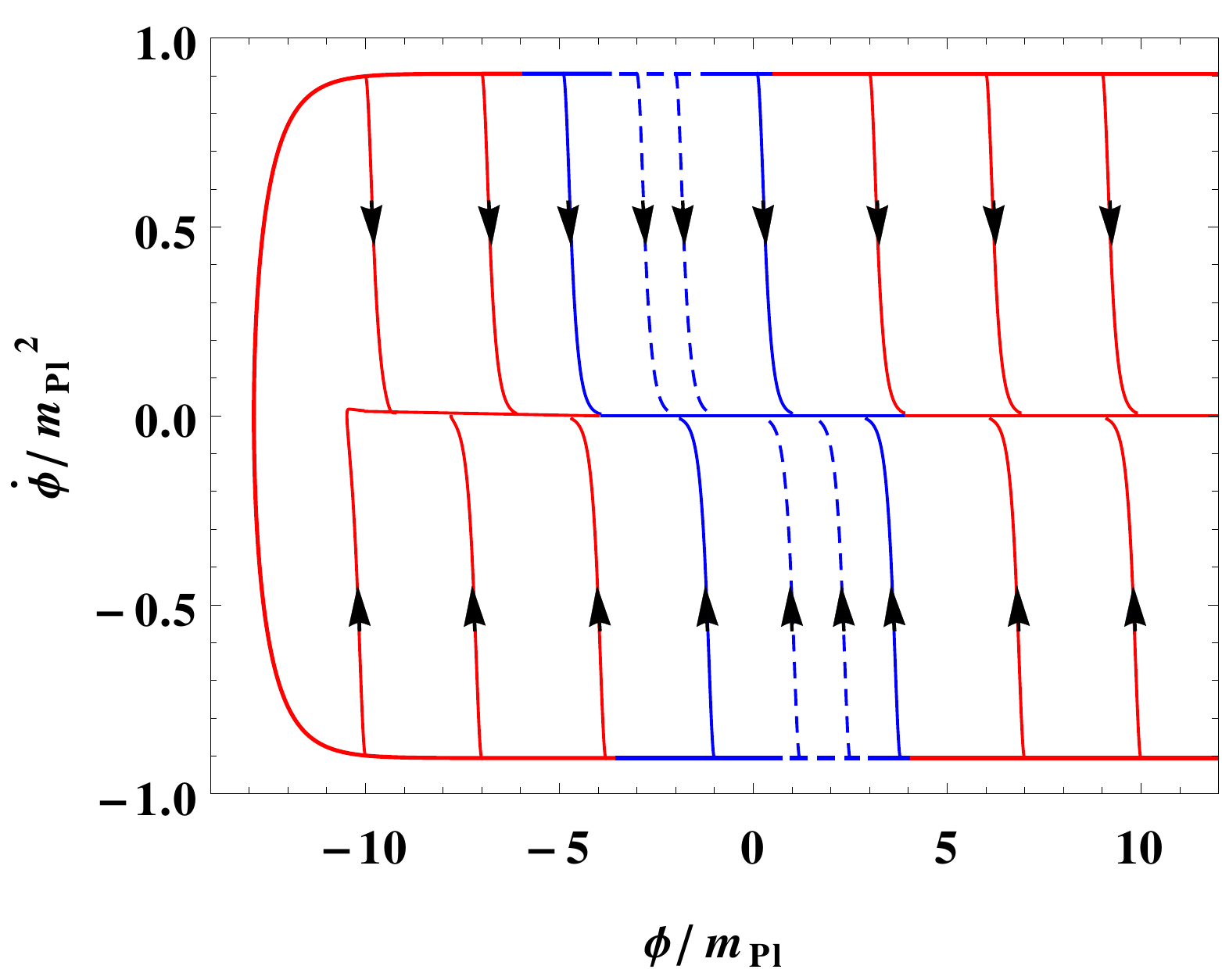}} 
\end{tabular}
\end{center}
\caption{This figure exhibits the phase portraits for $\alpha=0.5m_{Pl}^2$ (left) and $\alpha=5m_{Pl}^2$ (right), and similar to Fig. \ref{fig:Eport}, but for $\alpha-$attractor model with $n=2$.}
\label{fig:n2port}
\end{figure*}

Finally, we show the numerical evolutions of KE and PE in Fig. \ref{fig:n2rho}. The explanation of this figure is similar to Fig. \ref{fig:n1rho} of $E-model$.

\section{Phase space analysis and the desired slow-roll inflation}
\label{sec:phase}

In this section, we study the phase space analysis for the models   considered in the last sections. Let us first examine the symmetric $T-model$ with two different values of $\alpha$. In the case of  $\alpha=10 m_{Pl}^2$, the range of initial conditions having slow-roll/no slow-roll inflation, and consistent with observations are presented in  Eqs.(\ref{eq:TNphiB}),  (\ref{eq:TN60phiB}) and (\ref{eq:TNphiBsym}). Fig. \ref{fig:Tport} exhibits the evolution of few trajectories in $(\phi/m_{Pl}, \dot{\phi}/m_{Pl}^2)$ plane starting  from the quantum bounce (boundary curve without arrows). As mentioned in the subsection \ref{sec:Tmodel}, the initial data surface is not compressed: $| \dot{\phi}_B | < 0.91 m_{Pl}^2 $ and $\phi_B \rightarrow \pm \infty$. The dashed (blue) trajectories demonstrate the non-inflationary phase, while the solid (blue) ones show the inflationary phase but do not lead to the desired slow-roll inflation as they do not produce sufficient $e-$folds. Only red trajectories exhibit the desired slow-roll inflation which are compatible with observations. Similarly, dashed and solid (blue) parts of the initial  surface correspond to non-inflationary and a subset of inflationary phase that is not consistent with observations, whereas red part is compatible with observations. One can clearly see that the region of the non-inflationary phase and the part which does not provide the desired slow-roll inflation are almost negligible in comparison with the whole initial phase. Thus, a substantial fraction of the initial conditions generate a desired slow-roll inflation. For $\alpha=10^{10} m_{Pl}^2$, the range of initial conditions that are compatible with observations or not are presented in  Eqs.(\ref{eq:TNphiB2}), (\ref{eq:TN60phiB2}) and (\ref{eq:TNphiB2sym}), and the phase portrait is shown in Fig. \ref{fig:Tport}. In this case, the initial data surface is compact:  $| \dot{\phi}_B | < 0.91 m_{Pl}^2 $ and $\phi_B \rightarrow \pm 2.56 \times 10^5 m_{Pl}$ [see  Eq.(\ref{eq:TNphimax2})]. The rest is the same as in the case of $\alpha=10 m_{Pl}^2$. Note that, a small portion of the full initial conditions are shown in Fig. \ref{fig:Tport}.

Next, we carry out the phase analysis for $E-model$ with $\alpha=0.1 m_{Pl}^2$ and $5 m_{Pl}^2$. The phase portraits are shown in Fig. \ref{fig:Eport}. Let us first consider $\alpha=0.1 m_{Pl}^2$. In this case, the initial  surface is semi compact: $| \dot{\phi}_B | < 0.91 m_{Pl}^2 $ and $\phi_B \in (-3.64, \infty)$. Fig. \ref{fig:Eport} shows that the trajectories starting from the bounce   represent the slow-roll (red ones correspond to enough $e-$folds that are consistent with observations while solid blue ones are not) and without slow-roll inflation (dashed blue). Here, PED and a subset of KED initial conditions do not lead to the slow-roll inflation (blue; dashed and solid lines),  while KED initial values (except a small subset) provide (red). In a similar way, the dashed and solid (blue) parts of the bounce  (boundary curve) display the region of non-inflationary and inflationary phases (not compatible with observations), and the red part denotes the desired slow-roll inflation phase that is consistent with observations. The range of initial conditions are presented in  Eqs.(\ref{eq:EphiBP}), (\ref{eq:EphiBN}) and (\ref{eq:ENphiB}). Second, we take $\alpha=5 m_{Pl}^2$, here also the bounce surface is semi compact: $| \dot{\phi}_B | < 0.91 m_{Pl}^2 $ and $\phi_B \in (-25.65, \infty)$. The rest is the same as in the case of  $\alpha=0.1 m_{Pl}^2$ except PED initial conditions. Here, one can obtain the desired slow-roll inflation   with the KED (except a subset) and the PED initial values. However, it is not possible for the $\alpha=0.1 m_{Pl}^2$ case. The range of initial conditions is shown in  Eqs.(\ref{eq:E5phiBP}), (\ref{eq:E5phiBN}), (\ref{eq:E5NphiBP}) and (\ref{eq:E5NphiBN}). Notice that, a small portion of the whole initial conditions is displayed in Fig. \ref{fig:Eport}. 

Finally, we investigate the $\alpha-$attractor model with $n=2$. The phase portraits are presented in Fig. \ref{fig:n2port} for $\alpha=0.5 m_{Pl}^2$ (left) and $\alpha=5 m_{Pl}^2$ (right). In this case, the bouncing phase is also semi-finite: $| \dot{\phi}_B | < 0.91 m_{Pl}^2 $; $\phi_B \in (-4.08, \infty)$ for $\alpha=0.5 m_{Pl}^2$ and $\phi_B \in (-12.88, \infty)$ for $\alpha=5 m_{Pl}^2$. Similar to $E-model$, here also one value of $\alpha$ leads to the desired slow-roll inflation for both KED and PED initial conditions whereas it is not feasible for another value. In Fig. \ref{fig:n2port}, only a small part of initial values is shown. However, the entire range of initial conditions is given by  Eqs.(\ref{eq:n2phiBP})$-$(\ref{eq:n2N60phiB2}).

\section{Comparison with the power-law and the Starobinsky potentials}
\label{sec:compare}

In the literature, a large number of inflationary models have been studied that can be consistent with observations. In the case of a single field inflation, Planck 2015 results demonstrate that the quadratic potential is not favored compared to the power-law [$V(\phi) \propto \phi^n$ with $n<2$], the Starobinsky and $\alpha-$attractor models \cite{Planck2015}. Therefore, in this section, we shall compare our results with these known models  \cite{alam2017,Bonga2016}.

Let us first consider the results of the $T-model$ with the power-law and Starobinsky potentials. In the case of power-law potential, both KED and PED initial conditions produce the desired slow-roll inflation, and are consistent with observations in terms of the number of $e-$folds \cite{alam2017}. In the case of $T-model$ (with $\alpha=10^{10} m_{Pl}^2$), there is a small subset of KED initial conditions that does not generate the slow-roll inflation, and in terms of the number of $e-$folds, both KED (except for a very small subset) and PED initial values are consistent with observations, while the Starobinsky potential is observationally compatible only for KED (except for a very small subset) initial values and not for PED ones \cite{Bonga2016}.

Next, we consider the results of the $E-model$ with the power-law and Starobinsky potentials. In the case of power-law potential, both KED and PED initial conditions are compatible with observations, whereas the Starobinsky inflation is consistent only for KED ones. In the case of $E-model$, there is a subset of KED initial values that corresponds to the non-inflationary phase. For $\alpha=0.1 m_{Pl}^2$, our results are consistent with the Starobinsky model as both models (Starobinsky and $E-model$ with $\alpha=0.1 m_{Pl}^2$) lead to the desired slow-roll inflationary phase only for KED (except for a very small subset) initial values and not for PED ones. However, if we consider large values of $\alpha$ (say $\alpha=5 m_{Pl}^2$), the scenario will be different as in this case both KED (except a very small subset) and PED initial conditions lead to the desired slow-roll inflation and consistent with present observations in terms of the number of $e-$folds.

Finally, we consider the obtained results of $\alpha-$attractor model with $n=2$. Similar to $E-model$, here also the $\alpha-$model with $n=2$ is consistent with the Starobinsky model for small values of $\alpha$ (say $\alpha=0.5 m_{Pl}^2$), but for large values of $\alpha$ (say $\alpha=5 m_{Pl}^2$), both KED (except for a small subset) and PED initial values provide the desired slow-roll inflation, and compatible with the current observations as they all  produce enough $e-$folds.

\section{Conclusions}
\label{sec:conc}

In the context of LQC, in this paper we have systematically  investigated the preinflationary dynamics of the $T$, $E-models$ and  $\alpha-$attractor with $n=2$ for various  cases (PIV and NIV, also KED and PED). Our analysis bears resemblance with the study of the scalar field dynamics for the $\alpha-$attractor effective potential on RS brane with time like extra dimension \cite{shtanov}. We have chosen these models as they are favored by the Planck 2015 data \cite{Planck2015}. 

In particular,  we have first performed the detailed numerical analysis of the background evolution of the universe  for  $T-model$ with $\alpha=10m_{Pl}^2$ and $10^{10}m_{Pl}^2$. Due to the symmetry of $T-model$ potential, we have chosen only PIV at the quantum bounce. Further, initial conditions are divided into the KED and PED cases at the bounce. In the case of $\alpha=10m_{Pl}^2$, we have only KED initial conditions during the entire bouncing phase. However, to obtain both KED and PED initial values at the bounce, $\alpha$ should be very large (say $\alpha=10^{10}m_{Pl}^2$). The numerical results for $T-model$ are shown in Figs. \ref{fig:n0alpha10_dphp} and \ref{fig:n0alpha10p10_dphp}, in which the scale factor $a(t)$, EoS $w(\phi)$ and slow-roll parameter $\epsilon_H$ are shown for the same set of initial values of $\phi_B$. In the evolution of $w(\phi)$ and $\epsilon_H$, we have obtained inflationary and non-inflationary phases for the KED case. This implies that a small subset exists in which it does not give inflation, see Figs. \ref{fig:n0alpha10_dphp}, \ref{fig:n0alpha10p10_dphp} and Table \ref{tab:n012_alpha_dphi}. In the case of KED initial conditions (except for a very small a subset), the universe is always divided into three distinct phases prior to the reheating: {\em bouncing, transition and the slow-roll inflation}. In the bouncing phase, the evolution of the background is independent not only of  the wide ranges of initial values but also of the potentials. Specially, the numerical evolution of the expansion factor $a(t)$ has shown the universal feature and well approximated by the analytical solution (\ref{eq:a}), see upper panels of Fig. \ref{fig:n0alpha10p10_dphp}. During this phase, the EoS stays pegged at unity, $w(\phi) \simeq +1$. Though, in the transition phase, it decreases quickly from $w(\phi) \simeq +1$ to $w(\phi) \simeq -1$. The span of the transition phase is very short in comparison with  other two phases. Afterwards, the universe enters the slow-roll inflationary phase,   where $\epsilon_H$ is large initially, but soon  decreases to almost  zero, by which the slow-roll inflation starts, as exhibited in the upper panels of Fig. \ref{fig:n0alpha10p10_dphp}. During the slow-roll inflation, we also obtained the number of $e$-folds that is displayed in Table \ref{tab:n0_dphip}. In the case of PED initial values, the universality of the scale factor $a(t)$ is lost, and the bouncing phase no longer exists. However, the slow-roll inflation can still be acquired for a long period, and correspondingly one can obtain a large number of $e$-folds, as shown in the bottom panels of Fig. \ref{fig:n0alpha10p10_dphp} and Table \ref{tab:n0_dphip}.

Next, we have investigated  the evolution of  the background for $E-model$ with $\alpha=0.1m_{Pl}^2$ and $5m_{Pl}^2$. This model is not symmetric. Therefore, we have examined both PIV and NIV: the numerical evolution of the background is divided into the form of KED and PED initial conditions at the bounce. In LQC, the total energy density can not be larger than $\rho_c$. We have found that the KED evolution has a large range of $\phi_B$ than the PED ones, see Table \ref{tab:n012_alpha_dphi}. The numerical evolutions of $a(t)$, $w(\phi)$ and $\epsilon_H$ for $E-model$ with $\alpha=0.1m_{Pl}^2$ and $5m_{Pl}^2$ are shown in Figs. \ref{fig:n1alpha01_dphp}, \ref{fig:n1alpha01_dphn}, \ref{fig:n1alpha5_dphp} and \ref{fig:n1alpha5_dphn}. In the case of $\alpha=0.1m_{Pl}^2$, the entire range of $\phi_B$ does not give rise to the slow-roll inflation. In other words, a large range of KED (except a small subset) initial conditions provide the slow-roll inflation whereas a small subset of KED and the whole range of PED initial conditions do not. Similar results for the Starobinsky model were shown in  \cite{Bonga2016}. Although, in the case of $\alpha=5m_{Pl}^2$, both KED (except a small subset) and PED initial values provide slow-roll inflation as shown in Figs. \ref{fig:n1alpha5_dphp}, \ref{fig:n1alpha5_dphn} and Table \ref{tab:n012_alpha_dphi}. We have also found the number of $e$-folds which is exhibited in Table \ref{tab:n1_dphip} and \ref{tab:n1_dphin}.

Then, we have considered   the background evolution  of the $\alpha-$attractor model with $n=2$ for $\alpha=0.5m_{Pl}^2$ and $5m_{Pl}^2$. This model is also asymmetric, and the total energy density can not exceed $\rho_c$. The numerical results are shown in Figs. \ref{fig:n2alpha05_dphp}, \ref{fig:n2alpha05_dphn}, \ref{fig:n2alpha5_dphp} and \ref{fig:n2alpha5_dphn}. Similar to the $E-model$, here also, for small values of $\alpha$ (say $\alpha=0.5m_{Pl}^2$ ), we do not get slow-roll inflation for a small subset of KED and the entire range of PED initial conditions,  while a large range of KED initial values produces the slow-roll inflation. Though, for large  values of $\alpha$ (say $\alpha=5m_{Pl}^2$ ), both KED (except a small subset) and PED initial values are capable to produce the slow-roll inflationary phase. We have also obtained $N_{inf}$'s that are displayed in Tables \ref{tab:n2_dphip} and \ref{tab:n2_dphin}. Looking at both tables, physically viable initial conditions are identified that are consistent with the Planck data \cite{Planck2015}.

Finally, we have presented the phase space analysis for the above three models.  For $T-model$ with $\alpha=10 m_{Pl}^2$, the quantum bounce  is not compact: $| \dot{\phi}_B | < 0.91 m_{Pl}^2 $ and $\phi_B \rightarrow \pm \infty$ whereas for $\alpha=10^{10} m_{Pl}^2$, it is compact:  $| \dot{\phi}_B | < 0.91 m_{Pl}^2 $ and $\phi_B \rightarrow \pm 2.56 \times 10^5 m_{Pl}$. In the case of $E-model$ and $\alpha-$attractor with $n=2$, the initial surface is semi-finite: for $E-model$ with $\alpha=0.1 m_{Pl}^2$ and $\alpha=5 m_{Pl}^2$, it is $| \dot{\phi}_B | < 0.91 m_{Pl}^2 $; $\phi_B \in (-3.64, \infty)$ and $\phi_B \in (-25.65, \infty)$, respectively,  while for $\alpha-$attractor with $n=2$, this is given as $| \dot{\phi}_B | < 0.91 m_{Pl}^2 $; $\phi_B \in (-4.08, \infty)$ for $\alpha=0.5 m_{Pl}^2$ and $\phi_B \in (-12.88, \infty)$ for $\alpha=5 m_{Pl}^2$. The phase portraits for these models are shown in Figs. \ref{fig:Tport}, \ref{fig:Eport} and \ref{fig:n2port}, where dashed blue curves correspond to the cases without  slow-roll inflationary phase and solid curves (red and blue) provide slow-roll inflation. However, only the red curves are observationally consistent with the Planck 2015 data, not the blue ones \cite{Planck2015}.

\section*{Acknowledgements}
M.S. would like to thank  T. Zhu for fruitful  discussions. A.W. is supported in part by the National Natural Science Foundation of China (NNSFC) with the Grants Nos. 11375153 and 11675145.

\section*{Appendix A: Some Physical Quantities}
\label{sec:Append}
\renewcommand{\theequation}{A.\arabic{equation}}\setcounter{equation}{0}

From  Eq.(\ref{eq:Ninf}), one finds
\begin{eqnarray}
N_{inf} \simeq \int_{\phi_{end}}^{\phi_*} \frac{V(\phi)}{V'({\phi})} d\phi, 
\label{eq:Ninf2}
\end{eqnarray}
where $\phi_*$ and $\phi_{end}$ are the values of the inflaton field at the onset and end of the slow-roll inflation.

The slow-roll parameter $\epsilon_V$ is defined as
\begin{eqnarray}
\epsilon_V = \frac{M_{Pl}^2}{2} \left(\frac{V'(\phi)}{V(\phi)}\right)^2, 
\label{eq:ev}
\end{eqnarray}
where $M_{Pl}=m_{Pl}/\sqrt{8 \pi}$. At the end of the slow-roll inflation,  $\epsilon_V=1$. Hence, one can obtain $\phi_{end}$ from  Eq.(\ref{eq:ev}).

During the slow-roll inflation, $\dot{\phi}^2 \ll V(\phi)$ then  Eq.(\ref{eq:Hub}) becomes
\begin{eqnarray}
H_*{^2} \simeq \frac{8 \pi}{3 m_{Pl}^2}~V(\phi_*).
\label{eq:HubSR}
\end{eqnarray}
The upper bound on $H_*$ during the slow-roll inflation is given by \cite{Planck2015}
\begin{eqnarray}
\frac{H_*}{M_{Pl}} < 3.6 \times 10^{-5} ~~ (\text{95 \% Confidence level}).
\label{eq:H*}
\end{eqnarray}
In our analysis, we shall use $H_*{/M_{Pl}}=3.0 \times 10^{-5}$. Putting the value of $H_*{/M_{Pl}}$ in  Eq.(\ref{eq:HubSR}), one can get $\phi_*$.

Substituting the values of  $\phi_*$ and $\phi_{end}$ with $N_{inf}=60$ in  Eq.(\ref{eq:Ninf2}), we obtain the values of $\alpha$ and $c$ for $T$, $E-models$ and $\alpha-$attractor model with $n=2$.


\end{document}